%% file: main.tex
\documentclass[lettersize,journal]{IEEEtran}
\usepackage[utf8]{inputenc}
\usepackage{mathtools}   
\usepackage{amsfonts}       
\usepackage{amssymb}  
\usepackage{bm}  
\usepackage{url}
\usepackage{algorithm}  
\usepackage[noend]{algpseudocode}
\usepackage{pgfplots}  
\usepackage{booktabs}  
\usepackage{adjustbox}
\usepackage{subcaption}
\usepackage{cite}  
\usepackage{paralist}  
\usepackage{textcomp}  

\renewcommand{\vec}[1]{\bm{#1}}

\DeclareMathOperator*{\argmin}{argmin}
\DeclareMathOperator*{\argmax}{argmax}

\newcommand{\raw}{{\vec{w}}}  
\newcommand{\tot}{{\vec{t}}}  
\newcommand{\Nobj}{{N_\text{obj}}}  
\let\oldell\ell
\renewcommand{\ell}{\vec{\oldell}}  
\newcommand{\mats}{{\Omega}}
\newcommand{\Nmats}{{N_\text{mats}}}  
\newcommand{\Ne}{{N_\text{e}}}  
\newcommand{\x}{{\vec{x}}}  
\newcommand{\xgt}{{\vec{x}_\text{gt}}}  
\newcommand{\muvect}{{\vec{\mu}}}  
\renewcommand{\r}{{\vec{r}}}  

\newcommand{\Is}{{\vec{I}_s}}  
\newcommand{\Ip}{{\vec{I}_p}}  
\newcommand{\resp}{{\vec{g}}}  

\newcommand{\dir}{{\vec{d}}}
\newcommand{\sca}{{\vec{s}}}

\graphicspath{{./figures/}}

\begin{document}

\title{Material Identification From Radiographs \\ Without Energy Resolution}

\author{
Michael~T.~McCann,~\IEEEmembership{Member,~IEEE},
Elena~Guardincerri,
Samuel~M.~Gonzales,\\
Lauren~A.~Misurek,
Jennifer~L.~Schei, 
and
Marc~L.~Klasky%
    \thanks{All authors are with Los Alamos National Laboratory, Los Alamos, NM, 87545 USA.}%
	\thanks{
		M.~T.~McCann and M.~L.~Klasky and are with the Theoretical Division;
		S.~M~Gonzales and L.~A.~Misurek are with the Nuclear Engineering and Nonproliferation Division;
		E.~Guardincerri is with the Physics Division;
		and J.~L.~Schei is with the Integrated Weapons Experiments Division.
		}%
}

\markboth{}%
{}


\maketitle

\begin{abstract}
We propose a method for performing material identification
from radiographs without energy-resolved measurements.
Material identification has a wide variety of applications,
including in biomedical imaging, nondestructive testing, and security.
While existing techniques for radiographic material identification
make use of dual energy sources,
energy-resolving detectors,
or additional (e.g., neutron) measurements,
such setups are not always practical---%
requiring additional hardware and complicating imaging.
We tackle material identification without energy resolution,
allowing standard X-ray systems to provide
material identification information without requiring additional hardware.
Assuming a setting where the geometry of each object in the scene is known
and the materials come from a known set of possible materials,
we pose the problem as a combinatorial optimization with a loss function
that accounts for the presence of scatter
and an unknown gain 
and
propose a branch and bound algorithm to efficiently solve it.
We present experiments on both synthetic data and
real, experimental data with relevance to 
security applications---%
thick, dense objects imaged with MeV X-rays.
We show that material identification can be efficient and accurate,
for example,
in a scene with three shells (two copper, one aluminum),
our algorithm ran in six minutes on a consumer-level laptop
and identified the correct materials as being among the top 10 best matches
out of 8,000 possibilities.

\end{abstract}

\begin{IEEEkeywords}
X-ray radiography, material identification, combinatorial optimization, branch and bound algorithm.
\end{IEEEkeywords}

\section{Introduction and Related Work}

X-ray radiography is an important tool in many security applications,
including airport security~\cite{hussein_review_1998,macdonald_design_2001,wetter_imaging_2013,martin_learning_2015, mouton_materials_2015,zhang_determination_2016},
cargo inspection~\cite{chalmers_cargo_2007,yang_materials_2009,shikhaliev_megavoltage_2018,lee_efficient_2018},
and
in the emergency response to a suspicious item that may contain 
explosives or other hazards, i.e.,
for bomb squads~\cite{hill_su_2012,beckett_advances_2020}.
Automating aspects of X-ray-based screening is attractive
in terms of speed, cost, and safety
and is an active area of study; see \cite{akcay_towards_2022} for a recent review.
X-ray radiography provides both information about the internal geometry of what is being imaged
and about its material composition,
because different materials absorb X-rays to different degrees.
Both types of information can contribute to the security application.


Here, we are interested in using X-ray radiography for material identification.
There are several related yet distinct problems
that may be classified as radiographic material identification.
In one setting,
the possible materials come from a finite, discrete set,
e.g., \cite{yang_materials_2009} aims to identify 
whether a slab of material is plywood, polyethylene, aluminum, iron, copper, or lead.
In another setting,
the goal is to determine a continuous-valued material property
for each object (or each pixel in the radiograph)
that can be used to infer what materials are present,
e.g.
\cite{kimoto_precise_2017} estimates atomic number (as a continuous quantity)
in order to distinguish bone, aluminum, soft tissue, and acrylic.
Along similar lines,
some approaches attempt to determine the contribution of
multiple materials to each pixel/voxel of the reconstruction;
this is sometimes called \emph{material decomposition}~\cite{sukovle_basis_1999}.
Some works, e.g., \cite{lakshmanan_x_2014}, determine only \emph{whether} each material is present,
while others, e.g. \cite{martin_learning_2015}, determine \emph{where} each material is,
i.e., segment the radiograph.

Many approaches to material identification from X-ray measurements exist in the literature.
One class of approaches relies on hardware such as
dual energy scanners~\cite{ying_dual_2006,Bokun2010,long_multi_2014,modgil_material_2015,duvillier_inline_2018}
or photon-counting detectors~\cite{beldjoudi_optimised_2012,wu_hyperspectral_2017}
to collect energy-resolved measurements;
these methods date back to at least the 1970s~\cite{macovski_energy_1976}.
In these cases,
material identification can be achieved by
solving a system of equations that relate 
the energy-dependent measurements
to material properties,
e.g., Compton and photoelectric coefficients~\cite{yuan_robust_2016}.
Work exists \cite{tobias_limits_2019} that aims to elucidate the 
fundamental limits of dual-energy approaches.
Another, more recent approach, has been 
to use phase contrast X-ray imaging to decompose materials~\cite{braig_direct_2018,schaff_material_2020}.
Finally, another class of approaches
combines X-ray measurements with another modality,
e.g., neutron imaging~\cite{jones_material_1993,yang_materials_2009,cui_material_2020}.
Here, quantities such as the ratio of neutron to photo absorption
can be indicative of material identity.

In this work,
we focus on the problem of identifying the materials present in a radiograph
without energy-resolved measurements or additional imaging modalities,
thus allowing a standard MeV X-ray system to provide
material identification information without requiring additional hardware.
Solving this problem could provide an important new capability for security applications,
but,
to our knowledge, it has not been addressed in the literature.
Specifically, it could reduce the cost of security imaging systems
(because the source would only need to generate a single X-ray spectrum
and the detector would not need to provide energy resolution).
It could also
accelerate and simplify the imaging procedure
(because it requires only a single radiograph)
which is of particular importance in a suspicious package scenario
where radiography must be performed quickly, accurately, and potentially in uncontrolled settings, i.e., ``in the field''.
Finally,
it could help to reduce the incidental radiation exposure received by people near a threat device,
e.g., in the unfortunate scenario described in \cite{hill_su_2012}.

Material identification without energy resolution is a challenging,
underdetermined problem which is impossible to solve in general
(e.g., because different materials of different thicknesses can absorb exactly
the same amount of X-ray radiation);
in order to progress, we make a number of simplifying assumptions
that are nonetheless well-justified for security applications.
First, we treat each voxel as containing only one
material from a known list,
rather than a mixture;
this is reasonable because we expect to image man-made objects
and domain knowledge can inform what materials to look for.
Second, we assume that the geometry of the scene is already known;
this is also reasonable because the geometry can be determined
e.g., by inference from the measured radiograph based on domain knowledge,
previous interrogation with soft X-rays,
conventional photography,
or additional physical measurements.
Finally, our experiments use MeV X-rays and thick,
highly absorbing objects (see Figure~\ref{fig:radiographs}),
again this is justified because, in security applications,
we are looking for man-made, usually metallic objects.

Our specific contributions are \begin{enumerate}
\item the formulation of a novel and important material identification problem;
\item an algorithm to efficiently solve the resulting combinatorial problem,
allowing scenes with up to eight objects to be processed in minutes on a laptop computer;
and
\item experimental validation of the algorithm on real and synthetic data.
\end{enumerate}

The remainder of the manuscript is organized as follows.
In Section~\ref{sec:proposed} we formulate the material identification problem
and present our proposed algorithm for solving it.
In Section~\ref{sec:experiments},
we describe validation experiments on real and synthetic data.
We discuss these results in Section~\ref{sec:discussion}
and conclude in Section~\ref{sec:conclusions}.

\section{Proposed Approach} \label{sec:proposed}
We now formulate the material identification problem
and present our radiograph forward model,
branch and bound algorithm,
and preprocessing and calibration procedures.

\subsection{Problem Formulation} \label{sec:formulation}
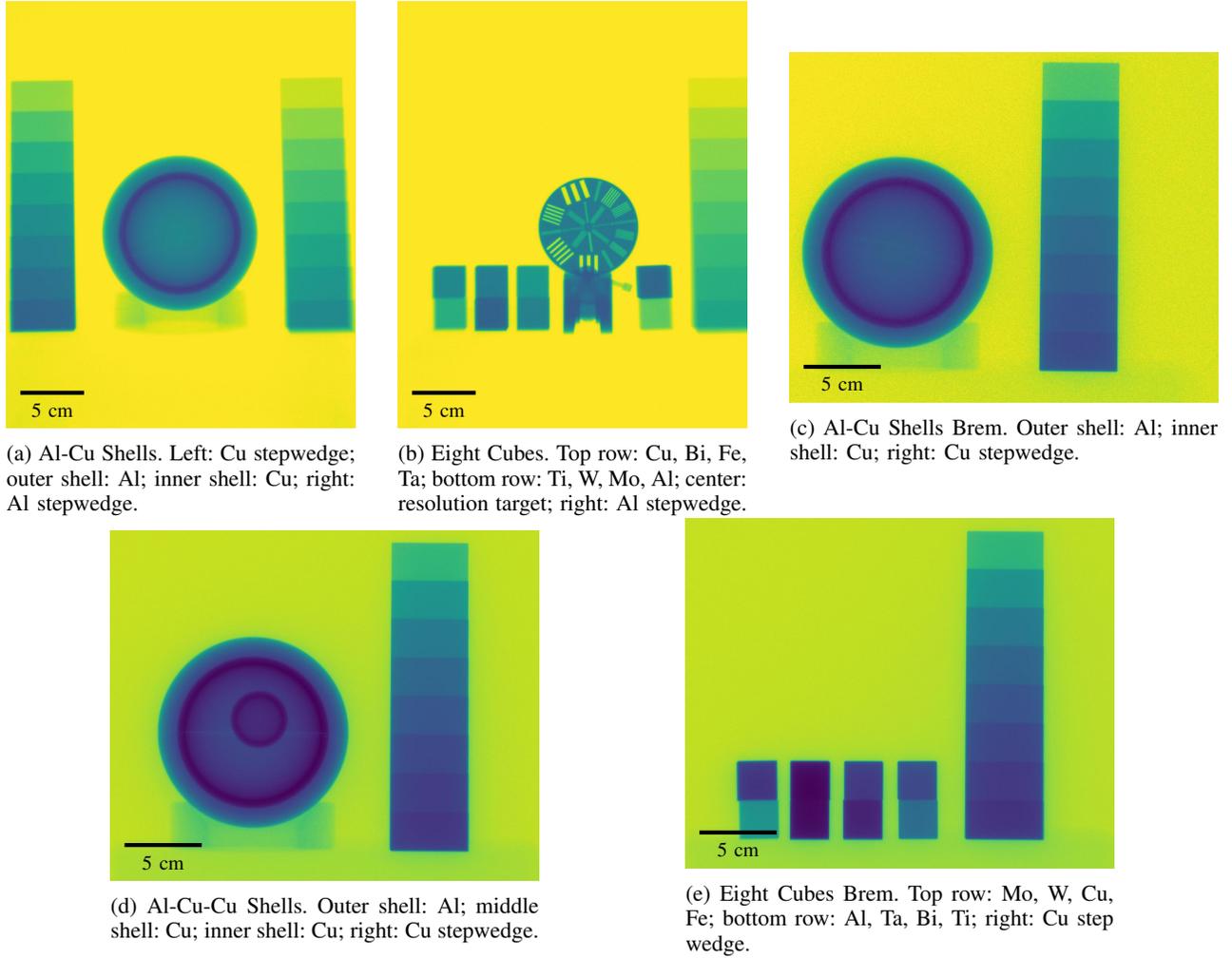
\begin{figure*}[htbp]
    \centering
    \begin{subfigure}{.269\linewidth}
        \begin{tikzpicture}
            \node[anchor=south west,inner sep=0] at (0,0) {\includegraphics[width=\linewidth]{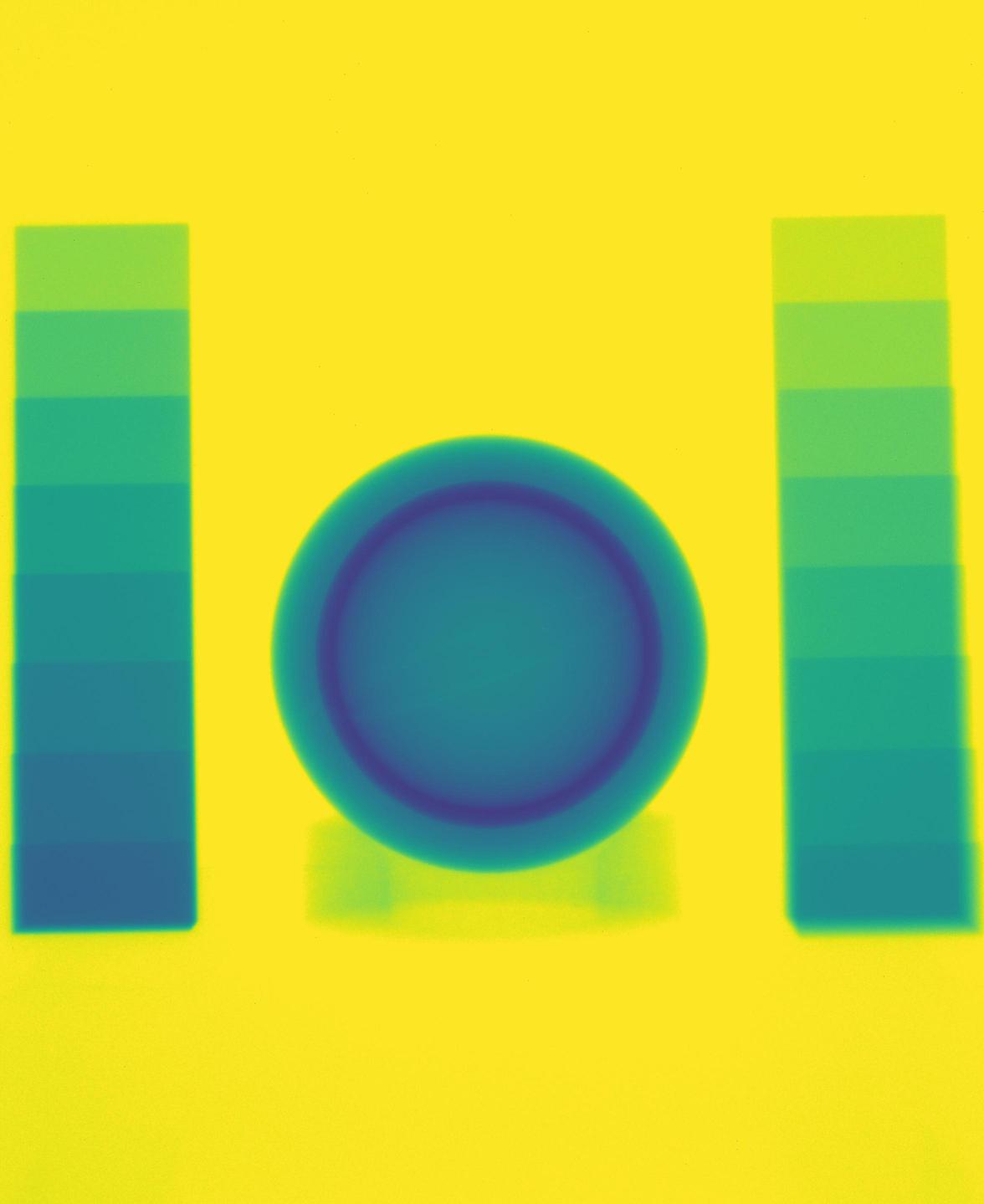}};
            \draw[ultra thick] (0.2,0.5) -- node[below, font=\footnotesize]{5 cm} (0.2+.89,0.5);
        \end{tikzpicture}
        \caption{Al-Cu Shells. Left: Cu stepwedge; outer shell: Al; inner shell: Cu; right: Al stepwedge. }
    \end{subfigure}\hfil
    \begin{subfigure}{.269\linewidth}
        \begin{tikzpicture}
            \node[anchor=south west,inner sep=0] at (0,0) {\includegraphics[width=\linewidth]{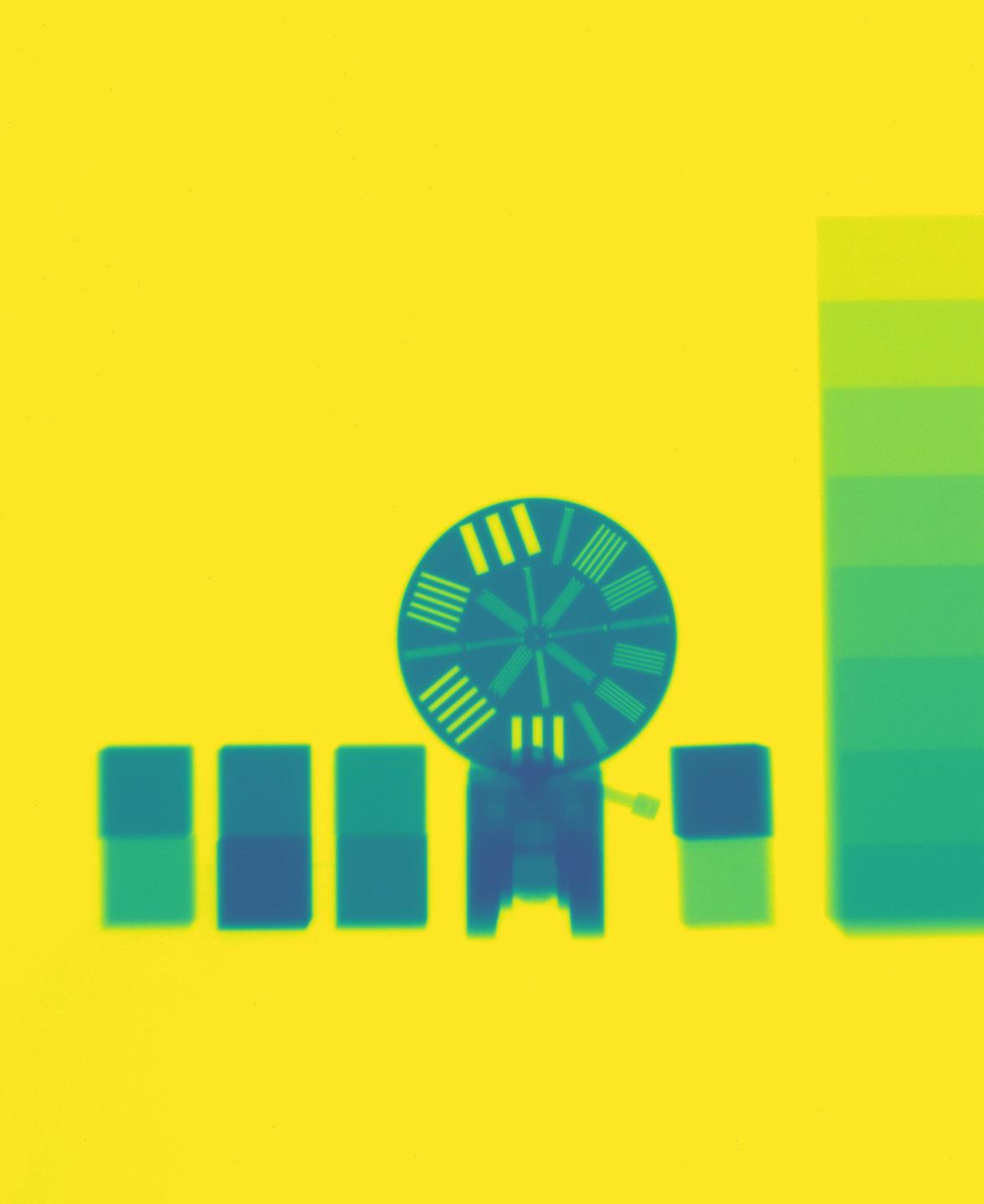}};
            \draw[ultra thick] (0.2,0.5) -- node[below, font=\footnotesize]{5 cm} (0.2+.89,0.5);
        \end{tikzpicture}
         \caption{Eight Cubes. Top row: Cu, Bi, Fe, Ta; bottom row: Ti, W, Mo, Al;
         center: resolution target; 
         right: Al stepwedge.
         }
    \end{subfigure}\hfil
    \begin{subfigure}{.330\linewidth}
        \begin{tikzpicture}
            \node[anchor=south west,inner sep=0] at (0,0) {\includegraphics[width=\linewidth]{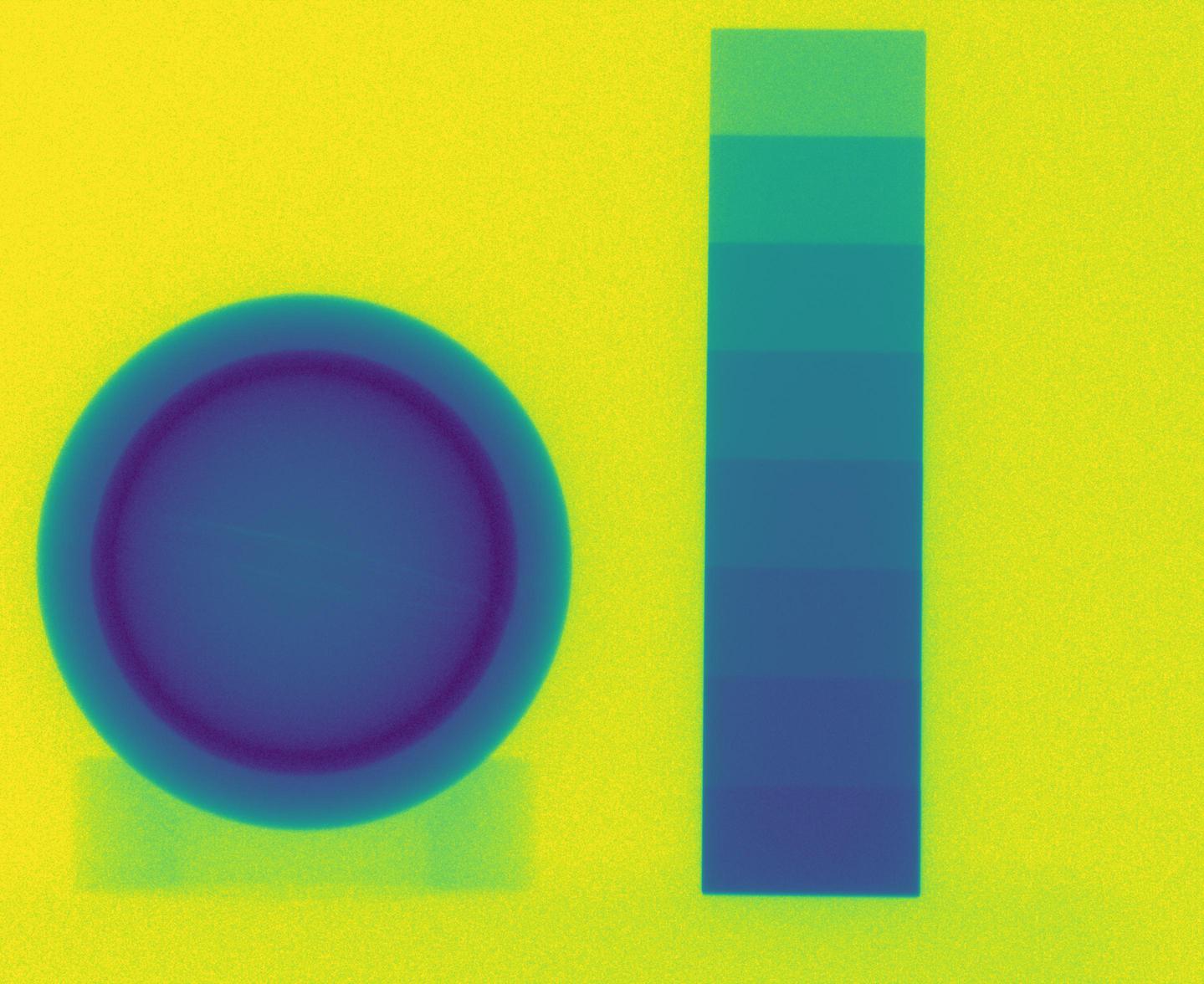}};
            \draw[ultra thick] (0.2,0.5) -- node[below, font=\footnotesize]{5 cm} (0.2 + 1.08,0.5);
        \end{tikzpicture}    
         \caption{Al-Cu Shells Brem. Outer shell: Al; inner shell: Cu; right:  Cu stepwedge.}
    \end{subfigure}\hfil\\
    \begin{subfigure}{.330\linewidth}
     \begin{tikzpicture}
            \node[anchor=south west,inner sep=0] at (0,0) {\includegraphics[width=\linewidth]{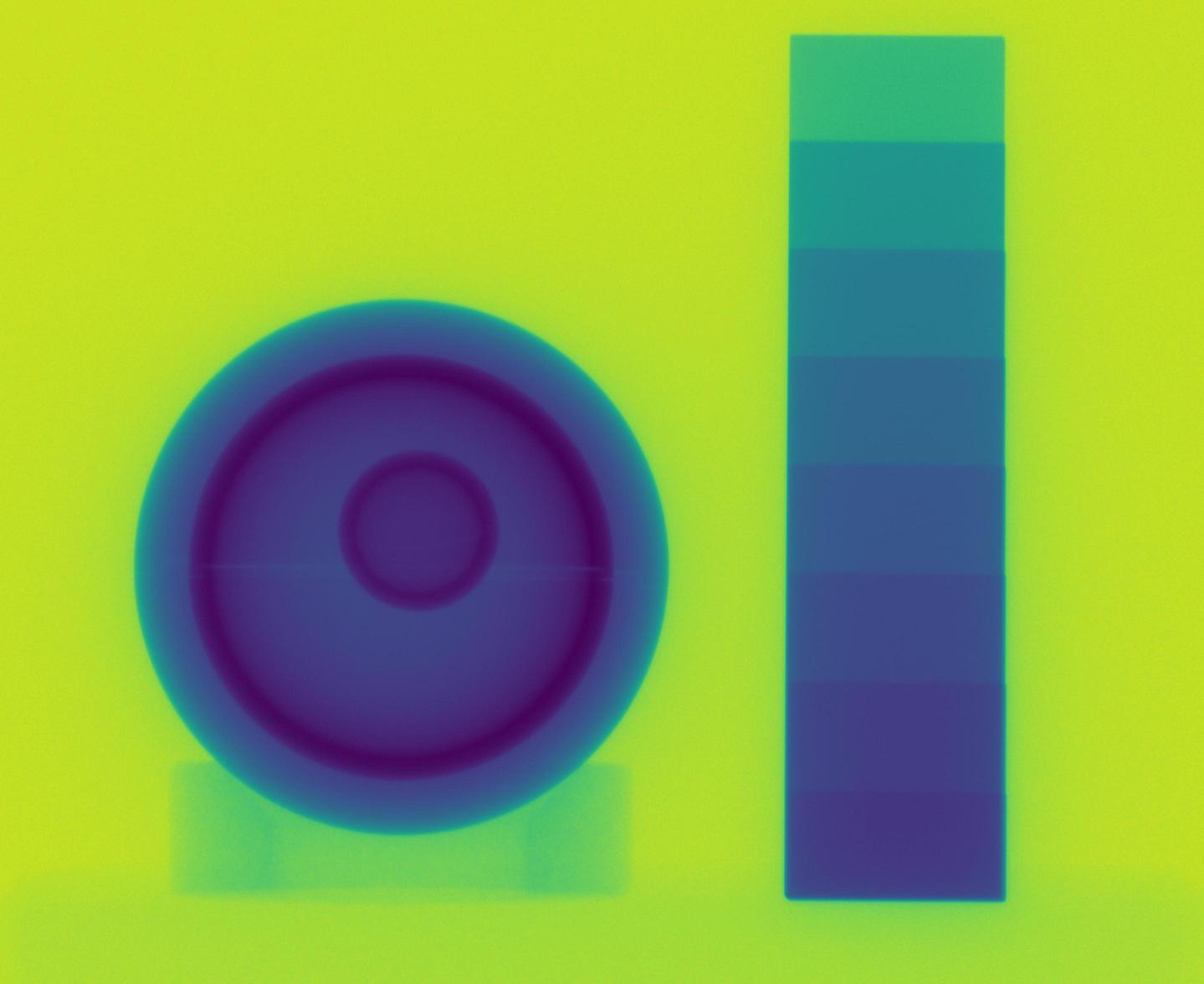}};
            \draw[ultra thick] (0.2,0.5) -- node[below, font=\footnotesize]{5 cm} (0.2 + 1.08,0.5);
        \end{tikzpicture}        
         \caption{Al-Cu-Cu Shells. Outer shell: Al; middle shell: Cu; inner shell: Cu; right: Cu stepwedge.}
    \end{subfigure}\hfil
    \begin{subfigure}{.330\linewidth}
     \begin{tikzpicture}
            \node[anchor=south west,inner sep=0] at (0,0) {\includegraphics[width=\linewidth]{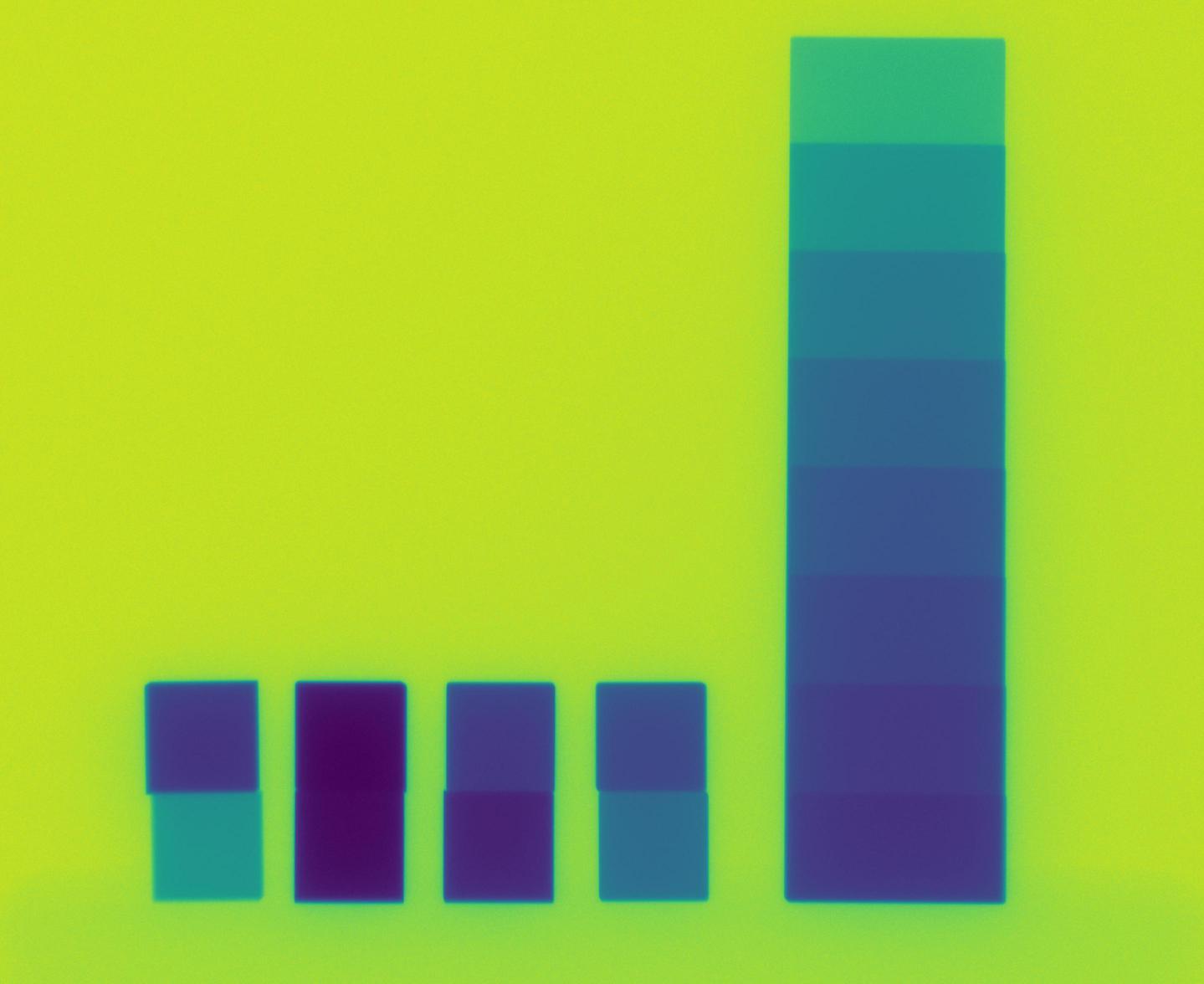}};
            \draw[ultra thick] (0.2,0.5) -- node[below, font=\footnotesize]{5 cm} (0.2 + 1.08,0.5);
        \end{tikzpicture}    
         \caption{Eight Cubes Brem. Top row: Mo, W, Cu, Fe; bottom row: Al, Ta, Bi, Ti; right: Cu step wedge.}
    \end{subfigure}%
    \caption{Radiographs (after preprocessing, see Section~\ref{sec:preprocessing}) of the scenes used for material identification. Images are $2352\times2878$.}
    \label{fig:radiographs}
\end{figure*}

Given a properly preprocessed radiograph $\tot \in \mathbb{R}^{V \times U}$,
where $V$ and $U$ denote the height and width in pixels, 
of a scene with $\Nobj$ single-material objects%
\footnote{Extending our approach to multiple radiographs is straightforward;
we present the single radiograph version here for brevity
and because our experiments use single radiographs.}
and 
a set of possible materials $\mats$ with $|\mats| = \Nmats$,
our goal is to
determine the material of each object: $\x \in \mats^\Nobj$.
For example radiographs, see Figure~\ref{fig:radiographs}.
Note by our definition an object is a connected region in space consisting of a single material:
an empty coffee cup is one object; a full coffee cup is two: mug and coffee.
We assume that we have the following ingredients of a radiographic forward model:
a raytrace for each object at each pixel, $\{\ell_n\}_{n=1}^\Nobj$ with $\ell_n \in \mathbb{R}^{V \times U}$,
the attenuation coefficients for each material at each of $\Ne$ energies
$\{\muvect_m\}_{m \in \mats}$ with $\muvect_m \in \mathbb{R}_{\ge 0}^\Ne$,
the spectrum of the source $\Is \in \mathbb{R}_{\ge 0}^\Ne$, and
the (energy-dependent) detector response $\resp \in \mathbb{R}_{\ge 0}^\Ne$.

Practically speaking,
it is useful for algorithms to return
a ranked list of guesses for $\x$,
thereby reducing the potentially large number of possible
material assignments to a tractable number.
For example, for a bomb squad investigating a suspicious package,
this would allow the team to focus their efforts on a handful of possible weapon designs.
Thus,
for performance evaluation,
we consider top-$N$ accuracy,
i.e., an algorithm is top-$N$ correct
if the true material assignment $\xgt$
is among the $N$ $\x$s returned.
We pose the material identification problem described above
in terms of minimizing the functional
\begin{equation} \label{eq:functional}
    J(\x) = \min_{\alpha, \theta} \| \tot - (\alpha \dir(\x) + \sca(\theta))\|_2^2,
\end{equation}
where $\tot$ is the measured radiograph,
$\alpha$ is a scalar,
$\dir$ is our model of the direct signal,
and $\sca(\theta)$ is our parametric scatter model
with parameters $\theta$
(which we detail later in \eqref{eq:poly_scatter_model}).
Seeking a set of top-$N$ material assignments amounts
to finding
\begin{equation} \label{eq:problem}
\begin{split}
    \x^*_1 &\in \argmin_{\x \in \mats^\Nobj} \ J(\x) \\
    \x^*_2 &\in \argmin_{\x \in \mats^\Nobj \setminus \x^*_1} J(\x)\\
    &\vdots\\
    \x^*_N &\in \argmin_{\x \in \mats^\Nobj \setminus \x^*_1,\x^*_2, \dots, \x^*_{N-1}} J(\x),
\end{split}
\end{equation}
where $\setminus$ denotes set difference.
We note that,
in general,
the problem \eqref{eq:problem} is underdetermined
in that several different material assignments may explain the measured radiograph equally
well;
we expect such cases to be rare in practice and, even if they do occur,
 it can still be valuable to find these assignments and rule out any others.

In the following sections,
we describe each element of the functional $\eqref{eq:functional}$
and how we solve the minimization problem \eqref{eq:problem}.

\subsection{Radiographic Forward Model} \label{sec:forward_model}
We model the signal entering the detector
as the sum of the direct and scattered signal.
The direct signal accounts for photons that travel
directly from the source to the detector;
objects create contrast in the direct signal because they remove photons from that straight path (either by absorption or scattering) according to Beer's law.
The scattered signal represents photons that 
scatter from the objects of interest as well as
the remainder of the scene (ground, walls, detector, etc.);
this signal is difficult to model accurately.
We now describe each in detail.

We model the direct signal entering the detector at pixel $\r$
for a scene with materials $\x$ as
\begin{equation} \label{eq:direct}
    \dir(\x)[\r] =  \sum_e \resp[e] \Is[e] \exp \left(-\sum_{n=1}^{\Nobj} \muvect_{\x[n]}[e] \ell_n[\r] \right),
\end{equation}
where $\resp$, $\Is$, $\{\muvect_m\}_{m \in \mats}$, and $\{\ell_n\}_{n=1}^\Nobj$
are known physical constants
and aspects of the scene as described in \ref{sec:formulation}.
This model is a standard discretized version of the polyenergetic Beer's law~\cite{kak_principles_2001};
for more detail on the discretization, see \cite{mccann_local_2021}.
The direct signal model \eqref{eq:direct} has no term to account
for spatial variations in the intensity of the source,
this is handled during preprocessing of the radiograph,
see Section~\ref{sec:preprocessing}

We do not directly model scatter;
instead, our cost functional $\eqref{eq:functional}$
is designed so that we fit a parametric function, $\sca$
to the difference between the prediction of the forward model $\dir(\x)$
and the measured radiograph $\tot$.
The effect of this fitting is that when the difference between $\dir(\x)$
and $\tot$ is in the (nonlinear) span of $\sca$,
the discrepancy between them is zero, i.e., $J(\x)=0$.
We emphasize that,
unlike other scatter modeling approaches~\cite{hansen_extraction_1997,ohnesorge_efficient_1999},
our scatter model need not be predictive.
Instead,
it needs to strike a balance between representing enough---%
so that it can bridge the gap between our model for the direct signal
and what is actually measured---%
and too much---%
so that it does not make incorrect material assignments appear to match the measured radiograph.

We use low-order polynomial fields to model the scatter, $\sca$.
These allow us to account for smoothly-varying scatter,
enable rapid fitting during the evaluation of the cost function \eqref{eq:functional},
and provide control over their complexity via changing the polynomial order.
Specifically,
\begin{equation} \label{eq:poly_scatter_model}
    \sca(\theta)[\r] = \sum_{m\ge0, n\ge0, m+n \le P}\theta_{m,n} u(\r)^m v(\r)^n,
\end{equation}
where $P$ is the degree of the polynomial,
$\theta_{m,n} \in \mathbb{R}$ is one element of the scatter model parameters, $\theta$,
and $u(\r)$ and $v(\r)$ give the $u$ and $v$ coordinate of the pixel $\r$
(where the coordinate system is arbitrary;
we place $(0,0)$ at the upper-left corner of the radiograph with positive $u$ to the right
and positive $v$ down).
For a demonstration that polynomial fields are a reasonable way to approximate scatter, see Section~\ref{sec:polynomial_scatter}.

\subsection{Branch and Bound Algorithm} \label{sec:branch_and_bound}
The optimization problem \eqref{eq:problem}
is a combinatorial problem:
the space of solutions is discrete and finite,
so the problem may be solved via exhaustive search,
however the size of the space grows exponentially with the number of objects
($\Nmats^\Nobj$).
For example, performing material identification among twenty materials on five objects 
requires evaluating the forward model $20^5 \approx 3.2$ million times.
For comparison, solving a convex problem would typically involve evaluating the forward model and its adjoint only hundreds of times.
While this search may be parallelized,
with even a few more materials it quickly becomes infeasible.

In order to efficiently solve the optimization problem \eqref{eq:problem},
we propose an algorithm in the branch and bound family.
Branch and bound algorithms rely on finding lower bounds on
the objective value of partial solutions to a combinatorial problem;
see \cite{parker_discrete_1988}, Chapter 5 for an introduction.
To design a branch and bound algorithm for our problem,
we need a way to express partial solutions and
bound their objective values.
We also need to extend the standard branch and bound framework
to return the top $N$ solutions rather than simply the best.
(Such extensions have been used before in other contexts, e.g., \cite{tao_branch_2007,poensgen_branch_2020}.)

\begin{algorithm}[htbp]
\caption{Branch and bound for top-$N$ optimization}
 \begin{algorithmic}[1]
 \State $topN \gets \{\}$, $J^* \gets \infty$, $cands \gets \{[\varnothing, \dots, \varnothing]\}$
 \While {$cands$ is not empty}
    \State Remove a candidate $\x$ from $cands$ \label{algo:branch_and_bound:select}
    \If{$\x$ is a full solution and $J(\x) < J^*$}
        \State Add $\x$ to $topN$
        \State if $|topN| > N$, remove $\argmax_{\x \in topN} J(\x)$
        \State $J^* \gets \max_{\x \in topN} J(\x)$
     \EndIf
    \If{$\x$ is a partial solution and $bound(\x) < J^*$}
        \State $cands \gets cands \cup branch(\x)$
    \EndIf
 \EndWhile
 \end{algorithmic}
 \label{algo:branch_and_bound}
\end{algorithm}%

We outline our branch and bound algorithm in Algorithm~\ref{algo:branch_and_bound}.
The approach follows a standard branch and bound structure 
with the exception that a list of solutions, $topN$, must be maintained
instead of a single best solution.
The $branch$ and $bound$ subroutines are novel
and adapted to the problem at hand.
Our bounding approach makes use of the fact
that, for typical radiographs,
only a subset of objects are present at each pixel;
thereby allowing us to compute a lower bound on the error at that pixel
for any partial solution that assigns materials to each of the objects that
are present.
We now describe the algorithm in detail.

To represent partial solutions,
we augment our list of possible materials with an additional symbol, $\varnothing$,
which we take to mean ``not yet assigned'';
partial solutions are then elements of  $(\mats \cup \varnothing)^\Nobj$
and full solutions are the elements of the same set that do not involve any $\varnothing$s.
We will say that a partial or full solution $\vec{y}$ is a descendent of a partial solution $\x$, written $\vec{y} \in \x$, when $\vec{y}$ matches $\x$ in every non-$\varnothing$ position of $\x$.
The list of candidate solutions, $cands$,
is initialized with a single partial solution that is full of $\varnothing$s,
this represents all possible material assignments.

The subroutine $branch(\x)$ creates new partial (or possibly full) solutions from a partial solution.
To do this, we replace the first $\varnothing$ in $\x$ with each material in the set of materials $\mats$.
For example, if $\mats = \{\text{Cu}, \text{Ni}, \text{Sn}\}$,
then $branch([\text{Cu}, \varnothing, \varnothing]) = \{[\text{Cu}, \text{Cu}, \varnothing], [\text{Cu}, \text{Ni}, \varnothing], [\text{Cu}, \text{Sn}, \varnothing]\}$.

The subroutine $bound(\x)$ produces a lower bound on the value of $J(\vec{y})$ for any full solution $\vec{y} \in \x$.
Good bounds are critical to the efficiency of the algorithm because 
they are what allow bad partial solutions (and all of their descendant full solutions) to be skipped.
To design $bound$, we make use of the fact that,
in many real-world scenarios,
only a subset of materials are present at each pixel in the radiograph.
Because of this,
when a partial solution fixes all the materials present at a set of pixels,
the error incurred at those pixels cannot be reduced by making further assignments
and therefore provides a lower bound on the total error for the partial solution.
Specifically, to compute $bound(\x)$ we note that 
\begin{equation}\label{eq:bound}
\begin{split}
        \min_{\vec{y} \in \x} \min_{\alpha, \theta} \| \tot - (\alpha \dir(\vec{y}) &+ \sca(\theta))\|_2^2\\
        &\ge \min_{\vec{y} \in \x} \min_{\alpha, \theta} \| \vec{M}_\x (\tot - (\alpha\dir(\vec{y}) + \sca(\theta)))\|_2^2 \\
        &= \min_{\alpha, \theta} \| \vec{M}_\x (\tot - (\alpha \dir(\vec{y}_0) + \sca(\theta)))\| _2^2,
\end{split}
\end{equation}
where $\vec{M}_\x$ is a binary masking operator chosen to mask away any pixels that depend on $\varnothing$ values in $\x$
and $\vec{y}_0$ is any full solution descendent of $\x$.
The inequality holds because the fact that $\|\cdot\|_2^2$ is a sum of squares means that, for all ($\vec{y}, \alpha, \theta$),
we have
 $\| \tot - (\alpha\dir(\vec{y}) + \sca(\theta))\|_2^2 \ge \|  \vec{M}_\x (\tot - (\alpha\dir(\vec{y}) + \sca(\theta)))\|_2^2$;
 thus the same relationship holds for the minimum values.
 The equality holds because
 the mask $\vec{M}_\x$ is defined to zero out exactly the pixels that would make two different full solution descents of $\vec{x}$ have different functional values.
The mask is recomputed for each partial solution $\x$,
but this operation is fast because it involves only pixelwise, boolean operations
on binary masks.

The final piece of our branch and bound algorithm is
to determine the order in which candidate solutions are evaluated,
i.e.,
the method for selecting a candidate from $cands$ in line~\ref{algo:branch_and_bound:select} of Algorithm~\ref{algo:branch_and_bound}.
Common options are depth-first (first in, last out), breadth-first (first in, first out), or best-first (least bound, first out).
In our context, we need to avoid breadth-first search because the $cands$ list would become huge (containing nearly every possible partial solution) before $J^*$ was ever updated.
Depth-first search has a different shortcoming,
in that it can spend a lot of time 
exploring solutions with poor initial material assignments.
Finally,
best-first search ends up being similar to breadth-first search because
we tend to give overly optimistic bounds to partial solutions with few assignments.
Our solution is use first in, last out ordering (depth-first search) but to also sort new candidates according to their $bound$ value when adding them to $cands$.
The result is a depth-first search that chooses the most promising candidate at each depth first.

\subsection{Preprocessing} \label{sec:preprocessing}
We have so far assumed that we have access to a properly preprocessed radiograph, $\tot$.
Here, we detail our preprocessing steps,
which include detector pixel calibration and flat-field correction.

To account for the fact that each detector pixel
may have its own gain and offset value (also called \emph{pedestal}),
we model a (uncorrected) radiograph, $\raw$ as
\begin{equation} \label{eq:raw}
\raw[\r] = t \alpha[\r] \tot[\r] + t\beta[\r] + \gamma[\r]
\end{equation}
where $\alpha$, $\beta$, and $\gamma$ represent pixelwise gain, dark count rate,
and offset values,
$\tot$ is the unnormalized total transmission,
and $t$ is the exposure time.
The goal of our preprocessing is to use calibration data
to estimate $\alpha$, $\beta$, and $\gamma$ in order to undo their effects
and to simultaneously remove the effect of spatial variation in the X-ray source intensity.

As calibration data, we use
a set of flat-field radiographs
(radiographs of an empty scene) taken using a high activity $^{60}Co$ radioactive source
and a set of dark-field radiographs (radiographs without an X-ray source). 
The reason for using a radioactive source as opposed to a betatron is that for the former the 
dose rate at the detector is known and constant in time, whereas for the latter this is not true.
We can assume that there is no transmission present in the dark-field radiographs, so $\tot_\text{dark}[r]=0$.
For the flat-field radiographs, we can write
\begin{equation} \label{eq:flat}
  \tot_\text{flat}[\r] =
    \Ip[\r] \sum_e \resp[e] \Is[e] 
    + \sca_\text{flat}[\r]
\end{equation}
where $\Ip$ is the spatial source profile,
$\resp$ is the energy-dependent detector response,
$\Is$ is the beam spectrum,
and
$\sca_\text{flat}$ is an unknown scatter term.
Note that \eqref{eq:flat} assumes that the beam spectrum
and detector response are multiplicatively separable with respect
to position, $\r$, and energy, $e$;
if the relative contribution of different energies change significantly
between pixels, it would not be accounted for by this calibration.
With the approximation that $\sca_\text{flat}[\r] = 0$
and the convention to scale $\resp$ and $\Is$ so that their inner product is one,
we have $\tot_\text{flat}=\Ip$.
Substituting into \eqref{eq:raw},
we have 
\begin{align} 
\raw_\text{dark}[\r] &= t\beta[\r] + \gamma[\r] \\
\raw_\text{flat}[\r] &= t \alpha[\r] \Ip[\r] + t\beta[\r] + \gamma[\r].
\end{align}
Therefore, using several dark- and flat-field radiographs,
we can estimate $\alpha[\r]\Ip[\r]$, $\beta[\r]$, and $\gamma[\r]$ by solving the linear system
\begin{equation}
    \begin{bmatrix}
    \raw_{\text{dark}, 1}[\r]\\
    \raw_{\text{dark}, 2}[\r]\\
    \vdots\\
    \raw_{\text{dark}, N_\text{dark}}[\r]\\
    \raw_{\text{flat}, 1}[\r]\\
    \raw_{\text{flat}, 2}[\r]\\
    \vdots\\
    \raw_{\text{flat}, N_\text{flat}}[\r]\\
    \end{bmatrix}
    =
    \begin{bmatrix}
    0 & t_{\text{dark}, 1} & 1 \\
    0 & t_{\text{dark}, 2} & 1 \\
\vdots & \vdots & \vdots\\
    0 & t_{\text{dark}, N_{\text{dark}}} & 1 \\
    t_{\text{flat}, 1} & t_{\text{flat}, 1} & 1 \\
    t_{\text{flat}, 2} & t_{\text{flat}, 2} & 1 \\
\vdots & \vdots & \vdots\\
    t_{\text{flat}, N_{\text{flat}}} & t_{\text{flat}, N_{\text{flat}}} & 1
    \end{bmatrix}
        \begin{bmatrix}
    \alpha[\r]\Ip[\r] \\
    \beta[\r] \\
    \gamma[\r]
    \end{bmatrix},
\end{equation}
To account for stuck or dead pixels on the detector,
we set a threshold on the R-squared value of the fit
and remove each pixel that falls below it.

Finally, to preprocess a radiograph with exposure time $t$, we use
\begin{equation} \label{eq:prep}
    \text{prep}(\raw)[\r] = \frac{\raw[\r] - t \beta[\r] - \gamma[\r]}{t \alpha[\r]\Ip[\r]}.
\end{equation}
Substituting \eqref{eq:prep} into \eqref{eq:raw} confirms that
\begin{equation}
    \text{prep}(\raw)[\r] = \frac{\tot[\r]}{\Ip[\r]}.
\end{equation}
In short, our preprocessing removes detector gain and pedestal
and also normalizes the transmission with respect to the flat-field image.

\subsection{Source Spectrum and Detector Response Calibration} \label{sec:calibration}
Our model depends on knowing the product of the source spectrum
and the detector response over the energies where they are nonzero.
In practice, there are challenges to knowing each of these:
The spectrum for a line source is known but affected by shielding around the source,
e.g., the container holding it.
A betatron generates, ideally,
a Bremsstrahlung spectrum,
but in practice
involves random fluctuations
as well as systematic changes as the source warms.
Detector responses may be provided by the manufacturer;
if they are not, they may be estimated from the materials used in the detector,
but both of these responses come with some degree of uncertainty
and are difficult to validate.

To overcome these challenges,
we propose to calibrate the quantity $\Is\resp$ from a calibration radiograph
by solving
\begin{equation} \label{eq:calibration}
    \argmin_{\Is\resp} J(\xgt),
\end{equation}
where the data fidelity term $J$ is given in \eqref{eq:functional}
and $\xgt$ is the ground truth material assignment.
Note that computing $J(\xgt)$ itself involves computing a linear fit
and that 
the problem is generally ill-posed and nonconvex.
To minimize \eqref{eq:calibration}
we used standard gradient descent,
where $\nabla J$ was computed using
automatic differentiation software~\cite{bradbury_jax_2021}.

\section{Experiments and Results} \label{sec:experiments}
We now describe our experiments on real and synthetic data.
All material identifications were performed on a consumer-level laptop
(MacBook Pro, 2.4 GHz processor, 32 GB of RAM).
Objects were identified as being one of the 20 materials listed in Table~\ref{tab:materials}.
Attenuation coefficients for these materials were obtained from the MCNP particle transport software~\cite{werner_mcnp6.2_2018},
specifically the mcplib04 library, which derives from ENDF/B-VI.8.

\begin{table}[htbp]
    \caption{Materials list} 
    \centering
    \begin{tabular}{llll}
    \toprule
    air & aluminum & beryllium & bismuth   \\
    boron & carbon & copper & gold     \\
    high explosive & iron & lead      & lithium   \\
    molybdenum & plutonium & polycarbonate & tantalum \\
    tin & titanium & tungsten & uranium \\
    \bottomrule
    \end{tabular}
    \label{tab:materials}
\end{table}

\subsection{Material Identification From Experimental Data---Cobalt-60 Source}

As a first validation with experimental data,
we performed material identification on radiographs taken with a cobalt-60 source.
The source had an activity of 3.501 Ci at the time of the experiment.
All experiments were performed in the basement of at the Nonproliferation International Security Center at the Los Alamos National Laboratory;
all the images were collected using a Pixium 3543EZ digital readout panel.
For the source spectrum, we used the two emission lines of cobalt-60: 1.17 MeV and 1.33 MeV.
We assumed the detector response was the same for both lines
and did not perform further calibration.
The source-to-detector distance was 100 cm 
and the object-to-detector distance was 17.5 cm.

\textbf{Al-Cu Shells.}
The first scene consisted of concentric shells of aluminum (1 inch thick, 5 inch outer diameter)
and copper (0.5 inch thick, 4 inch outer diameter)
and aluminum and copper step wedges.
The scene was imaged with an exposure time of 100 seconds;
the radiograph is shown in Figure~\ref{fig:radiographs}.
The task was to identify the materials of the two shells from the list of possible materials in Table~\ref{tab:materials}.
We preprocessed the image (Section~\ref{sec:preprocessing}) and cropped out the table, step wedges, and stand.

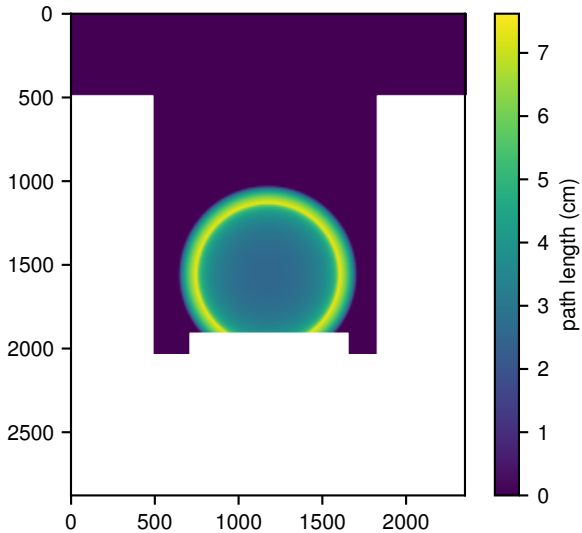
\begin{figure}[htbp]
    \centering
    \input{figures/AlCu_ell0.pgf}
    \caption{Path length of the aluminum shell in the Al-Cu Shells scene, $\ell_0$.}
    \label{fig:AlCu_ell0}
\end{figure}

In order to compute path lengths through the shells,
we assumed that the center of the sphere was on the optical axis
and used the known outer diameter of the aluminum shell,
source-to-shells distance,
source-to-detector distance,
and diameter of the sphere in pixels
to fix the rest of the scene geometry.
We then raytraced the shells analytically.
The result of this process was an image of the path length at each pixel
for each shell, $\ell_0$ and $\ell_1$;
Figure~\ref{fig:AlCu_ell0} shows $\ell_0$.

\begin{figure}[htbp]
    \centering
    \includegraphics[width=\linewidth]{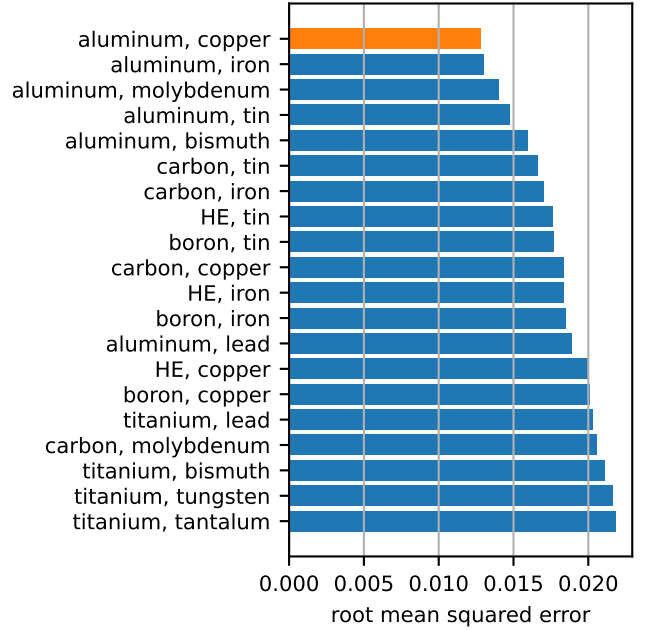}
    \caption{Results of material identification on the aluminum and copper shell scene with a cobalt-60 source.}
    \label{fig:AlCu_result}
\end{figure}

We ran the branch and bound algorithm (Section~\ref{sec:branch_and_bound})
with scatter model order $P=2$
to identify the top 20 material configurations for the scene.
The correct material assignment,  [aluminum, copper], was the one with the lowest fitting error (RMSE=0.0128),
i.e., it was the value $\x$ that minimized the loss functional \eqref{eq:functional}
out of all possible assignments.
meaning that our algorithm was able to correcting identify these materials;
the full ranking and RMSEs for the top 20 assignments is given in Figure~\ref{fig:AlCu_result}.

The total number of possible configurations was $20^2=400$
and the branch and bound algorithm evaluated all of them
plus an additional 20 partial solutions,
i.e., $[\text{air}, \varnothing,]$, $[\text{B}, \varnothing,]$, $[\text{high explosive}, \varnothing,]$, \dots.
This process took under three minutes on a consumer-level laptop.

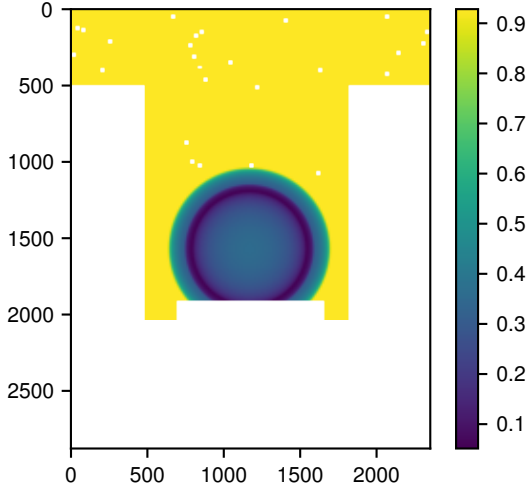
\begin{figure}
        \adjustbox{width=\linewidth}{\input{figures/AlCu_d.pgf}}
        \caption{Modeled direct radiograph.
        The isolated white pixels are those that were identified as dead or stuck pixels
        during preprocessing (Section~\ref{sec:preprocessing}).}
        \label{fig:AlCu_direct}
\end{figure}

\begin{figure}[htbp]
    \centering
    \input{figures/AlCu_lineouts_zoom.pgf}
    \caption{Zoomed in lineout across the diameter of the sphere comparing the measured radiograph to the first prediction (blue, dashed)
    and the second prediction (orange, dotted).}
    \label{fig:AlCu_lineout_zoomed}
\end{figure}
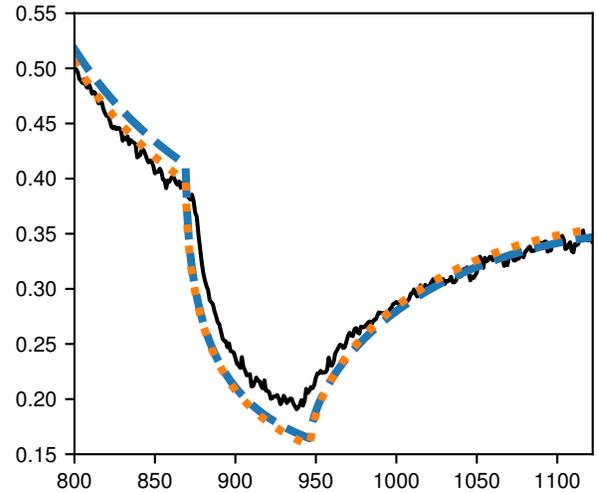

To further explore these results,
we plot the direct radiograph that our model predicted for the material assignment $\x = [\text{aluminum}, \text{copper}]$
(Figure~\ref{fig:AlCu_direct})
and the corresponding modeled scatter field (Figure~\ref{fig:AlCu_scatter}).
The fact that that the modeled direct only reaches a value around 0.9
in the background region
(instead of approximately 1.0, as direct theoretically should)
suggests that the gain parameter and the scatter field parameters ($\alpha$ and $\theta$ in \eqref{eq:functional})
work together to model scatter and any other aspect of the real imaging system that our forward model does not capture.
Finally, we plot a lineout across the diameter of the shells,
Figures~\ref{fig:AlCu_lineout} and \ref{fig:AlCu_lineout_zoomed}.
These show good agreement with the measured radiograph,
but also show evidence of systematic error
in that the predicted transmission consistently undershoots
the measured transmission.

\begin{figure}[htbp]
    \centering
    \input{figures/AlCu_s.pgf}
    \caption{Modeled scatter radiograph.}
    \label{fig:AlCu_scatter}
\end{figure}
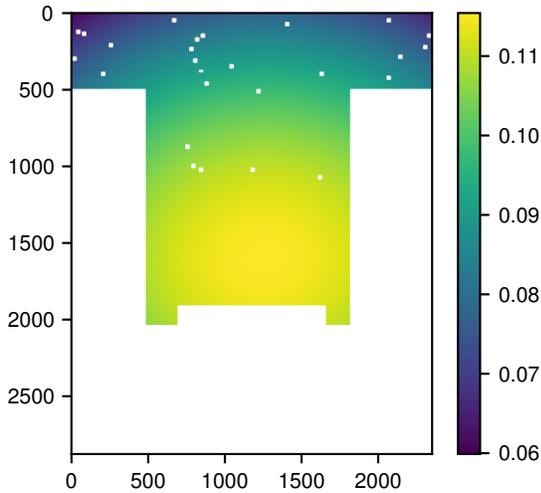

\begin{figure}[htbp]
    \centering
    \input{figures/AlCu_lineouts.pgf}
    \caption{Lineout across the diameter of the sphere comparing the measured radiograph to the first prediction (blue, dashed)
    and the second prediction (orange, dotted).}
    \label{fig:AlCu_lineout}
\end{figure}

\textbf{Eight Cubes.}
As a more challenging validation,
we radiographed a scene of eight, one-inch cubes made of copper, bismuth, iron, tantalum, titanium, tungsten, molybdenum, and aluminum (Figure~\ref{fig:radiographs}).
The scene was imaged with an exposure time of 100 seconds.
We set the path length through each cube at one inch,
neglecting the conebeam geometry;
this is justified because all the cubes are within a few inches of the optical axis,
which is small compared to the 82.5 cm source-to-object distance.

For this configuration,
we used a warm start technique:
we first ran a branch and bound search with only lithium, tin, and uranium
(low, medium, and high absorbing materials)
and used the top 20 assignments as an initialization for the $topN$ list and the bound $J^*$ in Algorithm~\ref{algo:branch_and_bound}.
This approach ensured a good initial value for the bound,
allowing the full search to finish more quickly.

\begin{figure}
    \centering
    \includegraphics[width=\linewidth]{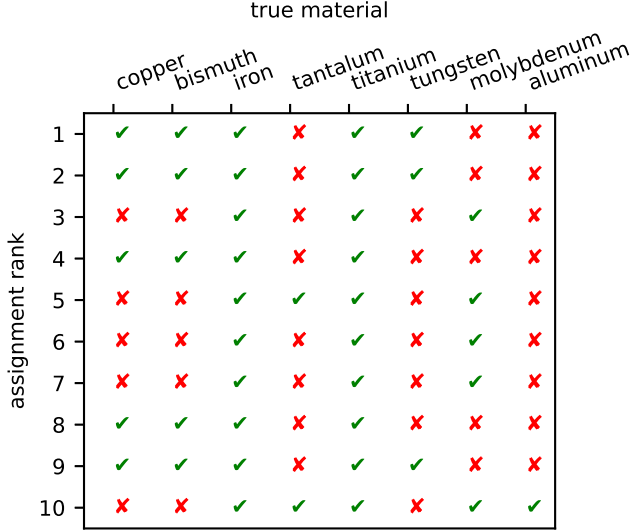}
    \caption{Material identification accuracy for the top ten assignments in for the Eight Cubes data.
    A check mark in row $n$, column $m$ indicates that the $n$th assignment correctly identified material $m$.
    }
    \label{fig:EightCubes_checks}
\end{figure}

The initial and full search took approximately 31 minutes on the same consumer-level laptop.
The correct material assignment had an RMSE of $6.79 \times 10^{-3}$
and did not rank among the top twenty assignments,
which had RMSEs between $6.24 \times 10^{-3}$ and $6.41 \times 10^{-3}$.
However all of the top ten assignments were partially correct (Figure~\ref{fig:EightCubes_checks})
and each cube was identified corrected in at least one of them.

Here, the benefits of the branch and bound algorithm were much more apparent.
There were $20^8 = 2.56 \times 10^{10}$ possible configurations,
and the full search used only 130,720 bounding operations and evaluated 12,640 complete configurations.
This means that the vast majority (99.99995\%) of the cases where excluded by the algorithm,
representing an approximately $1.8 \times 10^5$ times speedup over exhaustive search.

\subsection{Material Identification From Experimental Data---Betatron Source}
\begin{figure}[htbp]
    \centering
   \input{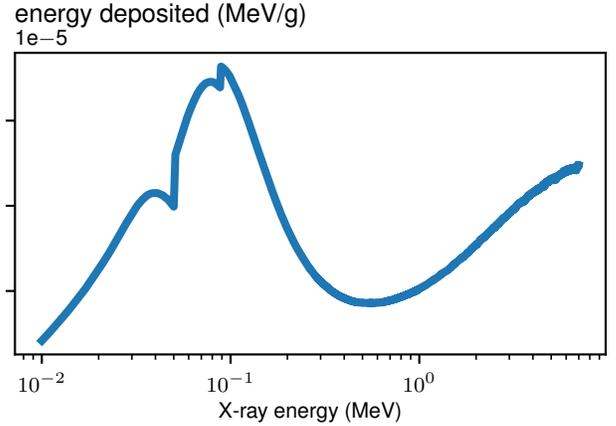}
    \caption{The response function of the Pixium 3543EZ digital readout panel calculated using MCNP.}
    \label{fig:detresponse}
\end{figure}

In a separate experiment, 
we collected radiographs using a SEA-7 betatron accelerator by Instauro where the energy of the primary electron beam 
was set to 7 MeV.
We modeled the source spectrum as a 7 MeV Bremsstrahlung spectrum
and used a simulation-based model for the detector response.
The response function of the digital readout panel was calculated using MCNP. 
The panel was modeled as a sequence of material layers as indicated by the manufacturer and shown in table \ref{tab:detmaterials}.
Multiple simulations were run to calculate the detector response.
In each of them, a monochromatic X-ray pencil beam
was shot on the panel, perpendicular to it, and the total energy deposited in the sensitive layers of the panel was obtained.
The energy of the incident X-rays was different in each simulation run and varied in the range (0 MeV, 7 MeV].
The resulting response function is shown in figure \ref{fig:detresponse}.
\begin{table}[]
    \centering
 \caption{Pixium 3543EZ digital readout panel}
    \begin{tabular}{llll}
    \toprule
    layer & thickness (mm)  & modeled as \\ \midrule
   front side housing  & 3.6 & plastic \\
    gadox scintillator layer & 0.4 & gadox   \\
    photodiode plate & 0.7 & silicon \\
    mechanical assembly & 2.3 & air \\
    lead & 1.0 & lead\\
    electronics & 7.25 & plastic \\
    back side housing & 1.5 & plastic \\
    \bottomrule
    \end{tabular}
    \label{tab:detmaterials}
\end{table}

\begin{figure}[htbp]
    \centering
   \input{figures/source_det.pgf}
    \caption{Product of the detector response and source spectrum before and after calibration.}
    \label{fig:source_det}
\end{figure}
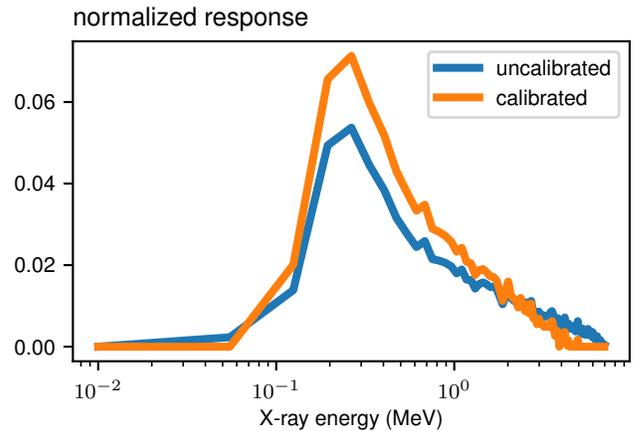

We subsequently calibrated the product of source spectrum and detector response on the Al-Cu Shells Brem data
using the method described in Section~\ref{sec:calibration},
see Figure~\ref{fig:source_det} for the results of the calibration.
Calibration emphasized energies between about 0.1 MeV and 1.0 MeV.
The source-to-detector distance was 300 cm 
and the object-to-detector distance was 100 cm.

\begin{figure*}[htbp]
    \centering
    \includegraphics[width=.33\linewidth]{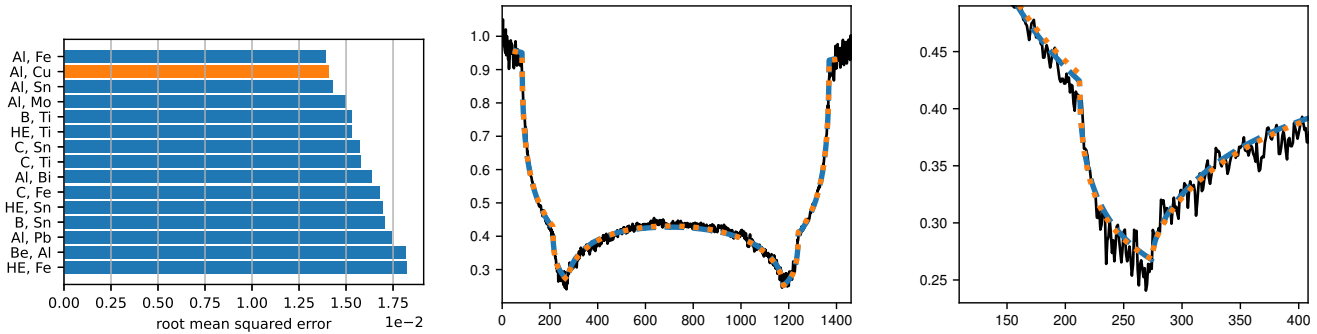}\hfill
    \adjustbox{width=0.33\linewidth}{\input{figures/AlCuShellsBrem_lineouts.pgf}}\hfill
    \adjustbox{width=0.33\linewidth}{\input{figures/AlCuShellsBrem_lineouts_zoom.pgf}}
    \caption{Left: results of material identification for Al-Cu Shells scene with a betatron source. 
    Center: Lineout across the diameter of the shells comparing the measured radiograph to the first prediction (blue, dashed)
    and the second prediction (orange, dotted).
    Right: Zoomed lineout.
  }
    \label{fig:AlCuShellsBrem_results}
\end{figure*}

\textbf{Al-Cu Shells Brem.}
We repeated the Al-Cu Shells scene with the betatron source,
with an exposure time of 8 seconds.
We again used the known shell diameters,
source-to-object distance,
and source-to-detector distance
to analytically compute the path lengths.
Branch and bound again reduced to exhaustive search
and took 7.2 minutes.
The correct assignment was at rank 2 (RMSE=0.014);
the error for the top 15 results and lineouts for the top 2 results are given in Figure~\ref{fig:AlCuShellsBrem_results}.

\begin{figure*}[htbp]
    \centering
    \includegraphics[width=.33\linewidth]{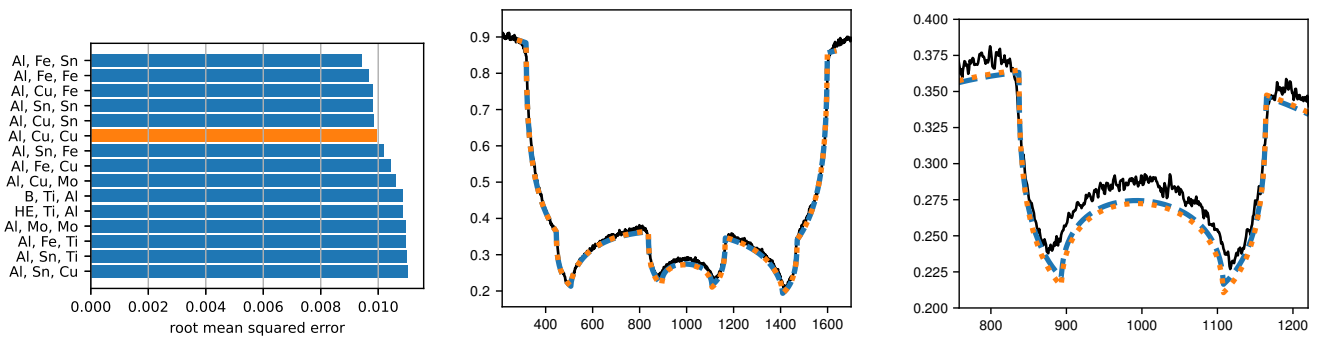}\hfill
    \adjustbox{width=0.33\linewidth}{\input{figures/AlCuCuShellsBrem_lineouts.pgf}}\hfill
    \adjustbox{width=0.33\linewidth}{\input{figures/AlCuCuShellsBrem_lineouts_zoom.pgf}}
    \caption{Left: results of material identification for the Al Cu Cu Shells with a betatron source. 
    Center: Lineout across the diameter of the sphere comparing the measured radiograph to the first prediction (blue, dashed)
    and the second prediction (orange, dotted).
    Right: Zoomed lineout.
    The disagreement between measurement and model can be reduced by increasing the polynomial scatter model order,
    but at the cost of decreased identification performance, see Figure~\ref{fig:polynomial_order}.
  }
    \label{fig:AlCuCu_results}
\end{figure*}

\textbf{Al-Cu-Cu Shells Brem.}
As a more challenging problem,
we performed material identification on a scene with three shells: aluminum, copper, copper
(Figure~\ref{fig:radiographs}).
For the outer shells,
we analytically computed the path lengths as in the previous experiments.
The center shell in this configuration is not on the optical axis,
which complicates the process of computing its path lengths.
We manually searched for a position (in 3D) for the shell such
that when we computed path lengths for a shell of the correct dimensions,
they aligned well with the measured radiograph.
The exposure time was 8 seconds.
The search for the top 20 assignments took under six minutes on the same consumer-level laptop.
The branch and bound algorithm computed error for 140 full assignments
and
excluded the remaining 7860 assignments by 
computing error for 180 partial assignments;
this represents an approximately $25$ times speedup over exhaustive search.
The ground truth was at position six (RMSE=$9.95 \times 10^{-3}$);
errors for the top 15 assignments and lineouts are shown in Figure~\ref{fig:AlCuCu_results}.

\begin{figure}
    \centering
    \includegraphics[width=\linewidth]{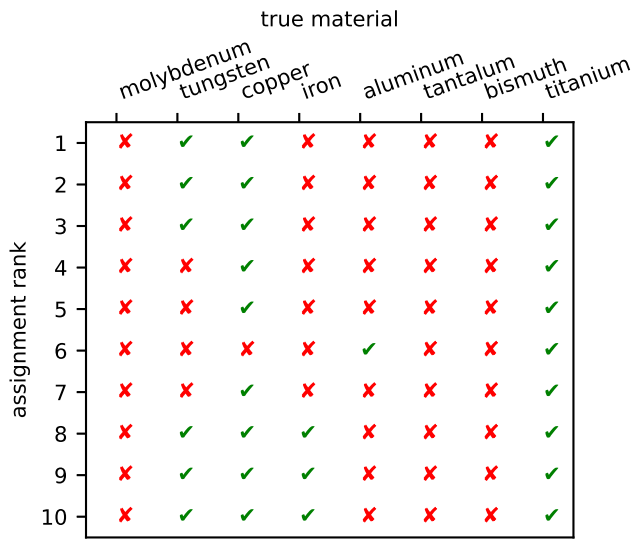}
    \caption{Material identification accuracy for the top ten assignments in for the Eight Cubes Brem data.
    A check mark in row $n$, column $m$ indicates that the $n$th assignment correctly identified material $m$.
    }
    \label{fig:EightCubesBrem_checks}
\end{figure}

\textbf{Eight Cubes Brem}
We repeated the eight cubes experiment with the betatron source
with an exposure time of 8 seconds.
We again set the path length through each cube to one inch,
ignoring conebeam effects.
Using the same warm start approach as described for the Eight Cubes data,
branch and bound took 5.3 hours.
Again, the correct assignment did not appear in the top 20 assignments.
The RMSE of the ground truth was $7.91 \times 10^{-3}$,
with the top 20 assignments having RMSEs between
 $5.36 \times 10^{-3}$
 and
  $5.77 \times 10^{-3}$.
The top results still had some correct assignments,
but several materials (molybdenum, tantalum, and bismuth)
were not identified correctly in any of the top 10 assignments,
shown in Figure~\ref{fig:EightCubesBrem_checks}.

\subsection{Effect of Material Thickness}
To evaluate the effect of material thickness on material identification performance,
we made a series of simulated radiographs with the MCNP Monte Carlo particle transport code~\cite{werner_mcnp6.2_2018}.
For all simulations,
we used a 6 MeV Bremsstrahlung source with an source-to-object distance of 4 m 
and a source-to-detector distance of 6.5 m
and a detector pixel size of 50 \textmu m.
In all simulations, the scene consisted of a ball of lithium
surrounded by shells of polycarbonate, boron, iron, and plutonium.
The simulations differed in that 
the radius of the ball and the thickness of the shells varied from 0.1 cm to 2 cm.

\begin{figure}[htbp]
  \centering
  \includegraphics{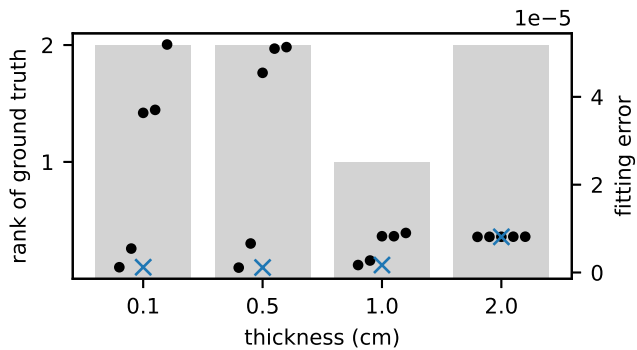}
  \caption{Rank (bars) and error (crosses) of the ground truth material assignment as a function of shell thickness,
    along with errors for the top five incorrect assignments (points, horizontally spread for visibility)
    on simulated data.
    The 2 cm shells are the most challenging in that more assignments have an error similar to that of the ground truth.
    }
  \label{fig:thickness}
  
\end{figure}

Figure~\ref{fig:thickness} shows the results of performing material identification on these simulations.
The ground truth ranked at the first position for the 1 cm simulation
and the second position for the rest,
suggesting that the contrast between materials is largest at intermediate lengths.
This makes sense because with large lengths, transmissions are always small no matter the material,
and vise versa for small lengths.
At the smaller lengths, confusions were between the boron shell and other materials with small attenuation coefficients:
high explosive and carbon.

\subsection{Effect of Noise and Path Length Error}
In this section,
we examine how noise and error in path length estimation
may impact material identification.
In general,
this is a complex question:
our algorithm returns a discrete answer,
meaning that performance cannot smoothly degrade
as noise or path length errors increase.
Further,
the effect of noise and errors will be scene-dependent,
limiting the value of any particular empirical study.

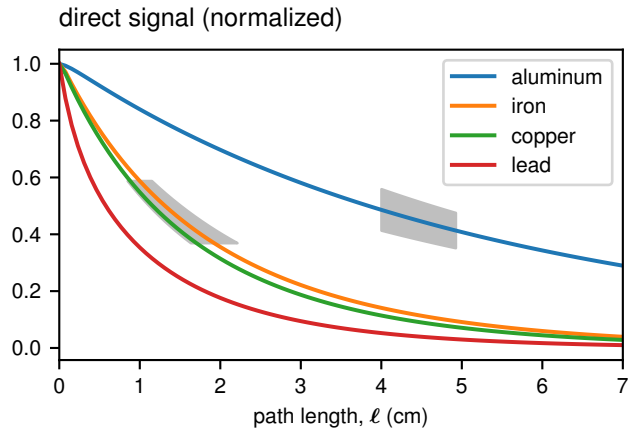
\begin{figure}[htbp]    
    \centering
   \input{figures/contrast.pgf}
    \caption{Expected direct signal for aluminum, iron, copper, and lead for path lengths up to 7 cm.
    At 15\% path length error for objects between 1 and 2 cm thick
    (left gray box) copper may be confused with iron,
    because the green line falls within the grey box.
    At 15\% measurement noise for objects between 4 and 5 cm thick (right gray box),
    aluminum is unlikely to be confused with any of the other materials.
    }
    \label{fig:contrast}
\end{figure}

As a partial solution to these challenges,
we analytically computed the value
of the forward model \eqref{eq:direct}
using our betatron source spectrum and detector response
for several materials over a range
of path lengths,
see Figure~\ref{fig:contrast}.
This analysis provides an analytical way of determining
what level of noise and path length error
is acceptable for a given application:
the smallest vertical cap between
a pair of curves puts an upper limit
on the acceptable measurement noise
before those two materials will be confused;
likewise the smallest horizontal gap 
limits the acceptable error in path length estimation.

\subsection{Justification for Polynomial Field Scatter Model}
\label{sec:polynomial_scatter}
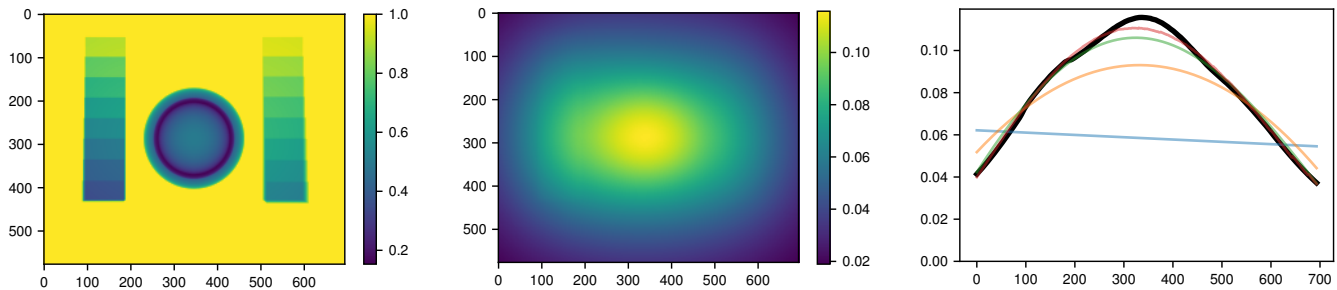
\begin{figure*}
  \centering
\begin{subfigure}{.33\linewidth}
    \adjustbox{width=\linewidth}{\input{figures/MCNP_t.pgf}}
\end{subfigure}\hfill
\begin{subfigure}{.33\linewidth}
    \adjustbox{width=\linewidth}{\input{figures/MCNP_s.pgf}}
  \end{subfigure}\hfill
  \begin{subfigure}{.33\linewidth}
    \adjustbox{width=\linewidth}{\input{figures/MCNP_lineouts.pgf}}
  \end{subfigure}
  \caption{Total signal (left), scatter signal (center), and scatter lineouts with polynomial fits (right) of orders one, two, four, and eight
    for a simulated scene.
      Scatter reaches over 10\% of the flat-field signal (note the different color scales)
    and can account for more than half of the total signal in areas where the direct is small.
  }
  \label{fig:scatter_fit}
\end{figure*}

To validate the appropriateness of our polynomial field scatter model
(Section~\ref{sec:forward_model}),
we performed a photon transport simulation similar to the the Al-Cu Shells experiment with the cobalt-60 source using MCNP~\cite{werner_mcnp6.2_2018}.
The resulting radiograph (including both direct and scattered photons) after flat-field correction
and the scatter radiograph (only scattered photons)
are shown in Figure~\ref{fig:scatter_fit}.
We observe that the scatter is mostly smooth,
although the effect of the left step-wedge is barely visible.

We fit the scatter field with polynomial fields of order one, two, four, and eight;
Figure~\ref{fig:scatter_fit} shows the resulting lineouts.
By order two, the polynomial fit is reasonably close to the scatter,
and by order eight, the maximum absolute error is less than 0.009 over the whole field of view.
In the context of material identification,
the advantage of using lower orders is that it is more computationally efficient
(fewer coefficients lead to faster solutions during the fitting in \eqref{eq:functional})
and is less prone to making incorrect material assignments match the data well.

While this experiment shows that polynomial fields are reasonable for representing scatter fields,
it is not straightforward to generalize these result to real data.
The simulation does not (and can not) represent scene scatter,
i.e. photons leaving the source, interacting with material in the scene but outside the field of view of the detector,
and deflecting back into the detector.
Nor have we accounted for scattering within the detector itself.
It is unlikely we will ever be able to precisely measure the scatter component of a real radiograph
to further validate scatter models.
However, we suspect most of the effects that are difficult to model would tend to make scatter more spatially smooth than our simulation,
again justifying the use of a smooth field to model scatter.

It is also important that the scatter model \emph{not}
be able to account for too much of the direct signal.
As an extreme case, if we allowed the scatter to be any image,
then scatter would account for the entire measured signal
and material identification using our method would be impossible.
This question boils down to how closely the scatter model
(in our case, a low-order polynomial)
can fit a direct signal,
which is the path length as a function of space
passed through the polyenergetic Beer's law \eqref{eq:direct},
which is a pointwise nonlinearity resembling a decaying exponential.
For the types of objects that we expect to encounter,
the path length increases quickly at the edge of the object,
creating a sharp decrease in the direct signal
(see, e.g., Figure~\ref{fig:AlCu_lineout})
that polynomials do not fit well.

\subsection{Effect of Polynomial Scatter Model Order}
\begin{figure}[htbp]
  \centering
  \includegraphics{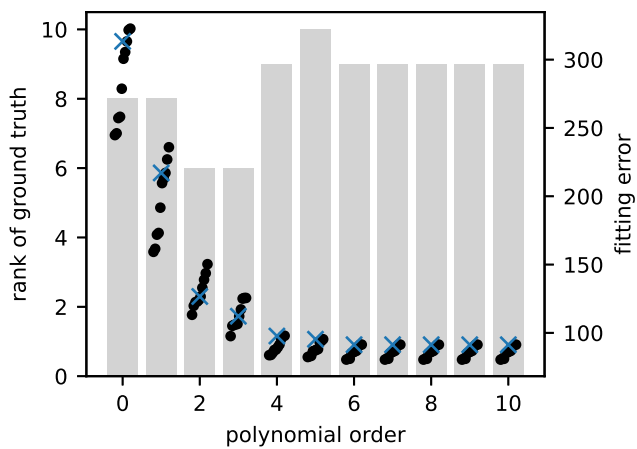}
  \caption{Rank (bars) and error (crosses) of the ground truth material assignment as a function of polynomial scatter model order,
    along with errors for the top ten incorrect assignments (points, horizontally spread for visibility)
    on the Al-Cu-Cu Shells Brem data.
  Model orders two and three are optimal for this data.}
  \label{fig:polynomial_order}
  
\end{figure}

To explore the effect of the order of the polynomial scatter model,
we ran the Al-Cu-Cu Shells Brem material identification
with model orders zero through ten.
The results of this experiment are summarized in Figure~\ref{fig:polynomial_order}.
The fitting error of all assignments decreases as the polynomial model order increases,
which must be the case because higher order polynomials fit strictly more of the residual than lower order ones.
What is more important is the error of the ground truth relative to other, incorrect assignments;
this value is at its minimum for orders two and three,
which place the correct assignment at rank six.

\section{Discussion} \label{sec:discussion}
Comparing experiments with the cobalt-60 source to those with the betatron source,
those with the betatron source resulted material identification
that was slower and 
less accurate.
One reason for the difference in speed
is that the forward model is slower to evaluate due to the larger number of energy bins
(101 versus 2).
Material identification in the betatron experiments was worse than for the cobalt-60 experiments,
(the Al-Cu Shells correct assignment moved from rank one to two
and the Eight Cubes top assignments changed from five of eight to three of eight correct).
We believe that 
this difference in accuracy
stems from the betatron source being relatively unstable,
making measurements less repeatable and flat-field correction difficult.
This effect is apparent in Figure~\ref{fig:radiographs},
where the background for the Bremsstrahlung datasets is notably nonuniform
even after flat-field correction.
While this effect can be partially mitigated by dividing by the open-beam value
for each radiograph individually,
doing so does not affect our algorithm's material assignments because
the scalar $\alpha$ in \eqref{eq:functional} makes the approach
invariant to rescaling the radiograph $\tot$.
It also appears in Figure~\ref{fig:AlCuCu_results},
where the transmission appears nonsymmetric,
i.e., the minimum transmission on the left side of the inner copper shell 
is higher than the minimum transmission on its right side
(although this effect may also be due to scatter
or blur).

The efficiency of branch and bound in our experiments
depended on the arrangement of the scene.
In the Al-Cu Shells scene, branch and bound was moderately less efficient than exhaustive search would have been
because none of the partial solution bounds resulted in exclusion of any full solutions.
This is probably because the shells overlap significantly in the radiograph
and therefore choosing the material of the outer shell only determines a small amount of the total error.
In the extreme case where all objects overlap completely in the radiograph,
e.g., a set of sheets perpendicular to the optical axis,
our branch and bound algorithm would never be faster than exhaustive search because all partial bounds would be zero.
However, we note that the overhead is modest (roughly log of the total number of solutions),
so that in situations where exhaustive search is computationally feasible,
branch and bound will be, also.
On the other hand,
our experiments on the Eight Cubes scene demonstrate that our algorithm can provide tremendous time savings over exhaustive search.
This happens because there are many partial solutions that are so inaccurate that no material assignment for the remaining objects can make them viable.
We also note that both our branch and bound algorithm and exhaustive search are simple to parallelize,
which should allow decreasing the runtimes reported here by several orders of magnitude 
when running on a moderately-sized cluster (100s of nodes).

Our experiments also showed that the accuracy of material identification depends on the arrangement of objects:
the algorithm performed better on scenes with concentric shells than on the cubes data.
We believe that the cubes data is especially challenging because it presents no diversity
of path lengths.
This means that identification must be performed on the basis of one number
(the transmission through the cube) per object.
Practically speaking, 
such arrangements of objects are unlikely because they require the objects to be perfectly aligned with
the detector.
We suspect that performance would have been better if the cubes had been rotated so as to break this alignment
and provide a range of path lengths.
We also note that,
because of the exponential growth of the number of possible solutions,
placing the ground truth in the top $N$ best fits is more difficult
when there are more objects.
For example, the top 20 assignments represent $0.25\%$ of the total possible assignments
when there are 3 objects,
but only $7.8 \times 10 ^{-8}\%$ of the total when there are 8 objects.
This effect may also explain why our method did not place the correct material assignment
into the top 20 on the cubes data.

Finally,
our experiments demonstrate how challenging it is to accurately model real radiographic data.
For example, even in the relatively simple Al-Cu Shells scene,
our best model showed deviations as large as ten percent of the measured signal (Figure~\ref{fig:AlCu_lineout_zoomed}).
This mismatch points to the need for more sophisticated scatter models
and/or further system modeling and calibration.
However,
our experiments also show that the method tolerates some level of approximation in modeling the data.

\section{Limitations}
We now discuss several limitations of the current work.

First,
we require significant knowledge of the scene:
a list of possible materials,
their attenuation coefficients,
the source spectrum,
the detector response,
and the object geometry.
Attenuation coefficients for many materials are freely available,
e.g., see the National Institute of Standards and Technology Photon Cross Sections Database
\cite{berger_xcom_2010}.
The challenge of producing a list of possible materials depends on the application,
but one can imagine security scenarios wherein only a handful of materials are relevant
and the rest can be safely covered by a few generic material definitions.
We have demonstrated that the source spectrum and detector response
can be accounted for via calibration,
which represents a relatively small time investment amortized
over the life of the system
(or time between calibrations, if done repeatedly).
The most restrictive assumption, then,
is the known object geometry.
Solving this challenge in general is beyond the scope of this paper,
but
we argue that it can be approached by performing a kind of tomographic reconstruction
using multiple radiographs.
This tomographic problem should be less challenging than a full tomographic reconstruction
because only object boundaries must be recovered,
rather than a map of internal densities.
In settings where the shape of the objects is known
but their 3D pose is not,
techniques from the computer vision community
for camera calibration
and pose detection/registration~\cite{forsyth_computer_2011}
will prove useful.

Second,
our model does not include source or detector blur,
and we have not performed the experiments needed to characterize the blur in our setups.
Our model could be extended to include blur (if properly characterized),
however this would likely incur significant computational cost.
Another approach would be to mitigate the effect of blur by excluding pixels near boundaries from the analysis.

Finally,
our experiments covered either relatively simple
(two or three spherical shells)
or very challenging (eight axis-aligned cubes)
scenarios.
So,
while our study both provides a proof of concept
and identifies a regime where performance degrades,
more experimentation is needed to elucidate the performance of the approach in intermediate settings.

\section{Conclusions and Future Work} \label{sec:conclusions}
We have presented a method for identifying materials from radiographs without energy-resolved measurements,
thus allowing standard X-ray systems to provide
material identification information without requiring additional hardware.
To solve the underlying combinatorial optimization problem,
we developed a branch and bound algorithm
with a bounding function specifically tailored to the radiographic setting.
Our experiments on metallic objects with MeV X-rays show that the method can successfully identify materials from real radiographs,
although its accuracy and runtime are scene-dependent.
This method could prove to be a useful new capability in a variety of security applications.

This method can be extended in several ways:
We could explore other scatter models (e.g., piecewise polynomials, smooth images)
to allow finer control of the representation power
or try physically-motivated kernel-based scatter models such as  \cite{hansen_extraction_1997,ohnesorge_efficient_1999}.
We could add a technique to automatically estimate scene geometry and/or pose
(which would presumably require multiple radiographs).
Finally, we could seek improved initialization, branching, and bounding algorithms
to speed up the optimization process,
allowing even larger problems to be tackled efficiently.

\bibliographystyle{IEEEtran}
\bibliography{refs_mike}

\end{document}

%% file: figures/AlCu_ell0.pgf
\begingroup%
\makeatletter%
\begin{pgfpicture}%
\pgfpathrectangle{\pgfpointorigin}{\pgfqpoint{3.500000in}{3.000000in}}%
\pgfusepath{use as bounding box, clip}%
\begin{pgfscope}%
\pgfsetbuttcap%
\pgfsetmiterjoin%
\definecolor{currentfill}{rgb}{1.000000,1.000000,1.000000}%
\pgfsetfillcolor{currentfill}%
\pgfsetlinewidth{0.000000pt}%
\definecolor{currentstroke}{rgb}{1.000000,1.000000,1.000000}%
\pgfsetstrokecolor{currentstroke}%
\pgfsetdash{}{0pt}%
\pgfpathmoveto{\pgfqpoint{0.000000in}{0.000000in}}%
\pgfpathlineto{\pgfqpoint{3.500000in}{0.000000in}}%
\pgfpathlineto{\pgfqpoint{3.500000in}{3.000000in}}%
\pgfpathlineto{\pgfqpoint{0.000000in}{3.000000in}}%
\pgfpathclose%
\pgfusepath{fill}%
\end{pgfscope}%
\begin{pgfscope}%
\pgfsetbuttcap%
\pgfsetmiterjoin%
\definecolor{currentfill}{rgb}{1.000000,1.000000,1.000000}%
\pgfsetfillcolor{currentfill}%
\pgfsetlinewidth{0.000000pt}%
\definecolor{currentstroke}{rgb}{0.000000,0.000000,0.000000}%
\pgfsetstrokecolor{currentstroke}%
\pgfsetstrokeopacity{0.000000}%
\pgfsetdash{}{0pt}%
\pgfpathmoveto{\pgfqpoint{0.659856in}{0.317222in}}%
\pgfpathlineto{\pgfqpoint{2.721884in}{0.317222in}}%
\pgfpathlineto{\pgfqpoint{2.721884in}{2.840401in}}%
\pgfpathlineto{\pgfqpoint{0.659856in}{2.840401in}}%
\pgfpathclose%
\pgfusepath{fill}%
\end{pgfscope}%
\begin{pgfscope}%
\pgfsys@transformshift{0.660000in}{1.060000in}%
\pgftext[left,bottom]{\includegraphics[interpolate=true,width=2.070000in,height=1.780000in]{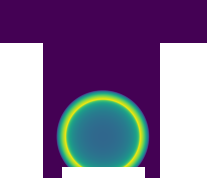}}%
\end{pgfscope}%
\begin{pgfscope}%
\pgfsetbuttcap%
\pgfsetroundjoin%
\definecolor{currentfill}{rgb}{0.000000,0.000000,0.000000}%
\pgfsetfillcolor{currentfill}%
\pgfsetlinewidth{0.803000pt}%
\definecolor{currentstroke}{rgb}{0.000000,0.000000,0.000000}%
\pgfsetstrokecolor{currentstroke}%
\pgfsetdash{}{0pt}%
\pgfsys@defobject{currentmarker}{\pgfqpoint{0.000000in}{-0.048611in}}{\pgfqpoint{0.000000in}{0.000000in}}{%
\pgfpathmoveto{\pgfqpoint{0.000000in}{0.000000in}}%
\pgfpathlineto{\pgfqpoint{0.000000in}{-0.048611in}}%
\pgfusepath{stroke,fill}%
}%
\begin{pgfscope}%
\pgfsys@transformshift{0.660295in}{0.317222in}%
\pgfsys@useobject{currentmarker}{}%
\end{pgfscope}%
\end{pgfscope}%
\begin{pgfscope}%
\definecolor{textcolor}{rgb}{0.000000,0.000000,0.000000}%
\pgfsetstrokecolor{textcolor}%
\pgfsetfillcolor{textcolor}%
\pgftext[x=0.660295in,y=0.220000in,,top]{\color{textcolor}\sffamily\fontsize{8.000000}{9.600000}\selectfont 0}%
\end{pgfscope}%
\begin{pgfscope}%
\pgfsetbuttcap%
\pgfsetroundjoin%
\definecolor{currentfill}{rgb}{0.000000,0.000000,0.000000}%
\pgfsetfillcolor{currentfill}%
\pgfsetlinewidth{0.803000pt}%
\definecolor{currentstroke}{rgb}{0.000000,0.000000,0.000000}%
\pgfsetstrokecolor{currentstroke}%
\pgfsetdash{}{0pt}%
\pgfsys@defobject{currentmarker}{\pgfqpoint{0.000000in}{-0.048611in}}{\pgfqpoint{0.000000in}{0.000000in}}{%
\pgfpathmoveto{\pgfqpoint{0.000000in}{0.000000in}}%
\pgfpathlineto{\pgfqpoint{0.000000in}{-0.048611in}}%
\pgfusepath{stroke,fill}%
}%
\begin{pgfscope}%
\pgfsys@transformshift{1.098651in}{0.317222in}%
\pgfsys@useobject{currentmarker}{}%
\end{pgfscope}%
\end{pgfscope}%
\begin{pgfscope}%
\definecolor{textcolor}{rgb}{0.000000,0.000000,0.000000}%
\pgfsetstrokecolor{textcolor}%
\pgfsetfillcolor{textcolor}%
\pgftext[x=1.098651in,y=0.220000in,,top]{\color{textcolor}\sffamily\fontsize{8.000000}{9.600000}\selectfont 500}%
\end{pgfscope}%
\begin{pgfscope}%
\pgfsetbuttcap%
\pgfsetroundjoin%
\definecolor{currentfill}{rgb}{0.000000,0.000000,0.000000}%
\pgfsetfillcolor{currentfill}%
\pgfsetlinewidth{0.803000pt}%
\definecolor{currentstroke}{rgb}{0.000000,0.000000,0.000000}%
\pgfsetstrokecolor{currentstroke}%
\pgfsetdash{}{0pt}%
\pgfsys@defobject{currentmarker}{\pgfqpoint{0.000000in}{-0.048611in}}{\pgfqpoint{0.000000in}{0.000000in}}{%
\pgfpathmoveto{\pgfqpoint{0.000000in}{0.000000in}}%
\pgfpathlineto{\pgfqpoint{0.000000in}{-0.048611in}}%
\pgfusepath{stroke,fill}%
}%
\begin{pgfscope}%
\pgfsys@transformshift{1.537007in}{0.317222in}%
\pgfsys@useobject{currentmarker}{}%
\end{pgfscope}%
\end{pgfscope}%
\begin{pgfscope}%
\definecolor{textcolor}{rgb}{0.000000,0.000000,0.000000}%
\pgfsetstrokecolor{textcolor}%
\pgfsetfillcolor{textcolor}%
\pgftext[x=1.537007in,y=0.220000in,,top]{\color{textcolor}\sffamily\fontsize{8.000000}{9.600000}\selectfont 1000}%
\end{pgfscope}%
\begin{pgfscope}%
\pgfsetbuttcap%
\pgfsetroundjoin%
\definecolor{currentfill}{rgb}{0.000000,0.000000,0.000000}%
\pgfsetfillcolor{currentfill}%
\pgfsetlinewidth{0.803000pt}%
\definecolor{currentstroke}{rgb}{0.000000,0.000000,0.000000}%
\pgfsetstrokecolor{currentstroke}%
\pgfsetdash{}{0pt}%
\pgfsys@defobject{currentmarker}{\pgfqpoint{0.000000in}{-0.048611in}}{\pgfqpoint{0.000000in}{0.000000in}}{%
\pgfpathmoveto{\pgfqpoint{0.000000in}{0.000000in}}%
\pgfpathlineto{\pgfqpoint{0.000000in}{-0.048611in}}%
\pgfusepath{stroke,fill}%
}%
\begin{pgfscope}%
\pgfsys@transformshift{1.975364in}{0.317222in}%
\pgfsys@useobject{currentmarker}{}%
\end{pgfscope}%
\end{pgfscope}%
\begin{pgfscope}%
\definecolor{textcolor}{rgb}{0.000000,0.000000,0.000000}%
\pgfsetstrokecolor{textcolor}%
\pgfsetfillcolor{textcolor}%
\pgftext[x=1.975364in,y=0.220000in,,top]{\color{textcolor}\sffamily\fontsize{8.000000}{9.600000}\selectfont 1500}%
\end{pgfscope}%
\begin{pgfscope}%
\pgfsetbuttcap%
\pgfsetroundjoin%
\definecolor{currentfill}{rgb}{0.000000,0.000000,0.000000}%
\pgfsetfillcolor{currentfill}%
\pgfsetlinewidth{0.803000pt}%
\definecolor{currentstroke}{rgb}{0.000000,0.000000,0.000000}%
\pgfsetstrokecolor{currentstroke}%
\pgfsetdash{}{0pt}%
\pgfsys@defobject{currentmarker}{\pgfqpoint{0.000000in}{-0.048611in}}{\pgfqpoint{0.000000in}{0.000000in}}{%
\pgfpathmoveto{\pgfqpoint{0.000000in}{0.000000in}}%
\pgfpathlineto{\pgfqpoint{0.000000in}{-0.048611in}}%
\pgfusepath{stroke,fill}%
}%
\begin{pgfscope}%
\pgfsys@transformshift{2.413720in}{0.317222in}%
\pgfsys@useobject{currentmarker}{}%
\end{pgfscope}%
\end{pgfscope}%
\begin{pgfscope}%
\definecolor{textcolor}{rgb}{0.000000,0.000000,0.000000}%
\pgfsetstrokecolor{textcolor}%
\pgfsetfillcolor{textcolor}%
\pgftext[x=2.413720in,y=0.220000in,,top]{\color{textcolor}\sffamily\fontsize{8.000000}{9.600000}\selectfont 2000}%
\end{pgfscope}%
\begin{pgfscope}%
\pgfsetbuttcap%
\pgfsetroundjoin%
\definecolor{currentfill}{rgb}{0.000000,0.000000,0.000000}%
\pgfsetfillcolor{currentfill}%
\pgfsetlinewidth{0.803000pt}%
\definecolor{currentstroke}{rgb}{0.000000,0.000000,0.000000}%
\pgfsetstrokecolor{currentstroke}%
\pgfsetdash{}{0pt}%
\pgfsys@defobject{currentmarker}{\pgfqpoint{-0.048611in}{0.000000in}}{\pgfqpoint{-0.000000in}{0.000000in}}{%
\pgfpathmoveto{\pgfqpoint{-0.000000in}{0.000000in}}%
\pgfpathlineto{\pgfqpoint{-0.048611in}{0.000000in}}%
\pgfusepath{stroke,fill}%
}%
\begin{pgfscope}%
\pgfsys@transformshift{0.659856in}{2.839963in}%
\pgfsys@useobject{currentmarker}{}%
\end{pgfscope}%
\end{pgfscope}%
\begin{pgfscope}%
\definecolor{textcolor}{rgb}{0.000000,0.000000,0.000000}%
\pgfsetstrokecolor{textcolor}%
\pgfsetfillcolor{textcolor}%
\pgftext[x=0.503634in, y=2.801407in, left, base]{\color{textcolor}\sffamily\fontsize{8.000000}{9.600000}\selectfont 0}%
\end{pgfscope}%
\begin{pgfscope}%
\pgfsetbuttcap%
\pgfsetroundjoin%
\definecolor{currentfill}{rgb}{0.000000,0.000000,0.000000}%
\pgfsetfillcolor{currentfill}%
\pgfsetlinewidth{0.803000pt}%
\definecolor{currentstroke}{rgb}{0.000000,0.000000,0.000000}%
\pgfsetstrokecolor{currentstroke}%
\pgfsetdash{}{0pt}%
\pgfsys@defobject{currentmarker}{\pgfqpoint{-0.048611in}{0.000000in}}{\pgfqpoint{-0.000000in}{0.000000in}}{%
\pgfpathmoveto{\pgfqpoint{-0.000000in}{0.000000in}}%
\pgfpathlineto{\pgfqpoint{-0.048611in}{0.000000in}}%
\pgfusepath{stroke,fill}%
}%
\begin{pgfscope}%
\pgfsys@transformshift{0.659856in}{2.401607in}%
\pgfsys@useobject{currentmarker}{}%
\end{pgfscope}%
\end{pgfscope}%
\begin{pgfscope}%
\definecolor{textcolor}{rgb}{0.000000,0.000000,0.000000}%
\pgfsetstrokecolor{textcolor}%
\pgfsetfillcolor{textcolor}%
\pgftext[x=0.385634in, y=2.363051in, left, base]{\color{textcolor}\sffamily\fontsize{8.000000}{9.600000}\selectfont 500}%
\end{pgfscope}%
\begin{pgfscope}%
\pgfsetbuttcap%
\pgfsetroundjoin%
\definecolor{currentfill}{rgb}{0.000000,0.000000,0.000000}%
\pgfsetfillcolor{currentfill}%
\pgfsetlinewidth{0.803000pt}%
\definecolor{currentstroke}{rgb}{0.000000,0.000000,0.000000}%
\pgfsetstrokecolor{currentstroke}%
\pgfsetdash{}{0pt}%
\pgfsys@defobject{currentmarker}{\pgfqpoint{-0.048611in}{0.000000in}}{\pgfqpoint{-0.000000in}{0.000000in}}{%
\pgfpathmoveto{\pgfqpoint{-0.000000in}{0.000000in}}%
\pgfpathlineto{\pgfqpoint{-0.048611in}{0.000000in}}%
\pgfusepath{stroke,fill}%
}%
\begin{pgfscope}%
\pgfsys@transformshift{0.659856in}{1.963250in}%
\pgfsys@useobject{currentmarker}{}%
\end{pgfscope}%
\end{pgfscope}%
\begin{pgfscope}%
\definecolor{textcolor}{rgb}{0.000000,0.000000,0.000000}%
\pgfsetstrokecolor{textcolor}%
\pgfsetfillcolor{textcolor}%
\pgftext[x=0.326634in, y=1.924695in, left, base]{\color{textcolor}\sffamily\fontsize{8.000000}{9.600000}\selectfont 1000}%
\end{pgfscope}%
\begin{pgfscope}%
\pgfsetbuttcap%
\pgfsetroundjoin%
\definecolor{currentfill}{rgb}{0.000000,0.000000,0.000000}%
\pgfsetfillcolor{currentfill}%
\pgfsetlinewidth{0.803000pt}%
\definecolor{currentstroke}{rgb}{0.000000,0.000000,0.000000}%
\pgfsetstrokecolor{currentstroke}%
\pgfsetdash{}{0pt}%
\pgfsys@defobject{currentmarker}{\pgfqpoint{-0.048611in}{0.000000in}}{\pgfqpoint{-0.000000in}{0.000000in}}{%
\pgfpathmoveto{\pgfqpoint{-0.000000in}{0.000000in}}%
\pgfpathlineto{\pgfqpoint{-0.048611in}{0.000000in}}%
\pgfusepath{stroke,fill}%
}%
\begin{pgfscope}%
\pgfsys@transformshift{0.659856in}{1.524894in}%
\pgfsys@useobject{currentmarker}{}%
\end{pgfscope}%
\end{pgfscope}%
\begin{pgfscope}%
\definecolor{textcolor}{rgb}{0.000000,0.000000,0.000000}%
\pgfsetstrokecolor{textcolor}%
\pgfsetfillcolor{textcolor}%
\pgftext[x=0.326634in, y=1.486338in, left, base]{\color{textcolor}\sffamily\fontsize{8.000000}{9.600000}\selectfont 1500}%
\end{pgfscope}%
\begin{pgfscope}%
\pgfsetbuttcap%
\pgfsetroundjoin%
\definecolor{currentfill}{rgb}{0.000000,0.000000,0.000000}%
\pgfsetfillcolor{currentfill}%
\pgfsetlinewidth{0.803000pt}%
\definecolor{currentstroke}{rgb}{0.000000,0.000000,0.000000}%
\pgfsetstrokecolor{currentstroke}%
\pgfsetdash{}{0pt}%
\pgfsys@defobject{currentmarker}{\pgfqpoint{-0.048611in}{0.000000in}}{\pgfqpoint{-0.000000in}{0.000000in}}{%
\pgfpathmoveto{\pgfqpoint{-0.000000in}{0.000000in}}%
\pgfpathlineto{\pgfqpoint{-0.048611in}{0.000000in}}%
\pgfusepath{stroke,fill}%
}%
\begin{pgfscope}%
\pgfsys@transformshift{0.659856in}{1.086538in}%
\pgfsys@useobject{currentmarker}{}%
\end{pgfscope}%
\end{pgfscope}%
\begin{pgfscope}%
\definecolor{textcolor}{rgb}{0.000000,0.000000,0.000000}%
\pgfsetstrokecolor{textcolor}%
\pgfsetfillcolor{textcolor}%
\pgftext[x=0.326634in, y=1.047982in, left, base]{\color{textcolor}\sffamily\fontsize{8.000000}{9.600000}\selectfont 2000}%
\end{pgfscope}%
\begin{pgfscope}%
\pgfsetbuttcap%
\pgfsetroundjoin%
\definecolor{currentfill}{rgb}{0.000000,0.000000,0.000000}%
\pgfsetfillcolor{currentfill}%
\pgfsetlinewidth{0.803000pt}%
\definecolor{currentstroke}{rgb}{0.000000,0.000000,0.000000}%
\pgfsetstrokecolor{currentstroke}%
\pgfsetdash{}{0pt}%
\pgfsys@defobject{currentmarker}{\pgfqpoint{-0.048611in}{0.000000in}}{\pgfqpoint{-0.000000in}{0.000000in}}{%
\pgfpathmoveto{\pgfqpoint{-0.000000in}{0.000000in}}%
\pgfpathlineto{\pgfqpoint{-0.048611in}{0.000000in}}%
\pgfusepath{stroke,fill}%
}%
\begin{pgfscope}%
\pgfsys@transformshift{0.659856in}{0.648181in}%
\pgfsys@useobject{currentmarker}{}%
\end{pgfscope}%
\end{pgfscope}%
\begin{pgfscope}%
\definecolor{textcolor}{rgb}{0.000000,0.000000,0.000000}%
\pgfsetstrokecolor{textcolor}%
\pgfsetfillcolor{textcolor}%
\pgftext[x=0.326634in, y=0.609626in, left, base]{\color{textcolor}\sffamily\fontsize{8.000000}{9.600000}\selectfont 2500}%
\end{pgfscope}%
\begin{pgfscope}%
\pgfsetrectcap%
\pgfsetmiterjoin%
\pgfsetlinewidth{0.803000pt}%
\definecolor{currentstroke}{rgb}{0.000000,0.000000,0.000000}%
\pgfsetstrokecolor{currentstroke}%
\pgfsetdash{}{0pt}%
\pgfpathmoveto{\pgfqpoint{0.659856in}{0.317222in}}%
\pgfpathlineto{\pgfqpoint{0.659856in}{2.840401in}}%
\pgfusepath{stroke}%
\end{pgfscope}%
\begin{pgfscope}%
\pgfsetrectcap%
\pgfsetmiterjoin%
\pgfsetlinewidth{0.803000pt}%
\definecolor{currentstroke}{rgb}{0.000000,0.000000,0.000000}%
\pgfsetstrokecolor{currentstroke}%
\pgfsetdash{}{0pt}%
\pgfpathmoveto{\pgfqpoint{2.721884in}{0.317222in}}%
\pgfpathlineto{\pgfqpoint{2.721884in}{2.840401in}}%
\pgfusepath{stroke}%
\end{pgfscope}%
\begin{pgfscope}%
\pgfsetrectcap%
\pgfsetmiterjoin%
\pgfsetlinewidth{0.803000pt}%
\definecolor{currentstroke}{rgb}{0.000000,0.000000,0.000000}%
\pgfsetstrokecolor{currentstroke}%
\pgfsetdash{}{0pt}%
\pgfpathmoveto{\pgfqpoint{0.659856in}{0.317222in}}%
\pgfpathlineto{\pgfqpoint{2.721884in}{0.317222in}}%
\pgfusepath{stroke}%
\end{pgfscope}%
\begin{pgfscope}%
\pgfsetrectcap%
\pgfsetmiterjoin%
\pgfsetlinewidth{0.803000pt}%
\definecolor{currentstroke}{rgb}{0.000000,0.000000,0.000000}%
\pgfsetstrokecolor{currentstroke}%
\pgfsetdash{}{0pt}%
\pgfpathmoveto{\pgfqpoint{0.659856in}{2.840401in}}%
\pgfpathlineto{\pgfqpoint{2.721884in}{2.840401in}}%
\pgfusepath{stroke}%
\end{pgfscope}%
\begin{pgfscope}%
\pgfsetbuttcap%
\pgfsetmiterjoin%
\definecolor{currentfill}{rgb}{1.000000,1.000000,1.000000}%
\pgfsetfillcolor{currentfill}%
\pgfsetlinewidth{0.000000pt}%
\definecolor{currentstroke}{rgb}{0.000000,0.000000,0.000000}%
\pgfsetstrokecolor{currentstroke}%
\pgfsetstrokeopacity{0.000000}%
\pgfsetdash{}{0pt}%
\pgfpathmoveto{\pgfqpoint{2.878250in}{0.317222in}}%
\pgfpathlineto{\pgfqpoint{3.004409in}{0.317222in}}%
\pgfpathlineto{\pgfqpoint{3.004409in}{2.840401in}}%
\pgfpathlineto{\pgfqpoint{2.878250in}{2.840401in}}%
\pgfpathclose%
\pgfusepath{fill}%
\end{pgfscope}%
\begin{pgfscope}%
\pgfpathrectangle{\pgfqpoint{2.878250in}{0.317222in}}{\pgfqpoint{0.126159in}{2.523179in}}%
\pgfusepath{clip}%
\pgfsetbuttcap%
\pgfsetmiterjoin%
\definecolor{currentfill}{rgb}{1.000000,1.000000,1.000000}%
\pgfsetfillcolor{currentfill}%
\pgfsetlinewidth{0.010037pt}%
\definecolor{currentstroke}{rgb}{1.000000,1.000000,1.000000}%
\pgfsetstrokecolor{currentstroke}%
\pgfsetdash{}{0pt}%
\pgfpathmoveto{\pgfqpoint{2.878250in}{0.317222in}}%
\pgfpathlineto{\pgfqpoint{2.878250in}{0.327078in}}%
\pgfpathlineto{\pgfqpoint{2.878250in}{2.830545in}}%
\pgfpathlineto{\pgfqpoint{2.878250in}{2.840401in}}%
\pgfpathlineto{\pgfqpoint{3.004409in}{2.840401in}}%
\pgfpathlineto{\pgfqpoint{3.004409in}{2.830545in}}%
\pgfpathlineto{\pgfqpoint{3.004409in}{0.327078in}}%
\pgfpathlineto{\pgfqpoint{3.004409in}{0.317222in}}%
\pgfpathlineto{\pgfqpoint{3.004409in}{0.317222in}}%
\pgfpathclose%
\pgfusepath{stroke,fill}%
\end{pgfscope}%
\begin{pgfscope}%
\pgfsys@transformshift{2.880000in}{0.320000in}%
\pgftext[left,bottom]{\includegraphics[interpolate=true,width=0.120000in,height=2.520000in]{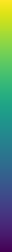}}%
\end{pgfscope}%
\begin{pgfscope}%
\pgfsetbuttcap%
\pgfsetroundjoin%
\definecolor{currentfill}{rgb}{0.000000,0.000000,0.000000}%
\pgfsetfillcolor{currentfill}%
\pgfsetlinewidth{0.803000pt}%
\definecolor{currentstroke}{rgb}{0.000000,0.000000,0.000000}%
\pgfsetstrokecolor{currentstroke}%
\pgfsetdash{}{0pt}%
\pgfsys@defobject{currentmarker}{\pgfqpoint{0.000000in}{0.000000in}}{\pgfqpoint{0.048611in}{0.000000in}}{%
\pgfpathmoveto{\pgfqpoint{0.000000in}{0.000000in}}%
\pgfpathlineto{\pgfqpoint{0.048611in}{0.000000in}}%
\pgfusepath{stroke,fill}%
}%
\begin{pgfscope}%
\pgfsys@transformshift{3.004409in}{0.317222in}%
\pgfsys@useobject{currentmarker}{}%
\end{pgfscope}%
\end{pgfscope}%
\begin{pgfscope}%
\definecolor{textcolor}{rgb}{0.000000,0.000000,0.000000}%
\pgfsetstrokecolor{textcolor}%
\pgfsetfillcolor{textcolor}%
\pgftext[x=3.101631in, y=0.278667in, left, base]{\color{textcolor}\sffamily\fontsize{8.000000}{9.600000}\selectfont 0}%
\end{pgfscope}%
\begin{pgfscope}%
\pgfsetbuttcap%
\pgfsetroundjoin%
\definecolor{currentfill}{rgb}{0.000000,0.000000,0.000000}%
\pgfsetfillcolor{currentfill}%
\pgfsetlinewidth{0.803000pt}%
\definecolor{currentstroke}{rgb}{0.000000,0.000000,0.000000}%
\pgfsetstrokecolor{currentstroke}%
\pgfsetdash{}{0pt}%
\pgfsys@defobject{currentmarker}{\pgfqpoint{0.000000in}{0.000000in}}{\pgfqpoint{0.048611in}{0.000000in}}{%
\pgfpathmoveto{\pgfqpoint{0.000000in}{0.000000in}}%
\pgfpathlineto{\pgfqpoint{0.048611in}{0.000000in}}%
\pgfusepath{stroke,fill}%
}%
\begin{pgfscope}%
\pgfsys@transformshift{3.004409in}{0.648349in}%
\pgfsys@useobject{currentmarker}{}%
\end{pgfscope}%
\end{pgfscope}%
\begin{pgfscope}%
\definecolor{textcolor}{rgb}{0.000000,0.000000,0.000000}%
\pgfsetstrokecolor{textcolor}%
\pgfsetfillcolor{textcolor}%
\pgftext[x=3.101631in, y=0.609794in, left, base]{\color{textcolor}\sffamily\fontsize{8.000000}{9.600000}\selectfont 1}%
\end{pgfscope}%
\begin{pgfscope}%
\pgfsetbuttcap%
\pgfsetroundjoin%
\definecolor{currentfill}{rgb}{0.000000,0.000000,0.000000}%
\pgfsetfillcolor{currentfill}%
\pgfsetlinewidth{0.803000pt}%
\definecolor{currentstroke}{rgb}{0.000000,0.000000,0.000000}%
\pgfsetstrokecolor{currentstroke}%
\pgfsetdash{}{0pt}%
\pgfsys@defobject{currentmarker}{\pgfqpoint{0.000000in}{0.000000in}}{\pgfqpoint{0.048611in}{0.000000in}}{%
\pgfpathmoveto{\pgfqpoint{0.000000in}{0.000000in}}%
\pgfpathlineto{\pgfqpoint{0.048611in}{0.000000in}}%
\pgfusepath{stroke,fill}%
}%
\begin{pgfscope}%
\pgfsys@transformshift{3.004409in}{0.979476in}%
\pgfsys@useobject{currentmarker}{}%
\end{pgfscope}%
\end{pgfscope}%
\begin{pgfscope}%
\definecolor{textcolor}{rgb}{0.000000,0.000000,0.000000}%
\pgfsetstrokecolor{textcolor}%
\pgfsetfillcolor{textcolor}%
\pgftext[x=3.101631in, y=0.940921in, left, base]{\color{textcolor}\sffamily\fontsize{8.000000}{9.600000}\selectfont 2}%
\end{pgfscope}%
\begin{pgfscope}%
\pgfsetbuttcap%
\pgfsetroundjoin%
\definecolor{currentfill}{rgb}{0.000000,0.000000,0.000000}%
\pgfsetfillcolor{currentfill}%
\pgfsetlinewidth{0.803000pt}%
\definecolor{currentstroke}{rgb}{0.000000,0.000000,0.000000}%
\pgfsetstrokecolor{currentstroke}%
\pgfsetdash{}{0pt}%
\pgfsys@defobject{currentmarker}{\pgfqpoint{0.000000in}{0.000000in}}{\pgfqpoint{0.048611in}{0.000000in}}{%
\pgfpathmoveto{\pgfqpoint{0.000000in}{0.000000in}}%
\pgfpathlineto{\pgfqpoint{0.048611in}{0.000000in}}%
\pgfusepath{stroke,fill}%
}%
\begin{pgfscope}%
\pgfsys@transformshift{3.004409in}{1.310603in}%
\pgfsys@useobject{currentmarker}{}%
\end{pgfscope}%
\end{pgfscope}%
\begin{pgfscope}%
\definecolor{textcolor}{rgb}{0.000000,0.000000,0.000000}%
\pgfsetstrokecolor{textcolor}%
\pgfsetfillcolor{textcolor}%
\pgftext[x=3.101631in, y=1.272048in, left, base]{\color{textcolor}\sffamily\fontsize{8.000000}{9.600000}\selectfont 3}%
\end{pgfscope}%
\begin{pgfscope}%
\pgfsetbuttcap%
\pgfsetroundjoin%
\definecolor{currentfill}{rgb}{0.000000,0.000000,0.000000}%
\pgfsetfillcolor{currentfill}%
\pgfsetlinewidth{0.803000pt}%
\definecolor{currentstroke}{rgb}{0.000000,0.000000,0.000000}%
\pgfsetstrokecolor{currentstroke}%
\pgfsetdash{}{0pt}%
\pgfsys@defobject{currentmarker}{\pgfqpoint{0.000000in}{0.000000in}}{\pgfqpoint{0.048611in}{0.000000in}}{%
\pgfpathmoveto{\pgfqpoint{0.000000in}{0.000000in}}%
\pgfpathlineto{\pgfqpoint{0.048611in}{0.000000in}}%
\pgfusepath{stroke,fill}%
}%
\begin{pgfscope}%
\pgfsys@transformshift{3.004409in}{1.641730in}%
\pgfsys@useobject{currentmarker}{}%
\end{pgfscope}%
\end{pgfscope}%
\begin{pgfscope}%
\definecolor{textcolor}{rgb}{0.000000,0.000000,0.000000}%
\pgfsetstrokecolor{textcolor}%
\pgfsetfillcolor{textcolor}%
\pgftext[x=3.101631in, y=1.603175in, left, base]{\color{textcolor}\sffamily\fontsize{8.000000}{9.600000}\selectfont 4}%
\end{pgfscope}%
\begin{pgfscope}%
\pgfsetbuttcap%
\pgfsetroundjoin%
\definecolor{currentfill}{rgb}{0.000000,0.000000,0.000000}%
\pgfsetfillcolor{currentfill}%
\pgfsetlinewidth{0.803000pt}%
\definecolor{currentstroke}{rgb}{0.000000,0.000000,0.000000}%
\pgfsetstrokecolor{currentstroke}%
\pgfsetdash{}{0pt}%
\pgfsys@defobject{currentmarker}{\pgfqpoint{0.000000in}{0.000000in}}{\pgfqpoint{0.048611in}{0.000000in}}{%
\pgfpathmoveto{\pgfqpoint{0.000000in}{0.000000in}}%
\pgfpathlineto{\pgfqpoint{0.048611in}{0.000000in}}%
\pgfusepath{stroke,fill}%
}%
\begin{pgfscope}%
\pgfsys@transformshift{3.004409in}{1.972857in}%
\pgfsys@useobject{currentmarker}{}%
\end{pgfscope}%
\end{pgfscope}%
\begin{pgfscope}%
\definecolor{textcolor}{rgb}{0.000000,0.000000,0.000000}%
\pgfsetstrokecolor{textcolor}%
\pgfsetfillcolor{textcolor}%
\pgftext[x=3.101631in, y=1.934302in, left, base]{\color{textcolor}\sffamily\fontsize{8.000000}{9.600000}\selectfont 5}%
\end{pgfscope}%
\begin{pgfscope}%
\pgfsetbuttcap%
\pgfsetroundjoin%
\definecolor{currentfill}{rgb}{0.000000,0.000000,0.000000}%
\pgfsetfillcolor{currentfill}%
\pgfsetlinewidth{0.803000pt}%
\definecolor{currentstroke}{rgb}{0.000000,0.000000,0.000000}%
\pgfsetstrokecolor{currentstroke}%
\pgfsetdash{}{0pt}%
\pgfsys@defobject{currentmarker}{\pgfqpoint{0.000000in}{0.000000in}}{\pgfqpoint{0.048611in}{0.000000in}}{%
\pgfpathmoveto{\pgfqpoint{0.000000in}{0.000000in}}%
\pgfpathlineto{\pgfqpoint{0.048611in}{0.000000in}}%
\pgfusepath{stroke,fill}%
}%
\begin{pgfscope}%
\pgfsys@transformshift{3.004409in}{2.303984in}%
\pgfsys@useobject{currentmarker}{}%
\end{pgfscope}%
\end{pgfscope}%
\begin{pgfscope}%
\definecolor{textcolor}{rgb}{0.000000,0.000000,0.000000}%
\pgfsetstrokecolor{textcolor}%
\pgfsetfillcolor{textcolor}%
\pgftext[x=3.101631in, y=2.265429in, left, base]{\color{textcolor}\sffamily\fontsize{8.000000}{9.600000}\selectfont 6}%
\end{pgfscope}%
\begin{pgfscope}%
\pgfsetbuttcap%
\pgfsetroundjoin%
\definecolor{currentfill}{rgb}{0.000000,0.000000,0.000000}%
\pgfsetfillcolor{currentfill}%
\pgfsetlinewidth{0.803000pt}%
\definecolor{currentstroke}{rgb}{0.000000,0.000000,0.000000}%
\pgfsetstrokecolor{currentstroke}%
\pgfsetdash{}{0pt}%
\pgfsys@defobject{currentmarker}{\pgfqpoint{0.000000in}{0.000000in}}{\pgfqpoint{0.048611in}{0.000000in}}{%
\pgfpathmoveto{\pgfqpoint{0.000000in}{0.000000in}}%
\pgfpathlineto{\pgfqpoint{0.048611in}{0.000000in}}%
\pgfusepath{stroke,fill}%
}%
\begin{pgfscope}%
\pgfsys@transformshift{3.004409in}{2.635111in}%
\pgfsys@useobject{currentmarker}{}%
\end{pgfscope}%
\end{pgfscope}%
\begin{pgfscope}%
\definecolor{textcolor}{rgb}{0.000000,0.000000,0.000000}%
\pgfsetstrokecolor{textcolor}%
\pgfsetfillcolor{textcolor}%
\pgftext[x=3.101631in, y=2.596556in, left, base]{\color{textcolor}\sffamily\fontsize{8.000000}{9.600000}\selectfont 7}%
\end{pgfscope}%
\begin{pgfscope}%
\definecolor{textcolor}{rgb}{0.000000,0.000000,0.000000}%
\pgfsetstrokecolor{textcolor}%
\pgfsetfillcolor{textcolor}%
\pgftext[x=3.216187in,y=1.578812in,,top,rotate=90.000000]{\color{textcolor}\sffamily\fontsize{8.000000}{9.600000}\selectfont path length (cm)}%
\end{pgfscope}%
\begin{pgfscope}%
\pgfsetrectcap%
\pgfsetmiterjoin%
\pgfsetlinewidth{0.803000pt}%
\definecolor{currentstroke}{rgb}{0.000000,0.000000,0.000000}%
\pgfsetstrokecolor{currentstroke}%
\pgfsetdash{}{0pt}%
\pgfpathmoveto{\pgfqpoint{2.878250in}{0.317222in}}%
\pgfpathlineto{\pgfqpoint{2.878250in}{0.327078in}}%
\pgfpathlineto{\pgfqpoint{2.878250in}{2.830545in}}%
\pgfpathlineto{\pgfqpoint{2.878250in}{2.840401in}}%
\pgfpathlineto{\pgfqpoint{3.004409in}{2.840401in}}%
\pgfpathlineto{\pgfqpoint{3.004409in}{2.830545in}}%
\pgfpathlineto{\pgfqpoint{3.004409in}{0.327078in}}%
\pgfpathlineto{\pgfqpoint{3.004409in}{0.317222in}}%
\pgfpathclose%
\pgfusepath{stroke}%
\end{pgfscope}%
\end{pgfpicture}%
\makeatother%
\endgroup%

%% file: figures/AlCu_d.pgf
\begingroup%
\makeatletter%
\begin{pgfpicture}%
\pgfpathrectangle{\pgfpointorigin}{\pgfqpoint{3.500000in}{3.000000in}}%
\pgfusepath{use as bounding box, clip}%
\begin{pgfscope}%
\pgfsetbuttcap%
\pgfsetmiterjoin%
\definecolor{currentfill}{rgb}{1.000000,1.000000,1.000000}%
\pgfsetfillcolor{currentfill}%
\pgfsetlinewidth{0.000000pt}%
\definecolor{currentstroke}{rgb}{1.000000,1.000000,1.000000}%
\pgfsetstrokecolor{currentstroke}%
\pgfsetdash{}{0pt}%
\pgfpathmoveto{\pgfqpoint{0.000000in}{0.000000in}}%
\pgfpathlineto{\pgfqpoint{3.500000in}{0.000000in}}%
\pgfpathlineto{\pgfqpoint{3.500000in}{3.000000in}}%
\pgfpathlineto{\pgfqpoint{0.000000in}{3.000000in}}%
\pgfpathclose%
\pgfusepath{fill}%
\end{pgfscope}%
\begin{pgfscope}%
\pgfsetbuttcap%
\pgfsetmiterjoin%
\definecolor{currentfill}{rgb}{1.000000,1.000000,1.000000}%
\pgfsetfillcolor{currentfill}%
\pgfsetlinewidth{0.000000pt}%
\definecolor{currentstroke}{rgb}{0.000000,0.000000,0.000000}%
\pgfsetstrokecolor{currentstroke}%
\pgfsetstrokeopacity{0.000000}%
\pgfsetdash{}{0pt}%
\pgfpathmoveto{\pgfqpoint{0.719689in}{0.330000in}}%
\pgfpathlineto{\pgfqpoint{2.607500in}{0.330000in}}%
\pgfpathlineto{\pgfqpoint{2.607500in}{2.640000in}}%
\pgfpathlineto{\pgfqpoint{0.719689in}{2.640000in}}%
\pgfpathclose%
\pgfusepath{fill}%
\end{pgfscope}%
\begin{pgfscope}%
\pgfsys@transformshift{0.720000in}{1.010000in}%
\pgftext[left,bottom]{\includegraphics[interpolate=true,width=1.890000in,height=1.630000in]{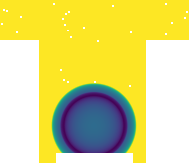}}%
\end{pgfscope}%
\begin{pgfscope}%
\pgfsetbuttcap%
\pgfsetroundjoin%
\definecolor{currentfill}{rgb}{0.000000,0.000000,0.000000}%
\pgfsetfillcolor{currentfill}%
\pgfsetlinewidth{0.803000pt}%
\definecolor{currentstroke}{rgb}{0.000000,0.000000,0.000000}%
\pgfsetstrokecolor{currentstroke}%
\pgfsetdash{}{0pt}%
\pgfsys@defobject{currentmarker}{\pgfqpoint{0.000000in}{-0.048611in}}{\pgfqpoint{0.000000in}{0.000000in}}{%
\pgfpathmoveto{\pgfqpoint{0.000000in}{0.000000in}}%
\pgfpathlineto{\pgfqpoint{0.000000in}{-0.048611in}}%
\pgfusepath{stroke,fill}%
}%
\begin{pgfscope}%
\pgfsys@transformshift{0.720090in}{0.330000in}%
\pgfsys@useobject{currentmarker}{}%
\end{pgfscope}%
\end{pgfscope}%
\begin{pgfscope}%
\definecolor{textcolor}{rgb}{0.000000,0.000000,0.000000}%
\pgfsetstrokecolor{textcolor}%
\pgfsetfillcolor{textcolor}%
\pgftext[x=0.720090in,y=0.232778in,,top]{\color{textcolor}\sffamily\fontsize{8.000000}{9.600000}\selectfont 0}%
\end{pgfscope}%
\begin{pgfscope}%
\pgfsetbuttcap%
\pgfsetroundjoin%
\definecolor{currentfill}{rgb}{0.000000,0.000000,0.000000}%
\pgfsetfillcolor{currentfill}%
\pgfsetlinewidth{0.803000pt}%
\definecolor{currentstroke}{rgb}{0.000000,0.000000,0.000000}%
\pgfsetstrokecolor{currentstroke}%
\pgfsetdash{}{0pt}%
\pgfsys@defobject{currentmarker}{\pgfqpoint{0.000000in}{-0.048611in}}{\pgfqpoint{0.000000in}{0.000000in}}{%
\pgfpathmoveto{\pgfqpoint{0.000000in}{0.000000in}}%
\pgfpathlineto{\pgfqpoint{0.000000in}{-0.048611in}}%
\pgfusepath{stroke,fill}%
}%
\begin{pgfscope}%
\pgfsys@transformshift{1.121411in}{0.330000in}%
\pgfsys@useobject{currentmarker}{}%
\end{pgfscope}%
\end{pgfscope}%
\begin{pgfscope}%
\definecolor{textcolor}{rgb}{0.000000,0.000000,0.000000}%
\pgfsetstrokecolor{textcolor}%
\pgfsetfillcolor{textcolor}%
\pgftext[x=1.121411in,y=0.232778in,,top]{\color{textcolor}\sffamily\fontsize{8.000000}{9.600000}\selectfont 500}%
\end{pgfscope}%
\begin{pgfscope}%
\pgfsetbuttcap%
\pgfsetroundjoin%
\definecolor{currentfill}{rgb}{0.000000,0.000000,0.000000}%
\pgfsetfillcolor{currentfill}%
\pgfsetlinewidth{0.803000pt}%
\definecolor{currentstroke}{rgb}{0.000000,0.000000,0.000000}%
\pgfsetstrokecolor{currentstroke}%
\pgfsetdash{}{0pt}%
\pgfsys@defobject{currentmarker}{\pgfqpoint{0.000000in}{-0.048611in}}{\pgfqpoint{0.000000in}{0.000000in}}{%
\pgfpathmoveto{\pgfqpoint{0.000000in}{0.000000in}}%
\pgfpathlineto{\pgfqpoint{0.000000in}{-0.048611in}}%
\pgfusepath{stroke,fill}%
}%
\begin{pgfscope}%
\pgfsys@transformshift{1.522731in}{0.330000in}%
\pgfsys@useobject{currentmarker}{}%
\end{pgfscope}%
\end{pgfscope}%
\begin{pgfscope}%
\definecolor{textcolor}{rgb}{0.000000,0.000000,0.000000}%
\pgfsetstrokecolor{textcolor}%
\pgfsetfillcolor{textcolor}%
\pgftext[x=1.522731in,y=0.232778in,,top]{\color{textcolor}\sffamily\fontsize{8.000000}{9.600000}\selectfont 1000}%
\end{pgfscope}%
\begin{pgfscope}%
\pgfsetbuttcap%
\pgfsetroundjoin%
\definecolor{currentfill}{rgb}{0.000000,0.000000,0.000000}%
\pgfsetfillcolor{currentfill}%
\pgfsetlinewidth{0.803000pt}%
\definecolor{currentstroke}{rgb}{0.000000,0.000000,0.000000}%
\pgfsetstrokecolor{currentstroke}%
\pgfsetdash{}{0pt}%
\pgfsys@defobject{currentmarker}{\pgfqpoint{0.000000in}{-0.048611in}}{\pgfqpoint{0.000000in}{0.000000in}}{%
\pgfpathmoveto{\pgfqpoint{0.000000in}{0.000000in}}%
\pgfpathlineto{\pgfqpoint{0.000000in}{-0.048611in}}%
\pgfusepath{stroke,fill}%
}%
\begin{pgfscope}%
\pgfsys@transformshift{1.924051in}{0.330000in}%
\pgfsys@useobject{currentmarker}{}%
\end{pgfscope}%
\end{pgfscope}%
\begin{pgfscope}%
\definecolor{textcolor}{rgb}{0.000000,0.000000,0.000000}%
\pgfsetstrokecolor{textcolor}%
\pgfsetfillcolor{textcolor}%
\pgftext[x=1.924051in,y=0.232778in,,top]{\color{textcolor}\sffamily\fontsize{8.000000}{9.600000}\selectfont 1500}%
\end{pgfscope}%
\begin{pgfscope}%
\pgfsetbuttcap%
\pgfsetroundjoin%
\definecolor{currentfill}{rgb}{0.000000,0.000000,0.000000}%
\pgfsetfillcolor{currentfill}%
\pgfsetlinewidth{0.803000pt}%
\definecolor{currentstroke}{rgb}{0.000000,0.000000,0.000000}%
\pgfsetstrokecolor{currentstroke}%
\pgfsetdash{}{0pt}%
\pgfsys@defobject{currentmarker}{\pgfqpoint{0.000000in}{-0.048611in}}{\pgfqpoint{0.000000in}{0.000000in}}{%
\pgfpathmoveto{\pgfqpoint{0.000000in}{0.000000in}}%
\pgfpathlineto{\pgfqpoint{0.000000in}{-0.048611in}}%
\pgfusepath{stroke,fill}%
}%
\begin{pgfscope}%
\pgfsys@transformshift{2.325372in}{0.330000in}%
\pgfsys@useobject{currentmarker}{}%
\end{pgfscope}%
\end{pgfscope}%
\begin{pgfscope}%
\definecolor{textcolor}{rgb}{0.000000,0.000000,0.000000}%
\pgfsetstrokecolor{textcolor}%
\pgfsetfillcolor{textcolor}%
\pgftext[x=2.325372in,y=0.232778in,,top]{\color{textcolor}\sffamily\fontsize{8.000000}{9.600000}\selectfont 2000}%
\end{pgfscope}%
\begin{pgfscope}%
\pgfsetbuttcap%
\pgfsetroundjoin%
\definecolor{currentfill}{rgb}{0.000000,0.000000,0.000000}%
\pgfsetfillcolor{currentfill}%
\pgfsetlinewidth{0.803000pt}%
\definecolor{currentstroke}{rgb}{0.000000,0.000000,0.000000}%
\pgfsetstrokecolor{currentstroke}%
\pgfsetdash{}{0pt}%
\pgfsys@defobject{currentmarker}{\pgfqpoint{-0.048611in}{0.000000in}}{\pgfqpoint{-0.000000in}{0.000000in}}{%
\pgfpathmoveto{\pgfqpoint{-0.000000in}{0.000000in}}%
\pgfpathlineto{\pgfqpoint{-0.048611in}{0.000000in}}%
\pgfusepath{stroke,fill}%
}%
\begin{pgfscope}%
\pgfsys@transformshift{0.719689in}{2.639599in}%
\pgfsys@useobject{currentmarker}{}%
\end{pgfscope}%
\end{pgfscope}%
\begin{pgfscope}%
\definecolor{textcolor}{rgb}{0.000000,0.000000,0.000000}%
\pgfsetstrokecolor{textcolor}%
\pgfsetfillcolor{textcolor}%
\pgftext[x=0.563467in, y=2.601043in, left, base]{\color{textcolor}\sffamily\fontsize{8.000000}{9.600000}\selectfont 0}%
\end{pgfscope}%
\begin{pgfscope}%
\pgfsetbuttcap%
\pgfsetroundjoin%
\definecolor{currentfill}{rgb}{0.000000,0.000000,0.000000}%
\pgfsetfillcolor{currentfill}%
\pgfsetlinewidth{0.803000pt}%
\definecolor{currentstroke}{rgb}{0.000000,0.000000,0.000000}%
\pgfsetstrokecolor{currentstroke}%
\pgfsetdash{}{0pt}%
\pgfsys@defobject{currentmarker}{\pgfqpoint{-0.048611in}{0.000000in}}{\pgfqpoint{-0.000000in}{0.000000in}}{%
\pgfpathmoveto{\pgfqpoint{-0.000000in}{0.000000in}}%
\pgfpathlineto{\pgfqpoint{-0.048611in}{0.000000in}}%
\pgfusepath{stroke,fill}%
}%
\begin{pgfscope}%
\pgfsys@transformshift{0.719689in}{2.238278in}%
\pgfsys@useobject{currentmarker}{}%
\end{pgfscope}%
\end{pgfscope}%
\begin{pgfscope}%
\definecolor{textcolor}{rgb}{0.000000,0.000000,0.000000}%
\pgfsetstrokecolor{textcolor}%
\pgfsetfillcolor{textcolor}%
\pgftext[x=0.445467in, y=2.199723in, left, base]{\color{textcolor}\sffamily\fontsize{8.000000}{9.600000}\selectfont 500}%
\end{pgfscope}%
\begin{pgfscope}%
\pgfsetbuttcap%
\pgfsetroundjoin%
\definecolor{currentfill}{rgb}{0.000000,0.000000,0.000000}%
\pgfsetfillcolor{currentfill}%
\pgfsetlinewidth{0.803000pt}%
\definecolor{currentstroke}{rgb}{0.000000,0.000000,0.000000}%
\pgfsetstrokecolor{currentstroke}%
\pgfsetdash{}{0pt}%
\pgfsys@defobject{currentmarker}{\pgfqpoint{-0.048611in}{0.000000in}}{\pgfqpoint{-0.000000in}{0.000000in}}{%
\pgfpathmoveto{\pgfqpoint{-0.000000in}{0.000000in}}%
\pgfpathlineto{\pgfqpoint{-0.048611in}{0.000000in}}%
\pgfusepath{stroke,fill}%
}%
\begin{pgfscope}%
\pgfsys@transformshift{0.719689in}{1.836958in}%
\pgfsys@useobject{currentmarker}{}%
\end{pgfscope}%
\end{pgfscope}%
\begin{pgfscope}%
\definecolor{textcolor}{rgb}{0.000000,0.000000,0.000000}%
\pgfsetstrokecolor{textcolor}%
\pgfsetfillcolor{textcolor}%
\pgftext[x=0.386467in, y=1.798402in, left, base]{\color{textcolor}\sffamily\fontsize{8.000000}{9.600000}\selectfont 1000}%
\end{pgfscope}%
\begin{pgfscope}%
\pgfsetbuttcap%
\pgfsetroundjoin%
\definecolor{currentfill}{rgb}{0.000000,0.000000,0.000000}%
\pgfsetfillcolor{currentfill}%
\pgfsetlinewidth{0.803000pt}%
\definecolor{currentstroke}{rgb}{0.000000,0.000000,0.000000}%
\pgfsetstrokecolor{currentstroke}%
\pgfsetdash{}{0pt}%
\pgfsys@defobject{currentmarker}{\pgfqpoint{-0.048611in}{0.000000in}}{\pgfqpoint{-0.000000in}{0.000000in}}{%
\pgfpathmoveto{\pgfqpoint{-0.000000in}{0.000000in}}%
\pgfpathlineto{\pgfqpoint{-0.048611in}{0.000000in}}%
\pgfusepath{stroke,fill}%
}%
\begin{pgfscope}%
\pgfsys@transformshift{0.719689in}{1.435638in}%
\pgfsys@useobject{currentmarker}{}%
\end{pgfscope}%
\end{pgfscope}%
\begin{pgfscope}%
\definecolor{textcolor}{rgb}{0.000000,0.000000,0.000000}%
\pgfsetstrokecolor{textcolor}%
\pgfsetfillcolor{textcolor}%
\pgftext[x=0.386467in, y=1.397082in, left, base]{\color{textcolor}\sffamily\fontsize{8.000000}{9.600000}\selectfont 1500}%
\end{pgfscope}%
\begin{pgfscope}%
\pgfsetbuttcap%
\pgfsetroundjoin%
\definecolor{currentfill}{rgb}{0.000000,0.000000,0.000000}%
\pgfsetfillcolor{currentfill}%
\pgfsetlinewidth{0.803000pt}%
\definecolor{currentstroke}{rgb}{0.000000,0.000000,0.000000}%
\pgfsetstrokecolor{currentstroke}%
\pgfsetdash{}{0pt}%
\pgfsys@defobject{currentmarker}{\pgfqpoint{-0.048611in}{0.000000in}}{\pgfqpoint{-0.000000in}{0.000000in}}{%
\pgfpathmoveto{\pgfqpoint{-0.000000in}{0.000000in}}%
\pgfpathlineto{\pgfqpoint{-0.048611in}{0.000000in}}%
\pgfusepath{stroke,fill}%
}%
\begin{pgfscope}%
\pgfsys@transformshift{0.719689in}{1.034317in}%
\pgfsys@useobject{currentmarker}{}%
\end{pgfscope}%
\end{pgfscope}%
\begin{pgfscope}%
\definecolor{textcolor}{rgb}{0.000000,0.000000,0.000000}%
\pgfsetstrokecolor{textcolor}%
\pgfsetfillcolor{textcolor}%
\pgftext[x=0.386467in, y=0.995762in, left, base]{\color{textcolor}\sffamily\fontsize{8.000000}{9.600000}\selectfont 2000}%
\end{pgfscope}%
\begin{pgfscope}%
\pgfsetbuttcap%
\pgfsetroundjoin%
\definecolor{currentfill}{rgb}{0.000000,0.000000,0.000000}%
\pgfsetfillcolor{currentfill}%
\pgfsetlinewidth{0.803000pt}%
\definecolor{currentstroke}{rgb}{0.000000,0.000000,0.000000}%
\pgfsetstrokecolor{currentstroke}%
\pgfsetdash{}{0pt}%
\pgfsys@defobject{currentmarker}{\pgfqpoint{-0.048611in}{0.000000in}}{\pgfqpoint{-0.000000in}{0.000000in}}{%
\pgfpathmoveto{\pgfqpoint{-0.000000in}{0.000000in}}%
\pgfpathlineto{\pgfqpoint{-0.048611in}{0.000000in}}%
\pgfusepath{stroke,fill}%
}%
\begin{pgfscope}%
\pgfsys@transformshift{0.719689in}{0.632997in}%
\pgfsys@useobject{currentmarker}{}%
\end{pgfscope}%
\end{pgfscope}%
\begin{pgfscope}%
\definecolor{textcolor}{rgb}{0.000000,0.000000,0.000000}%
\pgfsetstrokecolor{textcolor}%
\pgfsetfillcolor{textcolor}%
\pgftext[x=0.386467in, y=0.594441in, left, base]{\color{textcolor}\sffamily\fontsize{8.000000}{9.600000}\selectfont 2500}%
\end{pgfscope}%
\begin{pgfscope}%
\pgfsetrectcap%
\pgfsetmiterjoin%
\pgfsetlinewidth{0.803000pt}%
\definecolor{currentstroke}{rgb}{0.000000,0.000000,0.000000}%
\pgfsetstrokecolor{currentstroke}%
\pgfsetdash{}{0pt}%
\pgfpathmoveto{\pgfqpoint{0.719689in}{0.330000in}}%
\pgfpathlineto{\pgfqpoint{0.719689in}{2.640000in}}%
\pgfusepath{stroke}%
\end{pgfscope}%
\begin{pgfscope}%
\pgfsetrectcap%
\pgfsetmiterjoin%
\pgfsetlinewidth{0.803000pt}%
\definecolor{currentstroke}{rgb}{0.000000,0.000000,0.000000}%
\pgfsetstrokecolor{currentstroke}%
\pgfsetdash{}{0pt}%
\pgfpathmoveto{\pgfqpoint{2.607500in}{0.330000in}}%
\pgfpathlineto{\pgfqpoint{2.607500in}{2.640000in}}%
\pgfusepath{stroke}%
\end{pgfscope}%
\begin{pgfscope}%
\pgfsetrectcap%
\pgfsetmiterjoin%
\pgfsetlinewidth{0.803000pt}%
\definecolor{currentstroke}{rgb}{0.000000,0.000000,0.000000}%
\pgfsetstrokecolor{currentstroke}%
\pgfsetdash{}{0pt}%
\pgfpathmoveto{\pgfqpoint{0.719689in}{0.330000in}}%
\pgfpathlineto{\pgfqpoint{2.607500in}{0.330000in}}%
\pgfusepath{stroke}%
\end{pgfscope}%
\begin{pgfscope}%
\pgfsetrectcap%
\pgfsetmiterjoin%
\pgfsetlinewidth{0.803000pt}%
\definecolor{currentstroke}{rgb}{0.000000,0.000000,0.000000}%
\pgfsetstrokecolor{currentstroke}%
\pgfsetdash{}{0pt}%
\pgfpathmoveto{\pgfqpoint{0.719689in}{2.640000in}}%
\pgfpathlineto{\pgfqpoint{2.607500in}{2.640000in}}%
\pgfusepath{stroke}%
\end{pgfscope}%
\begin{pgfscope}%
\pgfsetbuttcap%
\pgfsetmiterjoin%
\definecolor{currentfill}{rgb}{1.000000,1.000000,1.000000}%
\pgfsetfillcolor{currentfill}%
\pgfsetlinewidth{0.000000pt}%
\definecolor{currentstroke}{rgb}{0.000000,0.000000,0.000000}%
\pgfsetstrokecolor{currentstroke}%
\pgfsetstrokeopacity{0.000000}%
\pgfsetdash{}{0pt}%
\pgfpathmoveto{\pgfqpoint{2.743125in}{0.330000in}}%
\pgfpathlineto{\pgfqpoint{2.858625in}{0.330000in}}%
\pgfpathlineto{\pgfqpoint{2.858625in}{2.640000in}}%
\pgfpathlineto{\pgfqpoint{2.743125in}{2.640000in}}%
\pgfpathclose%
\pgfusepath{fill}%
\end{pgfscope}%
\begin{pgfscope}%
\pgfpathrectangle{\pgfqpoint{2.743125in}{0.330000in}}{\pgfqpoint{0.115500in}{2.310000in}}%
\pgfusepath{clip}%
\pgfsetbuttcap%
\pgfsetmiterjoin%
\definecolor{currentfill}{rgb}{1.000000,1.000000,1.000000}%
\pgfsetfillcolor{currentfill}%
\pgfsetlinewidth{0.010037pt}%
\definecolor{currentstroke}{rgb}{1.000000,1.000000,1.000000}%
\pgfsetstrokecolor{currentstroke}%
\pgfsetdash{}{0pt}%
\pgfpathmoveto{\pgfqpoint{2.743125in}{0.330000in}}%
\pgfpathlineto{\pgfqpoint{2.743125in}{0.339023in}}%
\pgfpathlineto{\pgfqpoint{2.743125in}{2.630977in}}%
\pgfpathlineto{\pgfqpoint{2.743125in}{2.640000in}}%
\pgfpathlineto{\pgfqpoint{2.858625in}{2.640000in}}%
\pgfpathlineto{\pgfqpoint{2.858625in}{2.630977in}}%
\pgfpathlineto{\pgfqpoint{2.858625in}{0.339023in}}%
\pgfpathlineto{\pgfqpoint{2.858625in}{0.330000in}}%
\pgfpathlineto{\pgfqpoint{2.858625in}{0.330000in}}%
\pgfpathclose%
\pgfusepath{stroke,fill}%
\end{pgfscope}%
\begin{pgfscope}%
\pgfsys@transformshift{2.740000in}{0.330000in}%
\pgftext[left,bottom]{\includegraphics[interpolate=true,width=0.120000in,height=2.310000in]{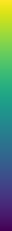}}%
\end{pgfscope}%
\begin{pgfscope}%
\pgfsetbuttcap%
\pgfsetroundjoin%
\definecolor{currentfill}{rgb}{0.000000,0.000000,0.000000}%
\pgfsetfillcolor{currentfill}%
\pgfsetlinewidth{0.803000pt}%
\definecolor{currentstroke}{rgb}{0.000000,0.000000,0.000000}%
\pgfsetstrokecolor{currentstroke}%
\pgfsetdash{}{0pt}%
\pgfsys@defobject{currentmarker}{\pgfqpoint{0.000000in}{0.000000in}}{\pgfqpoint{0.048611in}{0.000000in}}{%
\pgfpathmoveto{\pgfqpoint{0.000000in}{0.000000in}}%
\pgfpathlineto{\pgfqpoint{0.048611in}{0.000000in}}%
\pgfusepath{stroke,fill}%
}%
\begin{pgfscope}%
\pgfsys@transformshift{2.858625in}{0.458078in}%
\pgfsys@useobject{currentmarker}{}%
\end{pgfscope}%
\end{pgfscope}%
\begin{pgfscope}%
\definecolor{textcolor}{rgb}{0.000000,0.000000,0.000000}%
\pgfsetstrokecolor{textcolor}%
\pgfsetfillcolor{textcolor}%
\pgftext[x=2.955847in, y=0.419523in, left, base]{\color{textcolor}\sffamily\fontsize{8.000000}{9.600000}\selectfont 0.1}%
\end{pgfscope}%
\begin{pgfscope}%
\pgfsetbuttcap%
\pgfsetroundjoin%
\definecolor{currentfill}{rgb}{0.000000,0.000000,0.000000}%
\pgfsetfillcolor{currentfill}%
\pgfsetlinewidth{0.803000pt}%
\definecolor{currentstroke}{rgb}{0.000000,0.000000,0.000000}%
\pgfsetstrokecolor{currentstroke}%
\pgfsetdash{}{0pt}%
\pgfsys@defobject{currentmarker}{\pgfqpoint{0.000000in}{0.000000in}}{\pgfqpoint{0.048611in}{0.000000in}}{%
\pgfpathmoveto{\pgfqpoint{0.000000in}{0.000000in}}%
\pgfpathlineto{\pgfqpoint{0.048611in}{0.000000in}}%
\pgfusepath{stroke,fill}%
}%
\begin{pgfscope}%
\pgfsys@transformshift{2.858625in}{0.721532in}%
\pgfsys@useobject{currentmarker}{}%
\end{pgfscope}%
\end{pgfscope}%
\begin{pgfscope}%
\definecolor{textcolor}{rgb}{0.000000,0.000000,0.000000}%
\pgfsetstrokecolor{textcolor}%
\pgfsetfillcolor{textcolor}%
\pgftext[x=2.955847in, y=0.682977in, left, base]{\color{textcolor}\sffamily\fontsize{8.000000}{9.600000}\selectfont 0.2}%
\end{pgfscope}%
\begin{pgfscope}%
\pgfsetbuttcap%
\pgfsetroundjoin%
\definecolor{currentfill}{rgb}{0.000000,0.000000,0.000000}%
\pgfsetfillcolor{currentfill}%
\pgfsetlinewidth{0.803000pt}%
\definecolor{currentstroke}{rgb}{0.000000,0.000000,0.000000}%
\pgfsetstrokecolor{currentstroke}%
\pgfsetdash{}{0pt}%
\pgfsys@defobject{currentmarker}{\pgfqpoint{0.000000in}{0.000000in}}{\pgfqpoint{0.048611in}{0.000000in}}{%
\pgfpathmoveto{\pgfqpoint{0.000000in}{0.000000in}}%
\pgfpathlineto{\pgfqpoint{0.048611in}{0.000000in}}%
\pgfusepath{stroke,fill}%
}%
\begin{pgfscope}%
\pgfsys@transformshift{2.858625in}{0.984986in}%
\pgfsys@useobject{currentmarker}{}%
\end{pgfscope}%
\end{pgfscope}%
\begin{pgfscope}%
\definecolor{textcolor}{rgb}{0.000000,0.000000,0.000000}%
\pgfsetstrokecolor{textcolor}%
\pgfsetfillcolor{textcolor}%
\pgftext[x=2.955847in, y=0.946431in, left, base]{\color{textcolor}\sffamily\fontsize{8.000000}{9.600000}\selectfont 0.3}%
\end{pgfscope}%
\begin{pgfscope}%
\pgfsetbuttcap%
\pgfsetroundjoin%
\definecolor{currentfill}{rgb}{0.000000,0.000000,0.000000}%
\pgfsetfillcolor{currentfill}%
\pgfsetlinewidth{0.803000pt}%
\definecolor{currentstroke}{rgb}{0.000000,0.000000,0.000000}%
\pgfsetstrokecolor{currentstroke}%
\pgfsetdash{}{0pt}%
\pgfsys@defobject{currentmarker}{\pgfqpoint{0.000000in}{0.000000in}}{\pgfqpoint{0.048611in}{0.000000in}}{%
\pgfpathmoveto{\pgfqpoint{0.000000in}{0.000000in}}%
\pgfpathlineto{\pgfqpoint{0.048611in}{0.000000in}}%
\pgfusepath{stroke,fill}%
}%
\begin{pgfscope}%
\pgfsys@transformshift{2.858625in}{1.248440in}%
\pgfsys@useobject{currentmarker}{}%
\end{pgfscope}%
\end{pgfscope}%
\begin{pgfscope}%
\definecolor{textcolor}{rgb}{0.000000,0.000000,0.000000}%
\pgfsetstrokecolor{textcolor}%
\pgfsetfillcolor{textcolor}%
\pgftext[x=2.955847in, y=1.209885in, left, base]{\color{textcolor}\sffamily\fontsize{8.000000}{9.600000}\selectfont 0.4}%
\end{pgfscope}%
\begin{pgfscope}%
\pgfsetbuttcap%
\pgfsetroundjoin%
\definecolor{currentfill}{rgb}{0.000000,0.000000,0.000000}%
\pgfsetfillcolor{currentfill}%
\pgfsetlinewidth{0.803000pt}%
\definecolor{currentstroke}{rgb}{0.000000,0.000000,0.000000}%
\pgfsetstrokecolor{currentstroke}%
\pgfsetdash{}{0pt}%
\pgfsys@defobject{currentmarker}{\pgfqpoint{0.000000in}{0.000000in}}{\pgfqpoint{0.048611in}{0.000000in}}{%
\pgfpathmoveto{\pgfqpoint{0.000000in}{0.000000in}}%
\pgfpathlineto{\pgfqpoint{0.048611in}{0.000000in}}%
\pgfusepath{stroke,fill}%
}%
\begin{pgfscope}%
\pgfsys@transformshift{2.858625in}{1.511894in}%
\pgfsys@useobject{currentmarker}{}%
\end{pgfscope}%
\end{pgfscope}%
\begin{pgfscope}%
\definecolor{textcolor}{rgb}{0.000000,0.000000,0.000000}%
\pgfsetstrokecolor{textcolor}%
\pgfsetfillcolor{textcolor}%
\pgftext[x=2.955847in, y=1.473339in, left, base]{\color{textcolor}\sffamily\fontsize{8.000000}{9.600000}\selectfont 0.5}%
\end{pgfscope}%
\begin{pgfscope}%
\pgfsetbuttcap%
\pgfsetroundjoin%
\definecolor{currentfill}{rgb}{0.000000,0.000000,0.000000}%
\pgfsetfillcolor{currentfill}%
\pgfsetlinewidth{0.803000pt}%
\definecolor{currentstroke}{rgb}{0.000000,0.000000,0.000000}%
\pgfsetstrokecolor{currentstroke}%
\pgfsetdash{}{0pt}%
\pgfsys@defobject{currentmarker}{\pgfqpoint{0.000000in}{0.000000in}}{\pgfqpoint{0.048611in}{0.000000in}}{%
\pgfpathmoveto{\pgfqpoint{0.000000in}{0.000000in}}%
\pgfpathlineto{\pgfqpoint{0.048611in}{0.000000in}}%
\pgfusepath{stroke,fill}%
}%
\begin{pgfscope}%
\pgfsys@transformshift{2.858625in}{1.775349in}%
\pgfsys@useobject{currentmarker}{}%
\end{pgfscope}%
\end{pgfscope}%
\begin{pgfscope}%
\definecolor{textcolor}{rgb}{0.000000,0.000000,0.000000}%
\pgfsetstrokecolor{textcolor}%
\pgfsetfillcolor{textcolor}%
\pgftext[x=2.955847in, y=1.736793in, left, base]{\color{textcolor}\sffamily\fontsize{8.000000}{9.600000}\selectfont 0.6}%
\end{pgfscope}%
\begin{pgfscope}%
\pgfsetbuttcap%
\pgfsetroundjoin%
\definecolor{currentfill}{rgb}{0.000000,0.000000,0.000000}%
\pgfsetfillcolor{currentfill}%
\pgfsetlinewidth{0.803000pt}%
\definecolor{currentstroke}{rgb}{0.000000,0.000000,0.000000}%
\pgfsetstrokecolor{currentstroke}%
\pgfsetdash{}{0pt}%
\pgfsys@defobject{currentmarker}{\pgfqpoint{0.000000in}{0.000000in}}{\pgfqpoint{0.048611in}{0.000000in}}{%
\pgfpathmoveto{\pgfqpoint{0.000000in}{0.000000in}}%
\pgfpathlineto{\pgfqpoint{0.048611in}{0.000000in}}%
\pgfusepath{stroke,fill}%
}%
\begin{pgfscope}%
\pgfsys@transformshift{2.858625in}{2.038803in}%
\pgfsys@useobject{currentmarker}{}%
\end{pgfscope}%
\end{pgfscope}%
\begin{pgfscope}%
\definecolor{textcolor}{rgb}{0.000000,0.000000,0.000000}%
\pgfsetstrokecolor{textcolor}%
\pgfsetfillcolor{textcolor}%
\pgftext[x=2.955847in, y=2.000247in, left, base]{\color{textcolor}\sffamily\fontsize{8.000000}{9.600000}\selectfont 0.7}%
\end{pgfscope}%
\begin{pgfscope}%
\pgfsetbuttcap%
\pgfsetroundjoin%
\definecolor{currentfill}{rgb}{0.000000,0.000000,0.000000}%
\pgfsetfillcolor{currentfill}%
\pgfsetlinewidth{0.803000pt}%
\definecolor{currentstroke}{rgb}{0.000000,0.000000,0.000000}%
\pgfsetstrokecolor{currentstroke}%
\pgfsetdash{}{0pt}%
\pgfsys@defobject{currentmarker}{\pgfqpoint{0.000000in}{0.000000in}}{\pgfqpoint{0.048611in}{0.000000in}}{%
\pgfpathmoveto{\pgfqpoint{0.000000in}{0.000000in}}%
\pgfpathlineto{\pgfqpoint{0.048611in}{0.000000in}}%
\pgfusepath{stroke,fill}%
}%
\begin{pgfscope}%
\pgfsys@transformshift{2.858625in}{2.302257in}%
\pgfsys@useobject{currentmarker}{}%
\end{pgfscope}%
\end{pgfscope}%
\begin{pgfscope}%
\definecolor{textcolor}{rgb}{0.000000,0.000000,0.000000}%
\pgfsetstrokecolor{textcolor}%
\pgfsetfillcolor{textcolor}%
\pgftext[x=2.955847in, y=2.263701in, left, base]{\color{textcolor}\sffamily\fontsize{8.000000}{9.600000}\selectfont 0.8}%
\end{pgfscope}%
\begin{pgfscope}%
\pgfsetbuttcap%
\pgfsetroundjoin%
\definecolor{currentfill}{rgb}{0.000000,0.000000,0.000000}%
\pgfsetfillcolor{currentfill}%
\pgfsetlinewidth{0.803000pt}%
\definecolor{currentstroke}{rgb}{0.000000,0.000000,0.000000}%
\pgfsetstrokecolor{currentstroke}%
\pgfsetdash{}{0pt}%
\pgfsys@defobject{currentmarker}{\pgfqpoint{0.000000in}{0.000000in}}{\pgfqpoint{0.048611in}{0.000000in}}{%
\pgfpathmoveto{\pgfqpoint{0.000000in}{0.000000in}}%
\pgfpathlineto{\pgfqpoint{0.048611in}{0.000000in}}%
\pgfusepath{stroke,fill}%
}%
\begin{pgfscope}%
\pgfsys@transformshift{2.858625in}{2.565711in}%
\pgfsys@useobject{currentmarker}{}%
\end{pgfscope}%
\end{pgfscope}%
\begin{pgfscope}%
\definecolor{textcolor}{rgb}{0.000000,0.000000,0.000000}%
\pgfsetstrokecolor{textcolor}%
\pgfsetfillcolor{textcolor}%
\pgftext[x=2.955847in, y=2.527155in, left, base]{\color{textcolor}\sffamily\fontsize{8.000000}{9.600000}\selectfont 0.9}%
\end{pgfscope}%
\begin{pgfscope}%
\pgfsetrectcap%
\pgfsetmiterjoin%
\pgfsetlinewidth{0.803000pt}%
\definecolor{currentstroke}{rgb}{0.000000,0.000000,0.000000}%
\pgfsetstrokecolor{currentstroke}%
\pgfsetdash{}{0pt}%
\pgfpathmoveto{\pgfqpoint{2.743125in}{0.330000in}}%
\pgfpathlineto{\pgfqpoint{2.743125in}{0.339023in}}%
\pgfpathlineto{\pgfqpoint{2.743125in}{2.630977in}}%
\pgfpathlineto{\pgfqpoint{2.743125in}{2.640000in}}%
\pgfpathlineto{\pgfqpoint{2.858625in}{2.640000in}}%
\pgfpathlineto{\pgfqpoint{2.858625in}{2.630977in}}%
\pgfpathlineto{\pgfqpoint{2.858625in}{0.339023in}}%
\pgfpathlineto{\pgfqpoint{2.858625in}{0.330000in}}%
\pgfpathclose%
\pgfusepath{stroke}%
\end{pgfscope}%
\end{pgfpicture}%
\makeatother%
\endgroup%

%% file: figures/AlCu_s.pgf
\begingroup%
\makeatletter%
\begin{pgfpicture}%
\pgfpathrectangle{\pgfpointorigin}{\pgfqpoint{3.500000in}{3.000000in}}%
\pgfusepath{use as bounding box, clip}%
\begin{pgfscope}%
\pgfsetbuttcap%
\pgfsetmiterjoin%
\definecolor{currentfill}{rgb}{1.000000,1.000000,1.000000}%
\pgfsetfillcolor{currentfill}%
\pgfsetlinewidth{0.000000pt}%
\definecolor{currentstroke}{rgb}{1.000000,1.000000,1.000000}%
\pgfsetstrokecolor{currentstroke}%
\pgfsetdash{}{0pt}%
\pgfpathmoveto{\pgfqpoint{0.000000in}{0.000000in}}%
\pgfpathlineto{\pgfqpoint{3.500000in}{0.000000in}}%
\pgfpathlineto{\pgfqpoint{3.500000in}{3.000000in}}%
\pgfpathlineto{\pgfqpoint{0.000000in}{3.000000in}}%
\pgfpathclose%
\pgfusepath{fill}%
\end{pgfscope}%
\begin{pgfscope}%
\pgfsetbuttcap%
\pgfsetmiterjoin%
\definecolor{currentfill}{rgb}{1.000000,1.000000,1.000000}%
\pgfsetfillcolor{currentfill}%
\pgfsetlinewidth{0.000000pt}%
\definecolor{currentstroke}{rgb}{0.000000,0.000000,0.000000}%
\pgfsetstrokecolor{currentstroke}%
\pgfsetstrokeopacity{0.000000}%
\pgfsetdash{}{0pt}%
\pgfpathmoveto{\pgfqpoint{0.719689in}{0.330000in}}%
\pgfpathlineto{\pgfqpoint{2.607500in}{0.330000in}}%
\pgfpathlineto{\pgfqpoint{2.607500in}{2.640000in}}%
\pgfpathlineto{\pgfqpoint{0.719689in}{2.640000in}}%
\pgfpathclose%
\pgfusepath{fill}%
\end{pgfscope}%
\begin{pgfscope}%
\pgfsys@transformshift{0.720000in}{1.010000in}%
\pgftext[left,bottom]{\includegraphics[interpolate=true,width=1.890000in,height=1.630000in]{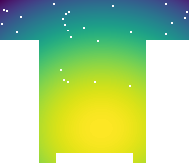}}%
\end{pgfscope}%
\begin{pgfscope}%
\pgfsetbuttcap%
\pgfsetroundjoin%
\definecolor{currentfill}{rgb}{0.000000,0.000000,0.000000}%
\pgfsetfillcolor{currentfill}%
\pgfsetlinewidth{0.803000pt}%
\definecolor{currentstroke}{rgb}{0.000000,0.000000,0.000000}%
\pgfsetstrokecolor{currentstroke}%
\pgfsetdash{}{0pt}%
\pgfsys@defobject{currentmarker}{\pgfqpoint{0.000000in}{-0.048611in}}{\pgfqpoint{0.000000in}{0.000000in}}{%
\pgfpathmoveto{\pgfqpoint{0.000000in}{0.000000in}}%
\pgfpathlineto{\pgfqpoint{0.000000in}{-0.048611in}}%
\pgfusepath{stroke,fill}%
}%
\begin{pgfscope}%
\pgfsys@transformshift{0.720090in}{0.330000in}%
\pgfsys@useobject{currentmarker}{}%
\end{pgfscope}%
\end{pgfscope}%
\begin{pgfscope}%
\definecolor{textcolor}{rgb}{0.000000,0.000000,0.000000}%
\pgfsetstrokecolor{textcolor}%
\pgfsetfillcolor{textcolor}%
\pgftext[x=0.720090in,y=0.232778in,,top]{\color{textcolor}\sffamily\fontsize{8.000000}{9.600000}\selectfont 0}%
\end{pgfscope}%
\begin{pgfscope}%
\pgfsetbuttcap%
\pgfsetroundjoin%
\definecolor{currentfill}{rgb}{0.000000,0.000000,0.000000}%
\pgfsetfillcolor{currentfill}%
\pgfsetlinewidth{0.803000pt}%
\definecolor{currentstroke}{rgb}{0.000000,0.000000,0.000000}%
\pgfsetstrokecolor{currentstroke}%
\pgfsetdash{}{0pt}%
\pgfsys@defobject{currentmarker}{\pgfqpoint{0.000000in}{-0.048611in}}{\pgfqpoint{0.000000in}{0.000000in}}{%
\pgfpathmoveto{\pgfqpoint{0.000000in}{0.000000in}}%
\pgfpathlineto{\pgfqpoint{0.000000in}{-0.048611in}}%
\pgfusepath{stroke,fill}%
}%
\begin{pgfscope}%
\pgfsys@transformshift{1.121411in}{0.330000in}%
\pgfsys@useobject{currentmarker}{}%
\end{pgfscope}%
\end{pgfscope}%
\begin{pgfscope}%
\definecolor{textcolor}{rgb}{0.000000,0.000000,0.000000}%
\pgfsetstrokecolor{textcolor}%
\pgfsetfillcolor{textcolor}%
\pgftext[x=1.121411in,y=0.232778in,,top]{\color{textcolor}\sffamily\fontsize{8.000000}{9.600000}\selectfont 500}%
\end{pgfscope}%
\begin{pgfscope}%
\pgfsetbuttcap%
\pgfsetroundjoin%
\definecolor{currentfill}{rgb}{0.000000,0.000000,0.000000}%
\pgfsetfillcolor{currentfill}%
\pgfsetlinewidth{0.803000pt}%
\definecolor{currentstroke}{rgb}{0.000000,0.000000,0.000000}%
\pgfsetstrokecolor{currentstroke}%
\pgfsetdash{}{0pt}%
\pgfsys@defobject{currentmarker}{\pgfqpoint{0.000000in}{-0.048611in}}{\pgfqpoint{0.000000in}{0.000000in}}{%
\pgfpathmoveto{\pgfqpoint{0.000000in}{0.000000in}}%
\pgfpathlineto{\pgfqpoint{0.000000in}{-0.048611in}}%
\pgfusepath{stroke,fill}%
}%
\begin{pgfscope}%
\pgfsys@transformshift{1.522731in}{0.330000in}%
\pgfsys@useobject{currentmarker}{}%
\end{pgfscope}%
\end{pgfscope}%
\begin{pgfscope}%
\definecolor{textcolor}{rgb}{0.000000,0.000000,0.000000}%
\pgfsetstrokecolor{textcolor}%
\pgfsetfillcolor{textcolor}%
\pgftext[x=1.522731in,y=0.232778in,,top]{\color{textcolor}\sffamily\fontsize{8.000000}{9.600000}\selectfont 1000}%
\end{pgfscope}%
\begin{pgfscope}%
\pgfsetbuttcap%
\pgfsetroundjoin%
\definecolor{currentfill}{rgb}{0.000000,0.000000,0.000000}%
\pgfsetfillcolor{currentfill}%
\pgfsetlinewidth{0.803000pt}%
\definecolor{currentstroke}{rgb}{0.000000,0.000000,0.000000}%
\pgfsetstrokecolor{currentstroke}%
\pgfsetdash{}{0pt}%
\pgfsys@defobject{currentmarker}{\pgfqpoint{0.000000in}{-0.048611in}}{\pgfqpoint{0.000000in}{0.000000in}}{%
\pgfpathmoveto{\pgfqpoint{0.000000in}{0.000000in}}%
\pgfpathlineto{\pgfqpoint{0.000000in}{-0.048611in}}%
\pgfusepath{stroke,fill}%
}%
\begin{pgfscope}%
\pgfsys@transformshift{1.924051in}{0.330000in}%
\pgfsys@useobject{currentmarker}{}%
\end{pgfscope}%
\end{pgfscope}%
\begin{pgfscope}%
\definecolor{textcolor}{rgb}{0.000000,0.000000,0.000000}%
\pgfsetstrokecolor{textcolor}%
\pgfsetfillcolor{textcolor}%
\pgftext[x=1.924051in,y=0.232778in,,top]{\color{textcolor}\sffamily\fontsize{8.000000}{9.600000}\selectfont 1500}%
\end{pgfscope}%
\begin{pgfscope}%
\pgfsetbuttcap%
\pgfsetroundjoin%
\definecolor{currentfill}{rgb}{0.000000,0.000000,0.000000}%
\pgfsetfillcolor{currentfill}%
\pgfsetlinewidth{0.803000pt}%
\definecolor{currentstroke}{rgb}{0.000000,0.000000,0.000000}%
\pgfsetstrokecolor{currentstroke}%
\pgfsetdash{}{0pt}%
\pgfsys@defobject{currentmarker}{\pgfqpoint{0.000000in}{-0.048611in}}{\pgfqpoint{0.000000in}{0.000000in}}{%
\pgfpathmoveto{\pgfqpoint{0.000000in}{0.000000in}}%
\pgfpathlineto{\pgfqpoint{0.000000in}{-0.048611in}}%
\pgfusepath{stroke,fill}%
}%
\begin{pgfscope}%
\pgfsys@transformshift{2.325372in}{0.330000in}%
\pgfsys@useobject{currentmarker}{}%
\end{pgfscope}%
\end{pgfscope}%
\begin{pgfscope}%
\definecolor{textcolor}{rgb}{0.000000,0.000000,0.000000}%
\pgfsetstrokecolor{textcolor}%
\pgfsetfillcolor{textcolor}%
\pgftext[x=2.325372in,y=0.232778in,,top]{\color{textcolor}\sffamily\fontsize{8.000000}{9.600000}\selectfont 2000}%
\end{pgfscope}%
\begin{pgfscope}%
\pgfsetbuttcap%
\pgfsetroundjoin%
\definecolor{currentfill}{rgb}{0.000000,0.000000,0.000000}%
\pgfsetfillcolor{currentfill}%
\pgfsetlinewidth{0.803000pt}%
\definecolor{currentstroke}{rgb}{0.000000,0.000000,0.000000}%
\pgfsetstrokecolor{currentstroke}%
\pgfsetdash{}{0pt}%
\pgfsys@defobject{currentmarker}{\pgfqpoint{-0.048611in}{0.000000in}}{\pgfqpoint{-0.000000in}{0.000000in}}{%
\pgfpathmoveto{\pgfqpoint{-0.000000in}{0.000000in}}%
\pgfpathlineto{\pgfqpoint{-0.048611in}{0.000000in}}%
\pgfusepath{stroke,fill}%
}%
\begin{pgfscope}%
\pgfsys@transformshift{0.719689in}{2.639599in}%
\pgfsys@useobject{currentmarker}{}%
\end{pgfscope}%
\end{pgfscope}%
\begin{pgfscope}%
\definecolor{textcolor}{rgb}{0.000000,0.000000,0.000000}%
\pgfsetstrokecolor{textcolor}%
\pgfsetfillcolor{textcolor}%
\pgftext[x=0.563467in, y=2.601043in, left, base]{\color{textcolor}\sffamily\fontsize{8.000000}{9.600000}\selectfont 0}%
\end{pgfscope}%
\begin{pgfscope}%
\pgfsetbuttcap%
\pgfsetroundjoin%
\definecolor{currentfill}{rgb}{0.000000,0.000000,0.000000}%
\pgfsetfillcolor{currentfill}%
\pgfsetlinewidth{0.803000pt}%
\definecolor{currentstroke}{rgb}{0.000000,0.000000,0.000000}%
\pgfsetstrokecolor{currentstroke}%
\pgfsetdash{}{0pt}%
\pgfsys@defobject{currentmarker}{\pgfqpoint{-0.048611in}{0.000000in}}{\pgfqpoint{-0.000000in}{0.000000in}}{%
\pgfpathmoveto{\pgfqpoint{-0.000000in}{0.000000in}}%
\pgfpathlineto{\pgfqpoint{-0.048611in}{0.000000in}}%
\pgfusepath{stroke,fill}%
}%
\begin{pgfscope}%
\pgfsys@transformshift{0.719689in}{2.238278in}%
\pgfsys@useobject{currentmarker}{}%
\end{pgfscope}%
\end{pgfscope}%
\begin{pgfscope}%
\definecolor{textcolor}{rgb}{0.000000,0.000000,0.000000}%
\pgfsetstrokecolor{textcolor}%
\pgfsetfillcolor{textcolor}%
\pgftext[x=0.445467in, y=2.199723in, left, base]{\color{textcolor}\sffamily\fontsize{8.000000}{9.600000}\selectfont 500}%
\end{pgfscope}%
\begin{pgfscope}%
\pgfsetbuttcap%
\pgfsetroundjoin%
\definecolor{currentfill}{rgb}{0.000000,0.000000,0.000000}%
\pgfsetfillcolor{currentfill}%
\pgfsetlinewidth{0.803000pt}%
\definecolor{currentstroke}{rgb}{0.000000,0.000000,0.000000}%
\pgfsetstrokecolor{currentstroke}%
\pgfsetdash{}{0pt}%
\pgfsys@defobject{currentmarker}{\pgfqpoint{-0.048611in}{0.000000in}}{\pgfqpoint{-0.000000in}{0.000000in}}{%
\pgfpathmoveto{\pgfqpoint{-0.000000in}{0.000000in}}%
\pgfpathlineto{\pgfqpoint{-0.048611in}{0.000000in}}%
\pgfusepath{stroke,fill}%
}%
\begin{pgfscope}%
\pgfsys@transformshift{0.719689in}{1.836958in}%
\pgfsys@useobject{currentmarker}{}%
\end{pgfscope}%
\end{pgfscope}%
\begin{pgfscope}%
\definecolor{textcolor}{rgb}{0.000000,0.000000,0.000000}%
\pgfsetstrokecolor{textcolor}%
\pgfsetfillcolor{textcolor}%
\pgftext[x=0.386467in, y=1.798402in, left, base]{\color{textcolor}\sffamily\fontsize{8.000000}{9.600000}\selectfont 1000}%
\end{pgfscope}%
\begin{pgfscope}%
\pgfsetbuttcap%
\pgfsetroundjoin%
\definecolor{currentfill}{rgb}{0.000000,0.000000,0.000000}%
\pgfsetfillcolor{currentfill}%
\pgfsetlinewidth{0.803000pt}%
\definecolor{currentstroke}{rgb}{0.000000,0.000000,0.000000}%
\pgfsetstrokecolor{currentstroke}%
\pgfsetdash{}{0pt}%
\pgfsys@defobject{currentmarker}{\pgfqpoint{-0.048611in}{0.000000in}}{\pgfqpoint{-0.000000in}{0.000000in}}{%
\pgfpathmoveto{\pgfqpoint{-0.000000in}{0.000000in}}%
\pgfpathlineto{\pgfqpoint{-0.048611in}{0.000000in}}%
\pgfusepath{stroke,fill}%
}%
\begin{pgfscope}%
\pgfsys@transformshift{0.719689in}{1.435638in}%
\pgfsys@useobject{currentmarker}{}%
\end{pgfscope}%
\end{pgfscope}%
\begin{pgfscope}%
\definecolor{textcolor}{rgb}{0.000000,0.000000,0.000000}%
\pgfsetstrokecolor{textcolor}%
\pgfsetfillcolor{textcolor}%
\pgftext[x=0.386467in, y=1.397082in, left, base]{\color{textcolor}\sffamily\fontsize{8.000000}{9.600000}\selectfont 1500}%
\end{pgfscope}%
\begin{pgfscope}%
\pgfsetbuttcap%
\pgfsetroundjoin%
\definecolor{currentfill}{rgb}{0.000000,0.000000,0.000000}%
\pgfsetfillcolor{currentfill}%
\pgfsetlinewidth{0.803000pt}%
\definecolor{currentstroke}{rgb}{0.000000,0.000000,0.000000}%
\pgfsetstrokecolor{currentstroke}%
\pgfsetdash{}{0pt}%
\pgfsys@defobject{currentmarker}{\pgfqpoint{-0.048611in}{0.000000in}}{\pgfqpoint{-0.000000in}{0.000000in}}{%
\pgfpathmoveto{\pgfqpoint{-0.000000in}{0.000000in}}%
\pgfpathlineto{\pgfqpoint{-0.048611in}{0.000000in}}%
\pgfusepath{stroke,fill}%
}%
\begin{pgfscope}%
\pgfsys@transformshift{0.719689in}{1.034317in}%
\pgfsys@useobject{currentmarker}{}%
\end{pgfscope}%
\end{pgfscope}%
\begin{pgfscope}%
\definecolor{textcolor}{rgb}{0.000000,0.000000,0.000000}%
\pgfsetstrokecolor{textcolor}%
\pgfsetfillcolor{textcolor}%
\pgftext[x=0.386467in, y=0.995762in, left, base]{\color{textcolor}\sffamily\fontsize{8.000000}{9.600000}\selectfont 2000}%
\end{pgfscope}%
\begin{pgfscope}%
\pgfsetbuttcap%
\pgfsetroundjoin%
\definecolor{currentfill}{rgb}{0.000000,0.000000,0.000000}%
\pgfsetfillcolor{currentfill}%
\pgfsetlinewidth{0.803000pt}%
\definecolor{currentstroke}{rgb}{0.000000,0.000000,0.000000}%
\pgfsetstrokecolor{currentstroke}%
\pgfsetdash{}{0pt}%
\pgfsys@defobject{currentmarker}{\pgfqpoint{-0.048611in}{0.000000in}}{\pgfqpoint{-0.000000in}{0.000000in}}{%
\pgfpathmoveto{\pgfqpoint{-0.000000in}{0.000000in}}%
\pgfpathlineto{\pgfqpoint{-0.048611in}{0.000000in}}%
\pgfusepath{stroke,fill}%
}%
\begin{pgfscope}%
\pgfsys@transformshift{0.719689in}{0.632997in}%
\pgfsys@useobject{currentmarker}{}%
\end{pgfscope}%
\end{pgfscope}%
\begin{pgfscope}%
\definecolor{textcolor}{rgb}{0.000000,0.000000,0.000000}%
\pgfsetstrokecolor{textcolor}%
\pgfsetfillcolor{textcolor}%
\pgftext[x=0.386467in, y=0.594441in, left, base]{\color{textcolor}\sffamily\fontsize{8.000000}{9.600000}\selectfont 2500}%
\end{pgfscope}%
\begin{pgfscope}%
\pgfsetrectcap%
\pgfsetmiterjoin%
\pgfsetlinewidth{0.803000pt}%
\definecolor{currentstroke}{rgb}{0.000000,0.000000,0.000000}%
\pgfsetstrokecolor{currentstroke}%
\pgfsetdash{}{0pt}%
\pgfpathmoveto{\pgfqpoint{0.719689in}{0.330000in}}%
\pgfpathlineto{\pgfqpoint{0.719689in}{2.640000in}}%
\pgfusepath{stroke}%
\end{pgfscope}%
\begin{pgfscope}%
\pgfsetrectcap%
\pgfsetmiterjoin%
\pgfsetlinewidth{0.803000pt}%
\definecolor{currentstroke}{rgb}{0.000000,0.000000,0.000000}%
\pgfsetstrokecolor{currentstroke}%
\pgfsetdash{}{0pt}%
\pgfpathmoveto{\pgfqpoint{2.607500in}{0.330000in}}%
\pgfpathlineto{\pgfqpoint{2.607500in}{2.640000in}}%
\pgfusepath{stroke}%
\end{pgfscope}%
\begin{pgfscope}%
\pgfsetrectcap%
\pgfsetmiterjoin%
\pgfsetlinewidth{0.803000pt}%
\definecolor{currentstroke}{rgb}{0.000000,0.000000,0.000000}%
\pgfsetstrokecolor{currentstroke}%
\pgfsetdash{}{0pt}%
\pgfpathmoveto{\pgfqpoint{0.719689in}{0.330000in}}%
\pgfpathlineto{\pgfqpoint{2.607500in}{0.330000in}}%
\pgfusepath{stroke}%
\end{pgfscope}%
\begin{pgfscope}%
\pgfsetrectcap%
\pgfsetmiterjoin%
\pgfsetlinewidth{0.803000pt}%
\definecolor{currentstroke}{rgb}{0.000000,0.000000,0.000000}%
\pgfsetstrokecolor{currentstroke}%
\pgfsetdash{}{0pt}%
\pgfpathmoveto{\pgfqpoint{0.719689in}{2.640000in}}%
\pgfpathlineto{\pgfqpoint{2.607500in}{2.640000in}}%
\pgfusepath{stroke}%
\end{pgfscope}%
\begin{pgfscope}%
\pgfsetbuttcap%
\pgfsetmiterjoin%
\definecolor{currentfill}{rgb}{1.000000,1.000000,1.000000}%
\pgfsetfillcolor{currentfill}%
\pgfsetlinewidth{0.000000pt}%
\definecolor{currentstroke}{rgb}{0.000000,0.000000,0.000000}%
\pgfsetstrokecolor{currentstroke}%
\pgfsetstrokeopacity{0.000000}%
\pgfsetdash{}{0pt}%
\pgfpathmoveto{\pgfqpoint{2.743125in}{0.330000in}}%
\pgfpathlineto{\pgfqpoint{2.858625in}{0.330000in}}%
\pgfpathlineto{\pgfqpoint{2.858625in}{2.640000in}}%
\pgfpathlineto{\pgfqpoint{2.743125in}{2.640000in}}%
\pgfpathclose%
\pgfusepath{fill}%
\end{pgfscope}%
\begin{pgfscope}%
\pgfpathrectangle{\pgfqpoint{2.743125in}{0.330000in}}{\pgfqpoint{0.115500in}{2.310000in}}%
\pgfusepath{clip}%
\pgfsetbuttcap%
\pgfsetmiterjoin%
\definecolor{currentfill}{rgb}{1.000000,1.000000,1.000000}%
\pgfsetfillcolor{currentfill}%
\pgfsetlinewidth{0.010037pt}%
\definecolor{currentstroke}{rgb}{1.000000,1.000000,1.000000}%
\pgfsetstrokecolor{currentstroke}%
\pgfsetdash{}{0pt}%
\pgfpathmoveto{\pgfqpoint{2.743125in}{0.330000in}}%
\pgfpathlineto{\pgfqpoint{2.743125in}{0.339023in}}%
\pgfpathlineto{\pgfqpoint{2.743125in}{2.630977in}}%
\pgfpathlineto{\pgfqpoint{2.743125in}{2.640000in}}%
\pgfpathlineto{\pgfqpoint{2.858625in}{2.640000in}}%
\pgfpathlineto{\pgfqpoint{2.858625in}{2.630977in}}%
\pgfpathlineto{\pgfqpoint{2.858625in}{0.339023in}}%
\pgfpathlineto{\pgfqpoint{2.858625in}{0.330000in}}%
\pgfpathlineto{\pgfqpoint{2.858625in}{0.330000in}}%
\pgfpathclose%
\pgfusepath{stroke,fill}%
\end{pgfscope}%
\begin{pgfscope}%
\pgfsys@transformshift{2.740000in}{0.330000in}%
\pgftext[left,bottom]{\includegraphics[interpolate=true,width=0.120000in,height=2.310000in]{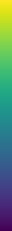}}%
\end{pgfscope}%
\begin{pgfscope}%
\pgfsetbuttcap%
\pgfsetroundjoin%
\definecolor{currentfill}{rgb}{0.000000,0.000000,0.000000}%
\pgfsetfillcolor{currentfill}%
\pgfsetlinewidth{0.803000pt}%
\definecolor{currentstroke}{rgb}{0.000000,0.000000,0.000000}%
\pgfsetstrokecolor{currentstroke}%
\pgfsetdash{}{0pt}%
\pgfsys@defobject{currentmarker}{\pgfqpoint{0.000000in}{0.000000in}}{\pgfqpoint{0.048611in}{0.000000in}}{%
\pgfpathmoveto{\pgfqpoint{0.000000in}{0.000000in}}%
\pgfpathlineto{\pgfqpoint{0.048611in}{0.000000in}}%
\pgfusepath{stroke,fill}%
}%
\begin{pgfscope}%
\pgfsys@transformshift{2.858625in}{0.334419in}%
\pgfsys@useobject{currentmarker}{}%
\end{pgfscope}%
\end{pgfscope}%
\begin{pgfscope}%
\definecolor{textcolor}{rgb}{0.000000,0.000000,0.000000}%
\pgfsetstrokecolor{textcolor}%
\pgfsetfillcolor{textcolor}%
\pgftext[x=2.955847in, y=0.295864in, left, base]{\color{textcolor}\sffamily\fontsize{8.000000}{9.600000}\selectfont 0.06}%
\end{pgfscope}%
\begin{pgfscope}%
\pgfsetbuttcap%
\pgfsetroundjoin%
\definecolor{currentfill}{rgb}{0.000000,0.000000,0.000000}%
\pgfsetfillcolor{currentfill}%
\pgfsetlinewidth{0.803000pt}%
\definecolor{currentstroke}{rgb}{0.000000,0.000000,0.000000}%
\pgfsetstrokecolor{currentstroke}%
\pgfsetdash{}{0pt}%
\pgfsys@defobject{currentmarker}{\pgfqpoint{0.000000in}{0.000000in}}{\pgfqpoint{0.048611in}{0.000000in}}{%
\pgfpathmoveto{\pgfqpoint{0.000000in}{0.000000in}}%
\pgfpathlineto{\pgfqpoint{0.048611in}{0.000000in}}%
\pgfusepath{stroke,fill}%
}%
\begin{pgfscope}%
\pgfsys@transformshift{2.858625in}{0.750462in}%
\pgfsys@useobject{currentmarker}{}%
\end{pgfscope}%
\end{pgfscope}%
\begin{pgfscope}%
\definecolor{textcolor}{rgb}{0.000000,0.000000,0.000000}%
\pgfsetstrokecolor{textcolor}%
\pgfsetfillcolor{textcolor}%
\pgftext[x=2.955847in, y=0.711906in, left, base]{\color{textcolor}\sffamily\fontsize{8.000000}{9.600000}\selectfont 0.07}%
\end{pgfscope}%
\begin{pgfscope}%
\pgfsetbuttcap%
\pgfsetroundjoin%
\definecolor{currentfill}{rgb}{0.000000,0.000000,0.000000}%
\pgfsetfillcolor{currentfill}%
\pgfsetlinewidth{0.803000pt}%
\definecolor{currentstroke}{rgb}{0.000000,0.000000,0.000000}%
\pgfsetstrokecolor{currentstroke}%
\pgfsetdash{}{0pt}%
\pgfsys@defobject{currentmarker}{\pgfqpoint{0.000000in}{0.000000in}}{\pgfqpoint{0.048611in}{0.000000in}}{%
\pgfpathmoveto{\pgfqpoint{0.000000in}{0.000000in}}%
\pgfpathlineto{\pgfqpoint{0.048611in}{0.000000in}}%
\pgfusepath{stroke,fill}%
}%
\begin{pgfscope}%
\pgfsys@transformshift{2.858625in}{1.166505in}%
\pgfsys@useobject{currentmarker}{}%
\end{pgfscope}%
\end{pgfscope}%
\begin{pgfscope}%
\definecolor{textcolor}{rgb}{0.000000,0.000000,0.000000}%
\pgfsetstrokecolor{textcolor}%
\pgfsetfillcolor{textcolor}%
\pgftext[x=2.955847in, y=1.127949in, left, base]{\color{textcolor}\sffamily\fontsize{8.000000}{9.600000}\selectfont 0.08}%
\end{pgfscope}%
\begin{pgfscope}%
\pgfsetbuttcap%
\pgfsetroundjoin%
\definecolor{currentfill}{rgb}{0.000000,0.000000,0.000000}%
\pgfsetfillcolor{currentfill}%
\pgfsetlinewidth{0.803000pt}%
\definecolor{currentstroke}{rgb}{0.000000,0.000000,0.000000}%
\pgfsetstrokecolor{currentstroke}%
\pgfsetdash{}{0pt}%
\pgfsys@defobject{currentmarker}{\pgfqpoint{0.000000in}{0.000000in}}{\pgfqpoint{0.048611in}{0.000000in}}{%
\pgfpathmoveto{\pgfqpoint{0.000000in}{0.000000in}}%
\pgfpathlineto{\pgfqpoint{0.048611in}{0.000000in}}%
\pgfusepath{stroke,fill}%
}%
\begin{pgfscope}%
\pgfsys@transformshift{2.858625in}{1.582547in}%
\pgfsys@useobject{currentmarker}{}%
\end{pgfscope}%
\end{pgfscope}%
\begin{pgfscope}%
\definecolor{textcolor}{rgb}{0.000000,0.000000,0.000000}%
\pgfsetstrokecolor{textcolor}%
\pgfsetfillcolor{textcolor}%
\pgftext[x=2.955847in, y=1.543992in, left, base]{\color{textcolor}\sffamily\fontsize{8.000000}{9.600000}\selectfont 0.09}%
\end{pgfscope}%
\begin{pgfscope}%
\pgfsetbuttcap%
\pgfsetroundjoin%
\definecolor{currentfill}{rgb}{0.000000,0.000000,0.000000}%
\pgfsetfillcolor{currentfill}%
\pgfsetlinewidth{0.803000pt}%
\definecolor{currentstroke}{rgb}{0.000000,0.000000,0.000000}%
\pgfsetstrokecolor{currentstroke}%
\pgfsetdash{}{0pt}%
\pgfsys@defobject{currentmarker}{\pgfqpoint{0.000000in}{0.000000in}}{\pgfqpoint{0.048611in}{0.000000in}}{%
\pgfpathmoveto{\pgfqpoint{0.000000in}{0.000000in}}%
\pgfpathlineto{\pgfqpoint{0.048611in}{0.000000in}}%
\pgfusepath{stroke,fill}%
}%
\begin{pgfscope}%
\pgfsys@transformshift{2.858625in}{1.998590in}%
\pgfsys@useobject{currentmarker}{}%
\end{pgfscope}%
\end{pgfscope}%
\begin{pgfscope}%
\definecolor{textcolor}{rgb}{0.000000,0.000000,0.000000}%
\pgfsetstrokecolor{textcolor}%
\pgfsetfillcolor{textcolor}%
\pgftext[x=2.955847in, y=1.960034in, left, base]{\color{textcolor}\sffamily\fontsize{8.000000}{9.600000}\selectfont 0.10}%
\end{pgfscope}%
\begin{pgfscope}%
\pgfsetbuttcap%
\pgfsetroundjoin%
\definecolor{currentfill}{rgb}{0.000000,0.000000,0.000000}%
\pgfsetfillcolor{currentfill}%
\pgfsetlinewidth{0.803000pt}%
\definecolor{currentstroke}{rgb}{0.000000,0.000000,0.000000}%
\pgfsetstrokecolor{currentstroke}%
\pgfsetdash{}{0pt}%
\pgfsys@defobject{currentmarker}{\pgfqpoint{0.000000in}{0.000000in}}{\pgfqpoint{0.048611in}{0.000000in}}{%
\pgfpathmoveto{\pgfqpoint{0.000000in}{0.000000in}}%
\pgfpathlineto{\pgfqpoint{0.048611in}{0.000000in}}%
\pgfusepath{stroke,fill}%
}%
\begin{pgfscope}%
\pgfsys@transformshift{2.858625in}{2.414632in}%
\pgfsys@useobject{currentmarker}{}%
\end{pgfscope}%
\end{pgfscope}%
\begin{pgfscope}%
\definecolor{textcolor}{rgb}{0.000000,0.000000,0.000000}%
\pgfsetstrokecolor{textcolor}%
\pgfsetfillcolor{textcolor}%
\pgftext[x=2.955847in, y=2.376077in, left, base]{\color{textcolor}\sffamily\fontsize{8.000000}{9.600000}\selectfont 0.11}%
\end{pgfscope}%
\begin{pgfscope}%
\pgfsetrectcap%
\pgfsetmiterjoin%
\pgfsetlinewidth{0.803000pt}%
\definecolor{currentstroke}{rgb}{0.000000,0.000000,0.000000}%
\pgfsetstrokecolor{currentstroke}%
\pgfsetdash{}{0pt}%
\pgfpathmoveto{\pgfqpoint{2.743125in}{0.330000in}}%
\pgfpathlineto{\pgfqpoint{2.743125in}{0.339023in}}%
\pgfpathlineto{\pgfqpoint{2.743125in}{2.630977in}}%
\pgfpathlineto{\pgfqpoint{2.743125in}{2.640000in}}%
\pgfpathlineto{\pgfqpoint{2.858625in}{2.640000in}}%
\pgfpathlineto{\pgfqpoint{2.858625in}{2.630977in}}%
\pgfpathlineto{\pgfqpoint{2.858625in}{0.339023in}}%
\pgfpathlineto{\pgfqpoint{2.858625in}{0.330000in}}%
\pgfpathclose%
\pgfusepath{stroke}%
\end{pgfscope}%
\end{pgfpicture}%
\makeatother%
\endgroup%

%% file: figures/AlCu_lineouts_zoom.pgf
\begingroup%
\makeatletter%
\begin{pgfpicture}%
\pgfpathrectangle{\pgfpointorigin}{\pgfqpoint{3.500000in}{3.000000in}}%
\pgfusepath{use as bounding box, clip}%
\begin{pgfscope}%
\pgfsetbuttcap%
\pgfsetmiterjoin%
\definecolor{currentfill}{rgb}{1.000000,1.000000,1.000000}%
\pgfsetfillcolor{currentfill}%
\pgfsetlinewidth{0.000000pt}%
\definecolor{currentstroke}{rgb}{1.000000,1.000000,1.000000}%
\pgfsetstrokecolor{currentstroke}%
\pgfsetdash{}{0pt}%
\pgfpathmoveto{\pgfqpoint{0.000000in}{0.000000in}}%
\pgfpathlineto{\pgfqpoint{3.500000in}{0.000000in}}%
\pgfpathlineto{\pgfqpoint{3.500000in}{3.000000in}}%
\pgfpathlineto{\pgfqpoint{0.000000in}{3.000000in}}%
\pgfpathclose%
\pgfusepath{fill}%
\end{pgfscope}%
\begin{pgfscope}%
\pgfsetbuttcap%
\pgfsetmiterjoin%
\definecolor{currentfill}{rgb}{1.000000,1.000000,1.000000}%
\pgfsetfillcolor{currentfill}%
\pgfsetlinewidth{0.000000pt}%
\definecolor{currentstroke}{rgb}{0.000000,0.000000,0.000000}%
\pgfsetstrokecolor{currentstroke}%
\pgfsetstrokeopacity{0.000000}%
\pgfsetdash{}{0pt}%
\pgfpathmoveto{\pgfqpoint{0.437500in}{0.330000in}}%
\pgfpathlineto{\pgfqpoint{3.150000in}{0.330000in}}%
\pgfpathlineto{\pgfqpoint{3.150000in}{2.640000in}}%
\pgfpathlineto{\pgfqpoint{0.437500in}{2.640000in}}%
\pgfpathclose%
\pgfusepath{fill}%
\end{pgfscope}%
\begin{pgfscope}%
\pgfsetbuttcap%
\pgfsetroundjoin%
\definecolor{currentfill}{rgb}{0.000000,0.000000,0.000000}%
\pgfsetfillcolor{currentfill}%
\pgfsetlinewidth{0.803000pt}%
\definecolor{currentstroke}{rgb}{0.000000,0.000000,0.000000}%
\pgfsetstrokecolor{currentstroke}%
\pgfsetdash{}{0pt}%
\pgfsys@defobject{currentmarker}{\pgfqpoint{0.000000in}{-0.048611in}}{\pgfqpoint{0.000000in}{0.000000in}}{%
\pgfpathmoveto{\pgfqpoint{0.000000in}{0.000000in}}%
\pgfpathlineto{\pgfqpoint{0.000000in}{-0.048611in}}%
\pgfusepath{stroke,fill}%
}%
\begin{pgfscope}%
\pgfsys@transformshift{0.437500in}{0.330000in}%
\pgfsys@useobject{currentmarker}{}%
\end{pgfscope}%
\end{pgfscope}%
\begin{pgfscope}%
\definecolor{textcolor}{rgb}{0.000000,0.000000,0.000000}%
\pgfsetstrokecolor{textcolor}%
\pgfsetfillcolor{textcolor}%
\pgftext[x=0.437500in,y=0.232778in,,top]{\color{textcolor}\sffamily\fontsize{8.000000}{9.600000}\selectfont 800}%
\end{pgfscope}%
\begin{pgfscope}%
\pgfsetbuttcap%
\pgfsetroundjoin%
\definecolor{currentfill}{rgb}{0.000000,0.000000,0.000000}%
\pgfsetfillcolor{currentfill}%
\pgfsetlinewidth{0.803000pt}%
\definecolor{currentstroke}{rgb}{0.000000,0.000000,0.000000}%
\pgfsetstrokecolor{currentstroke}%
\pgfsetdash{}{0pt}%
\pgfsys@defobject{currentmarker}{\pgfqpoint{0.000000in}{-0.048611in}}{\pgfqpoint{0.000000in}{0.000000in}}{%
\pgfpathmoveto{\pgfqpoint{0.000000in}{0.000000in}}%
\pgfpathlineto{\pgfqpoint{0.000000in}{-0.048611in}}%
\pgfusepath{stroke,fill}%
}%
\begin{pgfscope}%
\pgfsys@transformshift{0.858696in}{0.330000in}%
\pgfsys@useobject{currentmarker}{}%
\end{pgfscope}%
\end{pgfscope}%
\begin{pgfscope}%
\definecolor{textcolor}{rgb}{0.000000,0.000000,0.000000}%
\pgfsetstrokecolor{textcolor}%
\pgfsetfillcolor{textcolor}%
\pgftext[x=0.858696in,y=0.232778in,,top]{\color{textcolor}\sffamily\fontsize{8.000000}{9.600000}\selectfont 850}%
\end{pgfscope}%
\begin{pgfscope}%
\pgfsetbuttcap%
\pgfsetroundjoin%
\definecolor{currentfill}{rgb}{0.000000,0.000000,0.000000}%
\pgfsetfillcolor{currentfill}%
\pgfsetlinewidth{0.803000pt}%
\definecolor{currentstroke}{rgb}{0.000000,0.000000,0.000000}%
\pgfsetstrokecolor{currentstroke}%
\pgfsetdash{}{0pt}%
\pgfsys@defobject{currentmarker}{\pgfqpoint{0.000000in}{-0.048611in}}{\pgfqpoint{0.000000in}{0.000000in}}{%
\pgfpathmoveto{\pgfqpoint{0.000000in}{0.000000in}}%
\pgfpathlineto{\pgfqpoint{0.000000in}{-0.048611in}}%
\pgfusepath{stroke,fill}%
}%
\begin{pgfscope}%
\pgfsys@transformshift{1.279891in}{0.330000in}%
\pgfsys@useobject{currentmarker}{}%
\end{pgfscope}%
\end{pgfscope}%
\begin{pgfscope}%
\definecolor{textcolor}{rgb}{0.000000,0.000000,0.000000}%
\pgfsetstrokecolor{textcolor}%
\pgfsetfillcolor{textcolor}%
\pgftext[x=1.279891in,y=0.232778in,,top]{\color{textcolor}\sffamily\fontsize{8.000000}{9.600000}\selectfont 900}%
\end{pgfscope}%
\begin{pgfscope}%
\pgfsetbuttcap%
\pgfsetroundjoin%
\definecolor{currentfill}{rgb}{0.000000,0.000000,0.000000}%
\pgfsetfillcolor{currentfill}%
\pgfsetlinewidth{0.803000pt}%
\definecolor{currentstroke}{rgb}{0.000000,0.000000,0.000000}%
\pgfsetstrokecolor{currentstroke}%
\pgfsetdash{}{0pt}%
\pgfsys@defobject{currentmarker}{\pgfqpoint{0.000000in}{-0.048611in}}{\pgfqpoint{0.000000in}{0.000000in}}{%
\pgfpathmoveto{\pgfqpoint{0.000000in}{0.000000in}}%
\pgfpathlineto{\pgfqpoint{0.000000in}{-0.048611in}}%
\pgfusepath{stroke,fill}%
}%
\begin{pgfscope}%
\pgfsys@transformshift{1.701087in}{0.330000in}%
\pgfsys@useobject{currentmarker}{}%
\end{pgfscope}%
\end{pgfscope}%
\begin{pgfscope}%
\definecolor{textcolor}{rgb}{0.000000,0.000000,0.000000}%
\pgfsetstrokecolor{textcolor}%
\pgfsetfillcolor{textcolor}%
\pgftext[x=1.701087in,y=0.232778in,,top]{\color{textcolor}\sffamily\fontsize{8.000000}{9.600000}\selectfont 950}%
\end{pgfscope}%
\begin{pgfscope}%
\pgfsetbuttcap%
\pgfsetroundjoin%
\definecolor{currentfill}{rgb}{0.000000,0.000000,0.000000}%
\pgfsetfillcolor{currentfill}%
\pgfsetlinewidth{0.803000pt}%
\definecolor{currentstroke}{rgb}{0.000000,0.000000,0.000000}%
\pgfsetstrokecolor{currentstroke}%
\pgfsetdash{}{0pt}%
\pgfsys@defobject{currentmarker}{\pgfqpoint{0.000000in}{-0.048611in}}{\pgfqpoint{0.000000in}{0.000000in}}{%
\pgfpathmoveto{\pgfqpoint{0.000000in}{0.000000in}}%
\pgfpathlineto{\pgfqpoint{0.000000in}{-0.048611in}}%
\pgfusepath{stroke,fill}%
}%
\begin{pgfscope}%
\pgfsys@transformshift{2.122283in}{0.330000in}%
\pgfsys@useobject{currentmarker}{}%
\end{pgfscope}%
\end{pgfscope}%
\begin{pgfscope}%
\definecolor{textcolor}{rgb}{0.000000,0.000000,0.000000}%
\pgfsetstrokecolor{textcolor}%
\pgfsetfillcolor{textcolor}%
\pgftext[x=2.122283in,y=0.232778in,,top]{\color{textcolor}\sffamily\fontsize{8.000000}{9.600000}\selectfont 1000}%
\end{pgfscope}%
\begin{pgfscope}%
\pgfsetbuttcap%
\pgfsetroundjoin%
\definecolor{currentfill}{rgb}{0.000000,0.000000,0.000000}%
\pgfsetfillcolor{currentfill}%
\pgfsetlinewidth{0.803000pt}%
\definecolor{currentstroke}{rgb}{0.000000,0.000000,0.000000}%
\pgfsetstrokecolor{currentstroke}%
\pgfsetdash{}{0pt}%
\pgfsys@defobject{currentmarker}{\pgfqpoint{0.000000in}{-0.048611in}}{\pgfqpoint{0.000000in}{0.000000in}}{%
\pgfpathmoveto{\pgfqpoint{0.000000in}{0.000000in}}%
\pgfpathlineto{\pgfqpoint{0.000000in}{-0.048611in}}%
\pgfusepath{stroke,fill}%
}%
\begin{pgfscope}%
\pgfsys@transformshift{2.543478in}{0.330000in}%
\pgfsys@useobject{currentmarker}{}%
\end{pgfscope}%
\end{pgfscope}%
\begin{pgfscope}%
\definecolor{textcolor}{rgb}{0.000000,0.000000,0.000000}%
\pgfsetstrokecolor{textcolor}%
\pgfsetfillcolor{textcolor}%
\pgftext[x=2.543478in,y=0.232778in,,top]{\color{textcolor}\sffamily\fontsize{8.000000}{9.600000}\selectfont 1050}%
\end{pgfscope}%
\begin{pgfscope}%
\pgfsetbuttcap%
\pgfsetroundjoin%
\definecolor{currentfill}{rgb}{0.000000,0.000000,0.000000}%
\pgfsetfillcolor{currentfill}%
\pgfsetlinewidth{0.803000pt}%
\definecolor{currentstroke}{rgb}{0.000000,0.000000,0.000000}%
\pgfsetstrokecolor{currentstroke}%
\pgfsetdash{}{0pt}%
\pgfsys@defobject{currentmarker}{\pgfqpoint{0.000000in}{-0.048611in}}{\pgfqpoint{0.000000in}{0.000000in}}{%
\pgfpathmoveto{\pgfqpoint{0.000000in}{0.000000in}}%
\pgfpathlineto{\pgfqpoint{0.000000in}{-0.048611in}}%
\pgfusepath{stroke,fill}%
}%
\begin{pgfscope}%
\pgfsys@transformshift{2.964674in}{0.330000in}%
\pgfsys@useobject{currentmarker}{}%
\end{pgfscope}%
\end{pgfscope}%
\begin{pgfscope}%
\definecolor{textcolor}{rgb}{0.000000,0.000000,0.000000}%
\pgfsetstrokecolor{textcolor}%
\pgfsetfillcolor{textcolor}%
\pgftext[x=2.964674in,y=0.232778in,,top]{\color{textcolor}\sffamily\fontsize{8.000000}{9.600000}\selectfont 1100}%
\end{pgfscope}%
\begin{pgfscope}%
\pgfsetbuttcap%
\pgfsetroundjoin%
\definecolor{currentfill}{rgb}{0.000000,0.000000,0.000000}%
\pgfsetfillcolor{currentfill}%
\pgfsetlinewidth{0.803000pt}%
\definecolor{currentstroke}{rgb}{0.000000,0.000000,0.000000}%
\pgfsetstrokecolor{currentstroke}%
\pgfsetdash{}{0pt}%
\pgfsys@defobject{currentmarker}{\pgfqpoint{-0.048611in}{0.000000in}}{\pgfqpoint{-0.000000in}{0.000000in}}{%
\pgfpathmoveto{\pgfqpoint{-0.000000in}{0.000000in}}%
\pgfpathlineto{\pgfqpoint{-0.048611in}{0.000000in}}%
\pgfusepath{stroke,fill}%
}%
\begin{pgfscope}%
\pgfsys@transformshift{0.437500in}{0.330000in}%
\pgfsys@useobject{currentmarker}{}%
\end{pgfscope}%
\end{pgfscope}%
\begin{pgfscope}%
\definecolor{textcolor}{rgb}{0.000000,0.000000,0.000000}%
\pgfsetstrokecolor{textcolor}%
\pgfsetfillcolor{textcolor}%
\pgftext[x=0.130500in, y=0.291444in, left, base]{\color{textcolor}\sffamily\fontsize{8.000000}{9.600000}\selectfont 0.15}%
\end{pgfscope}%
\begin{pgfscope}%
\pgfsetbuttcap%
\pgfsetroundjoin%
\definecolor{currentfill}{rgb}{0.000000,0.000000,0.000000}%
\pgfsetfillcolor{currentfill}%
\pgfsetlinewidth{0.803000pt}%
\definecolor{currentstroke}{rgb}{0.000000,0.000000,0.000000}%
\pgfsetstrokecolor{currentstroke}%
\pgfsetdash{}{0pt}%
\pgfsys@defobject{currentmarker}{\pgfqpoint{-0.048611in}{0.000000in}}{\pgfqpoint{-0.000000in}{0.000000in}}{%
\pgfpathmoveto{\pgfqpoint{-0.000000in}{0.000000in}}%
\pgfpathlineto{\pgfqpoint{-0.048611in}{0.000000in}}%
\pgfusepath{stroke,fill}%
}%
\begin{pgfscope}%
\pgfsys@transformshift{0.437500in}{0.618750in}%
\pgfsys@useobject{currentmarker}{}%
\end{pgfscope}%
\end{pgfscope}%
\begin{pgfscope}%
\definecolor{textcolor}{rgb}{0.000000,0.000000,0.000000}%
\pgfsetstrokecolor{textcolor}%
\pgfsetfillcolor{textcolor}%
\pgftext[x=0.130500in, y=0.580194in, left, base]{\color{textcolor}\sffamily\fontsize{8.000000}{9.600000}\selectfont 0.20}%
\end{pgfscope}%
\begin{pgfscope}%
\pgfsetbuttcap%
\pgfsetroundjoin%
\definecolor{currentfill}{rgb}{0.000000,0.000000,0.000000}%
\pgfsetfillcolor{currentfill}%
\pgfsetlinewidth{0.803000pt}%
\definecolor{currentstroke}{rgb}{0.000000,0.000000,0.000000}%
\pgfsetstrokecolor{currentstroke}%
\pgfsetdash{}{0pt}%
\pgfsys@defobject{currentmarker}{\pgfqpoint{-0.048611in}{0.000000in}}{\pgfqpoint{-0.000000in}{0.000000in}}{%
\pgfpathmoveto{\pgfqpoint{-0.000000in}{0.000000in}}%
\pgfpathlineto{\pgfqpoint{-0.048611in}{0.000000in}}%
\pgfusepath{stroke,fill}%
}%
\begin{pgfscope}%
\pgfsys@transformshift{0.437500in}{0.907500in}%
\pgfsys@useobject{currentmarker}{}%
\end{pgfscope}%
\end{pgfscope}%
\begin{pgfscope}%
\definecolor{textcolor}{rgb}{0.000000,0.000000,0.000000}%
\pgfsetstrokecolor{textcolor}%
\pgfsetfillcolor{textcolor}%
\pgftext[x=0.130500in, y=0.868944in, left, base]{\color{textcolor}\sffamily\fontsize{8.000000}{9.600000}\selectfont 0.25}%
\end{pgfscope}%
\begin{pgfscope}%
\pgfsetbuttcap%
\pgfsetroundjoin%
\definecolor{currentfill}{rgb}{0.000000,0.000000,0.000000}%
\pgfsetfillcolor{currentfill}%
\pgfsetlinewidth{0.803000pt}%
\definecolor{currentstroke}{rgb}{0.000000,0.000000,0.000000}%
\pgfsetstrokecolor{currentstroke}%
\pgfsetdash{}{0pt}%
\pgfsys@defobject{currentmarker}{\pgfqpoint{-0.048611in}{0.000000in}}{\pgfqpoint{-0.000000in}{0.000000in}}{%
\pgfpathmoveto{\pgfqpoint{-0.000000in}{0.000000in}}%
\pgfpathlineto{\pgfqpoint{-0.048611in}{0.000000in}}%
\pgfusepath{stroke,fill}%
}%
\begin{pgfscope}%
\pgfsys@transformshift{0.437500in}{1.196250in}%
\pgfsys@useobject{currentmarker}{}%
\end{pgfscope}%
\end{pgfscope}%
\begin{pgfscope}%
\definecolor{textcolor}{rgb}{0.000000,0.000000,0.000000}%
\pgfsetstrokecolor{textcolor}%
\pgfsetfillcolor{textcolor}%
\pgftext[x=0.130500in, y=1.157694in, left, base]{\color{textcolor}\sffamily\fontsize{8.000000}{9.600000}\selectfont 0.30}%
\end{pgfscope}%
\begin{pgfscope}%
\pgfsetbuttcap%
\pgfsetroundjoin%
\definecolor{currentfill}{rgb}{0.000000,0.000000,0.000000}%
\pgfsetfillcolor{currentfill}%
\pgfsetlinewidth{0.803000pt}%
\definecolor{currentstroke}{rgb}{0.000000,0.000000,0.000000}%
\pgfsetstrokecolor{currentstroke}%
\pgfsetdash{}{0pt}%
\pgfsys@defobject{currentmarker}{\pgfqpoint{-0.048611in}{0.000000in}}{\pgfqpoint{-0.000000in}{0.000000in}}{%
\pgfpathmoveto{\pgfqpoint{-0.000000in}{0.000000in}}%
\pgfpathlineto{\pgfqpoint{-0.048611in}{0.000000in}}%
\pgfusepath{stroke,fill}%
}%
\begin{pgfscope}%
\pgfsys@transformshift{0.437500in}{1.485000in}%
\pgfsys@useobject{currentmarker}{}%
\end{pgfscope}%
\end{pgfscope}%
\begin{pgfscope}%
\definecolor{textcolor}{rgb}{0.000000,0.000000,0.000000}%
\pgfsetstrokecolor{textcolor}%
\pgfsetfillcolor{textcolor}%
\pgftext[x=0.130500in, y=1.446444in, left, base]{\color{textcolor}\sffamily\fontsize{8.000000}{9.600000}\selectfont 0.35}%
\end{pgfscope}%
\begin{pgfscope}%
\pgfsetbuttcap%
\pgfsetroundjoin%
\definecolor{currentfill}{rgb}{0.000000,0.000000,0.000000}%
\pgfsetfillcolor{currentfill}%
\pgfsetlinewidth{0.803000pt}%
\definecolor{currentstroke}{rgb}{0.000000,0.000000,0.000000}%
\pgfsetstrokecolor{currentstroke}%
\pgfsetdash{}{0pt}%
\pgfsys@defobject{currentmarker}{\pgfqpoint{-0.048611in}{0.000000in}}{\pgfqpoint{-0.000000in}{0.000000in}}{%
\pgfpathmoveto{\pgfqpoint{-0.000000in}{0.000000in}}%
\pgfpathlineto{\pgfqpoint{-0.048611in}{0.000000in}}%
\pgfusepath{stroke,fill}%
}%
\begin{pgfscope}%
\pgfsys@transformshift{0.437500in}{1.773750in}%
\pgfsys@useobject{currentmarker}{}%
\end{pgfscope}%
\end{pgfscope}%
\begin{pgfscope}%
\definecolor{textcolor}{rgb}{0.000000,0.000000,0.000000}%
\pgfsetstrokecolor{textcolor}%
\pgfsetfillcolor{textcolor}%
\pgftext[x=0.130500in, y=1.735194in, left, base]{\color{textcolor}\sffamily\fontsize{8.000000}{9.600000}\selectfont 0.40}%
\end{pgfscope}%
\begin{pgfscope}%
\pgfsetbuttcap%
\pgfsetroundjoin%
\definecolor{currentfill}{rgb}{0.000000,0.000000,0.000000}%
\pgfsetfillcolor{currentfill}%
\pgfsetlinewidth{0.803000pt}%
\definecolor{currentstroke}{rgb}{0.000000,0.000000,0.000000}%
\pgfsetstrokecolor{currentstroke}%
\pgfsetdash{}{0pt}%
\pgfsys@defobject{currentmarker}{\pgfqpoint{-0.048611in}{0.000000in}}{\pgfqpoint{-0.000000in}{0.000000in}}{%
\pgfpathmoveto{\pgfqpoint{-0.000000in}{0.000000in}}%
\pgfpathlineto{\pgfqpoint{-0.048611in}{0.000000in}}%
\pgfusepath{stroke,fill}%
}%
\begin{pgfscope}%
\pgfsys@transformshift{0.437500in}{2.062500in}%
\pgfsys@useobject{currentmarker}{}%
\end{pgfscope}%
\end{pgfscope}%
\begin{pgfscope}%
\definecolor{textcolor}{rgb}{0.000000,0.000000,0.000000}%
\pgfsetstrokecolor{textcolor}%
\pgfsetfillcolor{textcolor}%
\pgftext[x=0.130500in, y=2.023944in, left, base]{\color{textcolor}\sffamily\fontsize{8.000000}{9.600000}\selectfont 0.45}%
\end{pgfscope}%
\begin{pgfscope}%
\pgfsetbuttcap%
\pgfsetroundjoin%
\definecolor{currentfill}{rgb}{0.000000,0.000000,0.000000}%
\pgfsetfillcolor{currentfill}%
\pgfsetlinewidth{0.803000pt}%
\definecolor{currentstroke}{rgb}{0.000000,0.000000,0.000000}%
\pgfsetstrokecolor{currentstroke}%
\pgfsetdash{}{0pt}%
\pgfsys@defobject{currentmarker}{\pgfqpoint{-0.048611in}{0.000000in}}{\pgfqpoint{-0.000000in}{0.000000in}}{%
\pgfpathmoveto{\pgfqpoint{-0.000000in}{0.000000in}}%
\pgfpathlineto{\pgfqpoint{-0.048611in}{0.000000in}}%
\pgfusepath{stroke,fill}%
}%
\begin{pgfscope}%
\pgfsys@transformshift{0.437500in}{2.351250in}%
\pgfsys@useobject{currentmarker}{}%
\end{pgfscope}%
\end{pgfscope}%
\begin{pgfscope}%
\definecolor{textcolor}{rgb}{0.000000,0.000000,0.000000}%
\pgfsetstrokecolor{textcolor}%
\pgfsetfillcolor{textcolor}%
\pgftext[x=0.130500in, y=2.312694in, left, base]{\color{textcolor}\sffamily\fontsize{8.000000}{9.600000}\selectfont 0.50}%
\end{pgfscope}%
\begin{pgfscope}%
\pgfsetbuttcap%
\pgfsetroundjoin%
\definecolor{currentfill}{rgb}{0.000000,0.000000,0.000000}%
\pgfsetfillcolor{currentfill}%
\pgfsetlinewidth{0.803000pt}%
\definecolor{currentstroke}{rgb}{0.000000,0.000000,0.000000}%
\pgfsetstrokecolor{currentstroke}%
\pgfsetdash{}{0pt}%
\pgfsys@defobject{currentmarker}{\pgfqpoint{-0.048611in}{0.000000in}}{\pgfqpoint{-0.000000in}{0.000000in}}{%
\pgfpathmoveto{\pgfqpoint{-0.000000in}{0.000000in}}%
\pgfpathlineto{\pgfqpoint{-0.048611in}{0.000000in}}%
\pgfusepath{stroke,fill}%
}%
\begin{pgfscope}%
\pgfsys@transformshift{0.437500in}{2.640000in}%
\pgfsys@useobject{currentmarker}{}%
\end{pgfscope}%
\end{pgfscope}%
\begin{pgfscope}%
\definecolor{textcolor}{rgb}{0.000000,0.000000,0.000000}%
\pgfsetstrokecolor{textcolor}%
\pgfsetfillcolor{textcolor}%
\pgftext[x=0.130500in, y=2.601444in, left, base]{\color{textcolor}\sffamily\fontsize{8.000000}{9.600000}\selectfont 0.55}%
\end{pgfscope}%
\begin{pgfscope}%
\pgfpathrectangle{\pgfqpoint{0.437500in}{0.330000in}}{\pgfqpoint{2.712500in}{2.310000in}}%
\pgfusepath{clip}%
\pgfsetrectcap%
\pgfsetroundjoin%
\pgfsetlinewidth{1.505625pt}%
\definecolor{currentstroke}{rgb}{0.000000,0.000000,0.000000}%
\pgfsetstrokecolor{currentstroke}%
\pgfsetdash{}{0pt}%
\pgfpathmoveto{\pgfqpoint{0.429076in}{2.347650in}}%
\pgfpathlineto{\pgfqpoint{0.445924in}{2.346913in}}%
\pgfpathlineto{\pgfqpoint{0.454348in}{2.332649in}}%
\pgfpathlineto{\pgfqpoint{0.462772in}{2.296936in}}%
\pgfpathlineto{\pgfqpoint{0.471196in}{2.286663in}}%
\pgfpathlineto{\pgfqpoint{0.479620in}{2.274151in}}%
\pgfpathlineto{\pgfqpoint{0.488043in}{2.293676in}}%
\pgfpathlineto{\pgfqpoint{0.496467in}{2.263206in}}%
\pgfpathlineto{\pgfqpoint{0.504891in}{2.249234in}}%
\pgfpathlineto{\pgfqpoint{0.513315in}{2.268072in}}%
\pgfpathlineto{\pgfqpoint{0.521739in}{2.246909in}}%
\pgfpathlineto{\pgfqpoint{0.530163in}{2.218298in}}%
\pgfpathlineto{\pgfqpoint{0.538587in}{2.228191in}}%
\pgfpathlineto{\pgfqpoint{0.547011in}{2.220542in}}%
\pgfpathlineto{\pgfqpoint{0.563859in}{2.167192in}}%
\pgfpathlineto{\pgfqpoint{0.572283in}{2.176863in}}%
\pgfpathlineto{\pgfqpoint{0.580707in}{2.173211in}}%
\pgfpathlineto{\pgfqpoint{0.589130in}{2.137684in}}%
\pgfpathlineto{\pgfqpoint{0.597554in}{2.116205in}}%
\pgfpathlineto{\pgfqpoint{0.605978in}{2.102061in}}%
\pgfpathlineto{\pgfqpoint{0.614402in}{2.104478in}}%
\pgfpathlineto{\pgfqpoint{0.631250in}{2.045833in}}%
\pgfpathlineto{\pgfqpoint{0.639674in}{2.036806in}}%
\pgfpathlineto{\pgfqpoint{0.648098in}{2.045348in}}%
\pgfpathlineto{\pgfqpoint{0.656522in}{2.032703in}}%
\pgfpathlineto{\pgfqpoint{0.664946in}{2.039220in}}%
\pgfpathlineto{\pgfqpoint{0.673370in}{2.037398in}}%
\pgfpathlineto{\pgfqpoint{0.681793in}{2.020981in}}%
\pgfpathlineto{\pgfqpoint{0.690217in}{1.972007in}}%
\pgfpathlineto{\pgfqpoint{0.698641in}{1.989488in}}%
\pgfpathlineto{\pgfqpoint{0.707065in}{1.979222in}}%
\pgfpathlineto{\pgfqpoint{0.715489in}{1.994395in}}%
\pgfpathlineto{\pgfqpoint{0.723913in}{1.953344in}}%
\pgfpathlineto{\pgfqpoint{0.732337in}{1.975690in}}%
\pgfpathlineto{\pgfqpoint{0.740761in}{1.966579in}}%
\pgfpathlineto{\pgfqpoint{0.749185in}{1.962154in}}%
\pgfpathlineto{\pgfqpoint{0.757609in}{1.953623in}}%
\pgfpathlineto{\pgfqpoint{0.766033in}{1.884632in}}%
\pgfpathlineto{\pgfqpoint{0.774457in}{1.891013in}}%
\pgfpathlineto{\pgfqpoint{0.782880in}{1.899072in}}%
\pgfpathlineto{\pgfqpoint{0.791304in}{1.929381in}}%
\pgfpathlineto{\pgfqpoint{0.799728in}{1.922646in}}%
\pgfpathlineto{\pgfqpoint{0.808152in}{1.901742in}}%
\pgfpathlineto{\pgfqpoint{0.816576in}{1.877752in}}%
\pgfpathlineto{\pgfqpoint{0.825000in}{1.858761in}}%
\pgfpathlineto{\pgfqpoint{0.833424in}{1.869338in}}%
\pgfpathlineto{\pgfqpoint{0.841848in}{1.861407in}}%
\pgfpathlineto{\pgfqpoint{0.850272in}{1.839986in}}%
\pgfpathlineto{\pgfqpoint{0.858696in}{1.801028in}}%
\pgfpathlineto{\pgfqpoint{0.867120in}{1.810742in}}%
\pgfpathlineto{\pgfqpoint{0.875543in}{1.808787in}}%
\pgfpathlineto{\pgfqpoint{0.883967in}{1.826156in}}%
\pgfpathlineto{\pgfqpoint{0.892391in}{1.810437in}}%
\pgfpathlineto{\pgfqpoint{0.900815in}{1.768909in}}%
\pgfpathlineto{\pgfqpoint{0.909239in}{1.764786in}}%
\pgfpathlineto{\pgfqpoint{0.917663in}{1.723351in}}%
\pgfpathlineto{\pgfqpoint{0.926087in}{1.763797in}}%
\pgfpathlineto{\pgfqpoint{0.934511in}{1.757186in}}%
\pgfpathlineto{\pgfqpoint{0.942935in}{1.774419in}}%
\pgfpathlineto{\pgfqpoint{0.951359in}{1.785953in}}%
\pgfpathlineto{\pgfqpoint{0.959783in}{1.776409in}}%
\pgfpathlineto{\pgfqpoint{0.968207in}{1.752227in}}%
\pgfpathlineto{\pgfqpoint{0.976630in}{1.741760in}}%
\pgfpathlineto{\pgfqpoint{0.985054in}{1.759168in}}%
\pgfpathlineto{\pgfqpoint{0.993478in}{1.725668in}}%
\pgfpathlineto{\pgfqpoint{1.001902in}{1.751743in}}%
\pgfpathlineto{\pgfqpoint{1.010326in}{1.728559in}}%
\pgfpathlineto{\pgfqpoint{1.027174in}{1.687214in}}%
\pgfpathlineto{\pgfqpoint{1.035598in}{1.674291in}}%
\pgfpathlineto{\pgfqpoint{1.044022in}{1.664100in}}%
\pgfpathlineto{\pgfqpoint{1.052446in}{1.690024in}}%
\pgfpathlineto{\pgfqpoint{1.060870in}{1.657039in}}%
\pgfpathlineto{\pgfqpoint{1.069293in}{1.595181in}}%
\pgfpathlineto{\pgfqpoint{1.077717in}{1.573767in}}%
\pgfpathlineto{\pgfqpoint{1.086141in}{1.484036in}}%
\pgfpathlineto{\pgfqpoint{1.102989in}{1.338867in}}%
\pgfpathlineto{\pgfqpoint{1.111413in}{1.280748in}}%
\pgfpathlineto{\pgfqpoint{1.119837in}{1.241612in}}%
\pgfpathlineto{\pgfqpoint{1.128261in}{1.211780in}}%
\pgfpathlineto{\pgfqpoint{1.136685in}{1.189917in}}%
\pgfpathlineto{\pgfqpoint{1.145109in}{1.137427in}}%
\pgfpathlineto{\pgfqpoint{1.161957in}{1.116310in}}%
\pgfpathlineto{\pgfqpoint{1.170380in}{1.079011in}}%
\pgfpathlineto{\pgfqpoint{1.178804in}{1.052462in}}%
\pgfpathlineto{\pgfqpoint{1.187228in}{1.021128in}}%
\pgfpathlineto{\pgfqpoint{1.195652in}{1.011441in}}%
\pgfpathlineto{\pgfqpoint{1.212500in}{0.936609in}}%
\pgfpathlineto{\pgfqpoint{1.220924in}{0.923678in}}%
\pgfpathlineto{\pgfqpoint{1.229348in}{0.921734in}}%
\pgfpathlineto{\pgfqpoint{1.237772in}{0.890354in}}%
\pgfpathlineto{\pgfqpoint{1.246196in}{0.911432in}}%
\pgfpathlineto{\pgfqpoint{1.254620in}{0.869988in}}%
\pgfpathlineto{\pgfqpoint{1.263043in}{0.867294in}}%
\pgfpathlineto{\pgfqpoint{1.271467in}{0.862854in}}%
\pgfpathlineto{\pgfqpoint{1.288315in}{0.803549in}}%
\pgfpathlineto{\pgfqpoint{1.296739in}{0.794788in}}%
\pgfpathlineto{\pgfqpoint{1.305163in}{0.781755in}}%
\pgfpathlineto{\pgfqpoint{1.313587in}{0.838970in}}%
\pgfpathlineto{\pgfqpoint{1.322011in}{0.784851in}}%
\pgfpathlineto{\pgfqpoint{1.330435in}{0.768854in}}%
\pgfpathlineto{\pgfqpoint{1.338859in}{0.743225in}}%
\pgfpathlineto{\pgfqpoint{1.347283in}{0.747295in}}%
\pgfpathlineto{\pgfqpoint{1.355707in}{0.749536in}}%
\pgfpathlineto{\pgfqpoint{1.364130in}{0.685443in}}%
\pgfpathlineto{\pgfqpoint{1.372554in}{0.687193in}}%
\pgfpathlineto{\pgfqpoint{1.389402in}{0.731374in}}%
\pgfpathlineto{\pgfqpoint{1.397826in}{0.715347in}}%
\pgfpathlineto{\pgfqpoint{1.406250in}{0.712784in}}%
\pgfpathlineto{\pgfqpoint{1.414674in}{0.686431in}}%
\pgfpathlineto{\pgfqpoint{1.423098in}{0.675058in}}%
\pgfpathlineto{\pgfqpoint{1.431522in}{0.717479in}}%
\pgfpathlineto{\pgfqpoint{1.439946in}{0.662671in}}%
\pgfpathlineto{\pgfqpoint{1.448370in}{0.662016in}}%
\pgfpathlineto{\pgfqpoint{1.456793in}{0.654822in}}%
\pgfpathlineto{\pgfqpoint{1.465217in}{0.620110in}}%
\pgfpathlineto{\pgfqpoint{1.473641in}{0.624122in}}%
\pgfpathlineto{\pgfqpoint{1.482065in}{0.633765in}}%
\pgfpathlineto{\pgfqpoint{1.490489in}{0.633085in}}%
\pgfpathlineto{\pgfqpoint{1.498913in}{0.640586in}}%
\pgfpathlineto{\pgfqpoint{1.507337in}{0.620047in}}%
\pgfpathlineto{\pgfqpoint{1.515761in}{0.592305in}}%
\pgfpathlineto{\pgfqpoint{1.524185in}{0.604728in}}%
\pgfpathlineto{\pgfqpoint{1.532609in}{0.636552in}}%
\pgfpathlineto{\pgfqpoint{1.541033in}{0.622713in}}%
\pgfpathlineto{\pgfqpoint{1.549457in}{0.597917in}}%
\pgfpathlineto{\pgfqpoint{1.566304in}{0.621733in}}%
\pgfpathlineto{\pgfqpoint{1.574728in}{0.620000in}}%
\pgfpathlineto{\pgfqpoint{1.583152in}{0.595341in}}%
\pgfpathlineto{\pgfqpoint{1.591576in}{0.581767in}}%
\pgfpathlineto{\pgfqpoint{1.600000in}{0.565539in}}%
\pgfpathlineto{\pgfqpoint{1.608424in}{0.574127in}}%
\pgfpathlineto{\pgfqpoint{1.616848in}{0.588187in}}%
\pgfpathlineto{\pgfqpoint{1.625272in}{0.613024in}}%
\pgfpathlineto{\pgfqpoint{1.633696in}{0.584579in}}%
\pgfpathlineto{\pgfqpoint{1.642120in}{0.608336in}}%
\pgfpathlineto{\pgfqpoint{1.650543in}{0.672694in}}%
\pgfpathlineto{\pgfqpoint{1.658967in}{0.650351in}}%
\pgfpathlineto{\pgfqpoint{1.667391in}{0.662450in}}%
\pgfpathlineto{\pgfqpoint{1.675815in}{0.682024in}}%
\pgfpathlineto{\pgfqpoint{1.684239in}{0.676090in}}%
\pgfpathlineto{\pgfqpoint{1.692663in}{0.692911in}}%
\pgfpathlineto{\pgfqpoint{1.701087in}{0.730559in}}%
\pgfpathlineto{\pgfqpoint{1.709511in}{0.753538in}}%
\pgfpathlineto{\pgfqpoint{1.717935in}{0.753569in}}%
\pgfpathlineto{\pgfqpoint{1.726359in}{0.766899in}}%
\pgfpathlineto{\pgfqpoint{1.734783in}{0.755903in}}%
\pgfpathlineto{\pgfqpoint{1.743207in}{0.758607in}}%
\pgfpathlineto{\pgfqpoint{1.751630in}{0.759124in}}%
\pgfpathlineto{\pgfqpoint{1.760054in}{0.772558in}}%
\pgfpathlineto{\pgfqpoint{1.768478in}{0.836151in}}%
\pgfpathlineto{\pgfqpoint{1.776902in}{0.796690in}}%
\pgfpathlineto{\pgfqpoint{1.785326in}{0.794435in}}%
\pgfpathlineto{\pgfqpoint{1.793750in}{0.812180in}}%
\pgfpathlineto{\pgfqpoint{1.802174in}{0.856601in}}%
\pgfpathlineto{\pgfqpoint{1.827446in}{0.904044in}}%
\pgfpathlineto{\pgfqpoint{1.835870in}{0.900403in}}%
\pgfpathlineto{\pgfqpoint{1.844293in}{0.886230in}}%
\pgfpathlineto{\pgfqpoint{1.852717in}{0.890398in}}%
\pgfpathlineto{\pgfqpoint{1.869565in}{0.950511in}}%
\pgfpathlineto{\pgfqpoint{1.877989in}{0.966172in}}%
\pgfpathlineto{\pgfqpoint{1.886413in}{0.951613in}}%
\pgfpathlineto{\pgfqpoint{1.894837in}{0.948393in}}%
\pgfpathlineto{\pgfqpoint{1.903261in}{0.985537in}}%
\pgfpathlineto{\pgfqpoint{1.911685in}{0.990727in}}%
\pgfpathlineto{\pgfqpoint{1.920109in}{0.967193in}}%
\pgfpathlineto{\pgfqpoint{1.928533in}{0.961986in}}%
\pgfpathlineto{\pgfqpoint{1.936957in}{0.998762in}}%
\pgfpathlineto{\pgfqpoint{1.953804in}{0.994732in}}%
\pgfpathlineto{\pgfqpoint{1.979076in}{1.024456in}}%
\pgfpathlineto{\pgfqpoint{1.987500in}{1.024816in}}%
\pgfpathlineto{\pgfqpoint{1.995924in}{1.014985in}}%
\pgfpathlineto{\pgfqpoint{2.004348in}{1.032795in}}%
\pgfpathlineto{\pgfqpoint{2.012772in}{1.035085in}}%
\pgfpathlineto{\pgfqpoint{2.021196in}{1.050533in}}%
\pgfpathlineto{\pgfqpoint{2.029620in}{1.072666in}}%
\pgfpathlineto{\pgfqpoint{2.046467in}{1.068778in}}%
\pgfpathlineto{\pgfqpoint{2.054891in}{1.069582in}}%
\pgfpathlineto{\pgfqpoint{2.063315in}{1.061979in}}%
\pgfpathlineto{\pgfqpoint{2.071739in}{1.057435in}}%
\pgfpathlineto{\pgfqpoint{2.080163in}{1.064249in}}%
\pgfpathlineto{\pgfqpoint{2.088587in}{1.090713in}}%
\pgfpathlineto{\pgfqpoint{2.097011in}{1.098005in}}%
\pgfpathlineto{\pgfqpoint{2.105435in}{1.116580in}}%
\pgfpathlineto{\pgfqpoint{2.113859in}{1.121384in}}%
\pgfpathlineto{\pgfqpoint{2.122283in}{1.127695in}}%
\pgfpathlineto{\pgfqpoint{2.130707in}{1.122592in}}%
\pgfpathlineto{\pgfqpoint{2.139130in}{1.111944in}}%
\pgfpathlineto{\pgfqpoint{2.147554in}{1.135875in}}%
\pgfpathlineto{\pgfqpoint{2.155978in}{1.096130in}}%
\pgfpathlineto{\pgfqpoint{2.164402in}{1.120172in}}%
\pgfpathlineto{\pgfqpoint{2.181250in}{1.132845in}}%
\pgfpathlineto{\pgfqpoint{2.189674in}{1.145289in}}%
\pgfpathlineto{\pgfqpoint{2.198098in}{1.171454in}}%
\pgfpathlineto{\pgfqpoint{2.206522in}{1.174070in}}%
\pgfpathlineto{\pgfqpoint{2.214946in}{1.193567in}}%
\pgfpathlineto{\pgfqpoint{2.223370in}{1.169875in}}%
\pgfpathlineto{\pgfqpoint{2.231793in}{1.155882in}}%
\pgfpathlineto{\pgfqpoint{2.240217in}{1.166667in}}%
\pgfpathlineto{\pgfqpoint{2.248641in}{1.214965in}}%
\pgfpathlineto{\pgfqpoint{2.265489in}{1.171237in}}%
\pgfpathlineto{\pgfqpoint{2.273913in}{1.217346in}}%
\pgfpathlineto{\pgfqpoint{2.282337in}{1.201567in}}%
\pgfpathlineto{\pgfqpoint{2.290761in}{1.205491in}}%
\pgfpathlineto{\pgfqpoint{2.316033in}{1.248149in}}%
\pgfpathlineto{\pgfqpoint{2.324457in}{1.215872in}}%
\pgfpathlineto{\pgfqpoint{2.332880in}{1.248831in}}%
\pgfpathlineto{\pgfqpoint{2.341304in}{1.239161in}}%
\pgfpathlineto{\pgfqpoint{2.349728in}{1.219130in}}%
\pgfpathlineto{\pgfqpoint{2.358152in}{1.243576in}}%
\pgfpathlineto{\pgfqpoint{2.366576in}{1.247568in}}%
\pgfpathlineto{\pgfqpoint{2.375000in}{1.220090in}}%
\pgfpathlineto{\pgfqpoint{2.383424in}{1.238184in}}%
\pgfpathlineto{\pgfqpoint{2.391848in}{1.235745in}}%
\pgfpathlineto{\pgfqpoint{2.400272in}{1.243468in}}%
\pgfpathlineto{\pgfqpoint{2.408696in}{1.259492in}}%
\pgfpathlineto{\pgfqpoint{2.417120in}{1.230638in}}%
\pgfpathlineto{\pgfqpoint{2.425543in}{1.230397in}}%
\pgfpathlineto{\pgfqpoint{2.433967in}{1.238536in}}%
\pgfpathlineto{\pgfqpoint{2.442391in}{1.263567in}}%
\pgfpathlineto{\pgfqpoint{2.459239in}{1.260400in}}%
\pgfpathlineto{\pgfqpoint{2.467663in}{1.288893in}}%
\pgfpathlineto{\pgfqpoint{2.476087in}{1.299764in}}%
\pgfpathlineto{\pgfqpoint{2.484511in}{1.296227in}}%
\pgfpathlineto{\pgfqpoint{2.492935in}{1.309078in}}%
\pgfpathlineto{\pgfqpoint{2.501359in}{1.277092in}}%
\pgfpathlineto{\pgfqpoint{2.509783in}{1.240351in}}%
\pgfpathlineto{\pgfqpoint{2.518207in}{1.261142in}}%
\pgfpathlineto{\pgfqpoint{2.526630in}{1.253032in}}%
\pgfpathlineto{\pgfqpoint{2.543478in}{1.282179in}}%
\pgfpathlineto{\pgfqpoint{2.551902in}{1.336609in}}%
\pgfpathlineto{\pgfqpoint{2.560326in}{1.365035in}}%
\pgfpathlineto{\pgfqpoint{2.568750in}{1.341797in}}%
\pgfpathlineto{\pgfqpoint{2.577174in}{1.305719in}}%
\pgfpathlineto{\pgfqpoint{2.585598in}{1.359138in}}%
\pgfpathlineto{\pgfqpoint{2.594022in}{1.340197in}}%
\pgfpathlineto{\pgfqpoint{2.602446in}{1.345398in}}%
\pgfpathlineto{\pgfqpoint{2.610870in}{1.364297in}}%
\pgfpathlineto{\pgfqpoint{2.619293in}{1.361131in}}%
\pgfpathlineto{\pgfqpoint{2.627717in}{1.345194in}}%
\pgfpathlineto{\pgfqpoint{2.636141in}{1.380115in}}%
\pgfpathlineto{\pgfqpoint{2.644565in}{1.369159in}}%
\pgfpathlineto{\pgfqpoint{2.652989in}{1.364941in}}%
\pgfpathlineto{\pgfqpoint{2.661413in}{1.366776in}}%
\pgfpathlineto{\pgfqpoint{2.669837in}{1.354965in}}%
\pgfpathlineto{\pgfqpoint{2.678261in}{1.355762in}}%
\pgfpathlineto{\pgfqpoint{2.686685in}{1.316313in}}%
\pgfpathlineto{\pgfqpoint{2.695109in}{1.340406in}}%
\pgfpathlineto{\pgfqpoint{2.703533in}{1.319562in}}%
\pgfpathlineto{\pgfqpoint{2.711957in}{1.361141in}}%
\pgfpathlineto{\pgfqpoint{2.720380in}{1.383966in}}%
\pgfpathlineto{\pgfqpoint{2.737228in}{1.344441in}}%
\pgfpathlineto{\pgfqpoint{2.745652in}{1.345065in}}%
\pgfpathlineto{\pgfqpoint{2.754076in}{1.358638in}}%
\pgfpathlineto{\pgfqpoint{2.762500in}{1.353773in}}%
\pgfpathlineto{\pgfqpoint{2.770924in}{1.371962in}}%
\pgfpathlineto{\pgfqpoint{2.779348in}{1.370760in}}%
\pgfpathlineto{\pgfqpoint{2.787772in}{1.360940in}}%
\pgfpathlineto{\pgfqpoint{2.796196in}{1.389661in}}%
\pgfpathlineto{\pgfqpoint{2.804620in}{1.397301in}}%
\pgfpathlineto{\pgfqpoint{2.813043in}{1.411409in}}%
\pgfpathlineto{\pgfqpoint{2.821467in}{1.417657in}}%
\pgfpathlineto{\pgfqpoint{2.829891in}{1.393211in}}%
\pgfpathlineto{\pgfqpoint{2.846739in}{1.396473in}}%
\pgfpathlineto{\pgfqpoint{2.855163in}{1.403432in}}%
\pgfpathlineto{\pgfqpoint{2.863587in}{1.420501in}}%
\pgfpathlineto{\pgfqpoint{2.880435in}{1.437232in}}%
\pgfpathlineto{\pgfqpoint{2.888859in}{1.428180in}}%
\pgfpathlineto{\pgfqpoint{2.897283in}{1.443148in}}%
\pgfpathlineto{\pgfqpoint{2.905707in}{1.444167in}}%
\pgfpathlineto{\pgfqpoint{2.914130in}{1.391016in}}%
\pgfpathlineto{\pgfqpoint{2.922554in}{1.396296in}}%
\pgfpathlineto{\pgfqpoint{2.930978in}{1.397593in}}%
\pgfpathlineto{\pgfqpoint{2.939402in}{1.427688in}}%
\pgfpathlineto{\pgfqpoint{2.947826in}{1.443392in}}%
\pgfpathlineto{\pgfqpoint{2.956250in}{1.425167in}}%
\pgfpathlineto{\pgfqpoint{2.964674in}{1.461927in}}%
\pgfpathlineto{\pgfqpoint{2.973098in}{1.469489in}}%
\pgfpathlineto{\pgfqpoint{2.981522in}{1.445031in}}%
\pgfpathlineto{\pgfqpoint{2.989946in}{1.473819in}}%
\pgfpathlineto{\pgfqpoint{2.998370in}{1.455780in}}%
\pgfpathlineto{\pgfqpoint{3.006793in}{1.404565in}}%
\pgfpathlineto{\pgfqpoint{3.015217in}{1.409380in}}%
\pgfpathlineto{\pgfqpoint{3.023641in}{1.428054in}}%
\pgfpathlineto{\pgfqpoint{3.032065in}{1.440374in}}%
\pgfpathlineto{\pgfqpoint{3.040489in}{1.393921in}}%
\pgfpathlineto{\pgfqpoint{3.048913in}{1.386799in}}%
\pgfpathlineto{\pgfqpoint{3.057337in}{1.431853in}}%
\pgfpathlineto{\pgfqpoint{3.065761in}{1.425436in}}%
\pgfpathlineto{\pgfqpoint{3.074185in}{1.431198in}}%
\pgfpathlineto{\pgfqpoint{3.082609in}{1.455232in}}%
\pgfpathlineto{\pgfqpoint{3.091033in}{1.458350in}}%
\pgfpathlineto{\pgfqpoint{3.099457in}{1.501780in}}%
\pgfpathlineto{\pgfqpoint{3.107880in}{1.451279in}}%
\pgfpathlineto{\pgfqpoint{3.116304in}{1.457689in}}%
\pgfpathlineto{\pgfqpoint{3.124728in}{1.450148in}}%
\pgfpathlineto{\pgfqpoint{3.141576in}{1.456528in}}%
\pgfpathlineto{\pgfqpoint{3.150000in}{1.438230in}}%
\pgfpathlineto{\pgfqpoint{3.158424in}{1.473037in}}%
\pgfpathlineto{\pgfqpoint{3.158424in}{1.473037in}}%
\pgfusepath{stroke}%
\end{pgfscope}%
\begin{pgfscope}%
\pgfpathrectangle{\pgfqpoint{0.437500in}{0.330000in}}{\pgfqpoint{2.712500in}{2.310000in}}%
\pgfusepath{clip}%
\pgfsetbuttcap%
\pgfsetroundjoin%
\pgfsetlinewidth{3.011250pt}%
\definecolor{currentstroke}{rgb}{0.121569,0.466667,0.705882}%
\pgfsetstrokecolor{currentstroke}%
\pgfsetdash{{11.100000pt}{4.800000pt}}{0.000000pt}%
\pgfpathmoveto{\pgfqpoint{0.429076in}{2.465251in}}%
\pgfpathlineto{\pgfqpoint{0.471196in}{2.400609in}}%
\pgfpathlineto{\pgfqpoint{0.513315in}{2.340539in}}%
\pgfpathlineto{\pgfqpoint{0.555435in}{2.284531in}}%
\pgfpathlineto{\pgfqpoint{0.597554in}{2.232156in}}%
\pgfpathlineto{\pgfqpoint{0.639674in}{2.183050in}}%
\pgfpathlineto{\pgfqpoint{0.681793in}{2.136901in}}%
\pgfpathlineto{\pgfqpoint{0.732337in}{2.085048in}}%
\pgfpathlineto{\pgfqpoint{0.782880in}{2.036667in}}%
\pgfpathlineto{\pgfqpoint{0.833424in}{1.991416in}}%
\pgfpathlineto{\pgfqpoint{0.883967in}{1.949003in}}%
\pgfpathlineto{\pgfqpoint{0.934511in}{1.909173in}}%
\pgfpathlineto{\pgfqpoint{0.993478in}{1.865675in}}%
\pgfpathlineto{\pgfqpoint{1.018750in}{1.847939in}}%
\pgfpathlineto{\pgfqpoint{1.027174in}{1.600285in}}%
\pgfpathlineto{\pgfqpoint{1.035598in}{1.465590in}}%
\pgfpathlineto{\pgfqpoint{1.044022in}{1.375976in}}%
\pgfpathlineto{\pgfqpoint{1.052446in}{1.306290in}}%
\pgfpathlineto{\pgfqpoint{1.060870in}{1.248518in}}%
\pgfpathlineto{\pgfqpoint{1.077717in}{1.155217in}}%
\pgfpathlineto{\pgfqpoint{1.094565in}{1.080887in}}%
\pgfpathlineto{\pgfqpoint{1.111413in}{1.018972in}}%
\pgfpathlineto{\pgfqpoint{1.128261in}{0.965936in}}%
\pgfpathlineto{\pgfqpoint{1.145109in}{0.919622in}}%
\pgfpathlineto{\pgfqpoint{1.161957in}{0.878596in}}%
\pgfpathlineto{\pgfqpoint{1.178804in}{0.841855in}}%
\pgfpathlineto{\pgfqpoint{1.204076in}{0.793218in}}%
\pgfpathlineto{\pgfqpoint{1.229348in}{0.750817in}}%
\pgfpathlineto{\pgfqpoint{1.254620in}{0.713410in}}%
\pgfpathlineto{\pgfqpoint{1.279891in}{0.680091in}}%
\pgfpathlineto{\pgfqpoint{1.305163in}{0.650183in}}%
\pgfpathlineto{\pgfqpoint{1.330435in}{0.623159in}}%
\pgfpathlineto{\pgfqpoint{1.364130in}{0.590912in}}%
\pgfpathlineto{\pgfqpoint{1.397826in}{0.562300in}}%
\pgfpathlineto{\pgfqpoint{1.431522in}{0.536736in}}%
\pgfpathlineto{\pgfqpoint{1.465217in}{0.513759in}}%
\pgfpathlineto{\pgfqpoint{1.507337in}{0.488121in}}%
\pgfpathlineto{\pgfqpoint{1.549457in}{0.465384in}}%
\pgfpathlineto{\pgfqpoint{1.591576in}{0.445099in}}%
\pgfpathlineto{\pgfqpoint{1.642120in}{0.423496in}}%
\pgfpathlineto{\pgfqpoint{1.667391in}{0.413668in}}%
\pgfpathlineto{\pgfqpoint{1.675815in}{0.450705in}}%
\pgfpathlineto{\pgfqpoint{1.684239in}{0.498149in}}%
\pgfpathlineto{\pgfqpoint{1.692663in}{0.529742in}}%
\pgfpathlineto{\pgfqpoint{1.709511in}{0.578518in}}%
\pgfpathlineto{\pgfqpoint{1.726359in}{0.618211in}}%
\pgfpathlineto{\pgfqpoint{1.743207in}{0.652830in}}%
\pgfpathlineto{\pgfqpoint{1.768478in}{0.698642in}}%
\pgfpathlineto{\pgfqpoint{1.793750in}{0.739363in}}%
\pgfpathlineto{\pgfqpoint{1.819022in}{0.776367in}}%
\pgfpathlineto{\pgfqpoint{1.852717in}{0.821285in}}%
\pgfpathlineto{\pgfqpoint{1.886413in}{0.862208in}}%
\pgfpathlineto{\pgfqpoint{1.920109in}{0.899897in}}%
\pgfpathlineto{\pgfqpoint{1.962228in}{0.943242in}}%
\pgfpathlineto{\pgfqpoint{2.004348in}{0.983064in}}%
\pgfpathlineto{\pgfqpoint{2.046467in}{1.019879in}}%
\pgfpathlineto{\pgfqpoint{2.088587in}{1.054077in}}%
\pgfpathlineto{\pgfqpoint{2.139130in}{1.092077in}}%
\pgfpathlineto{\pgfqpoint{2.189674in}{1.127148in}}%
\pgfpathlineto{\pgfqpoint{2.240217in}{1.159612in}}%
\pgfpathlineto{\pgfqpoint{2.290761in}{1.189728in}}%
\pgfpathlineto{\pgfqpoint{2.349728in}{1.222181in}}%
\pgfpathlineto{\pgfqpoint{2.408696in}{1.252009in}}%
\pgfpathlineto{\pgfqpoint{2.467663in}{1.279443in}}%
\pgfpathlineto{\pgfqpoint{2.526630in}{1.304676in}}%
\pgfpathlineto{\pgfqpoint{2.594022in}{1.331024in}}%
\pgfpathlineto{\pgfqpoint{2.661413in}{1.354914in}}%
\pgfpathlineto{\pgfqpoint{2.728804in}{1.376517in}}%
\pgfpathlineto{\pgfqpoint{2.796196in}{1.395976in}}%
\pgfpathlineto{\pgfqpoint{2.872011in}{1.415460in}}%
\pgfpathlineto{\pgfqpoint{2.947826in}{1.432534in}}%
\pgfpathlineto{\pgfqpoint{3.023641in}{1.447320in}}%
\pgfpathlineto{\pgfqpoint{3.099457in}{1.459921in}}%
\pgfpathlineto{\pgfqpoint{3.158424in}{1.468264in}}%
\pgfpathlineto{\pgfqpoint{3.158424in}{1.468264in}}%
\pgfusepath{stroke}%
\end{pgfscope}%
\begin{pgfscope}%
\pgfpathrectangle{\pgfqpoint{0.437500in}{0.330000in}}{\pgfqpoint{2.712500in}{2.310000in}}%
\pgfusepath{clip}%
\pgfsetbuttcap%
\pgfsetroundjoin%
\pgfsetlinewidth{3.011250pt}%
\definecolor{currentstroke}{rgb}{1.000000,0.498039,0.054902}%
\pgfsetstrokecolor{currentstroke}%
\pgfsetdash{{3.000000pt}{4.950000pt}}{0.000000pt}%
\pgfpathmoveto{\pgfqpoint{0.429076in}{2.394346in}}%
\pgfpathlineto{\pgfqpoint{0.471196in}{2.328124in}}%
\pgfpathlineto{\pgfqpoint{0.513315in}{2.266584in}}%
\pgfpathlineto{\pgfqpoint{0.555435in}{2.209205in}}%
\pgfpathlineto{\pgfqpoint{0.597554in}{2.155546in}}%
\pgfpathlineto{\pgfqpoint{0.639674in}{2.105234in}}%
\pgfpathlineto{\pgfqpoint{0.681793in}{2.057951in}}%
\pgfpathlineto{\pgfqpoint{0.732337in}{2.004822in}}%
\pgfpathlineto{\pgfqpoint{0.782880in}{1.955250in}}%
\pgfpathlineto{\pgfqpoint{0.833424in}{1.908883in}}%
\pgfpathlineto{\pgfqpoint{0.883967in}{1.865423in}}%
\pgfpathlineto{\pgfqpoint{0.934511in}{1.824609in}}%
\pgfpathlineto{\pgfqpoint{0.993478in}{1.780033in}}%
\pgfpathlineto{\pgfqpoint{1.018750in}{1.761858in}}%
\pgfpathlineto{\pgfqpoint{1.027174in}{1.544270in}}%
\pgfpathlineto{\pgfqpoint{1.035598in}{1.423366in}}%
\pgfpathlineto{\pgfqpoint{1.044022in}{1.341741in}}%
\pgfpathlineto{\pgfqpoint{1.052446in}{1.277566in}}%
\pgfpathlineto{\pgfqpoint{1.060870in}{1.223871in}}%
\pgfpathlineto{\pgfqpoint{1.077717in}{1.136155in}}%
\pgfpathlineto{\pgfqpoint{1.094565in}{1.065330in}}%
\pgfpathlineto{\pgfqpoint{1.111413in}{1.005649in}}%
\pgfpathlineto{\pgfqpoint{1.128261in}{0.953999in}}%
\pgfpathlineto{\pgfqpoint{1.145109in}{0.908470in}}%
\pgfpathlineto{\pgfqpoint{1.161957in}{0.867793in}}%
\pgfpathlineto{\pgfqpoint{1.187228in}{0.813983in}}%
\pgfpathlineto{\pgfqpoint{1.212500in}{0.767023in}}%
\pgfpathlineto{\pgfqpoint{1.237772in}{0.725494in}}%
\pgfpathlineto{\pgfqpoint{1.263043in}{0.688387in}}%
\pgfpathlineto{\pgfqpoint{1.288315in}{0.654956in}}%
\pgfpathlineto{\pgfqpoint{1.313587in}{0.624629in}}%
\pgfpathlineto{\pgfqpoint{1.347283in}{0.588266in}}%
\pgfpathlineto{\pgfqpoint{1.380978in}{0.555819in}}%
\pgfpathlineto{\pgfqpoint{1.414674in}{0.526664in}}%
\pgfpathlineto{\pgfqpoint{1.448370in}{0.500310in}}%
\pgfpathlineto{\pgfqpoint{1.482065in}{0.476367in}}%
\pgfpathlineto{\pgfqpoint{1.524185in}{0.449353in}}%
\pgfpathlineto{\pgfqpoint{1.566304in}{0.425118in}}%
\pgfpathlineto{\pgfqpoint{1.608424in}{0.403266in}}%
\pgfpathlineto{\pgfqpoint{1.658967in}{0.379740in}}%
\pgfpathlineto{\pgfqpoint{1.667391in}{0.376074in}}%
\pgfpathlineto{\pgfqpoint{1.675815in}{0.419000in}}%
\pgfpathlineto{\pgfqpoint{1.684239in}{0.473266in}}%
\pgfpathlineto{\pgfqpoint{1.692663in}{0.509045in}}%
\pgfpathlineto{\pgfqpoint{1.709511in}{0.563796in}}%
\pgfpathlineto{\pgfqpoint{1.726359in}{0.607965in}}%
\pgfpathlineto{\pgfqpoint{1.743207in}{0.646237in}}%
\pgfpathlineto{\pgfqpoint{1.768478in}{0.696558in}}%
\pgfpathlineto{\pgfqpoint{1.793750in}{0.741007in}}%
\pgfpathlineto{\pgfqpoint{1.819022in}{0.781192in}}%
\pgfpathlineto{\pgfqpoint{1.852717in}{0.829729in}}%
\pgfpathlineto{\pgfqpoint{1.886413in}{0.873738in}}%
\pgfpathlineto{\pgfqpoint{1.920109in}{0.914105in}}%
\pgfpathlineto{\pgfqpoint{1.953804in}{0.951438in}}%
\pgfpathlineto{\pgfqpoint{1.995924in}{0.994497in}}%
\pgfpathlineto{\pgfqpoint{2.038043in}{1.034131in}}%
\pgfpathlineto{\pgfqpoint{2.080163in}{1.070811in}}%
\pgfpathlineto{\pgfqpoint{2.122283in}{1.104898in}}%
\pgfpathlineto{\pgfqpoint{2.172826in}{1.142779in}}%
\pgfpathlineto{\pgfqpoint{2.223370in}{1.177732in}}%
\pgfpathlineto{\pgfqpoint{2.273913in}{1.210069in}}%
\pgfpathlineto{\pgfqpoint{2.324457in}{1.240048in}}%
\pgfpathlineto{\pgfqpoint{2.383424in}{1.272321in}}%
\pgfpathlineto{\pgfqpoint{2.442391in}{1.301948in}}%
\pgfpathlineto{\pgfqpoint{2.501359in}{1.329160in}}%
\pgfpathlineto{\pgfqpoint{2.560326in}{1.354148in}}%
\pgfpathlineto{\pgfqpoint{2.627717in}{1.380191in}}%
\pgfpathlineto{\pgfqpoint{2.695109in}{1.403747in}}%
\pgfpathlineto{\pgfqpoint{2.762500in}{1.424988in}}%
\pgfpathlineto{\pgfqpoint{2.829891in}{1.444059in}}%
\pgfpathlineto{\pgfqpoint{2.905707in}{1.463073in}}%
\pgfpathlineto{\pgfqpoint{2.981522in}{1.479642in}}%
\pgfpathlineto{\pgfqpoint{3.057337in}{1.493889in}}%
\pgfpathlineto{\pgfqpoint{3.133152in}{1.505913in}}%
\pgfpathlineto{\pgfqpoint{3.158424in}{1.509441in}}%
\pgfpathlineto{\pgfqpoint{3.158424in}{1.509441in}}%
\pgfusepath{stroke}%
\end{pgfscope}%
\begin{pgfscope}%
\pgfsetrectcap%
\pgfsetmiterjoin%
\pgfsetlinewidth{0.803000pt}%
\definecolor{currentstroke}{rgb}{0.000000,0.000000,0.000000}%
\pgfsetstrokecolor{currentstroke}%
\pgfsetdash{}{0pt}%
\pgfpathmoveto{\pgfqpoint{0.437500in}{0.330000in}}%
\pgfpathlineto{\pgfqpoint{0.437500in}{2.640000in}}%
\pgfusepath{stroke}%
\end{pgfscope}%
\begin{pgfscope}%
\pgfsetrectcap%
\pgfsetmiterjoin%
\pgfsetlinewidth{0.803000pt}%
\definecolor{currentstroke}{rgb}{0.000000,0.000000,0.000000}%
\pgfsetstrokecolor{currentstroke}%
\pgfsetdash{}{0pt}%
\pgfpathmoveto{\pgfqpoint{3.150000in}{0.330000in}}%
\pgfpathlineto{\pgfqpoint{3.150000in}{2.640000in}}%
\pgfusepath{stroke}%
\end{pgfscope}%
\begin{pgfscope}%
\pgfsetrectcap%
\pgfsetmiterjoin%
\pgfsetlinewidth{0.803000pt}%
\definecolor{currentstroke}{rgb}{0.000000,0.000000,0.000000}%
\pgfsetstrokecolor{currentstroke}%
\pgfsetdash{}{0pt}%
\pgfpathmoveto{\pgfqpoint{0.437500in}{0.330000in}}%
\pgfpathlineto{\pgfqpoint{3.150000in}{0.330000in}}%
\pgfusepath{stroke}%
\end{pgfscope}%
\begin{pgfscope}%
\pgfsetrectcap%
\pgfsetmiterjoin%
\pgfsetlinewidth{0.803000pt}%
\definecolor{currentstroke}{rgb}{0.000000,0.000000,0.000000}%
\pgfsetstrokecolor{currentstroke}%
\pgfsetdash{}{0pt}%
\pgfpathmoveto{\pgfqpoint{0.437500in}{2.640000in}}%
\pgfpathlineto{\pgfqpoint{3.150000in}{2.640000in}}%
\pgfusepath{stroke}%
\end{pgfscope}%
\end{pgfpicture}%
\makeatother%
\endgroup%

%% file: figures/source_det.pgf
\begingroup%
\makeatletter%
\begin{pgfpicture}%
\pgfpathrectangle{\pgfpointorigin}{\pgfqpoint{3.500000in}{2.500000in}}%
\pgfusepath{use as bounding box, clip}%
\begin{pgfscope}%
\pgfsetbuttcap%
\pgfsetmiterjoin%
\definecolor{currentfill}{rgb}{1.000000,1.000000,1.000000}%
\pgfsetfillcolor{currentfill}%
\pgfsetlinewidth{0.000000pt}%
\definecolor{currentstroke}{rgb}{1.000000,1.000000,1.000000}%
\pgfsetstrokecolor{currentstroke}%
\pgfsetdash{}{0pt}%
\pgfpathmoveto{\pgfqpoint{0.000000in}{0.000000in}}%
\pgfpathlineto{\pgfqpoint{3.500000in}{0.000000in}}%
\pgfpathlineto{\pgfqpoint{3.500000in}{2.500000in}}%
\pgfpathlineto{\pgfqpoint{0.000000in}{2.500000in}}%
\pgfpathclose%
\pgfusepath{fill}%
\end{pgfscope}%
\begin{pgfscope}%
\pgfsetbuttcap%
\pgfsetmiterjoin%
\definecolor{currentfill}{rgb}{1.000000,1.000000,1.000000}%
\pgfsetfillcolor{currentfill}%
\pgfsetlinewidth{0.000000pt}%
\definecolor{currentstroke}{rgb}{0.000000,0.000000,0.000000}%
\pgfsetstrokecolor{currentstroke}%
\pgfsetstrokeopacity{0.000000}%
\pgfsetdash{}{0pt}%
\pgfpathmoveto{\pgfqpoint{0.465972in}{0.510778in}}%
\pgfpathlineto{\pgfqpoint{3.380000in}{0.510778in}}%
\pgfpathlineto{\pgfqpoint{3.380000in}{2.186667in}}%
\pgfpathlineto{\pgfqpoint{0.465972in}{2.186667in}}%
\pgfpathclose%
\pgfusepath{fill}%
\end{pgfscope}%
\begin{pgfscope}%
\pgfsetbuttcap%
\pgfsetroundjoin%
\definecolor{currentfill}{rgb}{0.000000,0.000000,0.000000}%
\pgfsetfillcolor{currentfill}%
\pgfsetlinewidth{0.803000pt}%
\definecolor{currentstroke}{rgb}{0.000000,0.000000,0.000000}%
\pgfsetstrokecolor{currentstroke}%
\pgfsetdash{}{0pt}%
\pgfsys@defobject{currentmarker}{\pgfqpoint{0.000000in}{-0.048611in}}{\pgfqpoint{0.000000in}{0.000000in}}{%
\pgfpathmoveto{\pgfqpoint{0.000000in}{0.000000in}}%
\pgfpathlineto{\pgfqpoint{0.000000in}{-0.048611in}}%
\pgfusepath{stroke,fill}%
}%
\begin{pgfscope}%
\pgfsys@transformshift{0.598428in}{0.510778in}%
\pgfsys@useobject{currentmarker}{}%
\end{pgfscope}%
\end{pgfscope}%
\begin{pgfscope}%
\definecolor{textcolor}{rgb}{0.000000,0.000000,0.000000}%
\pgfsetstrokecolor{textcolor}%
\pgfsetfillcolor{textcolor}%
\pgftext[x=0.598428in,y=0.413556in,,top]{\color{textcolor}\sffamily\fontsize{8.000000}{9.600000}\selectfont \(\displaystyle {10^{-2}}\)}%
\end{pgfscope}%
\begin{pgfscope}%
\pgfsetbuttcap%
\pgfsetroundjoin%
\definecolor{currentfill}{rgb}{0.000000,0.000000,0.000000}%
\pgfsetfillcolor{currentfill}%
\pgfsetlinewidth{0.803000pt}%
\definecolor{currentstroke}{rgb}{0.000000,0.000000,0.000000}%
\pgfsetstrokecolor{currentstroke}%
\pgfsetdash{}{0pt}%
\pgfsys@defobject{currentmarker}{\pgfqpoint{0.000000in}{-0.048611in}}{\pgfqpoint{0.000000in}{0.000000in}}{%
\pgfpathmoveto{\pgfqpoint{0.000000in}{0.000000in}}%
\pgfpathlineto{\pgfqpoint{0.000000in}{-0.048611in}}%
\pgfusepath{stroke,fill}%
}%
\begin{pgfscope}%
\pgfsys@transformshift{1.530255in}{0.510778in}%
\pgfsys@useobject{currentmarker}{}%
\end{pgfscope}%
\end{pgfscope}%
\begin{pgfscope}%
\definecolor{textcolor}{rgb}{0.000000,0.000000,0.000000}%
\pgfsetstrokecolor{textcolor}%
\pgfsetfillcolor{textcolor}%
\pgftext[x=1.530255in,y=0.413556in,,top]{\color{textcolor}\sffamily\fontsize{8.000000}{9.600000}\selectfont \(\displaystyle {10^{-1}}\)}%
\end{pgfscope}%
\begin{pgfscope}%
\pgfsetbuttcap%
\pgfsetroundjoin%
\definecolor{currentfill}{rgb}{0.000000,0.000000,0.000000}%
\pgfsetfillcolor{currentfill}%
\pgfsetlinewidth{0.803000pt}%
\definecolor{currentstroke}{rgb}{0.000000,0.000000,0.000000}%
\pgfsetstrokecolor{currentstroke}%
\pgfsetdash{}{0pt}%
\pgfsys@defobject{currentmarker}{\pgfqpoint{0.000000in}{-0.048611in}}{\pgfqpoint{0.000000in}{0.000000in}}{%
\pgfpathmoveto{\pgfqpoint{0.000000in}{0.000000in}}%
\pgfpathlineto{\pgfqpoint{0.000000in}{-0.048611in}}%
\pgfusepath{stroke,fill}%
}%
\begin{pgfscope}%
\pgfsys@transformshift{2.462082in}{0.510778in}%
\pgfsys@useobject{currentmarker}{}%
\end{pgfscope}%
\end{pgfscope}%
\begin{pgfscope}%
\definecolor{textcolor}{rgb}{0.000000,0.000000,0.000000}%
\pgfsetstrokecolor{textcolor}%
\pgfsetfillcolor{textcolor}%
\pgftext[x=2.462082in,y=0.413556in,,top]{\color{textcolor}\sffamily\fontsize{8.000000}{9.600000}\selectfont \(\displaystyle {10^{0}}\)}%
\end{pgfscope}%
\begin{pgfscope}%
\pgfsetbuttcap%
\pgfsetroundjoin%
\definecolor{currentfill}{rgb}{0.000000,0.000000,0.000000}%
\pgfsetfillcolor{currentfill}%
\pgfsetlinewidth{0.602250pt}%
\definecolor{currentstroke}{rgb}{0.000000,0.000000,0.000000}%
\pgfsetstrokecolor{currentstroke}%
\pgfsetdash{}{0pt}%
\pgfsys@defobject{currentmarker}{\pgfqpoint{0.000000in}{-0.027778in}}{\pgfqpoint{0.000000in}{0.000000in}}{%
\pgfpathmoveto{\pgfqpoint{0.000000in}{0.000000in}}%
\pgfpathlineto{\pgfqpoint{0.000000in}{-0.027778in}}%
\pgfusepath{stroke,fill}%
}%
\begin{pgfscope}%
\pgfsys@transformshift{0.508125in}{0.510778in}%
\pgfsys@useobject{currentmarker}{}%
\end{pgfscope}%
\end{pgfscope}%
\begin{pgfscope}%
\pgfsetbuttcap%
\pgfsetroundjoin%
\definecolor{currentfill}{rgb}{0.000000,0.000000,0.000000}%
\pgfsetfillcolor{currentfill}%
\pgfsetlinewidth{0.602250pt}%
\definecolor{currentstroke}{rgb}{0.000000,0.000000,0.000000}%
\pgfsetstrokecolor{currentstroke}%
\pgfsetdash{}{0pt}%
\pgfsys@defobject{currentmarker}{\pgfqpoint{0.000000in}{-0.027778in}}{\pgfqpoint{0.000000in}{0.000000in}}{%
\pgfpathmoveto{\pgfqpoint{0.000000in}{0.000000in}}%
\pgfpathlineto{\pgfqpoint{0.000000in}{-0.027778in}}%
\pgfusepath{stroke,fill}%
}%
\begin{pgfscope}%
\pgfsys@transformshift{0.555790in}{0.510778in}%
\pgfsys@useobject{currentmarker}{}%
\end{pgfscope}%
\end{pgfscope}%
\begin{pgfscope}%
\pgfsetbuttcap%
\pgfsetroundjoin%
\definecolor{currentfill}{rgb}{0.000000,0.000000,0.000000}%
\pgfsetfillcolor{currentfill}%
\pgfsetlinewidth{0.602250pt}%
\definecolor{currentstroke}{rgb}{0.000000,0.000000,0.000000}%
\pgfsetstrokecolor{currentstroke}%
\pgfsetdash{}{0pt}%
\pgfsys@defobject{currentmarker}{\pgfqpoint{0.000000in}{-0.027778in}}{\pgfqpoint{0.000000in}{0.000000in}}{%
\pgfpathmoveto{\pgfqpoint{0.000000in}{0.000000in}}%
\pgfpathlineto{\pgfqpoint{0.000000in}{-0.027778in}}%
\pgfusepath{stroke,fill}%
}%
\begin{pgfscope}%
\pgfsys@transformshift{0.878936in}{0.510778in}%
\pgfsys@useobject{currentmarker}{}%
\end{pgfscope}%
\end{pgfscope}%
\begin{pgfscope}%
\pgfsetbuttcap%
\pgfsetroundjoin%
\definecolor{currentfill}{rgb}{0.000000,0.000000,0.000000}%
\pgfsetfillcolor{currentfill}%
\pgfsetlinewidth{0.602250pt}%
\definecolor{currentstroke}{rgb}{0.000000,0.000000,0.000000}%
\pgfsetstrokecolor{currentstroke}%
\pgfsetdash{}{0pt}%
\pgfsys@defobject{currentmarker}{\pgfqpoint{0.000000in}{-0.027778in}}{\pgfqpoint{0.000000in}{0.000000in}}{%
\pgfpathmoveto{\pgfqpoint{0.000000in}{0.000000in}}%
\pgfpathlineto{\pgfqpoint{0.000000in}{-0.027778in}}%
\pgfusepath{stroke,fill}%
}%
\begin{pgfscope}%
\pgfsys@transformshift{1.043022in}{0.510778in}%
\pgfsys@useobject{currentmarker}{}%
\end{pgfscope}%
\end{pgfscope}%
\begin{pgfscope}%
\pgfsetbuttcap%
\pgfsetroundjoin%
\definecolor{currentfill}{rgb}{0.000000,0.000000,0.000000}%
\pgfsetfillcolor{currentfill}%
\pgfsetlinewidth{0.602250pt}%
\definecolor{currentstroke}{rgb}{0.000000,0.000000,0.000000}%
\pgfsetstrokecolor{currentstroke}%
\pgfsetdash{}{0pt}%
\pgfsys@defobject{currentmarker}{\pgfqpoint{0.000000in}{-0.027778in}}{\pgfqpoint{0.000000in}{0.000000in}}{%
\pgfpathmoveto{\pgfqpoint{0.000000in}{0.000000in}}%
\pgfpathlineto{\pgfqpoint{0.000000in}{-0.027778in}}%
\pgfusepath{stroke,fill}%
}%
\begin{pgfscope}%
\pgfsys@transformshift{1.159444in}{0.510778in}%
\pgfsys@useobject{currentmarker}{}%
\end{pgfscope}%
\end{pgfscope}%
\begin{pgfscope}%
\pgfsetbuttcap%
\pgfsetroundjoin%
\definecolor{currentfill}{rgb}{0.000000,0.000000,0.000000}%
\pgfsetfillcolor{currentfill}%
\pgfsetlinewidth{0.602250pt}%
\definecolor{currentstroke}{rgb}{0.000000,0.000000,0.000000}%
\pgfsetstrokecolor{currentstroke}%
\pgfsetdash{}{0pt}%
\pgfsys@defobject{currentmarker}{\pgfqpoint{0.000000in}{-0.027778in}}{\pgfqpoint{0.000000in}{0.000000in}}{%
\pgfpathmoveto{\pgfqpoint{0.000000in}{0.000000in}}%
\pgfpathlineto{\pgfqpoint{0.000000in}{-0.027778in}}%
\pgfusepath{stroke,fill}%
}%
\begin{pgfscope}%
\pgfsys@transformshift{1.249747in}{0.510778in}%
\pgfsys@useobject{currentmarker}{}%
\end{pgfscope}%
\end{pgfscope}%
\begin{pgfscope}%
\pgfsetbuttcap%
\pgfsetroundjoin%
\definecolor{currentfill}{rgb}{0.000000,0.000000,0.000000}%
\pgfsetfillcolor{currentfill}%
\pgfsetlinewidth{0.602250pt}%
\definecolor{currentstroke}{rgb}{0.000000,0.000000,0.000000}%
\pgfsetstrokecolor{currentstroke}%
\pgfsetdash{}{0pt}%
\pgfsys@defobject{currentmarker}{\pgfqpoint{0.000000in}{-0.027778in}}{\pgfqpoint{0.000000in}{0.000000in}}{%
\pgfpathmoveto{\pgfqpoint{0.000000in}{0.000000in}}%
\pgfpathlineto{\pgfqpoint{0.000000in}{-0.027778in}}%
\pgfusepath{stroke,fill}%
}%
\begin{pgfscope}%
\pgfsys@transformshift{1.323530in}{0.510778in}%
\pgfsys@useobject{currentmarker}{}%
\end{pgfscope}%
\end{pgfscope}%
\begin{pgfscope}%
\pgfsetbuttcap%
\pgfsetroundjoin%
\definecolor{currentfill}{rgb}{0.000000,0.000000,0.000000}%
\pgfsetfillcolor{currentfill}%
\pgfsetlinewidth{0.602250pt}%
\definecolor{currentstroke}{rgb}{0.000000,0.000000,0.000000}%
\pgfsetstrokecolor{currentstroke}%
\pgfsetdash{}{0pt}%
\pgfsys@defobject{currentmarker}{\pgfqpoint{0.000000in}{-0.027778in}}{\pgfqpoint{0.000000in}{0.000000in}}{%
\pgfpathmoveto{\pgfqpoint{0.000000in}{0.000000in}}%
\pgfpathlineto{\pgfqpoint{0.000000in}{-0.027778in}}%
\pgfusepath{stroke,fill}%
}%
\begin{pgfscope}%
\pgfsys@transformshift{1.385913in}{0.510778in}%
\pgfsys@useobject{currentmarker}{}%
\end{pgfscope}%
\end{pgfscope}%
\begin{pgfscope}%
\pgfsetbuttcap%
\pgfsetroundjoin%
\definecolor{currentfill}{rgb}{0.000000,0.000000,0.000000}%
\pgfsetfillcolor{currentfill}%
\pgfsetlinewidth{0.602250pt}%
\definecolor{currentstroke}{rgb}{0.000000,0.000000,0.000000}%
\pgfsetstrokecolor{currentstroke}%
\pgfsetdash{}{0pt}%
\pgfsys@defobject{currentmarker}{\pgfqpoint{0.000000in}{-0.027778in}}{\pgfqpoint{0.000000in}{0.000000in}}{%
\pgfpathmoveto{\pgfqpoint{0.000000in}{0.000000in}}%
\pgfpathlineto{\pgfqpoint{0.000000in}{-0.027778in}}%
\pgfusepath{stroke,fill}%
}%
\begin{pgfscope}%
\pgfsys@transformshift{1.439952in}{0.510778in}%
\pgfsys@useobject{currentmarker}{}%
\end{pgfscope}%
\end{pgfscope}%
\begin{pgfscope}%
\pgfsetbuttcap%
\pgfsetroundjoin%
\definecolor{currentfill}{rgb}{0.000000,0.000000,0.000000}%
\pgfsetfillcolor{currentfill}%
\pgfsetlinewidth{0.602250pt}%
\definecolor{currentstroke}{rgb}{0.000000,0.000000,0.000000}%
\pgfsetstrokecolor{currentstroke}%
\pgfsetdash{}{0pt}%
\pgfsys@defobject{currentmarker}{\pgfqpoint{0.000000in}{-0.027778in}}{\pgfqpoint{0.000000in}{0.000000in}}{%
\pgfpathmoveto{\pgfqpoint{0.000000in}{0.000000in}}%
\pgfpathlineto{\pgfqpoint{0.000000in}{-0.027778in}}%
\pgfusepath{stroke,fill}%
}%
\begin{pgfscope}%
\pgfsys@transformshift{1.487617in}{0.510778in}%
\pgfsys@useobject{currentmarker}{}%
\end{pgfscope}%
\end{pgfscope}%
\begin{pgfscope}%
\pgfsetbuttcap%
\pgfsetroundjoin%
\definecolor{currentfill}{rgb}{0.000000,0.000000,0.000000}%
\pgfsetfillcolor{currentfill}%
\pgfsetlinewidth{0.602250pt}%
\definecolor{currentstroke}{rgb}{0.000000,0.000000,0.000000}%
\pgfsetstrokecolor{currentstroke}%
\pgfsetdash{}{0pt}%
\pgfsys@defobject{currentmarker}{\pgfqpoint{0.000000in}{-0.027778in}}{\pgfqpoint{0.000000in}{0.000000in}}{%
\pgfpathmoveto{\pgfqpoint{0.000000in}{0.000000in}}%
\pgfpathlineto{\pgfqpoint{0.000000in}{-0.027778in}}%
\pgfusepath{stroke,fill}%
}%
\begin{pgfscope}%
\pgfsys@transformshift{1.810763in}{0.510778in}%
\pgfsys@useobject{currentmarker}{}%
\end{pgfscope}%
\end{pgfscope}%
\begin{pgfscope}%
\pgfsetbuttcap%
\pgfsetroundjoin%
\definecolor{currentfill}{rgb}{0.000000,0.000000,0.000000}%
\pgfsetfillcolor{currentfill}%
\pgfsetlinewidth{0.602250pt}%
\definecolor{currentstroke}{rgb}{0.000000,0.000000,0.000000}%
\pgfsetstrokecolor{currentstroke}%
\pgfsetdash{}{0pt}%
\pgfsys@defobject{currentmarker}{\pgfqpoint{0.000000in}{-0.027778in}}{\pgfqpoint{0.000000in}{0.000000in}}{%
\pgfpathmoveto{\pgfqpoint{0.000000in}{0.000000in}}%
\pgfpathlineto{\pgfqpoint{0.000000in}{-0.027778in}}%
\pgfusepath{stroke,fill}%
}%
\begin{pgfscope}%
\pgfsys@transformshift{1.974849in}{0.510778in}%
\pgfsys@useobject{currentmarker}{}%
\end{pgfscope}%
\end{pgfscope}%
\begin{pgfscope}%
\pgfsetbuttcap%
\pgfsetroundjoin%
\definecolor{currentfill}{rgb}{0.000000,0.000000,0.000000}%
\pgfsetfillcolor{currentfill}%
\pgfsetlinewidth{0.602250pt}%
\definecolor{currentstroke}{rgb}{0.000000,0.000000,0.000000}%
\pgfsetstrokecolor{currentstroke}%
\pgfsetdash{}{0pt}%
\pgfsys@defobject{currentmarker}{\pgfqpoint{0.000000in}{-0.027778in}}{\pgfqpoint{0.000000in}{0.000000in}}{%
\pgfpathmoveto{\pgfqpoint{0.000000in}{0.000000in}}%
\pgfpathlineto{\pgfqpoint{0.000000in}{-0.027778in}}%
\pgfusepath{stroke,fill}%
}%
\begin{pgfscope}%
\pgfsys@transformshift{2.091271in}{0.510778in}%
\pgfsys@useobject{currentmarker}{}%
\end{pgfscope}%
\end{pgfscope}%
\begin{pgfscope}%
\pgfsetbuttcap%
\pgfsetroundjoin%
\definecolor{currentfill}{rgb}{0.000000,0.000000,0.000000}%
\pgfsetfillcolor{currentfill}%
\pgfsetlinewidth{0.602250pt}%
\definecolor{currentstroke}{rgb}{0.000000,0.000000,0.000000}%
\pgfsetstrokecolor{currentstroke}%
\pgfsetdash{}{0pt}%
\pgfsys@defobject{currentmarker}{\pgfqpoint{0.000000in}{-0.027778in}}{\pgfqpoint{0.000000in}{0.000000in}}{%
\pgfpathmoveto{\pgfqpoint{0.000000in}{0.000000in}}%
\pgfpathlineto{\pgfqpoint{0.000000in}{-0.027778in}}%
\pgfusepath{stroke,fill}%
}%
\begin{pgfscope}%
\pgfsys@transformshift{2.181574in}{0.510778in}%
\pgfsys@useobject{currentmarker}{}%
\end{pgfscope}%
\end{pgfscope}%
\begin{pgfscope}%
\pgfsetbuttcap%
\pgfsetroundjoin%
\definecolor{currentfill}{rgb}{0.000000,0.000000,0.000000}%
\pgfsetfillcolor{currentfill}%
\pgfsetlinewidth{0.602250pt}%
\definecolor{currentstroke}{rgb}{0.000000,0.000000,0.000000}%
\pgfsetstrokecolor{currentstroke}%
\pgfsetdash{}{0pt}%
\pgfsys@defobject{currentmarker}{\pgfqpoint{0.000000in}{-0.027778in}}{\pgfqpoint{0.000000in}{0.000000in}}{%
\pgfpathmoveto{\pgfqpoint{0.000000in}{0.000000in}}%
\pgfpathlineto{\pgfqpoint{0.000000in}{-0.027778in}}%
\pgfusepath{stroke,fill}%
}%
\begin{pgfscope}%
\pgfsys@transformshift{2.255357in}{0.510778in}%
\pgfsys@useobject{currentmarker}{}%
\end{pgfscope}%
\end{pgfscope}%
\begin{pgfscope}%
\pgfsetbuttcap%
\pgfsetroundjoin%
\definecolor{currentfill}{rgb}{0.000000,0.000000,0.000000}%
\pgfsetfillcolor{currentfill}%
\pgfsetlinewidth{0.602250pt}%
\definecolor{currentstroke}{rgb}{0.000000,0.000000,0.000000}%
\pgfsetstrokecolor{currentstroke}%
\pgfsetdash{}{0pt}%
\pgfsys@defobject{currentmarker}{\pgfqpoint{0.000000in}{-0.027778in}}{\pgfqpoint{0.000000in}{0.000000in}}{%
\pgfpathmoveto{\pgfqpoint{0.000000in}{0.000000in}}%
\pgfpathlineto{\pgfqpoint{0.000000in}{-0.027778in}}%
\pgfusepath{stroke,fill}%
}%
\begin{pgfscope}%
\pgfsys@transformshift{2.317740in}{0.510778in}%
\pgfsys@useobject{currentmarker}{}%
\end{pgfscope}%
\end{pgfscope}%
\begin{pgfscope}%
\pgfsetbuttcap%
\pgfsetroundjoin%
\definecolor{currentfill}{rgb}{0.000000,0.000000,0.000000}%
\pgfsetfillcolor{currentfill}%
\pgfsetlinewidth{0.602250pt}%
\definecolor{currentstroke}{rgb}{0.000000,0.000000,0.000000}%
\pgfsetstrokecolor{currentstroke}%
\pgfsetdash{}{0pt}%
\pgfsys@defobject{currentmarker}{\pgfqpoint{0.000000in}{-0.027778in}}{\pgfqpoint{0.000000in}{0.000000in}}{%
\pgfpathmoveto{\pgfqpoint{0.000000in}{0.000000in}}%
\pgfpathlineto{\pgfqpoint{0.000000in}{-0.027778in}}%
\pgfusepath{stroke,fill}%
}%
\begin{pgfscope}%
\pgfsys@transformshift{2.371778in}{0.510778in}%
\pgfsys@useobject{currentmarker}{}%
\end{pgfscope}%
\end{pgfscope}%
\begin{pgfscope}%
\pgfsetbuttcap%
\pgfsetroundjoin%
\definecolor{currentfill}{rgb}{0.000000,0.000000,0.000000}%
\pgfsetfillcolor{currentfill}%
\pgfsetlinewidth{0.602250pt}%
\definecolor{currentstroke}{rgb}{0.000000,0.000000,0.000000}%
\pgfsetstrokecolor{currentstroke}%
\pgfsetdash{}{0pt}%
\pgfsys@defobject{currentmarker}{\pgfqpoint{0.000000in}{-0.027778in}}{\pgfqpoint{0.000000in}{0.000000in}}{%
\pgfpathmoveto{\pgfqpoint{0.000000in}{0.000000in}}%
\pgfpathlineto{\pgfqpoint{0.000000in}{-0.027778in}}%
\pgfusepath{stroke,fill}%
}%
\begin{pgfscope}%
\pgfsys@transformshift{2.419444in}{0.510778in}%
\pgfsys@useobject{currentmarker}{}%
\end{pgfscope}%
\end{pgfscope}%
\begin{pgfscope}%
\pgfsetbuttcap%
\pgfsetroundjoin%
\definecolor{currentfill}{rgb}{0.000000,0.000000,0.000000}%
\pgfsetfillcolor{currentfill}%
\pgfsetlinewidth{0.602250pt}%
\definecolor{currentstroke}{rgb}{0.000000,0.000000,0.000000}%
\pgfsetstrokecolor{currentstroke}%
\pgfsetdash{}{0pt}%
\pgfsys@defobject{currentmarker}{\pgfqpoint{0.000000in}{-0.027778in}}{\pgfqpoint{0.000000in}{0.000000in}}{%
\pgfpathmoveto{\pgfqpoint{0.000000in}{0.000000in}}%
\pgfpathlineto{\pgfqpoint{0.000000in}{-0.027778in}}%
\pgfusepath{stroke,fill}%
}%
\begin{pgfscope}%
\pgfsys@transformshift{2.742590in}{0.510778in}%
\pgfsys@useobject{currentmarker}{}%
\end{pgfscope}%
\end{pgfscope}%
\begin{pgfscope}%
\pgfsetbuttcap%
\pgfsetroundjoin%
\definecolor{currentfill}{rgb}{0.000000,0.000000,0.000000}%
\pgfsetfillcolor{currentfill}%
\pgfsetlinewidth{0.602250pt}%
\definecolor{currentstroke}{rgb}{0.000000,0.000000,0.000000}%
\pgfsetstrokecolor{currentstroke}%
\pgfsetdash{}{0pt}%
\pgfsys@defobject{currentmarker}{\pgfqpoint{0.000000in}{-0.027778in}}{\pgfqpoint{0.000000in}{0.000000in}}{%
\pgfpathmoveto{\pgfqpoint{0.000000in}{0.000000in}}%
\pgfpathlineto{\pgfqpoint{0.000000in}{-0.027778in}}%
\pgfusepath{stroke,fill}%
}%
\begin{pgfscope}%
\pgfsys@transformshift{2.906676in}{0.510778in}%
\pgfsys@useobject{currentmarker}{}%
\end{pgfscope}%
\end{pgfscope}%
\begin{pgfscope}%
\pgfsetbuttcap%
\pgfsetroundjoin%
\definecolor{currentfill}{rgb}{0.000000,0.000000,0.000000}%
\pgfsetfillcolor{currentfill}%
\pgfsetlinewidth{0.602250pt}%
\definecolor{currentstroke}{rgb}{0.000000,0.000000,0.000000}%
\pgfsetstrokecolor{currentstroke}%
\pgfsetdash{}{0pt}%
\pgfsys@defobject{currentmarker}{\pgfqpoint{0.000000in}{-0.027778in}}{\pgfqpoint{0.000000in}{0.000000in}}{%
\pgfpathmoveto{\pgfqpoint{0.000000in}{0.000000in}}%
\pgfpathlineto{\pgfqpoint{0.000000in}{-0.027778in}}%
\pgfusepath{stroke,fill}%
}%
\begin{pgfscope}%
\pgfsys@transformshift{3.023098in}{0.510778in}%
\pgfsys@useobject{currentmarker}{}%
\end{pgfscope}%
\end{pgfscope}%
\begin{pgfscope}%
\pgfsetbuttcap%
\pgfsetroundjoin%
\definecolor{currentfill}{rgb}{0.000000,0.000000,0.000000}%
\pgfsetfillcolor{currentfill}%
\pgfsetlinewidth{0.602250pt}%
\definecolor{currentstroke}{rgb}{0.000000,0.000000,0.000000}%
\pgfsetstrokecolor{currentstroke}%
\pgfsetdash{}{0pt}%
\pgfsys@defobject{currentmarker}{\pgfqpoint{0.000000in}{-0.027778in}}{\pgfqpoint{0.000000in}{0.000000in}}{%
\pgfpathmoveto{\pgfqpoint{0.000000in}{0.000000in}}%
\pgfpathlineto{\pgfqpoint{0.000000in}{-0.027778in}}%
\pgfusepath{stroke,fill}%
}%
\begin{pgfscope}%
\pgfsys@transformshift{3.113401in}{0.510778in}%
\pgfsys@useobject{currentmarker}{}%
\end{pgfscope}%
\end{pgfscope}%
\begin{pgfscope}%
\pgfsetbuttcap%
\pgfsetroundjoin%
\definecolor{currentfill}{rgb}{0.000000,0.000000,0.000000}%
\pgfsetfillcolor{currentfill}%
\pgfsetlinewidth{0.602250pt}%
\definecolor{currentstroke}{rgb}{0.000000,0.000000,0.000000}%
\pgfsetstrokecolor{currentstroke}%
\pgfsetdash{}{0pt}%
\pgfsys@defobject{currentmarker}{\pgfqpoint{0.000000in}{-0.027778in}}{\pgfqpoint{0.000000in}{0.000000in}}{%
\pgfpathmoveto{\pgfqpoint{0.000000in}{0.000000in}}%
\pgfpathlineto{\pgfqpoint{0.000000in}{-0.027778in}}%
\pgfusepath{stroke,fill}%
}%
\begin{pgfscope}%
\pgfsys@transformshift{3.187184in}{0.510778in}%
\pgfsys@useobject{currentmarker}{}%
\end{pgfscope}%
\end{pgfscope}%
\begin{pgfscope}%
\pgfsetbuttcap%
\pgfsetroundjoin%
\definecolor{currentfill}{rgb}{0.000000,0.000000,0.000000}%
\pgfsetfillcolor{currentfill}%
\pgfsetlinewidth{0.602250pt}%
\definecolor{currentstroke}{rgb}{0.000000,0.000000,0.000000}%
\pgfsetstrokecolor{currentstroke}%
\pgfsetdash{}{0pt}%
\pgfsys@defobject{currentmarker}{\pgfqpoint{0.000000in}{-0.027778in}}{\pgfqpoint{0.000000in}{0.000000in}}{%
\pgfpathmoveto{\pgfqpoint{0.000000in}{0.000000in}}%
\pgfpathlineto{\pgfqpoint{0.000000in}{-0.027778in}}%
\pgfusepath{stroke,fill}%
}%
\begin{pgfscope}%
\pgfsys@transformshift{3.249567in}{0.510778in}%
\pgfsys@useobject{currentmarker}{}%
\end{pgfscope}%
\end{pgfscope}%
\begin{pgfscope}%
\pgfsetbuttcap%
\pgfsetroundjoin%
\definecolor{currentfill}{rgb}{0.000000,0.000000,0.000000}%
\pgfsetfillcolor{currentfill}%
\pgfsetlinewidth{0.602250pt}%
\definecolor{currentstroke}{rgb}{0.000000,0.000000,0.000000}%
\pgfsetstrokecolor{currentstroke}%
\pgfsetdash{}{0pt}%
\pgfsys@defobject{currentmarker}{\pgfqpoint{0.000000in}{-0.027778in}}{\pgfqpoint{0.000000in}{0.000000in}}{%
\pgfpathmoveto{\pgfqpoint{0.000000in}{0.000000in}}%
\pgfpathlineto{\pgfqpoint{0.000000in}{-0.027778in}}%
\pgfusepath{stroke,fill}%
}%
\begin{pgfscope}%
\pgfsys@transformshift{3.303605in}{0.510778in}%
\pgfsys@useobject{currentmarker}{}%
\end{pgfscope}%
\end{pgfscope}%
\begin{pgfscope}%
\pgfsetbuttcap%
\pgfsetroundjoin%
\definecolor{currentfill}{rgb}{0.000000,0.000000,0.000000}%
\pgfsetfillcolor{currentfill}%
\pgfsetlinewidth{0.602250pt}%
\definecolor{currentstroke}{rgb}{0.000000,0.000000,0.000000}%
\pgfsetstrokecolor{currentstroke}%
\pgfsetdash{}{0pt}%
\pgfsys@defobject{currentmarker}{\pgfqpoint{0.000000in}{-0.027778in}}{\pgfqpoint{0.000000in}{0.000000in}}{%
\pgfpathmoveto{\pgfqpoint{0.000000in}{0.000000in}}%
\pgfpathlineto{\pgfqpoint{0.000000in}{-0.027778in}}%
\pgfusepath{stroke,fill}%
}%
\begin{pgfscope}%
\pgfsys@transformshift{3.351271in}{0.510778in}%
\pgfsys@useobject{currentmarker}{}%
\end{pgfscope}%
\end{pgfscope}%
\begin{pgfscope}%
\definecolor{textcolor}{rgb}{0.000000,0.000000,0.000000}%
\pgfsetstrokecolor{textcolor}%
\pgfsetfillcolor{textcolor}%
\pgftext[x=1.922986in,y=0.258139in,,top]{\color{textcolor}\sffamily\fontsize{8.000000}{9.600000}\selectfont X-ray energy (MeV)}%
\end{pgfscope}%
\begin{pgfscope}%
\pgfsetbuttcap%
\pgfsetroundjoin%
\definecolor{currentfill}{rgb}{0.000000,0.000000,0.000000}%
\pgfsetfillcolor{currentfill}%
\pgfsetlinewidth{0.803000pt}%
\definecolor{currentstroke}{rgb}{0.000000,0.000000,0.000000}%
\pgfsetstrokecolor{currentstroke}%
\pgfsetdash{}{0pt}%
\pgfsys@defobject{currentmarker}{\pgfqpoint{-0.048611in}{0.000000in}}{\pgfqpoint{-0.000000in}{0.000000in}}{%
\pgfpathmoveto{\pgfqpoint{-0.000000in}{0.000000in}}%
\pgfpathlineto{\pgfqpoint{-0.048611in}{0.000000in}}%
\pgfusepath{stroke,fill}%
}%
\begin{pgfscope}%
\pgfsys@transformshift{0.465972in}{0.586955in}%
\pgfsys@useobject{currentmarker}{}%
\end{pgfscope}%
\end{pgfscope}%
\begin{pgfscope}%
\definecolor{textcolor}{rgb}{0.000000,0.000000,0.000000}%
\pgfsetstrokecolor{textcolor}%
\pgfsetfillcolor{textcolor}%
\pgftext[x=0.158972in, y=0.548399in, left, base]{\color{textcolor}\sffamily\fontsize{8.000000}{9.600000}\selectfont 0.00}%
\end{pgfscope}%
\begin{pgfscope}%
\pgfsetbuttcap%
\pgfsetroundjoin%
\definecolor{currentfill}{rgb}{0.000000,0.000000,0.000000}%
\pgfsetfillcolor{currentfill}%
\pgfsetlinewidth{0.803000pt}%
\definecolor{currentstroke}{rgb}{0.000000,0.000000,0.000000}%
\pgfsetstrokecolor{currentstroke}%
\pgfsetdash{}{0pt}%
\pgfsys@defobject{currentmarker}{\pgfqpoint{-0.048611in}{0.000000in}}{\pgfqpoint{-0.000000in}{0.000000in}}{%
\pgfpathmoveto{\pgfqpoint{-0.000000in}{0.000000in}}%
\pgfpathlineto{\pgfqpoint{-0.048611in}{0.000000in}}%
\pgfusepath{stroke,fill}%
}%
\begin{pgfscope}%
\pgfsys@transformshift{0.465972in}{1.014165in}%
\pgfsys@useobject{currentmarker}{}%
\end{pgfscope}%
\end{pgfscope}%
\begin{pgfscope}%
\definecolor{textcolor}{rgb}{0.000000,0.000000,0.000000}%
\pgfsetstrokecolor{textcolor}%
\pgfsetfillcolor{textcolor}%
\pgftext[x=0.158972in, y=0.975609in, left, base]{\color{textcolor}\sffamily\fontsize{8.000000}{9.600000}\selectfont 0.02}%
\end{pgfscope}%
\begin{pgfscope}%
\pgfsetbuttcap%
\pgfsetroundjoin%
\definecolor{currentfill}{rgb}{0.000000,0.000000,0.000000}%
\pgfsetfillcolor{currentfill}%
\pgfsetlinewidth{0.803000pt}%
\definecolor{currentstroke}{rgb}{0.000000,0.000000,0.000000}%
\pgfsetstrokecolor{currentstroke}%
\pgfsetdash{}{0pt}%
\pgfsys@defobject{currentmarker}{\pgfqpoint{-0.048611in}{0.000000in}}{\pgfqpoint{-0.000000in}{0.000000in}}{%
\pgfpathmoveto{\pgfqpoint{-0.000000in}{0.000000in}}%
\pgfpathlineto{\pgfqpoint{-0.048611in}{0.000000in}}%
\pgfusepath{stroke,fill}%
}%
\begin{pgfscope}%
\pgfsys@transformshift{0.465972in}{1.441376in}%
\pgfsys@useobject{currentmarker}{}%
\end{pgfscope}%
\end{pgfscope}%
\begin{pgfscope}%
\definecolor{textcolor}{rgb}{0.000000,0.000000,0.000000}%
\pgfsetstrokecolor{textcolor}%
\pgfsetfillcolor{textcolor}%
\pgftext[x=0.158972in, y=1.402820in, left, base]{\color{textcolor}\sffamily\fontsize{8.000000}{9.600000}\selectfont 0.04}%
\end{pgfscope}%
\begin{pgfscope}%
\pgfsetbuttcap%
\pgfsetroundjoin%
\definecolor{currentfill}{rgb}{0.000000,0.000000,0.000000}%
\pgfsetfillcolor{currentfill}%
\pgfsetlinewidth{0.803000pt}%
\definecolor{currentstroke}{rgb}{0.000000,0.000000,0.000000}%
\pgfsetstrokecolor{currentstroke}%
\pgfsetdash{}{0pt}%
\pgfsys@defobject{currentmarker}{\pgfqpoint{-0.048611in}{0.000000in}}{\pgfqpoint{-0.000000in}{0.000000in}}{%
\pgfpathmoveto{\pgfqpoint{-0.000000in}{0.000000in}}%
\pgfpathlineto{\pgfqpoint{-0.048611in}{0.000000in}}%
\pgfusepath{stroke,fill}%
}%
\begin{pgfscope}%
\pgfsys@transformshift{0.465972in}{1.868586in}%
\pgfsys@useobject{currentmarker}{}%
\end{pgfscope}%
\end{pgfscope}%
\begin{pgfscope}%
\definecolor{textcolor}{rgb}{0.000000,0.000000,0.000000}%
\pgfsetstrokecolor{textcolor}%
\pgfsetfillcolor{textcolor}%
\pgftext[x=0.158972in, y=1.830030in, left, base]{\color{textcolor}\sffamily\fontsize{8.000000}{9.600000}\selectfont 0.06}%
\end{pgfscope}%
\begin{pgfscope}%
\pgfpathrectangle{\pgfqpoint{0.465972in}{0.510778in}}{\pgfqpoint{2.914028in}{1.675889in}}%
\pgfusepath{clip}%
\pgfsetrectcap%
\pgfsetroundjoin%
\pgfsetlinewidth{3.011250pt}%
\definecolor{currentstroke}{rgb}{0.121569,0.466667,0.705882}%
\pgfsetstrokecolor{currentstroke}%
\pgfsetdash{}{0pt}%
\pgfpathmoveto{\pgfqpoint{0.598428in}{0.586955in}}%
\pgfpathlineto{\pgfqpoint{1.287581in}{0.636438in}}%
\pgfpathlineto{\pgfqpoint{1.619586in}{0.883458in}}%
\pgfpathlineto{\pgfqpoint{1.799478in}{1.641512in}}%
\pgfpathlineto{\pgfqpoint{1.923576in}{1.734536in}}%
\pgfpathlineto{\pgfqpoint{2.018417in}{1.536832in}}%
\pgfpathlineto{\pgfqpoint{2.095197in}{1.407779in}}%
\pgfpathlineto{\pgfqpoint{2.159707in}{1.260988in}}%
\pgfpathlineto{\pgfqpoint{2.215334in}{1.177578in}}%
\pgfpathlineto{\pgfqpoint{2.264230in}{1.108848in}}%
\pgfpathlineto{\pgfqpoint{2.307850in}{1.140234in}}%
\pgfpathlineto{\pgfqpoint{2.347222in}{1.045486in}}%
\pgfpathlineto{\pgfqpoint{2.383102in}{1.037807in}}%
\pgfpathlineto{\pgfqpoint{2.416057in}{1.025891in}}%
\pgfpathlineto{\pgfqpoint{2.446530in}{1.007674in}}%
\pgfpathlineto{\pgfqpoint{2.474868in}{0.970735in}}%
\pgfpathlineto{\pgfqpoint{2.501351in}{0.991934in}}%
\pgfpathlineto{\pgfqpoint{2.526207in}{0.938294in}}%
\pgfpathlineto{\pgfqpoint{2.549624in}{0.935613in}}%
\pgfpathlineto{\pgfqpoint{2.571760in}{0.891251in}}%
\pgfpathlineto{\pgfqpoint{2.592747in}{0.915770in}}%
\pgfpathlineto{\pgfqpoint{2.612700in}{0.923624in}}%
\pgfpathlineto{\pgfqpoint{2.631715in}{0.913478in}}%
\pgfpathlineto{\pgfqpoint{2.649876in}{0.899969in}}%
\pgfpathlineto{\pgfqpoint{2.667258in}{0.900762in}}%
\pgfpathlineto{\pgfqpoint{2.683923in}{0.889507in}}%
\pgfpathlineto{\pgfqpoint{2.699929in}{0.850035in}}%
\pgfpathlineto{\pgfqpoint{2.715326in}{0.809279in}}%
\pgfpathlineto{\pgfqpoint{2.730159in}{0.851072in}}%
\pgfpathlineto{\pgfqpoint{2.744467in}{0.896130in}}%
\pgfpathlineto{\pgfqpoint{2.758287in}{0.842546in}}%
\pgfpathlineto{\pgfqpoint{2.771650in}{0.829801in}}%
\pgfpathlineto{\pgfqpoint{2.784586in}{0.824731in}}%
\pgfpathlineto{\pgfqpoint{2.797121in}{0.797426in}}%
\pgfpathlineto{\pgfqpoint{2.809280in}{0.839249in}}%
\pgfpathlineto{\pgfqpoint{2.821084in}{0.814308in}}%
\pgfpathlineto{\pgfqpoint{2.832553in}{0.788673in}}%
\pgfpathlineto{\pgfqpoint{2.843706in}{0.819315in}}%
\pgfpathlineto{\pgfqpoint{2.854560in}{0.782878in}}%
\pgfpathlineto{\pgfqpoint{2.865131in}{0.826832in}}%
\pgfpathlineto{\pgfqpoint{2.875432in}{0.767265in}}%
\pgfpathlineto{\pgfqpoint{2.885478in}{0.770364in}}%
\pgfpathlineto{\pgfqpoint{2.895280in}{0.745601in}}%
\pgfpathlineto{\pgfqpoint{2.904851in}{0.770982in}}%
\pgfpathlineto{\pgfqpoint{2.914200in}{0.749126in}}%
\pgfpathlineto{\pgfqpoint{2.923339in}{0.753369in}}%
\pgfpathlineto{\pgfqpoint{2.932275in}{0.741504in}}%
\pgfpathlineto{\pgfqpoint{2.941019in}{0.747334in}}%
\pgfpathlineto{\pgfqpoint{2.949577in}{0.750561in}}%
\pgfpathlineto{\pgfqpoint{2.957959in}{0.744880in}}%
\pgfpathlineto{\pgfqpoint{2.966170in}{0.768007in}}%
\pgfpathlineto{\pgfqpoint{2.974218in}{0.773376in}}%
\pgfpathlineto{\pgfqpoint{2.982109in}{0.747025in}}%
\pgfpathlineto{\pgfqpoint{2.989849in}{0.714238in}}%
\pgfpathlineto{\pgfqpoint{2.997443in}{0.737168in}}%
\pgfpathlineto{\pgfqpoint{3.004898in}{0.735836in}}%
\pgfpathlineto{\pgfqpoint{3.012218in}{0.670844in}}%
\pgfpathlineto{\pgfqpoint{3.019408in}{0.713777in}}%
\pgfpathlineto{\pgfqpoint{3.026473in}{0.699937in}}%
\pgfpathlineto{\pgfqpoint{3.033416in}{0.705456in}}%
\pgfpathlineto{\pgfqpoint{3.040242in}{0.740260in}}%
\pgfpathlineto{\pgfqpoint{3.046955in}{0.695060in}}%
\pgfpathlineto{\pgfqpoint{3.053558in}{0.685442in}}%
\pgfpathlineto{\pgfqpoint{3.060056in}{0.707869in}}%
\pgfpathlineto{\pgfqpoint{3.066450in}{0.681686in}}%
\pgfpathlineto{\pgfqpoint{3.072745in}{0.681970in}}%
\pgfpathlineto{\pgfqpoint{3.078944in}{0.692749in}}%
\pgfpathlineto{\pgfqpoint{3.085049in}{0.680580in}}%
\pgfpathlineto{\pgfqpoint{3.091064in}{0.696739in}}%
\pgfpathlineto{\pgfqpoint{3.096990in}{0.666724in}}%
\pgfpathlineto{\pgfqpoint{3.102831in}{0.672429in}}%
\pgfpathlineto{\pgfqpoint{3.108589in}{0.721010in}}%
\pgfpathlineto{\pgfqpoint{3.114266in}{0.670288in}}%
\pgfpathlineto{\pgfqpoint{3.119864in}{0.673163in}}%
\pgfpathlineto{\pgfqpoint{3.125387in}{0.660810in}}%
\pgfpathlineto{\pgfqpoint{3.130834in}{0.680575in}}%
\pgfpathlineto{\pgfqpoint{3.136210in}{0.652745in}}%
\pgfpathlineto{\pgfqpoint{3.141515in}{0.678631in}}%
\pgfpathlineto{\pgfqpoint{3.146751in}{0.651169in}}%
\pgfpathlineto{\pgfqpoint{3.151920in}{0.653758in}}%
\pgfpathlineto{\pgfqpoint{3.157024in}{0.649006in}}%
\pgfpathlineto{\pgfqpoint{3.162065in}{0.680189in}}%
\pgfpathlineto{\pgfqpoint{3.167044in}{0.659754in}}%
\pgfpathlineto{\pgfqpoint{3.171962in}{0.668131in}}%
\pgfpathlineto{\pgfqpoint{3.176821in}{0.652195in}}%
\pgfpathlineto{\pgfqpoint{3.181622in}{0.639597in}}%
\pgfpathlineto{\pgfqpoint{3.186367in}{0.663312in}}%
\pgfpathlineto{\pgfqpoint{3.191057in}{0.634347in}}%
\pgfpathlineto{\pgfqpoint{3.195694in}{0.632084in}}%
\pgfpathlineto{\pgfqpoint{3.200277in}{0.624062in}}%
\pgfpathlineto{\pgfqpoint{3.204810in}{0.621513in}}%
\pgfpathlineto{\pgfqpoint{3.209292in}{0.642510in}}%
\pgfpathlineto{\pgfqpoint{3.213725in}{0.616111in}}%
\pgfpathlineto{\pgfqpoint{3.218111in}{0.605549in}}%
\pgfpathlineto{\pgfqpoint{3.222449in}{0.602911in}}%
\pgfpathlineto{\pgfqpoint{3.226741in}{0.621684in}}%
\pgfpathlineto{\pgfqpoint{3.230988in}{0.602964in}}%
\pgfpathlineto{\pgfqpoint{3.235191in}{0.605645in}}%
\pgfpathlineto{\pgfqpoint{3.239351in}{0.597676in}}%
\pgfpathlineto{\pgfqpoint{3.243468in}{0.600426in}}%
\pgfpathlineto{\pgfqpoint{3.247544in}{0.597731in}}%
\pgfusepath{stroke}%
\end{pgfscope}%
\begin{pgfscope}%
\pgfpathrectangle{\pgfqpoint{0.465972in}{0.510778in}}{\pgfqpoint{2.914028in}{1.675889in}}%
\pgfusepath{clip}%
\pgfsetrectcap%
\pgfsetroundjoin%
\pgfsetlinewidth{3.011250pt}%
\definecolor{currentstroke}{rgb}{1.000000,0.498039,0.054902}%
\pgfsetstrokecolor{currentstroke}%
\pgfsetdash{}{0pt}%
\pgfpathmoveto{\pgfqpoint{0.598428in}{0.586955in}}%
\pgfpathlineto{\pgfqpoint{1.287581in}{0.586955in}}%
\pgfpathlineto{\pgfqpoint{1.619586in}{1.016042in}}%
\pgfpathlineto{\pgfqpoint{1.799478in}{1.987828in}}%
\pgfpathlineto{\pgfqpoint{1.923576in}{2.110490in}}%
\pgfpathlineto{\pgfqpoint{2.018417in}{1.861609in}}%
\pgfpathlineto{\pgfqpoint{2.095197in}{1.695094in}}%
\pgfpathlineto{\pgfqpoint{2.159707in}{1.504685in}}%
\pgfpathlineto{\pgfqpoint{2.215334in}{1.392889in}}%
\pgfpathlineto{\pgfqpoint{2.264230in}{1.299300in}}%
\pgfpathlineto{\pgfqpoint{2.307850in}{1.330890in}}%
\pgfpathlineto{\pgfqpoint{2.347222in}{1.204989in}}%
\pgfpathlineto{\pgfqpoint{2.383102in}{1.188174in}}%
\pgfpathlineto{\pgfqpoint{2.416057in}{1.166320in}}%
\pgfpathlineto{\pgfqpoint{2.446530in}{1.136999in}}%
\pgfpathlineto{\pgfqpoint{2.474868in}{1.084533in}}%
\pgfpathlineto{\pgfqpoint{2.501351in}{1.105015in}}%
\pgfpathlineto{\pgfqpoint{2.526207in}{1.032279in}}%
\pgfpathlineto{\pgfqpoint{2.549624in}{1.023601in}}%
\pgfpathlineto{\pgfqpoint{2.571760in}{0.962917in}}%
\pgfpathlineto{\pgfqpoint{2.592747in}{0.988610in}}%
\pgfpathlineto{\pgfqpoint{2.612700in}{0.993769in}}%
\pgfpathlineto{\pgfqpoint{2.631715in}{0.976688in}}%
\pgfpathlineto{\pgfqpoint{2.649876in}{0.955582in}}%
\pgfpathlineto{\pgfqpoint{2.667258in}{0.952333in}}%
\pgfpathlineto{\pgfqpoint{2.683923in}{0.934352in}}%
\pgfpathlineto{\pgfqpoint{2.699929in}{0.881209in}}%
\pgfpathlineto{\pgfqpoint{2.715326in}{0.826561in}}%
\pgfpathlineto{\pgfqpoint{2.730159in}{0.875397in}}%
\pgfpathlineto{\pgfqpoint{2.744467in}{0.928387in}}%
\pgfpathlineto{\pgfqpoint{2.758287in}{0.858050in}}%
\pgfpathlineto{\pgfqpoint{2.771650in}{0.838933in}}%
\pgfpathlineto{\pgfqpoint{2.784586in}{0.829543in}}%
\pgfpathlineto{\pgfqpoint{2.797121in}{0.792457in}}%
\pgfpathlineto{\pgfqpoint{2.809280in}{0.841908in}}%
\pgfpathlineto{\pgfqpoint{2.821084in}{0.807986in}}%
\pgfpathlineto{\pgfqpoint{2.832553in}{0.773286in}}%
\pgfpathlineto{\pgfqpoint{2.843706in}{0.808958in}}%
\pgfpathlineto{\pgfqpoint{2.854560in}{0.760796in}}%
\pgfpathlineto{\pgfqpoint{2.865131in}{0.813258in}}%
\pgfpathlineto{\pgfqpoint{2.875432in}{0.736370in}}%
\pgfpathlineto{\pgfqpoint{2.885478in}{0.737926in}}%
\pgfpathlineto{\pgfqpoint{2.895280in}{0.704634in}}%
\pgfpathlineto{\pgfqpoint{2.904851in}{0.734141in}}%
\pgfpathlineto{\pgfqpoint{2.914200in}{0.704660in}}%
\pgfpathlineto{\pgfqpoint{2.923339in}{0.707883in}}%
\pgfpathlineto{\pgfqpoint{2.932275in}{0.690964in}}%
\pgfpathlineto{\pgfqpoint{2.941019in}{0.696224in}}%
\pgfpathlineto{\pgfqpoint{2.949577in}{0.698248in}}%
\pgfpathlineto{\pgfqpoint{2.957959in}{0.689167in}}%
\pgfpathlineto{\pgfqpoint{2.966170in}{0.716229in}}%
\pgfpathlineto{\pgfqpoint{2.974218in}{0.721085in}}%
\pgfpathlineto{\pgfqpoint{2.982109in}{0.686294in}}%
\pgfpathlineto{\pgfqpoint{2.989849in}{0.643525in}}%
\pgfpathlineto{\pgfqpoint{2.997443in}{0.670482in}}%
\pgfpathlineto{\pgfqpoint{3.004898in}{0.667123in}}%
\pgfpathlineto{\pgfqpoint{3.012218in}{0.586955in}}%
\pgfpathlineto{\pgfqpoint{3.019408in}{0.636185in}}%
\pgfpathlineto{\pgfqpoint{3.026473in}{0.617242in}}%
\pgfpathlineto{\pgfqpoint{3.033416in}{0.622668in}}%
\pgfpathlineto{\pgfqpoint{3.040242in}{0.664605in}}%
\pgfpathlineto{\pgfqpoint{3.046955in}{0.606456in}}%
\pgfpathlineto{\pgfqpoint{3.053558in}{0.592925in}}%
\pgfpathlineto{\pgfqpoint{3.060056in}{0.619448in}}%
\pgfpathlineto{\pgfqpoint{3.066450in}{0.586955in}}%
\pgfpathlineto{\pgfqpoint{3.072745in}{0.586955in}}%
\pgfpathlineto{\pgfqpoint{3.078944in}{0.596230in}}%
\pgfpathlineto{\pgfqpoint{3.085049in}{0.586955in}}%
\pgfpathlineto{\pgfqpoint{3.091064in}{0.598397in}}%
\pgfpathlineto{\pgfqpoint{3.096990in}{0.586955in}}%
\pgfpathlineto{\pgfqpoint{3.102831in}{0.586955in}}%
\pgfpathlineto{\pgfqpoint{3.108589in}{0.624794in}}%
\pgfpathlineto{\pgfqpoint{3.114266in}{0.586955in}}%
\pgfpathlineto{\pgfqpoint{3.119864in}{0.586955in}}%
\pgfpathlineto{\pgfqpoint{3.125387in}{0.586955in}}%
\pgfpathlineto{\pgfqpoint{3.130834in}{0.586955in}}%
\pgfpathlineto{\pgfqpoint{3.136210in}{0.586955in}}%
\pgfpathlineto{\pgfqpoint{3.141515in}{0.586955in}}%
\pgfpathlineto{\pgfqpoint{3.146751in}{0.586955in}}%
\pgfpathlineto{\pgfqpoint{3.151920in}{0.586955in}}%
\pgfpathlineto{\pgfqpoint{3.157024in}{0.586955in}}%
\pgfpathlineto{\pgfqpoint{3.162065in}{0.586955in}}%
\pgfpathlineto{\pgfqpoint{3.167044in}{0.586955in}}%
\pgfpathlineto{\pgfqpoint{3.171962in}{0.586955in}}%
\pgfpathlineto{\pgfqpoint{3.176821in}{0.586955in}}%
\pgfpathlineto{\pgfqpoint{3.181622in}{0.586955in}}%
\pgfpathlineto{\pgfqpoint{3.186367in}{0.586955in}}%
\pgfpathlineto{\pgfqpoint{3.191057in}{0.586955in}}%
\pgfpathlineto{\pgfqpoint{3.195694in}{0.586955in}}%
\pgfpathlineto{\pgfqpoint{3.200277in}{0.586955in}}%
\pgfpathlineto{\pgfqpoint{3.204810in}{0.586955in}}%
\pgfpathlineto{\pgfqpoint{3.209292in}{0.586955in}}%
\pgfpathlineto{\pgfqpoint{3.213725in}{0.586955in}}%
\pgfpathlineto{\pgfqpoint{3.218111in}{0.586955in}}%
\pgfpathlineto{\pgfqpoint{3.222449in}{0.586955in}}%
\pgfpathlineto{\pgfqpoint{3.226741in}{0.586955in}}%
\pgfpathlineto{\pgfqpoint{3.230988in}{0.586955in}}%
\pgfpathlineto{\pgfqpoint{3.235191in}{0.586955in}}%
\pgfpathlineto{\pgfqpoint{3.239351in}{0.586955in}}%
\pgfpathlineto{\pgfqpoint{3.243468in}{0.586955in}}%
\pgfpathlineto{\pgfqpoint{3.247544in}{0.586955in}}%
\pgfusepath{stroke}%
\end{pgfscope}%
\begin{pgfscope}%
\pgfsetrectcap%
\pgfsetmiterjoin%
\pgfsetlinewidth{0.803000pt}%
\definecolor{currentstroke}{rgb}{0.000000,0.000000,0.000000}%
\pgfsetstrokecolor{currentstroke}%
\pgfsetdash{}{0pt}%
\pgfpathmoveto{\pgfqpoint{0.465972in}{0.510778in}}%
\pgfpathlineto{\pgfqpoint{0.465972in}{2.186667in}}%
\pgfusepath{stroke}%
\end{pgfscope}%
\begin{pgfscope}%
\pgfsetrectcap%
\pgfsetmiterjoin%
\pgfsetlinewidth{0.803000pt}%
\definecolor{currentstroke}{rgb}{0.000000,0.000000,0.000000}%
\pgfsetstrokecolor{currentstroke}%
\pgfsetdash{}{0pt}%
\pgfpathmoveto{\pgfqpoint{3.380000in}{0.510778in}}%
\pgfpathlineto{\pgfqpoint{3.380000in}{2.186667in}}%
\pgfusepath{stroke}%
\end{pgfscope}%
\begin{pgfscope}%
\pgfsetrectcap%
\pgfsetmiterjoin%
\pgfsetlinewidth{0.803000pt}%
\definecolor{currentstroke}{rgb}{0.000000,0.000000,0.000000}%
\pgfsetstrokecolor{currentstroke}%
\pgfsetdash{}{0pt}%
\pgfpathmoveto{\pgfqpoint{0.465972in}{0.510778in}}%
\pgfpathlineto{\pgfqpoint{3.380000in}{0.510778in}}%
\pgfusepath{stroke}%
\end{pgfscope}%
\begin{pgfscope}%
\pgfsetrectcap%
\pgfsetmiterjoin%
\pgfsetlinewidth{0.803000pt}%
\definecolor{currentstroke}{rgb}{0.000000,0.000000,0.000000}%
\pgfsetstrokecolor{currentstroke}%
\pgfsetdash{}{0pt}%
\pgfpathmoveto{\pgfqpoint{0.465972in}{2.186667in}}%
\pgfpathlineto{\pgfqpoint{3.380000in}{2.186667in}}%
\pgfusepath{stroke}%
\end{pgfscope}%
\begin{pgfscope}%
\definecolor{textcolor}{rgb}{0.000000,0.000000,0.000000}%
\pgfsetstrokecolor{textcolor}%
\pgfsetfillcolor{textcolor}%
\pgftext[x=0.465972in,y=2.270000in,left,base]{\color{textcolor}\sffamily\fontsize{9.600000}{11.520000}\selectfont normalized response}%
\end{pgfscope}%
\begin{pgfscope}%
\pgfsetbuttcap%
\pgfsetmiterjoin%
\definecolor{currentfill}{rgb}{1.000000,1.000000,1.000000}%
\pgfsetfillcolor{currentfill}%
\pgfsetfillopacity{0.800000}%
\pgfsetlinewidth{1.003750pt}%
\definecolor{currentstroke}{rgb}{0.800000,0.800000,0.800000}%
\pgfsetstrokecolor{currentstroke}%
\pgfsetstrokeopacity{0.800000}%
\pgfsetdash{}{0pt}%
\pgfpathmoveto{\pgfqpoint{2.348556in}{1.788000in}}%
\pgfpathlineto{\pgfqpoint{3.302222in}{1.788000in}}%
\pgfpathquadraticcurveto{\pgfqpoint{3.324444in}{1.788000in}}{\pgfqpoint{3.324444in}{1.810223in}}%
\pgfpathlineto{\pgfqpoint{3.324444in}{2.108889in}}%
\pgfpathquadraticcurveto{\pgfqpoint{3.324444in}{2.131111in}}{\pgfqpoint{3.302222in}{2.131111in}}%
\pgfpathlineto{\pgfqpoint{2.348556in}{2.131111in}}%
\pgfpathquadraticcurveto{\pgfqpoint{2.326333in}{2.131111in}}{\pgfqpoint{2.326333in}{2.108889in}}%
\pgfpathlineto{\pgfqpoint{2.326333in}{1.810223in}}%
\pgfpathquadraticcurveto{\pgfqpoint{2.326333in}{1.788000in}}{\pgfqpoint{2.348556in}{1.788000in}}%
\pgfpathclose%
\pgfusepath{stroke,fill}%
\end{pgfscope}%
\begin{pgfscope}%
\pgfsetrectcap%
\pgfsetroundjoin%
\pgfsetlinewidth{3.011250pt}%
\definecolor{currentstroke}{rgb}{0.121569,0.466667,0.705882}%
\pgfsetstrokecolor{currentstroke}%
\pgfsetdash{}{0pt}%
\pgfpathmoveto{\pgfqpoint{2.370778in}{2.047778in}}%
\pgfpathlineto{\pgfqpoint{2.593000in}{2.047778in}}%
\pgfusepath{stroke}%
\end{pgfscope}%
\begin{pgfscope}%
\definecolor{textcolor}{rgb}{0.000000,0.000000,0.000000}%
\pgfsetstrokecolor{textcolor}%
\pgfsetfillcolor{textcolor}%
\pgftext[x=2.681889in,y=2.008889in,left,base]{\color{textcolor}\sffamily\fontsize{8.000000}{9.600000}\selectfont uncalibrated}%
\end{pgfscope}%
\begin{pgfscope}%
\pgfsetrectcap%
\pgfsetroundjoin%
\pgfsetlinewidth{3.011250pt}%
\definecolor{currentstroke}{rgb}{1.000000,0.498039,0.054902}%
\pgfsetstrokecolor{currentstroke}%
\pgfsetdash{}{0pt}%
\pgfpathmoveto{\pgfqpoint{2.370778in}{1.892889in}}%
\pgfpathlineto{\pgfqpoint{2.593000in}{1.892889in}}%
\pgfusepath{stroke}%
\end{pgfscope}%
\begin{pgfscope}%
\definecolor{textcolor}{rgb}{0.000000,0.000000,0.000000}%
\pgfsetstrokecolor{textcolor}%
\pgfsetfillcolor{textcolor}%
\pgftext[x=2.681889in,y=1.854000in,left,base]{\color{textcolor}\sffamily\fontsize{8.000000}{9.600000}\selectfont calibrated}%
\end{pgfscope}%
\end{pgfpicture}%
\makeatother%
\endgroup%

%% file: figures/AlCuShellsBrem_lineouts_zoom.pgf
\begingroup%
\makeatletter%
\begin{pgfpicture}%
\pgfpathrectangle{\pgfpointorigin}{\pgfqpoint{3.500000in}{3.000000in}}%
\pgfusepath{use as bounding box, clip}%
\begin{pgfscope}%
\pgfsetbuttcap%
\pgfsetmiterjoin%
\definecolor{currentfill}{rgb}{1.000000,1.000000,1.000000}%
\pgfsetfillcolor{currentfill}%
\pgfsetlinewidth{0.000000pt}%
\definecolor{currentstroke}{rgb}{1.000000,1.000000,1.000000}%
\pgfsetstrokecolor{currentstroke}%
\pgfsetdash{}{0pt}%
\pgfpathmoveto{\pgfqpoint{0.000000in}{0.000000in}}%
\pgfpathlineto{\pgfqpoint{3.500000in}{0.000000in}}%
\pgfpathlineto{\pgfqpoint{3.500000in}{3.000000in}}%
\pgfpathlineto{\pgfqpoint{0.000000in}{3.000000in}}%
\pgfpathclose%
\pgfusepath{fill}%
\end{pgfscope}%
\begin{pgfscope}%
\pgfsetbuttcap%
\pgfsetmiterjoin%
\definecolor{currentfill}{rgb}{1.000000,1.000000,1.000000}%
\pgfsetfillcolor{currentfill}%
\pgfsetlinewidth{0.000000pt}%
\definecolor{currentstroke}{rgb}{0.000000,0.000000,0.000000}%
\pgfsetstrokecolor{currentstroke}%
\pgfsetstrokeopacity{0.000000}%
\pgfsetdash{}{0pt}%
\pgfpathmoveto{\pgfqpoint{0.437500in}{0.330000in}}%
\pgfpathlineto{\pgfqpoint{3.150000in}{0.330000in}}%
\pgfpathlineto{\pgfqpoint{3.150000in}{2.640000in}}%
\pgfpathlineto{\pgfqpoint{0.437500in}{2.640000in}}%
\pgfpathclose%
\pgfusepath{fill}%
\end{pgfscope}%
\begin{pgfscope}%
\pgfsetbuttcap%
\pgfsetroundjoin%
\definecolor{currentfill}{rgb}{0.000000,0.000000,0.000000}%
\pgfsetfillcolor{currentfill}%
\pgfsetlinewidth{0.803000pt}%
\definecolor{currentstroke}{rgb}{0.000000,0.000000,0.000000}%
\pgfsetstrokecolor{currentstroke}%
\pgfsetdash{}{0pt}%
\pgfsys@defobject{currentmarker}{\pgfqpoint{0.000000in}{-0.048611in}}{\pgfqpoint{0.000000in}{0.000000in}}{%
\pgfpathmoveto{\pgfqpoint{0.000000in}{0.000000in}}%
\pgfpathlineto{\pgfqpoint{0.000000in}{-0.048611in}}%
\pgfusepath{stroke,fill}%
}%
\begin{pgfscope}%
\pgfsys@transformshift{0.809448in}{0.330000in}%
\pgfsys@useobject{currentmarker}{}%
\end{pgfscope}%
\end{pgfscope}%
\begin{pgfscope}%
\definecolor{textcolor}{rgb}{0.000000,0.000000,0.000000}%
\pgfsetstrokecolor{textcolor}%
\pgfsetfillcolor{textcolor}%
\pgftext[x=0.809448in,y=0.232778in,,top]{\color{textcolor}\sffamily\fontsize{8.000000}{9.600000}\selectfont 150}%
\end{pgfscope}%
\begin{pgfscope}%
\pgfsetbuttcap%
\pgfsetroundjoin%
\definecolor{currentfill}{rgb}{0.000000,0.000000,0.000000}%
\pgfsetfillcolor{currentfill}%
\pgfsetlinewidth{0.803000pt}%
\definecolor{currentstroke}{rgb}{0.000000,0.000000,0.000000}%
\pgfsetstrokecolor{currentstroke}%
\pgfsetdash{}{0pt}%
\pgfsys@defobject{currentmarker}{\pgfqpoint{0.000000in}{-0.048611in}}{\pgfqpoint{0.000000in}{0.000000in}}{%
\pgfpathmoveto{\pgfqpoint{0.000000in}{0.000000in}}%
\pgfpathlineto{\pgfqpoint{0.000000in}{-0.048611in}}%
\pgfusepath{stroke,fill}%
}%
\begin{pgfscope}%
\pgfsys@transformshift{1.263043in}{0.330000in}%
\pgfsys@useobject{currentmarker}{}%
\end{pgfscope}%
\end{pgfscope}%
\begin{pgfscope}%
\definecolor{textcolor}{rgb}{0.000000,0.000000,0.000000}%
\pgfsetstrokecolor{textcolor}%
\pgfsetfillcolor{textcolor}%
\pgftext[x=1.263043in,y=0.232778in,,top]{\color{textcolor}\sffamily\fontsize{8.000000}{9.600000}\selectfont 200}%
\end{pgfscope}%
\begin{pgfscope}%
\pgfsetbuttcap%
\pgfsetroundjoin%
\definecolor{currentfill}{rgb}{0.000000,0.000000,0.000000}%
\pgfsetfillcolor{currentfill}%
\pgfsetlinewidth{0.803000pt}%
\definecolor{currentstroke}{rgb}{0.000000,0.000000,0.000000}%
\pgfsetstrokecolor{currentstroke}%
\pgfsetdash{}{0pt}%
\pgfsys@defobject{currentmarker}{\pgfqpoint{0.000000in}{-0.048611in}}{\pgfqpoint{0.000000in}{0.000000in}}{%
\pgfpathmoveto{\pgfqpoint{0.000000in}{0.000000in}}%
\pgfpathlineto{\pgfqpoint{0.000000in}{-0.048611in}}%
\pgfusepath{stroke,fill}%
}%
\begin{pgfscope}%
\pgfsys@transformshift{1.716639in}{0.330000in}%
\pgfsys@useobject{currentmarker}{}%
\end{pgfscope}%
\end{pgfscope}%
\begin{pgfscope}%
\definecolor{textcolor}{rgb}{0.000000,0.000000,0.000000}%
\pgfsetstrokecolor{textcolor}%
\pgfsetfillcolor{textcolor}%
\pgftext[x=1.716639in,y=0.232778in,,top]{\color{textcolor}\sffamily\fontsize{8.000000}{9.600000}\selectfont 250}%
\end{pgfscope}%
\begin{pgfscope}%
\pgfsetbuttcap%
\pgfsetroundjoin%
\definecolor{currentfill}{rgb}{0.000000,0.000000,0.000000}%
\pgfsetfillcolor{currentfill}%
\pgfsetlinewidth{0.803000pt}%
\definecolor{currentstroke}{rgb}{0.000000,0.000000,0.000000}%
\pgfsetstrokecolor{currentstroke}%
\pgfsetdash{}{0pt}%
\pgfsys@defobject{currentmarker}{\pgfqpoint{0.000000in}{-0.048611in}}{\pgfqpoint{0.000000in}{0.000000in}}{%
\pgfpathmoveto{\pgfqpoint{0.000000in}{0.000000in}}%
\pgfpathlineto{\pgfqpoint{0.000000in}{-0.048611in}}%
\pgfusepath{stroke,fill}%
}%
\begin{pgfscope}%
\pgfsys@transformshift{2.170234in}{0.330000in}%
\pgfsys@useobject{currentmarker}{}%
\end{pgfscope}%
\end{pgfscope}%
\begin{pgfscope}%
\definecolor{textcolor}{rgb}{0.000000,0.000000,0.000000}%
\pgfsetstrokecolor{textcolor}%
\pgfsetfillcolor{textcolor}%
\pgftext[x=2.170234in,y=0.232778in,,top]{\color{textcolor}\sffamily\fontsize{8.000000}{9.600000}\selectfont 300}%
\end{pgfscope}%
\begin{pgfscope}%
\pgfsetbuttcap%
\pgfsetroundjoin%
\definecolor{currentfill}{rgb}{0.000000,0.000000,0.000000}%
\pgfsetfillcolor{currentfill}%
\pgfsetlinewidth{0.803000pt}%
\definecolor{currentstroke}{rgb}{0.000000,0.000000,0.000000}%
\pgfsetstrokecolor{currentstroke}%
\pgfsetdash{}{0pt}%
\pgfsys@defobject{currentmarker}{\pgfqpoint{0.000000in}{-0.048611in}}{\pgfqpoint{0.000000in}{0.000000in}}{%
\pgfpathmoveto{\pgfqpoint{0.000000in}{0.000000in}}%
\pgfpathlineto{\pgfqpoint{0.000000in}{-0.048611in}}%
\pgfusepath{stroke,fill}%
}%
\begin{pgfscope}%
\pgfsys@transformshift{2.623829in}{0.330000in}%
\pgfsys@useobject{currentmarker}{}%
\end{pgfscope}%
\end{pgfscope}%
\begin{pgfscope}%
\definecolor{textcolor}{rgb}{0.000000,0.000000,0.000000}%
\pgfsetstrokecolor{textcolor}%
\pgfsetfillcolor{textcolor}%
\pgftext[x=2.623829in,y=0.232778in,,top]{\color{textcolor}\sffamily\fontsize{8.000000}{9.600000}\selectfont 350}%
\end{pgfscope}%
\begin{pgfscope}%
\pgfsetbuttcap%
\pgfsetroundjoin%
\definecolor{currentfill}{rgb}{0.000000,0.000000,0.000000}%
\pgfsetfillcolor{currentfill}%
\pgfsetlinewidth{0.803000pt}%
\definecolor{currentstroke}{rgb}{0.000000,0.000000,0.000000}%
\pgfsetstrokecolor{currentstroke}%
\pgfsetdash{}{0pt}%
\pgfsys@defobject{currentmarker}{\pgfqpoint{0.000000in}{-0.048611in}}{\pgfqpoint{0.000000in}{0.000000in}}{%
\pgfpathmoveto{\pgfqpoint{0.000000in}{0.000000in}}%
\pgfpathlineto{\pgfqpoint{0.000000in}{-0.048611in}}%
\pgfusepath{stroke,fill}%
}%
\begin{pgfscope}%
\pgfsys@transformshift{3.077425in}{0.330000in}%
\pgfsys@useobject{currentmarker}{}%
\end{pgfscope}%
\end{pgfscope}%
\begin{pgfscope}%
\definecolor{textcolor}{rgb}{0.000000,0.000000,0.000000}%
\pgfsetstrokecolor{textcolor}%
\pgfsetfillcolor{textcolor}%
\pgftext[x=3.077425in,y=0.232778in,,top]{\color{textcolor}\sffamily\fontsize{8.000000}{9.600000}\selectfont 400}%
\end{pgfscope}%
\begin{pgfscope}%
\pgfsetbuttcap%
\pgfsetroundjoin%
\definecolor{currentfill}{rgb}{0.000000,0.000000,0.000000}%
\pgfsetfillcolor{currentfill}%
\pgfsetlinewidth{0.803000pt}%
\definecolor{currentstroke}{rgb}{0.000000,0.000000,0.000000}%
\pgfsetstrokecolor{currentstroke}%
\pgfsetdash{}{0pt}%
\pgfsys@defobject{currentmarker}{\pgfqpoint{-0.048611in}{0.000000in}}{\pgfqpoint{-0.000000in}{0.000000in}}{%
\pgfpathmoveto{\pgfqpoint{-0.000000in}{0.000000in}}%
\pgfpathlineto{\pgfqpoint{-0.048611in}{0.000000in}}%
\pgfusepath{stroke,fill}%
}%
\begin{pgfscope}%
\pgfsys@transformshift{0.437500in}{0.507692in}%
\pgfsys@useobject{currentmarker}{}%
\end{pgfscope}%
\end{pgfscope}%
\begin{pgfscope}%
\definecolor{textcolor}{rgb}{0.000000,0.000000,0.000000}%
\pgfsetstrokecolor{textcolor}%
\pgfsetfillcolor{textcolor}%
\pgftext[x=0.130500in, y=0.469137in, left, base]{\color{textcolor}\sffamily\fontsize{8.000000}{9.600000}\selectfont 0.25}%
\end{pgfscope}%
\begin{pgfscope}%
\pgfsetbuttcap%
\pgfsetroundjoin%
\definecolor{currentfill}{rgb}{0.000000,0.000000,0.000000}%
\pgfsetfillcolor{currentfill}%
\pgfsetlinewidth{0.803000pt}%
\definecolor{currentstroke}{rgb}{0.000000,0.000000,0.000000}%
\pgfsetstrokecolor{currentstroke}%
\pgfsetdash{}{0pt}%
\pgfsys@defobject{currentmarker}{\pgfqpoint{-0.048611in}{0.000000in}}{\pgfqpoint{-0.000000in}{0.000000in}}{%
\pgfpathmoveto{\pgfqpoint{-0.000000in}{0.000000in}}%
\pgfpathlineto{\pgfqpoint{-0.048611in}{0.000000in}}%
\pgfusepath{stroke,fill}%
}%
\begin{pgfscope}%
\pgfsys@transformshift{0.437500in}{0.951923in}%
\pgfsys@useobject{currentmarker}{}%
\end{pgfscope}%
\end{pgfscope}%
\begin{pgfscope}%
\definecolor{textcolor}{rgb}{0.000000,0.000000,0.000000}%
\pgfsetstrokecolor{textcolor}%
\pgfsetfillcolor{textcolor}%
\pgftext[x=0.130500in, y=0.913368in, left, base]{\color{textcolor}\sffamily\fontsize{8.000000}{9.600000}\selectfont 0.30}%
\end{pgfscope}%
\begin{pgfscope}%
\pgfsetbuttcap%
\pgfsetroundjoin%
\definecolor{currentfill}{rgb}{0.000000,0.000000,0.000000}%
\pgfsetfillcolor{currentfill}%
\pgfsetlinewidth{0.803000pt}%
\definecolor{currentstroke}{rgb}{0.000000,0.000000,0.000000}%
\pgfsetstrokecolor{currentstroke}%
\pgfsetdash{}{0pt}%
\pgfsys@defobject{currentmarker}{\pgfqpoint{-0.048611in}{0.000000in}}{\pgfqpoint{-0.000000in}{0.000000in}}{%
\pgfpathmoveto{\pgfqpoint{-0.000000in}{0.000000in}}%
\pgfpathlineto{\pgfqpoint{-0.048611in}{0.000000in}}%
\pgfusepath{stroke,fill}%
}%
\begin{pgfscope}%
\pgfsys@transformshift{0.437500in}{1.396154in}%
\pgfsys@useobject{currentmarker}{}%
\end{pgfscope}%
\end{pgfscope}%
\begin{pgfscope}%
\definecolor{textcolor}{rgb}{0.000000,0.000000,0.000000}%
\pgfsetstrokecolor{textcolor}%
\pgfsetfillcolor{textcolor}%
\pgftext[x=0.130500in, y=1.357598in, left, base]{\color{textcolor}\sffamily\fontsize{8.000000}{9.600000}\selectfont 0.35}%
\end{pgfscope}%
\begin{pgfscope}%
\pgfsetbuttcap%
\pgfsetroundjoin%
\definecolor{currentfill}{rgb}{0.000000,0.000000,0.000000}%
\pgfsetfillcolor{currentfill}%
\pgfsetlinewidth{0.803000pt}%
\definecolor{currentstroke}{rgb}{0.000000,0.000000,0.000000}%
\pgfsetstrokecolor{currentstroke}%
\pgfsetdash{}{0pt}%
\pgfsys@defobject{currentmarker}{\pgfqpoint{-0.048611in}{0.000000in}}{\pgfqpoint{-0.000000in}{0.000000in}}{%
\pgfpathmoveto{\pgfqpoint{-0.000000in}{0.000000in}}%
\pgfpathlineto{\pgfqpoint{-0.048611in}{0.000000in}}%
\pgfusepath{stroke,fill}%
}%
\begin{pgfscope}%
\pgfsys@transformshift{0.437500in}{1.840385in}%
\pgfsys@useobject{currentmarker}{}%
\end{pgfscope}%
\end{pgfscope}%
\begin{pgfscope}%
\definecolor{textcolor}{rgb}{0.000000,0.000000,0.000000}%
\pgfsetstrokecolor{textcolor}%
\pgfsetfillcolor{textcolor}%
\pgftext[x=0.130500in, y=1.801829in, left, base]{\color{textcolor}\sffamily\fontsize{8.000000}{9.600000}\selectfont 0.40}%
\end{pgfscope}%
\begin{pgfscope}%
\pgfsetbuttcap%
\pgfsetroundjoin%
\definecolor{currentfill}{rgb}{0.000000,0.000000,0.000000}%
\pgfsetfillcolor{currentfill}%
\pgfsetlinewidth{0.803000pt}%
\definecolor{currentstroke}{rgb}{0.000000,0.000000,0.000000}%
\pgfsetstrokecolor{currentstroke}%
\pgfsetdash{}{0pt}%
\pgfsys@defobject{currentmarker}{\pgfqpoint{-0.048611in}{0.000000in}}{\pgfqpoint{-0.000000in}{0.000000in}}{%
\pgfpathmoveto{\pgfqpoint{-0.000000in}{0.000000in}}%
\pgfpathlineto{\pgfqpoint{-0.048611in}{0.000000in}}%
\pgfusepath{stroke,fill}%
}%
\begin{pgfscope}%
\pgfsys@transformshift{0.437500in}{2.284615in}%
\pgfsys@useobject{currentmarker}{}%
\end{pgfscope}%
\end{pgfscope}%
\begin{pgfscope}%
\definecolor{textcolor}{rgb}{0.000000,0.000000,0.000000}%
\pgfsetstrokecolor{textcolor}%
\pgfsetfillcolor{textcolor}%
\pgftext[x=0.130500in, y=2.246060in, left, base]{\color{textcolor}\sffamily\fontsize{8.000000}{9.600000}\selectfont 0.45}%
\end{pgfscope}%
\begin{pgfscope}%
\pgfpathrectangle{\pgfqpoint{0.437500in}{0.330000in}}{\pgfqpoint{2.712500in}{2.310000in}}%
\pgfusepath{clip}%
\pgfsetrectcap%
\pgfsetroundjoin%
\pgfsetlinewidth{1.505625pt}%
\definecolor{currentstroke}{rgb}{0.000000,0.000000,0.000000}%
\pgfsetstrokecolor{currentstroke}%
\pgfsetdash{}{0pt}%
\pgfpathmoveto{\pgfqpoint{0.845535in}{2.650000in}}%
\pgfpathlineto{\pgfqpoint{0.845736in}{2.649233in}}%
\pgfpathlineto{\pgfqpoint{0.854808in}{2.642834in}}%
\pgfpathlineto{\pgfqpoint{0.863880in}{2.629227in}}%
\pgfpathlineto{\pgfqpoint{0.872952in}{2.628079in}}%
\pgfpathlineto{\pgfqpoint{0.882023in}{2.648894in}}%
\pgfpathlineto{\pgfqpoint{0.891095in}{2.643960in}}%
\pgfpathlineto{\pgfqpoint{0.896057in}{2.650000in}}%
\pgfpathmoveto{\pgfqpoint{0.904273in}{2.650000in}}%
\pgfpathlineto{\pgfqpoint{0.909239in}{2.643949in}}%
\pgfpathlineto{\pgfqpoint{0.917098in}{2.650000in}}%
\pgfpathmoveto{\pgfqpoint{0.918441in}{2.650000in}}%
\pgfpathlineto{\pgfqpoint{0.927383in}{2.585879in}}%
\pgfpathlineto{\pgfqpoint{0.936455in}{2.532007in}}%
\pgfpathlineto{\pgfqpoint{0.945527in}{2.530204in}}%
\pgfpathlineto{\pgfqpoint{0.954599in}{2.500933in}}%
\pgfpathlineto{\pgfqpoint{0.963671in}{2.487930in}}%
\pgfpathlineto{\pgfqpoint{0.981814in}{2.374638in}}%
\pgfpathlineto{\pgfqpoint{0.999958in}{2.504429in}}%
\pgfpathlineto{\pgfqpoint{1.009030in}{2.515179in}}%
\pgfpathlineto{\pgfqpoint{1.018102in}{2.531874in}}%
\pgfpathlineto{\pgfqpoint{1.027174in}{2.451133in}}%
\pgfpathlineto{\pgfqpoint{1.036246in}{2.394680in}}%
\pgfpathlineto{\pgfqpoint{1.045318in}{2.427474in}}%
\pgfpathlineto{\pgfqpoint{1.054390in}{2.388687in}}%
\pgfpathlineto{\pgfqpoint{1.063462in}{2.399394in}}%
\pgfpathlineto{\pgfqpoint{1.072533in}{2.343797in}}%
\pgfpathlineto{\pgfqpoint{1.081605in}{2.340727in}}%
\pgfpathlineto{\pgfqpoint{1.090677in}{2.237435in}}%
\pgfpathlineto{\pgfqpoint{1.099749in}{2.238336in}}%
\pgfpathlineto{\pgfqpoint{1.108821in}{2.295266in}}%
\pgfpathlineto{\pgfqpoint{1.117893in}{2.334482in}}%
\pgfpathlineto{\pgfqpoint{1.126965in}{2.299558in}}%
\pgfpathlineto{\pgfqpoint{1.136037in}{2.222447in}}%
\pgfpathlineto{\pgfqpoint{1.145109in}{2.160800in}}%
\pgfpathlineto{\pgfqpoint{1.154181in}{2.210349in}}%
\pgfpathlineto{\pgfqpoint{1.163253in}{2.284034in}}%
\pgfpathlineto{\pgfqpoint{1.172324in}{2.289960in}}%
\pgfpathlineto{\pgfqpoint{1.181396in}{2.314447in}}%
\pgfpathlineto{\pgfqpoint{1.190468in}{2.276011in}}%
\pgfpathlineto{\pgfqpoint{1.217684in}{2.070425in}}%
\pgfpathlineto{\pgfqpoint{1.226756in}{2.048717in}}%
\pgfpathlineto{\pgfqpoint{1.235828in}{2.073874in}}%
\pgfpathlineto{\pgfqpoint{1.244900in}{2.072659in}}%
\pgfpathlineto{\pgfqpoint{1.253972in}{2.035695in}}%
\pgfpathlineto{\pgfqpoint{1.263043in}{2.036372in}}%
\pgfpathlineto{\pgfqpoint{1.272115in}{2.112605in}}%
\pgfpathlineto{\pgfqpoint{1.281187in}{2.049729in}}%
\pgfpathlineto{\pgfqpoint{1.290259in}{2.082681in}}%
\pgfpathlineto{\pgfqpoint{1.299331in}{1.966933in}}%
\pgfpathlineto{\pgfqpoint{1.308403in}{2.026598in}}%
\pgfpathlineto{\pgfqpoint{1.317475in}{2.019192in}}%
\pgfpathlineto{\pgfqpoint{1.326547in}{1.881251in}}%
\pgfpathlineto{\pgfqpoint{1.335619in}{1.824342in}}%
\pgfpathlineto{\pgfqpoint{1.344691in}{1.973155in}}%
\pgfpathlineto{\pgfqpoint{1.353763in}{1.938626in}}%
\pgfpathlineto{\pgfqpoint{1.362834in}{1.948517in}}%
\pgfpathlineto{\pgfqpoint{1.371906in}{1.882204in}}%
\pgfpathlineto{\pgfqpoint{1.380978in}{1.906527in}}%
\pgfpathlineto{\pgfqpoint{1.390050in}{1.901844in}}%
\pgfpathlineto{\pgfqpoint{1.399122in}{1.745152in}}%
\pgfpathlineto{\pgfqpoint{1.408194in}{1.528654in}}%
\pgfpathlineto{\pgfqpoint{1.417266in}{1.529392in}}%
\pgfpathlineto{\pgfqpoint{1.426338in}{1.516625in}}%
\pgfpathlineto{\pgfqpoint{1.435410in}{1.472635in}}%
\pgfpathlineto{\pgfqpoint{1.444482in}{1.388554in}}%
\pgfpathlineto{\pgfqpoint{1.453554in}{1.368397in}}%
\pgfpathlineto{\pgfqpoint{1.462625in}{1.282197in}}%
\pgfpathlineto{\pgfqpoint{1.471697in}{1.285635in}}%
\pgfpathlineto{\pgfqpoint{1.480769in}{1.228603in}}%
\pgfpathlineto{\pgfqpoint{1.489841in}{1.219145in}}%
\pgfpathlineto{\pgfqpoint{1.498913in}{1.148133in}}%
\pgfpathlineto{\pgfqpoint{1.507985in}{0.993761in}}%
\pgfpathlineto{\pgfqpoint{1.517057in}{0.990452in}}%
\pgfpathlineto{\pgfqpoint{1.526129in}{1.068006in}}%
\pgfpathlineto{\pgfqpoint{1.535201in}{0.899712in}}%
\pgfpathlineto{\pgfqpoint{1.544273in}{0.936073in}}%
\pgfpathlineto{\pgfqpoint{1.553344in}{1.040170in}}%
\pgfpathlineto{\pgfqpoint{1.562416in}{1.068381in}}%
\pgfpathlineto{\pgfqpoint{1.571488in}{0.987957in}}%
\pgfpathlineto{\pgfqpoint{1.580560in}{0.632745in}}%
\pgfpathlineto{\pgfqpoint{1.589632in}{0.815575in}}%
\pgfpathlineto{\pgfqpoint{1.598704in}{0.844961in}}%
\pgfpathlineto{\pgfqpoint{1.607776in}{0.981743in}}%
\pgfpathlineto{\pgfqpoint{1.616848in}{0.870606in}}%
\pgfpathlineto{\pgfqpoint{1.625920in}{0.717948in}}%
\pgfpathlineto{\pgfqpoint{1.634992in}{0.816747in}}%
\pgfpathlineto{\pgfqpoint{1.644064in}{0.653103in}}%
\pgfpathlineto{\pgfqpoint{1.653135in}{0.771807in}}%
\pgfpathlineto{\pgfqpoint{1.662207in}{0.922833in}}%
\pgfpathlineto{\pgfqpoint{1.671279in}{0.936384in}}%
\pgfpathlineto{\pgfqpoint{1.680351in}{0.826500in}}%
\pgfpathlineto{\pgfqpoint{1.698495in}{0.644630in}}%
\pgfpathlineto{\pgfqpoint{1.707567in}{0.747797in}}%
\pgfpathlineto{\pgfqpoint{1.716639in}{0.884037in}}%
\pgfpathlineto{\pgfqpoint{1.725711in}{0.899691in}}%
\pgfpathlineto{\pgfqpoint{1.734783in}{0.713962in}}%
\pgfpathlineto{\pgfqpoint{1.743855in}{0.653169in}}%
\pgfpathlineto{\pgfqpoint{1.761998in}{0.768010in}}%
\pgfpathlineto{\pgfqpoint{1.771070in}{0.665985in}}%
\pgfpathlineto{\pgfqpoint{1.780142in}{0.541608in}}%
\pgfpathlineto{\pgfqpoint{1.789214in}{0.626188in}}%
\pgfpathlineto{\pgfqpoint{1.798286in}{0.898558in}}%
\pgfpathlineto{\pgfqpoint{1.807358in}{0.812883in}}%
\pgfpathlineto{\pgfqpoint{1.816430in}{0.889894in}}%
\pgfpathlineto{\pgfqpoint{1.825502in}{0.587067in}}%
\pgfpathlineto{\pgfqpoint{1.834574in}{0.465553in}}%
\pgfpathlineto{\pgfqpoint{1.843645in}{0.626199in}}%
\pgfpathlineto{\pgfqpoint{1.852717in}{0.660695in}}%
\pgfpathlineto{\pgfqpoint{1.870861in}{0.530169in}}%
\pgfpathlineto{\pgfqpoint{1.879933in}{0.599767in}}%
\pgfpathlineto{\pgfqpoint{1.889005in}{0.424968in}}%
\pgfpathlineto{\pgfqpoint{1.898077in}{0.471468in}}%
\pgfpathlineto{\pgfqpoint{1.907149in}{0.632780in}}%
\pgfpathlineto{\pgfqpoint{1.916221in}{0.569541in}}%
\pgfpathlineto{\pgfqpoint{1.925293in}{0.631419in}}%
\pgfpathlineto{\pgfqpoint{1.934365in}{0.593836in}}%
\pgfpathlineto{\pgfqpoint{1.943436in}{0.897196in}}%
\pgfpathlineto{\pgfqpoint{1.952508in}{0.853921in}}%
\pgfpathlineto{\pgfqpoint{1.961580in}{0.836163in}}%
\pgfpathlineto{\pgfqpoint{1.979724in}{0.835263in}}%
\pgfpathlineto{\pgfqpoint{1.988796in}{0.911032in}}%
\pgfpathlineto{\pgfqpoint{1.997868in}{0.881213in}}%
\pgfpathlineto{\pgfqpoint{2.006940in}{1.132014in}}%
\pgfpathlineto{\pgfqpoint{2.016012in}{1.103549in}}%
\pgfpathlineto{\pgfqpoint{2.025084in}{1.128049in}}%
\pgfpathlineto{\pgfqpoint{2.034156in}{1.061457in}}%
\pgfpathlineto{\pgfqpoint{2.043227in}{0.953006in}}%
\pgfpathlineto{\pgfqpoint{2.052299in}{1.067213in}}%
\pgfpathlineto{\pgfqpoint{2.061371in}{1.057614in}}%
\pgfpathlineto{\pgfqpoint{2.070443in}{1.038476in}}%
\pgfpathlineto{\pgfqpoint{2.079515in}{1.071702in}}%
\pgfpathlineto{\pgfqpoint{2.088587in}{0.945732in}}%
\pgfpathlineto{\pgfqpoint{2.097659in}{1.072902in}}%
\pgfpathlineto{\pgfqpoint{2.106731in}{1.148424in}}%
\pgfpathlineto{\pgfqpoint{2.115803in}{1.050450in}}%
\pgfpathlineto{\pgfqpoint{2.124875in}{1.077719in}}%
\pgfpathlineto{\pgfqpoint{2.133946in}{1.022232in}}%
\pgfpathlineto{\pgfqpoint{2.143018in}{0.983944in}}%
\pgfpathlineto{\pgfqpoint{2.152090in}{1.164289in}}%
\pgfpathlineto{\pgfqpoint{2.161162in}{1.208387in}}%
\pgfpathlineto{\pgfqpoint{2.170234in}{1.158910in}}%
\pgfpathlineto{\pgfqpoint{2.179306in}{1.161729in}}%
\pgfpathlineto{\pgfqpoint{2.188378in}{1.260808in}}%
\pgfpathlineto{\pgfqpoint{2.197450in}{1.246665in}}%
\pgfpathlineto{\pgfqpoint{2.206522in}{1.227127in}}%
\pgfpathlineto{\pgfqpoint{2.215594in}{1.069877in}}%
\pgfpathlineto{\pgfqpoint{2.224666in}{1.197634in}}%
\pgfpathlineto{\pgfqpoint{2.233737in}{1.275348in}}%
\pgfpathlineto{\pgfqpoint{2.242809in}{1.235659in}}%
\pgfpathlineto{\pgfqpoint{2.251881in}{1.220260in}}%
\pgfpathlineto{\pgfqpoint{2.260953in}{1.247299in}}%
\pgfpathlineto{\pgfqpoint{2.270025in}{1.198992in}}%
\pgfpathlineto{\pgfqpoint{2.279097in}{1.008147in}}%
\pgfpathlineto{\pgfqpoint{2.288169in}{1.105751in}}%
\pgfpathlineto{\pgfqpoint{2.297241in}{1.227964in}}%
\pgfpathlineto{\pgfqpoint{2.306313in}{1.335103in}}%
\pgfpathlineto{\pgfqpoint{2.315385in}{1.333955in}}%
\pgfpathlineto{\pgfqpoint{2.324457in}{1.199052in}}%
\pgfpathlineto{\pgfqpoint{2.333528in}{1.216481in}}%
\pgfpathlineto{\pgfqpoint{2.342600in}{1.286655in}}%
\pgfpathlineto{\pgfqpoint{2.351672in}{1.326483in}}%
\pgfpathlineto{\pgfqpoint{2.360744in}{1.400107in}}%
\pgfpathlineto{\pgfqpoint{2.369816in}{1.437243in}}%
\pgfpathlineto{\pgfqpoint{2.378888in}{1.301970in}}%
\pgfpathlineto{\pgfqpoint{2.387960in}{1.255784in}}%
\pgfpathlineto{\pgfqpoint{2.406104in}{1.373462in}}%
\pgfpathlineto{\pgfqpoint{2.415176in}{1.384342in}}%
\pgfpathlineto{\pgfqpoint{2.424247in}{1.376773in}}%
\pgfpathlineto{\pgfqpoint{2.433319in}{1.534886in}}%
\pgfpathlineto{\pgfqpoint{2.442391in}{1.506220in}}%
\pgfpathlineto{\pgfqpoint{2.451463in}{1.513625in}}%
\pgfpathlineto{\pgfqpoint{2.460535in}{1.499495in}}%
\pgfpathlineto{\pgfqpoint{2.469607in}{1.414101in}}%
\pgfpathlineto{\pgfqpoint{2.478679in}{1.391729in}}%
\pgfpathlineto{\pgfqpoint{2.487751in}{1.434325in}}%
\pgfpathlineto{\pgfqpoint{2.505895in}{1.464062in}}%
\pgfpathlineto{\pgfqpoint{2.514967in}{1.415753in}}%
\pgfpathlineto{\pgfqpoint{2.524038in}{1.333528in}}%
\pgfpathlineto{\pgfqpoint{2.533110in}{1.367281in}}%
\pgfpathlineto{\pgfqpoint{2.542182in}{1.422373in}}%
\pgfpathlineto{\pgfqpoint{2.551254in}{1.463013in}}%
\pgfpathlineto{\pgfqpoint{2.560326in}{1.490822in}}%
\pgfpathlineto{\pgfqpoint{2.569398in}{1.494297in}}%
\pgfpathlineto{\pgfqpoint{2.578470in}{1.509355in}}%
\pgfpathlineto{\pgfqpoint{2.587542in}{1.458137in}}%
\pgfpathlineto{\pgfqpoint{2.596614in}{1.484947in}}%
\pgfpathlineto{\pgfqpoint{2.605686in}{1.476141in}}%
\pgfpathlineto{\pgfqpoint{2.614758in}{1.544256in}}%
\pgfpathlineto{\pgfqpoint{2.623829in}{1.535712in}}%
\pgfpathlineto{\pgfqpoint{2.632901in}{1.550676in}}%
\pgfpathlineto{\pgfqpoint{2.641973in}{1.559228in}}%
\pgfpathlineto{\pgfqpoint{2.651045in}{1.617671in}}%
\pgfpathlineto{\pgfqpoint{2.660117in}{1.516326in}}%
\pgfpathlineto{\pgfqpoint{2.669189in}{1.511986in}}%
\pgfpathlineto{\pgfqpoint{2.678261in}{1.514011in}}%
\pgfpathlineto{\pgfqpoint{2.687333in}{1.475870in}}%
\pgfpathlineto{\pgfqpoint{2.705477in}{1.554462in}}%
\pgfpathlineto{\pgfqpoint{2.714548in}{1.539737in}}%
\pgfpathlineto{\pgfqpoint{2.723620in}{1.490578in}}%
\pgfpathlineto{\pgfqpoint{2.732692in}{1.469118in}}%
\pgfpathlineto{\pgfqpoint{2.741764in}{1.428535in}}%
\pgfpathlineto{\pgfqpoint{2.750836in}{1.570857in}}%
\pgfpathlineto{\pgfqpoint{2.759908in}{1.419917in}}%
\pgfpathlineto{\pgfqpoint{2.768980in}{1.354448in}}%
\pgfpathlineto{\pgfqpoint{2.778052in}{1.368717in}}%
\pgfpathlineto{\pgfqpoint{2.787124in}{1.458429in}}%
\pgfpathlineto{\pgfqpoint{2.796196in}{1.467490in}}%
\pgfpathlineto{\pgfqpoint{2.805268in}{1.462223in}}%
\pgfpathlineto{\pgfqpoint{2.814339in}{1.575311in}}%
\pgfpathlineto{\pgfqpoint{2.823411in}{1.620347in}}%
\pgfpathlineto{\pgfqpoint{2.832483in}{1.688653in}}%
\pgfpathlineto{\pgfqpoint{2.841555in}{1.694194in}}%
\pgfpathlineto{\pgfqpoint{2.850627in}{1.663510in}}%
\pgfpathlineto{\pgfqpoint{2.859699in}{1.610827in}}%
\pgfpathlineto{\pgfqpoint{2.868771in}{1.600615in}}%
\pgfpathlineto{\pgfqpoint{2.877843in}{1.645486in}}%
\pgfpathlineto{\pgfqpoint{2.886915in}{1.654946in}}%
\pgfpathlineto{\pgfqpoint{2.895987in}{1.679815in}}%
\pgfpathlineto{\pgfqpoint{2.905059in}{1.713547in}}%
\pgfpathlineto{\pgfqpoint{2.914130in}{1.678930in}}%
\pgfpathlineto{\pgfqpoint{2.923202in}{1.488265in}}%
\pgfpathlineto{\pgfqpoint{2.932274in}{1.488245in}}%
\pgfpathlineto{\pgfqpoint{2.941346in}{1.541478in}}%
\pgfpathlineto{\pgfqpoint{2.950418in}{1.631390in}}%
\pgfpathlineto{\pgfqpoint{2.959490in}{1.668403in}}%
\pgfpathlineto{\pgfqpoint{2.968562in}{1.599659in}}%
\pgfpathlineto{\pgfqpoint{2.977634in}{1.511141in}}%
\pgfpathlineto{\pgfqpoint{2.986706in}{1.445597in}}%
\pgfpathlineto{\pgfqpoint{2.995778in}{1.599120in}}%
\pgfpathlineto{\pgfqpoint{3.004849in}{1.612590in}}%
\pgfpathlineto{\pgfqpoint{3.013921in}{1.549735in}}%
\pgfpathlineto{\pgfqpoint{3.022993in}{1.547994in}}%
\pgfpathlineto{\pgfqpoint{3.032065in}{1.574896in}}%
\pgfpathlineto{\pgfqpoint{3.041137in}{1.728444in}}%
\pgfpathlineto{\pgfqpoint{3.059281in}{1.652281in}}%
\pgfpathlineto{\pgfqpoint{3.068353in}{1.607169in}}%
\pgfpathlineto{\pgfqpoint{3.077425in}{1.649509in}}%
\pgfpathlineto{\pgfqpoint{3.095569in}{1.750100in}}%
\pgfpathlineto{\pgfqpoint{3.104640in}{1.777307in}}%
\pgfpathlineto{\pgfqpoint{3.122784in}{1.674761in}}%
\pgfpathlineto{\pgfqpoint{3.131856in}{1.689691in}}%
\pgfpathlineto{\pgfqpoint{3.140928in}{1.579752in}}%
\pgfpathlineto{\pgfqpoint{3.150000in}{1.576965in}}%
\pgfpathlineto{\pgfqpoint{3.159072in}{1.654711in}}%
\pgfpathlineto{\pgfqpoint{3.159072in}{1.654711in}}%
\pgfusepath{stroke}%
\end{pgfscope}%
\begin{pgfscope}%
\pgfpathrectangle{\pgfqpoint{0.437500in}{0.330000in}}{\pgfqpoint{2.712500in}{2.310000in}}%
\pgfusepath{clip}%
\pgfsetbuttcap%
\pgfsetroundjoin%
\pgfsetlinewidth{3.011250pt}%
\definecolor{currentstroke}{rgb}{0.121569,0.466667,0.705882}%
\pgfsetstrokecolor{currentstroke}%
\pgfsetdash{{11.100000pt}{4.800000pt}}{0.000000pt}%
\pgfpathmoveto{\pgfqpoint{0.883268in}{2.650000in}}%
\pgfpathlineto{\pgfqpoint{0.927383in}{2.580553in}}%
\pgfpathlineto{\pgfqpoint{0.972742in}{2.513494in}}%
\pgfpathlineto{\pgfqpoint{1.018102in}{2.450367in}}%
\pgfpathlineto{\pgfqpoint{1.063462in}{2.390773in}}%
\pgfpathlineto{\pgfqpoint{1.108821in}{2.334372in}}%
\pgfpathlineto{\pgfqpoint{1.163253in}{2.270494in}}%
\pgfpathlineto{\pgfqpoint{1.217684in}{2.210355in}}%
\pgfpathlineto{\pgfqpoint{1.272115in}{2.153583in}}%
\pgfpathlineto{\pgfqpoint{1.326547in}{2.099859in}}%
\pgfpathlineto{\pgfqpoint{1.371906in}{2.057217in}}%
\pgfpathlineto{\pgfqpoint{1.380978in}{1.793710in}}%
\pgfpathlineto{\pgfqpoint{1.390050in}{1.676255in}}%
\pgfpathlineto{\pgfqpoint{1.399122in}{1.594848in}}%
\pgfpathlineto{\pgfqpoint{1.408194in}{1.530514in}}%
\pgfpathlineto{\pgfqpoint{1.417266in}{1.476629in}}%
\pgfpathlineto{\pgfqpoint{1.435410in}{1.388627in}}%
\pgfpathlineto{\pgfqpoint{1.453554in}{1.317620in}}%
\pgfpathlineto{\pgfqpoint{1.471697in}{1.257788in}}%
\pgfpathlineto{\pgfqpoint{1.489841in}{1.205968in}}%
\pgfpathlineto{\pgfqpoint{1.507985in}{1.160219in}}%
\pgfpathlineto{\pgfqpoint{1.526129in}{1.119251in}}%
\pgfpathlineto{\pgfqpoint{1.553344in}{1.064855in}}%
\pgfpathlineto{\pgfqpoint{1.580560in}{1.017118in}}%
\pgfpathlineto{\pgfqpoint{1.607776in}{0.974621in}}%
\pgfpathlineto{\pgfqpoint{1.634992in}{0.936359in}}%
\pgfpathlineto{\pgfqpoint{1.662207in}{0.901597in}}%
\pgfpathlineto{\pgfqpoint{1.689423in}{0.869776in}}%
\pgfpathlineto{\pgfqpoint{1.725711in}{0.831184in}}%
\pgfpathlineto{\pgfqpoint{1.761998in}{0.796263in}}%
\pgfpathlineto{\pgfqpoint{1.798286in}{0.764413in}}%
\pgfpathlineto{\pgfqpoint{1.834574in}{0.735166in}}%
\pgfpathlineto{\pgfqpoint{1.879933in}{0.701715in}}%
\pgfpathlineto{\pgfqpoint{1.925293in}{0.671192in}}%
\pgfpathlineto{\pgfqpoint{1.934365in}{0.665398in}}%
\pgfpathlineto{\pgfqpoint{1.943436in}{0.729583in}}%
\pgfpathlineto{\pgfqpoint{1.952508in}{0.776657in}}%
\pgfpathlineto{\pgfqpoint{1.961580in}{0.810769in}}%
\pgfpathlineto{\pgfqpoint{1.979724in}{0.864309in}}%
\pgfpathlineto{\pgfqpoint{1.997868in}{0.907975in}}%
\pgfpathlineto{\pgfqpoint{2.016012in}{0.945959in}}%
\pgfpathlineto{\pgfqpoint{2.043227in}{0.996009in}}%
\pgfpathlineto{\pgfqpoint{2.070443in}{1.040281in}}%
\pgfpathlineto{\pgfqpoint{2.097659in}{1.080346in}}%
\pgfpathlineto{\pgfqpoint{2.124875in}{1.117141in}}%
\pgfpathlineto{\pgfqpoint{2.161162in}{1.162147in}}%
\pgfpathlineto{\pgfqpoint{2.197450in}{1.203394in}}%
\pgfpathlineto{\pgfqpoint{2.233737in}{1.241546in}}%
\pgfpathlineto{\pgfqpoint{2.279097in}{1.285599in}}%
\pgfpathlineto{\pgfqpoint{2.324457in}{1.326236in}}%
\pgfpathlineto{\pgfqpoint{2.369816in}{1.363959in}}%
\pgfpathlineto{\pgfqpoint{2.415176in}{1.399152in}}%
\pgfpathlineto{\pgfqpoint{2.469607in}{1.438466in}}%
\pgfpathlineto{\pgfqpoint{2.524038in}{1.474985in}}%
\pgfpathlineto{\pgfqpoint{2.578470in}{1.509040in}}%
\pgfpathlineto{\pgfqpoint{2.641973in}{1.546013in}}%
\pgfpathlineto{\pgfqpoint{2.705477in}{1.580346in}}%
\pgfpathlineto{\pgfqpoint{2.768980in}{1.612326in}}%
\pgfpathlineto{\pgfqpoint{2.841555in}{1.646296in}}%
\pgfpathlineto{\pgfqpoint{2.914130in}{1.677799in}}%
\pgfpathlineto{\pgfqpoint{2.995778in}{1.710601in}}%
\pgfpathlineto{\pgfqpoint{3.077425in}{1.740892in}}%
\pgfpathlineto{\pgfqpoint{3.159072in}{1.768923in}}%
\pgfpathlineto{\pgfqpoint{3.159072in}{1.768923in}}%
\pgfusepath{stroke}%
\end{pgfscope}%
\begin{pgfscope}%
\pgfpathrectangle{\pgfqpoint{0.437500in}{0.330000in}}{\pgfqpoint{2.712500in}{2.310000in}}%
\pgfusepath{clip}%
\pgfsetbuttcap%
\pgfsetroundjoin%
\pgfsetlinewidth{3.011250pt}%
\definecolor{currentstroke}{rgb}{1.000000,0.498039,0.054902}%
\pgfsetstrokecolor{currentstroke}%
\pgfsetdash{{3.000000pt}{4.950000pt}}{0.000000pt}%
\pgfpathmoveto{\pgfqpoint{0.901568in}{2.650000in}}%
\pgfpathlineto{\pgfqpoint{0.945527in}{2.585053in}}%
\pgfpathlineto{\pgfqpoint{0.990886in}{2.522166in}}%
\pgfpathlineto{\pgfqpoint{1.036246in}{2.463026in}}%
\pgfpathlineto{\pgfqpoint{1.081605in}{2.407258in}}%
\pgfpathlineto{\pgfqpoint{1.126965in}{2.354541in}}%
\pgfpathlineto{\pgfqpoint{1.181396in}{2.294919in}}%
\pgfpathlineto{\pgfqpoint{1.235828in}{2.238877in}}%
\pgfpathlineto{\pgfqpoint{1.290259in}{2.186062in}}%
\pgfpathlineto{\pgfqpoint{1.344691in}{2.136168in}}%
\pgfpathlineto{\pgfqpoint{1.371906in}{2.112234in}}%
\pgfpathlineto{\pgfqpoint{1.380978in}{1.804192in}}%
\pgfpathlineto{\pgfqpoint{1.390050in}{1.671875in}}%
\pgfpathlineto{\pgfqpoint{1.399122in}{1.581987in}}%
\pgfpathlineto{\pgfqpoint{1.408194in}{1.512014in}}%
\pgfpathlineto{\pgfqpoint{1.417266in}{1.454133in}}%
\pgfpathlineto{\pgfqpoint{1.435410in}{1.361050in}}%
\pgfpathlineto{\pgfqpoint{1.453554in}{1.287270in}}%
\pgfpathlineto{\pgfqpoint{1.471697in}{1.226039in}}%
\pgfpathlineto{\pgfqpoint{1.489841in}{1.173711in}}%
\pgfpathlineto{\pgfqpoint{1.507985in}{1.128067in}}%
\pgfpathlineto{\pgfqpoint{1.526129in}{1.087644in}}%
\pgfpathlineto{\pgfqpoint{1.544273in}{1.051417in}}%
\pgfpathlineto{\pgfqpoint{1.571488in}{1.003371in}}%
\pgfpathlineto{\pgfqpoint{1.598704in}{0.961337in}}%
\pgfpathlineto{\pgfqpoint{1.625920in}{0.924078in}}%
\pgfpathlineto{\pgfqpoint{1.653135in}{0.890700in}}%
\pgfpathlineto{\pgfqpoint{1.680351in}{0.860538in}}%
\pgfpathlineto{\pgfqpoint{1.707567in}{0.833082in}}%
\pgfpathlineto{\pgfqpoint{1.743855in}{0.800003in}}%
\pgfpathlineto{\pgfqpoint{1.780142in}{0.770299in}}%
\pgfpathlineto{\pgfqpoint{1.816430in}{0.743410in}}%
\pgfpathlineto{\pgfqpoint{1.852717in}{0.718903in}}%
\pgfpathlineto{\pgfqpoint{1.898077in}{0.691098in}}%
\pgfpathlineto{\pgfqpoint{1.934365in}{0.670787in}}%
\pgfpathlineto{\pgfqpoint{1.943436in}{0.728161in}}%
\pgfpathlineto{\pgfqpoint{1.952508in}{0.770830in}}%
\pgfpathlineto{\pgfqpoint{1.961580in}{0.802063in}}%
\pgfpathlineto{\pgfqpoint{1.979724in}{0.851571in}}%
\pgfpathlineto{\pgfqpoint{1.997868in}{0.892360in}}%
\pgfpathlineto{\pgfqpoint{2.016012in}{0.928121in}}%
\pgfpathlineto{\pgfqpoint{2.043227in}{0.975621in}}%
\pgfpathlineto{\pgfqpoint{2.070443in}{1.017975in}}%
\pgfpathlineto{\pgfqpoint{2.097659in}{1.056570in}}%
\pgfpathlineto{\pgfqpoint{2.133946in}{1.103560in}}%
\pgfpathlineto{\pgfqpoint{2.170234in}{1.146526in}}%
\pgfpathlineto{\pgfqpoint{2.206522in}{1.186249in}}%
\pgfpathlineto{\pgfqpoint{2.251881in}{1.232150in}}%
\pgfpathlineto{\pgfqpoint{2.297241in}{1.274560in}}%
\pgfpathlineto{\pgfqpoint{2.342600in}{1.314015in}}%
\pgfpathlineto{\pgfqpoint{2.397032in}{1.358018in}}%
\pgfpathlineto{\pgfqpoint{2.451463in}{1.398867in}}%
\pgfpathlineto{\pgfqpoint{2.505895in}{1.436967in}}%
\pgfpathlineto{\pgfqpoint{2.560326in}{1.472640in}}%
\pgfpathlineto{\pgfqpoint{2.623829in}{1.511537in}}%
\pgfpathlineto{\pgfqpoint{2.687333in}{1.547822in}}%
\pgfpathlineto{\pgfqpoint{2.750836in}{1.581771in}}%
\pgfpathlineto{\pgfqpoint{2.823411in}{1.618005in}}%
\pgfpathlineto{\pgfqpoint{2.895987in}{1.651774in}}%
\pgfpathlineto{\pgfqpoint{2.977634in}{1.687116in}}%
\pgfpathlineto{\pgfqpoint{3.059281in}{1.719931in}}%
\pgfpathlineto{\pgfqpoint{3.140928in}{1.750459in}}%
\pgfpathlineto{\pgfqpoint{3.159072in}{1.756954in}}%
\pgfpathlineto{\pgfqpoint{3.159072in}{1.756954in}}%
\pgfusepath{stroke}%
\end{pgfscope}%
\begin{pgfscope}%
\pgfsetrectcap%
\pgfsetmiterjoin%
\pgfsetlinewidth{0.803000pt}%
\definecolor{currentstroke}{rgb}{0.000000,0.000000,0.000000}%
\pgfsetstrokecolor{currentstroke}%
\pgfsetdash{}{0pt}%
\pgfpathmoveto{\pgfqpoint{0.437500in}{0.330000in}}%
\pgfpathlineto{\pgfqpoint{0.437500in}{2.640000in}}%
\pgfusepath{stroke}%
\end{pgfscope}%
\begin{pgfscope}%
\pgfsetrectcap%
\pgfsetmiterjoin%
\pgfsetlinewidth{0.803000pt}%
\definecolor{currentstroke}{rgb}{0.000000,0.000000,0.000000}%
\pgfsetstrokecolor{currentstroke}%
\pgfsetdash{}{0pt}%
\pgfpathmoveto{\pgfqpoint{3.150000in}{0.330000in}}%
\pgfpathlineto{\pgfqpoint{3.150000in}{2.640000in}}%
\pgfusepath{stroke}%
\end{pgfscope}%
\begin{pgfscope}%
\pgfsetrectcap%
\pgfsetmiterjoin%
\pgfsetlinewidth{0.803000pt}%
\definecolor{currentstroke}{rgb}{0.000000,0.000000,0.000000}%
\pgfsetstrokecolor{currentstroke}%
\pgfsetdash{}{0pt}%
\pgfpathmoveto{\pgfqpoint{0.437500in}{0.330000in}}%
\pgfpathlineto{\pgfqpoint{3.150000in}{0.330000in}}%
\pgfusepath{stroke}%
\end{pgfscope}%
\begin{pgfscope}%
\pgfsetrectcap%
\pgfsetmiterjoin%
\pgfsetlinewidth{0.803000pt}%
\definecolor{currentstroke}{rgb}{0.000000,0.000000,0.000000}%
\pgfsetstrokecolor{currentstroke}%
\pgfsetdash{}{0pt}%
\pgfpathmoveto{\pgfqpoint{0.437500in}{2.640000in}}%
\pgfpathlineto{\pgfqpoint{3.150000in}{2.640000in}}%
\pgfusepath{stroke}%
\end{pgfscope}%
\end{pgfpicture}%
\makeatother%
\endgroup%

%% file: figures/contrast.pgf
\begingroup%
\makeatletter%
\begin{pgfpicture}%
\pgfpathrectangle{\pgfpointorigin}{\pgfqpoint{3.500000in}{2.500000in}}%
\pgfusepath{use as bounding box, clip}%
\begin{pgfscope}%
\pgfsetbuttcap%
\pgfsetmiterjoin%
\definecolor{currentfill}{rgb}{1.000000,1.000000,1.000000}%
\pgfsetfillcolor{currentfill}%
\pgfsetlinewidth{0.000000pt}%
\definecolor{currentstroke}{rgb}{1.000000,1.000000,1.000000}%
\pgfsetstrokecolor{currentstroke}%
\pgfsetdash{}{0pt}%
\pgfpathmoveto{\pgfqpoint{0.000000in}{0.000000in}}%
\pgfpathlineto{\pgfqpoint{3.500000in}{0.000000in}}%
\pgfpathlineto{\pgfqpoint{3.500000in}{2.500000in}}%
\pgfpathlineto{\pgfqpoint{0.000000in}{2.500000in}}%
\pgfpathclose%
\pgfusepath{fill}%
\end{pgfscope}%
\begin{pgfscope}%
\pgfsetbuttcap%
\pgfsetmiterjoin%
\definecolor{currentfill}{rgb}{1.000000,1.000000,1.000000}%
\pgfsetfillcolor{currentfill}%
\pgfsetlinewidth{0.000000pt}%
\definecolor{currentstroke}{rgb}{0.000000,0.000000,0.000000}%
\pgfsetstrokecolor{currentstroke}%
\pgfsetstrokeopacity{0.000000}%
\pgfsetdash{}{0pt}%
\pgfpathmoveto{\pgfqpoint{0.394722in}{0.512778in}}%
\pgfpathlineto{\pgfqpoint{3.344375in}{0.512778in}}%
\pgfpathlineto{\pgfqpoint{3.344375in}{2.138542in}}%
\pgfpathlineto{\pgfqpoint{0.394722in}{2.138542in}}%
\pgfpathclose%
\pgfusepath{fill}%
\end{pgfscope}%
\begin{pgfscope}%
\pgfpathrectangle{\pgfqpoint{0.394722in}{0.512778in}}{\pgfqpoint{2.949653in}{1.625764in}}%
\pgfusepath{clip}%
\pgfsetbuttcap%
\pgfsetroundjoin%
\definecolor{currentfill}{rgb}{0.501961,0.501961,0.501961}%
\pgfsetfillcolor{currentfill}%
\pgfsetfillopacity{0.500000}%
\pgfsetlinewidth{1.003750pt}%
\definecolor{currentstroke}{rgb}{0.501961,0.501961,0.501961}%
\pgfsetstrokecolor{currentstroke}%
\pgfsetstrokeopacity{0.500000}%
\pgfsetdash{}{0pt}%
\pgfsys@defobject{currentmarker}{\pgfqpoint{2.081256in}{1.098732in}}{\pgfqpoint{2.470456in}{1.407700in}}{%
\pgfpathmoveto{\pgfqpoint{2.081256in}{1.407700in}}%
\pgfpathlineto{\pgfqpoint{2.081256in}{1.190741in}}%
\pgfpathlineto{\pgfqpoint{2.113689in}{1.182429in}}%
\pgfpathlineto{\pgfqpoint{2.146122in}{1.174241in}}%
\pgfpathlineto{\pgfqpoint{2.178556in}{1.166174in}}%
\pgfpathlineto{\pgfqpoint{2.210989in}{1.158228in}}%
\pgfpathlineto{\pgfqpoint{2.243422in}{1.150399in}}%
\pgfpathlineto{\pgfqpoint{2.275856in}{1.142686in}}%
\pgfpathlineto{\pgfqpoint{2.308289in}{1.135086in}}%
\pgfpathlineto{\pgfqpoint{2.340722in}{1.127599in}}%
\pgfpathlineto{\pgfqpoint{2.373156in}{1.120221in}}%
\pgfpathlineto{\pgfqpoint{2.405589in}{1.112952in}}%
\pgfpathlineto{\pgfqpoint{2.438022in}{1.105790in}}%
\pgfpathlineto{\pgfqpoint{2.470456in}{1.098732in}}%
\pgfpathlineto{\pgfqpoint{2.470456in}{1.283218in}}%
\pgfpathlineto{\pgfqpoint{2.470456in}{1.283218in}}%
\pgfpathlineto{\pgfqpoint{2.438022in}{1.292767in}}%
\pgfpathlineto{\pgfqpoint{2.405589in}{1.302457in}}%
\pgfpathlineto{\pgfqpoint{2.373156in}{1.312292in}}%
\pgfpathlineto{\pgfqpoint{2.340722in}{1.322273in}}%
\pgfpathlineto{\pgfqpoint{2.308289in}{1.332403in}}%
\pgfpathlineto{\pgfqpoint{2.275856in}{1.342685in}}%
\pgfpathlineto{\pgfqpoint{2.243422in}{1.353120in}}%
\pgfpathlineto{\pgfqpoint{2.210989in}{1.363712in}}%
\pgfpathlineto{\pgfqpoint{2.178556in}{1.374464in}}%
\pgfpathlineto{\pgfqpoint{2.146122in}{1.385377in}}%
\pgfpathlineto{\pgfqpoint{2.113689in}{1.396455in}}%
\pgfpathlineto{\pgfqpoint{2.081256in}{1.407700in}}%
\pgfpathclose%
\pgfusepath{stroke,fill}%
}%
\begin{pgfscope}%
\pgfsys@transformshift{0.000000in}{0.000000in}%
\pgfsys@useobject{currentmarker}{}%
\end{pgfscope}%
\end{pgfscope}%
\begin{pgfscope}%
\pgfpathrectangle{\pgfqpoint{0.394722in}{0.512778in}}{\pgfqpoint{2.949653in}{1.625764in}}%
\pgfusepath{clip}%
\pgfsetbuttcap%
\pgfsetroundjoin%
\definecolor{currentfill}{rgb}{0.501961,0.501961,0.501961}%
\pgfsetfillcolor{currentfill}%
\pgfsetfillopacity{0.500000}%
\pgfsetlinewidth{1.003750pt}%
\definecolor{currentstroke}{rgb}{0.501961,0.501961,0.501961}%
\pgfsetstrokecolor{currentstroke}%
\pgfsetstrokeopacity{0.500000}%
\pgfsetdash{}{0pt}%
\pgfsys@defobject{currentmarker}{\pgfqpoint{0.753111in}{1.125054in}}{\pgfqpoint{1.327181in}{1.450184in}}{%
\pgfpathmoveto{\pgfqpoint{0.879601in}{1.450184in}}%
\pgfpathlineto{\pgfqpoint{0.753111in}{1.450184in}}%
\pgfpathlineto{\pgfqpoint{0.780679in}{1.415935in}}%
\pgfpathlineto{\pgfqpoint{0.808247in}{1.383223in}}%
\pgfpathlineto{\pgfqpoint{0.835816in}{1.351968in}}%
\pgfpathlineto{\pgfqpoint{0.863384in}{1.322092in}}%
\pgfpathlineto{\pgfqpoint{0.890952in}{1.293524in}}%
\pgfpathlineto{\pgfqpoint{0.918521in}{1.266199in}}%
\pgfpathlineto{\pgfqpoint{0.946089in}{1.240052in}}%
\pgfpathlineto{\pgfqpoint{0.973657in}{1.215025in}}%
\pgfpathlineto{\pgfqpoint{1.001226in}{1.191062in}}%
\pgfpathlineto{\pgfqpoint{1.028794in}{1.168112in}}%
\pgfpathlineto{\pgfqpoint{1.056362in}{1.146125in}}%
\pgfpathlineto{\pgfqpoint{1.083931in}{1.125054in}}%
\pgfpathlineto{\pgfqpoint{1.327181in}{1.125054in}}%
\pgfpathlineto{\pgfqpoint{1.327181in}{1.125054in}}%
\pgfpathlineto{\pgfqpoint{1.289882in}{1.146125in}}%
\pgfpathlineto{\pgfqpoint{1.252584in}{1.168112in}}%
\pgfpathlineto{\pgfqpoint{1.215286in}{1.191062in}}%
\pgfpathlineto{\pgfqpoint{1.177987in}{1.215025in}}%
\pgfpathlineto{\pgfqpoint{1.140689in}{1.240052in}}%
\pgfpathlineto{\pgfqpoint{1.103391in}{1.266199in}}%
\pgfpathlineto{\pgfqpoint{1.066092in}{1.293524in}}%
\pgfpathlineto{\pgfqpoint{1.028794in}{1.322092in}}%
\pgfpathlineto{\pgfqpoint{0.991496in}{1.351968in}}%
\pgfpathlineto{\pgfqpoint{0.954197in}{1.383223in}}%
\pgfpathlineto{\pgfqpoint{0.916899in}{1.415935in}}%
\pgfpathlineto{\pgfqpoint{0.879601in}{1.450184in}}%
\pgfpathclose%
\pgfusepath{stroke,fill}%
}%
\begin{pgfscope}%
\pgfsys@transformshift{0.000000in}{0.000000in}%
\pgfsys@useobject{currentmarker}{}%
\end{pgfscope}%
\end{pgfscope}%
\begin{pgfscope}%
\pgfsetbuttcap%
\pgfsetroundjoin%
\definecolor{currentfill}{rgb}{0.000000,0.000000,0.000000}%
\pgfsetfillcolor{currentfill}%
\pgfsetlinewidth{0.803000pt}%
\definecolor{currentstroke}{rgb}{0.000000,0.000000,0.000000}%
\pgfsetstrokecolor{currentstroke}%
\pgfsetdash{}{0pt}%
\pgfsys@defobject{currentmarker}{\pgfqpoint{0.000000in}{-0.048611in}}{\pgfqpoint{0.000000in}{0.000000in}}{%
\pgfpathmoveto{\pgfqpoint{0.000000in}{0.000000in}}%
\pgfpathlineto{\pgfqpoint{0.000000in}{-0.048611in}}%
\pgfusepath{stroke,fill}%
}%
\begin{pgfscope}%
\pgfsys@transformshift{0.394722in}{0.512778in}%
\pgfsys@useobject{currentmarker}{}%
\end{pgfscope}%
\end{pgfscope}%
\begin{pgfscope}%
\definecolor{textcolor}{rgb}{0.000000,0.000000,0.000000}%
\pgfsetstrokecolor{textcolor}%
\pgfsetfillcolor{textcolor}%
\pgftext[x=0.394722in,y=0.415556in,,top]{\color{textcolor}\sffamily\fontsize{8.000000}{9.600000}\selectfont 0}%
\end{pgfscope}%
\begin{pgfscope}%
\pgfsetbuttcap%
\pgfsetroundjoin%
\definecolor{currentfill}{rgb}{0.000000,0.000000,0.000000}%
\pgfsetfillcolor{currentfill}%
\pgfsetlinewidth{0.803000pt}%
\definecolor{currentstroke}{rgb}{0.000000,0.000000,0.000000}%
\pgfsetstrokecolor{currentstroke}%
\pgfsetdash{}{0pt}%
\pgfsys@defobject{currentmarker}{\pgfqpoint{0.000000in}{-0.048611in}}{\pgfqpoint{0.000000in}{0.000000in}}{%
\pgfpathmoveto{\pgfqpoint{0.000000in}{0.000000in}}%
\pgfpathlineto{\pgfqpoint{0.000000in}{-0.048611in}}%
\pgfusepath{stroke,fill}%
}%
\begin{pgfscope}%
\pgfsys@transformshift{0.816101in}{0.512778in}%
\pgfsys@useobject{currentmarker}{}%
\end{pgfscope}%
\end{pgfscope}%
\begin{pgfscope}%
\definecolor{textcolor}{rgb}{0.000000,0.000000,0.000000}%
\pgfsetstrokecolor{textcolor}%
\pgfsetfillcolor{textcolor}%
\pgftext[x=0.816101in,y=0.415556in,,top]{\color{textcolor}\sffamily\fontsize{8.000000}{9.600000}\selectfont 1}%
\end{pgfscope}%
\begin{pgfscope}%
\pgfsetbuttcap%
\pgfsetroundjoin%
\definecolor{currentfill}{rgb}{0.000000,0.000000,0.000000}%
\pgfsetfillcolor{currentfill}%
\pgfsetlinewidth{0.803000pt}%
\definecolor{currentstroke}{rgb}{0.000000,0.000000,0.000000}%
\pgfsetstrokecolor{currentstroke}%
\pgfsetdash{}{0pt}%
\pgfsys@defobject{currentmarker}{\pgfqpoint{0.000000in}{-0.048611in}}{\pgfqpoint{0.000000in}{0.000000in}}{%
\pgfpathmoveto{\pgfqpoint{0.000000in}{0.000000in}}%
\pgfpathlineto{\pgfqpoint{0.000000in}{-0.048611in}}%
\pgfusepath{stroke,fill}%
}%
\begin{pgfscope}%
\pgfsys@transformshift{1.237480in}{0.512778in}%
\pgfsys@useobject{currentmarker}{}%
\end{pgfscope}%
\end{pgfscope}%
\begin{pgfscope}%
\definecolor{textcolor}{rgb}{0.000000,0.000000,0.000000}%
\pgfsetstrokecolor{textcolor}%
\pgfsetfillcolor{textcolor}%
\pgftext[x=1.237480in,y=0.415556in,,top]{\color{textcolor}\sffamily\fontsize{8.000000}{9.600000}\selectfont 2}%
\end{pgfscope}%
\begin{pgfscope}%
\pgfsetbuttcap%
\pgfsetroundjoin%
\definecolor{currentfill}{rgb}{0.000000,0.000000,0.000000}%
\pgfsetfillcolor{currentfill}%
\pgfsetlinewidth{0.803000pt}%
\definecolor{currentstroke}{rgb}{0.000000,0.000000,0.000000}%
\pgfsetstrokecolor{currentstroke}%
\pgfsetdash{}{0pt}%
\pgfsys@defobject{currentmarker}{\pgfqpoint{0.000000in}{-0.048611in}}{\pgfqpoint{0.000000in}{0.000000in}}{%
\pgfpathmoveto{\pgfqpoint{0.000000in}{0.000000in}}%
\pgfpathlineto{\pgfqpoint{0.000000in}{-0.048611in}}%
\pgfusepath{stroke,fill}%
}%
\begin{pgfscope}%
\pgfsys@transformshift{1.658859in}{0.512778in}%
\pgfsys@useobject{currentmarker}{}%
\end{pgfscope}%
\end{pgfscope}%
\begin{pgfscope}%
\definecolor{textcolor}{rgb}{0.000000,0.000000,0.000000}%
\pgfsetstrokecolor{textcolor}%
\pgfsetfillcolor{textcolor}%
\pgftext[x=1.658859in,y=0.415556in,,top]{\color{textcolor}\sffamily\fontsize{8.000000}{9.600000}\selectfont 3}%
\end{pgfscope}%
\begin{pgfscope}%
\pgfsetbuttcap%
\pgfsetroundjoin%
\definecolor{currentfill}{rgb}{0.000000,0.000000,0.000000}%
\pgfsetfillcolor{currentfill}%
\pgfsetlinewidth{0.803000pt}%
\definecolor{currentstroke}{rgb}{0.000000,0.000000,0.000000}%
\pgfsetstrokecolor{currentstroke}%
\pgfsetdash{}{0pt}%
\pgfsys@defobject{currentmarker}{\pgfqpoint{0.000000in}{-0.048611in}}{\pgfqpoint{0.000000in}{0.000000in}}{%
\pgfpathmoveto{\pgfqpoint{0.000000in}{0.000000in}}%
\pgfpathlineto{\pgfqpoint{0.000000in}{-0.048611in}}%
\pgfusepath{stroke,fill}%
}%
\begin{pgfscope}%
\pgfsys@transformshift{2.080238in}{0.512778in}%
\pgfsys@useobject{currentmarker}{}%
\end{pgfscope}%
\end{pgfscope}%
\begin{pgfscope}%
\definecolor{textcolor}{rgb}{0.000000,0.000000,0.000000}%
\pgfsetstrokecolor{textcolor}%
\pgfsetfillcolor{textcolor}%
\pgftext[x=2.080238in,y=0.415556in,,top]{\color{textcolor}\sffamily\fontsize{8.000000}{9.600000}\selectfont 4}%
\end{pgfscope}%
\begin{pgfscope}%
\pgfsetbuttcap%
\pgfsetroundjoin%
\definecolor{currentfill}{rgb}{0.000000,0.000000,0.000000}%
\pgfsetfillcolor{currentfill}%
\pgfsetlinewidth{0.803000pt}%
\definecolor{currentstroke}{rgb}{0.000000,0.000000,0.000000}%
\pgfsetstrokecolor{currentstroke}%
\pgfsetdash{}{0pt}%
\pgfsys@defobject{currentmarker}{\pgfqpoint{0.000000in}{-0.048611in}}{\pgfqpoint{0.000000in}{0.000000in}}{%
\pgfpathmoveto{\pgfqpoint{0.000000in}{0.000000in}}%
\pgfpathlineto{\pgfqpoint{0.000000in}{-0.048611in}}%
\pgfusepath{stroke,fill}%
}%
\begin{pgfscope}%
\pgfsys@transformshift{2.501617in}{0.512778in}%
\pgfsys@useobject{currentmarker}{}%
\end{pgfscope}%
\end{pgfscope}%
\begin{pgfscope}%
\definecolor{textcolor}{rgb}{0.000000,0.000000,0.000000}%
\pgfsetstrokecolor{textcolor}%
\pgfsetfillcolor{textcolor}%
\pgftext[x=2.501617in,y=0.415556in,,top]{\color{textcolor}\sffamily\fontsize{8.000000}{9.600000}\selectfont 5}%
\end{pgfscope}%
\begin{pgfscope}%
\pgfsetbuttcap%
\pgfsetroundjoin%
\definecolor{currentfill}{rgb}{0.000000,0.000000,0.000000}%
\pgfsetfillcolor{currentfill}%
\pgfsetlinewidth{0.803000pt}%
\definecolor{currentstroke}{rgb}{0.000000,0.000000,0.000000}%
\pgfsetstrokecolor{currentstroke}%
\pgfsetdash{}{0pt}%
\pgfsys@defobject{currentmarker}{\pgfqpoint{0.000000in}{-0.048611in}}{\pgfqpoint{0.000000in}{0.000000in}}{%
\pgfpathmoveto{\pgfqpoint{0.000000in}{0.000000in}}%
\pgfpathlineto{\pgfqpoint{0.000000in}{-0.048611in}}%
\pgfusepath{stroke,fill}%
}%
\begin{pgfscope}%
\pgfsys@transformshift{2.922996in}{0.512778in}%
\pgfsys@useobject{currentmarker}{}%
\end{pgfscope}%
\end{pgfscope}%
\begin{pgfscope}%
\definecolor{textcolor}{rgb}{0.000000,0.000000,0.000000}%
\pgfsetstrokecolor{textcolor}%
\pgfsetfillcolor{textcolor}%
\pgftext[x=2.922996in,y=0.415556in,,top]{\color{textcolor}\sffamily\fontsize{8.000000}{9.600000}\selectfont 6}%
\end{pgfscope}%
\begin{pgfscope}%
\pgfsetbuttcap%
\pgfsetroundjoin%
\definecolor{currentfill}{rgb}{0.000000,0.000000,0.000000}%
\pgfsetfillcolor{currentfill}%
\pgfsetlinewidth{0.803000pt}%
\definecolor{currentstroke}{rgb}{0.000000,0.000000,0.000000}%
\pgfsetstrokecolor{currentstroke}%
\pgfsetdash{}{0pt}%
\pgfsys@defobject{currentmarker}{\pgfqpoint{0.000000in}{-0.048611in}}{\pgfqpoint{0.000000in}{0.000000in}}{%
\pgfpathmoveto{\pgfqpoint{0.000000in}{0.000000in}}%
\pgfpathlineto{\pgfqpoint{0.000000in}{-0.048611in}}%
\pgfusepath{stroke,fill}%
}%
\begin{pgfscope}%
\pgfsys@transformshift{3.344375in}{0.512778in}%
\pgfsys@useobject{currentmarker}{}%
\end{pgfscope}%
\end{pgfscope}%
\begin{pgfscope}%
\definecolor{textcolor}{rgb}{0.000000,0.000000,0.000000}%
\pgfsetstrokecolor{textcolor}%
\pgfsetfillcolor{textcolor}%
\pgftext[x=3.344375in,y=0.415556in,,top]{\color{textcolor}\sffamily\fontsize{8.000000}{9.600000}\selectfont 7}%
\end{pgfscope}%
\begin{pgfscope}%
\definecolor{textcolor}{rgb}{0.000000,0.000000,0.000000}%
\pgfsetstrokecolor{textcolor}%
\pgfsetfillcolor{textcolor}%
\pgftext[x=1.869549in,y=0.261333in,,top]{\color{textcolor}\sffamily\fontsize{8.000000}{9.600000}\selectfont path length, \(\displaystyle \ell\) (cm)}%
\end{pgfscope}%
\begin{pgfscope}%
\pgfsetbuttcap%
\pgfsetroundjoin%
\definecolor{currentfill}{rgb}{0.000000,0.000000,0.000000}%
\pgfsetfillcolor{currentfill}%
\pgfsetlinewidth{0.803000pt}%
\definecolor{currentstroke}{rgb}{0.000000,0.000000,0.000000}%
\pgfsetstrokecolor{currentstroke}%
\pgfsetdash{}{0pt}%
\pgfsys@defobject{currentmarker}{\pgfqpoint{-0.048611in}{0.000000in}}{\pgfqpoint{-0.000000in}{0.000000in}}{%
\pgfpathmoveto{\pgfqpoint{-0.000000in}{0.000000in}}%
\pgfpathlineto{\pgfqpoint{-0.048611in}{0.000000in}}%
\pgfusepath{stroke,fill}%
}%
\begin{pgfscope}%
\pgfsys@transformshift{0.394722in}{0.576021in}%
\pgfsys@useobject{currentmarker}{}%
\end{pgfscope}%
\end{pgfscope}%
\begin{pgfscope}%
\definecolor{textcolor}{rgb}{0.000000,0.000000,0.000000}%
\pgfsetstrokecolor{textcolor}%
\pgfsetfillcolor{textcolor}%
\pgftext[x=0.146722in, y=0.537466in, left, base]{\color{textcolor}\sffamily\fontsize{8.000000}{9.600000}\selectfont 0.0}%
\end{pgfscope}%
\begin{pgfscope}%
\pgfsetbuttcap%
\pgfsetroundjoin%
\definecolor{currentfill}{rgb}{0.000000,0.000000,0.000000}%
\pgfsetfillcolor{currentfill}%
\pgfsetlinewidth{0.803000pt}%
\definecolor{currentstroke}{rgb}{0.000000,0.000000,0.000000}%
\pgfsetstrokecolor{currentstroke}%
\pgfsetdash{}{0pt}%
\pgfsys@defobject{currentmarker}{\pgfqpoint{-0.048611in}{0.000000in}}{\pgfqpoint{-0.000000in}{0.000000in}}{%
\pgfpathmoveto{\pgfqpoint{-0.000000in}{0.000000in}}%
\pgfpathlineto{\pgfqpoint{-0.048611in}{0.000000in}}%
\pgfusepath{stroke,fill}%
}%
\begin{pgfscope}%
\pgfsys@transformshift{0.394722in}{0.873746in}%
\pgfsys@useobject{currentmarker}{}%
\end{pgfscope}%
\end{pgfscope}%
\begin{pgfscope}%
\definecolor{textcolor}{rgb}{0.000000,0.000000,0.000000}%
\pgfsetstrokecolor{textcolor}%
\pgfsetfillcolor{textcolor}%
\pgftext[x=0.146722in, y=0.835190in, left, base]{\color{textcolor}\sffamily\fontsize{8.000000}{9.600000}\selectfont 0.2}%
\end{pgfscope}%
\begin{pgfscope}%
\pgfsetbuttcap%
\pgfsetroundjoin%
\definecolor{currentfill}{rgb}{0.000000,0.000000,0.000000}%
\pgfsetfillcolor{currentfill}%
\pgfsetlinewidth{0.803000pt}%
\definecolor{currentstroke}{rgb}{0.000000,0.000000,0.000000}%
\pgfsetstrokecolor{currentstroke}%
\pgfsetdash{}{0pt}%
\pgfsys@defobject{currentmarker}{\pgfqpoint{-0.048611in}{0.000000in}}{\pgfqpoint{-0.000000in}{0.000000in}}{%
\pgfpathmoveto{\pgfqpoint{-0.000000in}{0.000000in}}%
\pgfpathlineto{\pgfqpoint{-0.048611in}{0.000000in}}%
\pgfusepath{stroke,fill}%
}%
\begin{pgfscope}%
\pgfsys@transformshift{0.394722in}{1.171470in}%
\pgfsys@useobject{currentmarker}{}%
\end{pgfscope}%
\end{pgfscope}%
\begin{pgfscope}%
\definecolor{textcolor}{rgb}{0.000000,0.000000,0.000000}%
\pgfsetstrokecolor{textcolor}%
\pgfsetfillcolor{textcolor}%
\pgftext[x=0.146722in, y=1.132915in, left, base]{\color{textcolor}\sffamily\fontsize{8.000000}{9.600000}\selectfont 0.4}%
\end{pgfscope}%
\begin{pgfscope}%
\pgfsetbuttcap%
\pgfsetroundjoin%
\definecolor{currentfill}{rgb}{0.000000,0.000000,0.000000}%
\pgfsetfillcolor{currentfill}%
\pgfsetlinewidth{0.803000pt}%
\definecolor{currentstroke}{rgb}{0.000000,0.000000,0.000000}%
\pgfsetstrokecolor{currentstroke}%
\pgfsetdash{}{0pt}%
\pgfsys@defobject{currentmarker}{\pgfqpoint{-0.048611in}{0.000000in}}{\pgfqpoint{-0.000000in}{0.000000in}}{%
\pgfpathmoveto{\pgfqpoint{-0.000000in}{0.000000in}}%
\pgfpathlineto{\pgfqpoint{-0.048611in}{0.000000in}}%
\pgfusepath{stroke,fill}%
}%
\begin{pgfscope}%
\pgfsys@transformshift{0.394722in}{1.469195in}%
\pgfsys@useobject{currentmarker}{}%
\end{pgfscope}%
\end{pgfscope}%
\begin{pgfscope}%
\definecolor{textcolor}{rgb}{0.000000,0.000000,0.000000}%
\pgfsetstrokecolor{textcolor}%
\pgfsetfillcolor{textcolor}%
\pgftext[x=0.146722in, y=1.430639in, left, base]{\color{textcolor}\sffamily\fontsize{8.000000}{9.600000}\selectfont 0.6}%
\end{pgfscope}%
\begin{pgfscope}%
\pgfsetbuttcap%
\pgfsetroundjoin%
\definecolor{currentfill}{rgb}{0.000000,0.000000,0.000000}%
\pgfsetfillcolor{currentfill}%
\pgfsetlinewidth{0.803000pt}%
\definecolor{currentstroke}{rgb}{0.000000,0.000000,0.000000}%
\pgfsetstrokecolor{currentstroke}%
\pgfsetdash{}{0pt}%
\pgfsys@defobject{currentmarker}{\pgfqpoint{-0.048611in}{0.000000in}}{\pgfqpoint{-0.000000in}{0.000000in}}{%
\pgfpathmoveto{\pgfqpoint{-0.000000in}{0.000000in}}%
\pgfpathlineto{\pgfqpoint{-0.048611in}{0.000000in}}%
\pgfusepath{stroke,fill}%
}%
\begin{pgfscope}%
\pgfsys@transformshift{0.394722in}{1.766919in}%
\pgfsys@useobject{currentmarker}{}%
\end{pgfscope}%
\end{pgfscope}%
\begin{pgfscope}%
\definecolor{textcolor}{rgb}{0.000000,0.000000,0.000000}%
\pgfsetstrokecolor{textcolor}%
\pgfsetfillcolor{textcolor}%
\pgftext[x=0.146722in, y=1.728363in, left, base]{\color{textcolor}\sffamily\fontsize{8.000000}{9.600000}\selectfont 0.8}%
\end{pgfscope}%
\begin{pgfscope}%
\pgfsetbuttcap%
\pgfsetroundjoin%
\definecolor{currentfill}{rgb}{0.000000,0.000000,0.000000}%
\pgfsetfillcolor{currentfill}%
\pgfsetlinewidth{0.803000pt}%
\definecolor{currentstroke}{rgb}{0.000000,0.000000,0.000000}%
\pgfsetstrokecolor{currentstroke}%
\pgfsetdash{}{0pt}%
\pgfsys@defobject{currentmarker}{\pgfqpoint{-0.048611in}{0.000000in}}{\pgfqpoint{-0.000000in}{0.000000in}}{%
\pgfpathmoveto{\pgfqpoint{-0.000000in}{0.000000in}}%
\pgfpathlineto{\pgfqpoint{-0.048611in}{0.000000in}}%
\pgfusepath{stroke,fill}%
}%
\begin{pgfscope}%
\pgfsys@transformshift{0.394722in}{2.064643in}%
\pgfsys@useobject{currentmarker}{}%
\end{pgfscope}%
\end{pgfscope}%
\begin{pgfscope}%
\definecolor{textcolor}{rgb}{0.000000,0.000000,0.000000}%
\pgfsetstrokecolor{textcolor}%
\pgfsetfillcolor{textcolor}%
\pgftext[x=0.146722in, y=2.026088in, left, base]{\color{textcolor}\sffamily\fontsize{8.000000}{9.600000}\selectfont 1.0}%
\end{pgfscope}%
\begin{pgfscope}%
\pgfpathrectangle{\pgfqpoint{0.394722in}{0.512778in}}{\pgfqpoint{2.949653in}{1.625764in}}%
\pgfusepath{clip}%
\pgfsetrectcap%
\pgfsetroundjoin%
\pgfsetlinewidth{1.505625pt}%
\definecolor{currentstroke}{rgb}{0.121569,0.466667,0.705882}%
\pgfsetstrokecolor{currentstroke}%
\pgfsetdash{}{0pt}%
\pgfpathmoveto{\pgfqpoint{0.394722in}{2.064643in}}%
\pgfpathlineto{\pgfqpoint{0.427156in}{2.054460in}}%
\pgfpathlineto{\pgfqpoint{0.459589in}{2.038493in}}%
\pgfpathlineto{\pgfqpoint{0.492022in}{2.019879in}}%
\pgfpathlineto{\pgfqpoint{0.524456in}{2.000146in}}%
\pgfpathlineto{\pgfqpoint{0.556889in}{1.980041in}}%
\pgfpathlineto{\pgfqpoint{0.589322in}{1.959924in}}%
\pgfpathlineto{\pgfqpoint{0.621756in}{1.939968in}}%
\pgfpathlineto{\pgfqpoint{0.654189in}{1.920255in}}%
\pgfpathlineto{\pgfqpoint{0.686622in}{1.900822in}}%
\pgfpathlineto{\pgfqpoint{0.719056in}{1.881684in}}%
\pgfpathlineto{\pgfqpoint{0.751489in}{1.862846in}}%
\pgfpathlineto{\pgfqpoint{0.783922in}{1.844307in}}%
\pgfpathlineto{\pgfqpoint{0.816356in}{1.826064in}}%
\pgfpathlineto{\pgfqpoint{0.848789in}{1.808113in}}%
\pgfpathlineto{\pgfqpoint{0.881222in}{1.790450in}}%
\pgfpathlineto{\pgfqpoint{0.913656in}{1.773068in}}%
\pgfpathlineto{\pgfqpoint{0.946089in}{1.755964in}}%
\pgfpathlineto{\pgfqpoint{0.978522in}{1.739132in}}%
\pgfpathlineto{\pgfqpoint{1.010956in}{1.722568in}}%
\pgfpathlineto{\pgfqpoint{1.043389in}{1.706267in}}%
\pgfpathlineto{\pgfqpoint{1.075822in}{1.690223in}}%
\pgfpathlineto{\pgfqpoint{1.108256in}{1.674434in}}%
\pgfpathlineto{\pgfqpoint{1.140689in}{1.658894in}}%
\pgfpathlineto{\pgfqpoint{1.173122in}{1.643598in}}%
\pgfpathlineto{\pgfqpoint{1.205556in}{1.628543in}}%
\pgfpathlineto{\pgfqpoint{1.237989in}{1.613724in}}%
\pgfpathlineto{\pgfqpoint{1.270422in}{1.599137in}}%
\pgfpathlineto{\pgfqpoint{1.302856in}{1.584777in}}%
\pgfpathlineto{\pgfqpoint{1.335289in}{1.570642in}}%
\pgfpathlineto{\pgfqpoint{1.367722in}{1.556727in}}%
\pgfpathlineto{\pgfqpoint{1.400156in}{1.543028in}}%
\pgfpathlineto{\pgfqpoint{1.432589in}{1.529541in}}%
\pgfpathlineto{\pgfqpoint{1.465022in}{1.516264in}}%
\pgfpathlineto{\pgfqpoint{1.497456in}{1.503191in}}%
\pgfpathlineto{\pgfqpoint{1.529889in}{1.490320in}}%
\pgfpathlineto{\pgfqpoint{1.562322in}{1.477647in}}%
\pgfpathlineto{\pgfqpoint{1.594756in}{1.465168in}}%
\pgfpathlineto{\pgfqpoint{1.627189in}{1.452881in}}%
\pgfpathlineto{\pgfqpoint{1.659622in}{1.440782in}}%
\pgfpathlineto{\pgfqpoint{1.692056in}{1.428868in}}%
\pgfpathlineto{\pgfqpoint{1.724489in}{1.417136in}}%
\pgfpathlineto{\pgfqpoint{1.756922in}{1.405582in}}%
\pgfpathlineto{\pgfqpoint{1.789356in}{1.394203in}}%
\pgfpathlineto{\pgfqpoint{1.821789in}{1.382997in}}%
\pgfpathlineto{\pgfqpoint{1.854222in}{1.371961in}}%
\pgfpathlineto{\pgfqpoint{1.886656in}{1.361091in}}%
\pgfpathlineto{\pgfqpoint{1.919089in}{1.350385in}}%
\pgfpathlineto{\pgfqpoint{1.951522in}{1.339841in}}%
\pgfpathlineto{\pgfqpoint{1.983956in}{1.329455in}}%
\pgfpathlineto{\pgfqpoint{2.016389in}{1.319224in}}%
\pgfpathlineto{\pgfqpoint{2.048822in}{1.309147in}}%
\pgfpathlineto{\pgfqpoint{2.081256in}{1.299220in}}%
\pgfpathlineto{\pgfqpoint{2.113689in}{1.289442in}}%
\pgfpathlineto{\pgfqpoint{2.146122in}{1.279809in}}%
\pgfpathlineto{\pgfqpoint{2.178556in}{1.270319in}}%
\pgfpathlineto{\pgfqpoint{2.210989in}{1.260970in}}%
\pgfpathlineto{\pgfqpoint{2.243422in}{1.251759in}}%
\pgfpathlineto{\pgfqpoint{2.275856in}{1.242685in}}%
\pgfpathlineto{\pgfqpoint{2.308289in}{1.233745in}}%
\pgfpathlineto{\pgfqpoint{2.340722in}{1.224936in}}%
\pgfpathlineto{\pgfqpoint{2.373156in}{1.216257in}}%
\pgfpathlineto{\pgfqpoint{2.405589in}{1.207705in}}%
\pgfpathlineto{\pgfqpoint{2.438022in}{1.199278in}}%
\pgfpathlineto{\pgfqpoint{2.470456in}{1.190975in}}%
\pgfpathlineto{\pgfqpoint{2.502889in}{1.182793in}}%
\pgfpathlineto{\pgfqpoint{2.535322in}{1.174731in}}%
\pgfpathlineto{\pgfqpoint{2.567756in}{1.166785in}}%
\pgfpathlineto{\pgfqpoint{2.600189in}{1.158955in}}%
\pgfpathlineto{\pgfqpoint{2.632622in}{1.151239in}}%
\pgfpathlineto{\pgfqpoint{2.665056in}{1.143635in}}%
\pgfpathlineto{\pgfqpoint{2.697489in}{1.136140in}}%
\pgfpathlineto{\pgfqpoint{2.729922in}{1.128753in}}%
\pgfpathlineto{\pgfqpoint{2.762356in}{1.121473in}}%
\pgfpathlineto{\pgfqpoint{2.794789in}{1.114298in}}%
\pgfpathlineto{\pgfqpoint{2.827222in}{1.107225in}}%
\pgfpathlineto{\pgfqpoint{2.859656in}{1.100254in}}%
\pgfpathlineto{\pgfqpoint{2.892089in}{1.093383in}}%
\pgfpathlineto{\pgfqpoint{2.924522in}{1.086610in}}%
\pgfpathlineto{\pgfqpoint{2.956956in}{1.079933in}}%
\pgfpathlineto{\pgfqpoint{2.989389in}{1.073352in}}%
\pgfpathlineto{\pgfqpoint{3.021822in}{1.066864in}}%
\pgfpathlineto{\pgfqpoint{3.054256in}{1.060468in}}%
\pgfpathlineto{\pgfqpoint{3.086689in}{1.054163in}}%
\pgfpathlineto{\pgfqpoint{3.119122in}{1.047947in}}%
\pgfpathlineto{\pgfqpoint{3.151556in}{1.041818in}}%
\pgfpathlineto{\pgfqpoint{3.183989in}{1.035776in}}%
\pgfpathlineto{\pgfqpoint{3.216422in}{1.029820in}}%
\pgfpathlineto{\pgfqpoint{3.248856in}{1.023947in}}%
\pgfpathlineto{\pgfqpoint{3.281289in}{1.018156in}}%
\pgfpathlineto{\pgfqpoint{3.313722in}{1.012446in}}%
\pgfpathlineto{\pgfqpoint{3.346156in}{1.006816in}}%
\pgfpathlineto{\pgfqpoint{3.354375in}{1.005410in}}%
\pgfusepath{stroke}%
\end{pgfscope}%
\begin{pgfscope}%
\pgfpathrectangle{\pgfqpoint{0.394722in}{0.512778in}}{\pgfqpoint{2.949653in}{1.625764in}}%
\pgfusepath{clip}%
\pgfsetrectcap%
\pgfsetroundjoin%
\pgfsetlinewidth{1.505625pt}%
\definecolor{currentstroke}{rgb}{1.000000,0.498039,0.054902}%
\pgfsetstrokecolor{currentstroke}%
\pgfsetdash{}{0pt}%
\pgfpathmoveto{\pgfqpoint{0.394722in}{2.064643in}}%
\pgfpathlineto{\pgfqpoint{0.427156in}{2.022130in}}%
\pgfpathlineto{\pgfqpoint{0.459589in}{1.959740in}}%
\pgfpathlineto{\pgfqpoint{0.492022in}{1.900708in}}%
\pgfpathlineto{\pgfqpoint{0.524456in}{1.844773in}}%
\pgfpathlineto{\pgfqpoint{0.556889in}{1.791707in}}%
\pgfpathlineto{\pgfqpoint{0.589322in}{1.741308in}}%
\pgfpathlineto{\pgfqpoint{0.621756in}{1.693397in}}%
\pgfpathlineto{\pgfqpoint{0.654189in}{1.647814in}}%
\pgfpathlineto{\pgfqpoint{0.686622in}{1.604415in}}%
\pgfpathlineto{\pgfqpoint{0.719056in}{1.563068in}}%
\pgfpathlineto{\pgfqpoint{0.751489in}{1.523652in}}%
\pgfpathlineto{\pgfqpoint{0.783922in}{1.486058in}}%
\pgfpathlineto{\pgfqpoint{0.816356in}{1.450184in}}%
\pgfpathlineto{\pgfqpoint{0.848789in}{1.415935in}}%
\pgfpathlineto{\pgfqpoint{0.881222in}{1.383223in}}%
\pgfpathlineto{\pgfqpoint{0.913656in}{1.351968in}}%
\pgfpathlineto{\pgfqpoint{0.946089in}{1.322092in}}%
\pgfpathlineto{\pgfqpoint{0.978522in}{1.293524in}}%
\pgfpathlineto{\pgfqpoint{1.010956in}{1.266199in}}%
\pgfpathlineto{\pgfqpoint{1.043389in}{1.240052in}}%
\pgfpathlineto{\pgfqpoint{1.075822in}{1.215025in}}%
\pgfpathlineto{\pgfqpoint{1.108256in}{1.191062in}}%
\pgfpathlineto{\pgfqpoint{1.140689in}{1.168112in}}%
\pgfpathlineto{\pgfqpoint{1.173122in}{1.146125in}}%
\pgfpathlineto{\pgfqpoint{1.205556in}{1.125054in}}%
\pgfpathlineto{\pgfqpoint{1.237989in}{1.104857in}}%
\pgfpathlineto{\pgfqpoint{1.270422in}{1.085490in}}%
\pgfpathlineto{\pgfqpoint{1.302856in}{1.066916in}}%
\pgfpathlineto{\pgfqpoint{1.335289in}{1.049098in}}%
\pgfpathlineto{\pgfqpoint{1.367722in}{1.031999in}}%
\pgfpathlineto{\pgfqpoint{1.400156in}{1.015588in}}%
\pgfpathlineto{\pgfqpoint{1.432589in}{0.999832in}}%
\pgfpathlineto{\pgfqpoint{1.465022in}{0.984702in}}%
\pgfpathlineto{\pgfqpoint{1.497456in}{0.970170in}}%
\pgfpathlineto{\pgfqpoint{1.529889in}{0.956209in}}%
\pgfpathlineto{\pgfqpoint{1.562322in}{0.942793in}}%
\pgfpathlineto{\pgfqpoint{1.594756in}{0.929899in}}%
\pgfpathlineto{\pgfqpoint{1.627189in}{0.917503in}}%
\pgfpathlineto{\pgfqpoint{1.659622in}{0.905585in}}%
\pgfpathlineto{\pgfqpoint{1.692056in}{0.894122in}}%
\pgfpathlineto{\pgfqpoint{1.724489in}{0.883096in}}%
\pgfpathlineto{\pgfqpoint{1.756922in}{0.872488in}}%
\pgfpathlineto{\pgfqpoint{1.789356in}{0.862280in}}%
\pgfpathlineto{\pgfqpoint{1.821789in}{0.852455in}}%
\pgfpathlineto{\pgfqpoint{1.854222in}{0.842997in}}%
\pgfpathlineto{\pgfqpoint{1.886656in}{0.833891in}}%
\pgfpathlineto{\pgfqpoint{1.919089in}{0.825121in}}%
\pgfpathlineto{\pgfqpoint{1.951522in}{0.816675in}}%
\pgfpathlineto{\pgfqpoint{1.983956in}{0.808539in}}%
\pgfpathlineto{\pgfqpoint{2.016389in}{0.800701in}}%
\pgfpathlineto{\pgfqpoint{2.048822in}{0.793147in}}%
\pgfpathlineto{\pgfqpoint{2.081256in}{0.785867in}}%
\pgfpathlineto{\pgfqpoint{2.113689in}{0.778850in}}%
\pgfpathlineto{\pgfqpoint{2.146122in}{0.772084in}}%
\pgfpathlineto{\pgfqpoint{2.178556in}{0.765561in}}%
\pgfpathlineto{\pgfqpoint{2.210989in}{0.759271in}}%
\pgfpathlineto{\pgfqpoint{2.243422in}{0.753204in}}%
\pgfpathlineto{\pgfqpoint{2.275856in}{0.747351in}}%
\pgfpathlineto{\pgfqpoint{2.308289in}{0.741705in}}%
\pgfpathlineto{\pgfqpoint{2.340722in}{0.736257in}}%
\pgfpathlineto{\pgfqpoint{2.373156in}{0.731000in}}%
\pgfpathlineto{\pgfqpoint{2.405589in}{0.725926in}}%
\pgfpathlineto{\pgfqpoint{2.438022in}{0.721029in}}%
\pgfpathlineto{\pgfqpoint{2.470456in}{0.716301in}}%
\pgfpathlineto{\pgfqpoint{2.502889in}{0.711736in}}%
\pgfpathlineto{\pgfqpoint{2.535322in}{0.707328in}}%
\pgfpathlineto{\pgfqpoint{2.567756in}{0.703072in}}%
\pgfpathlineto{\pgfqpoint{2.600189in}{0.698961in}}%
\pgfpathlineto{\pgfqpoint{2.632622in}{0.694990in}}%
\pgfpathlineto{\pgfqpoint{2.665056in}{0.691154in}}%
\pgfpathlineto{\pgfqpoint{2.697489in}{0.687448in}}%
\pgfpathlineto{\pgfqpoint{2.729922in}{0.683867in}}%
\pgfpathlineto{\pgfqpoint{2.762356in}{0.680406in}}%
\pgfpathlineto{\pgfqpoint{2.794789in}{0.677062in}}%
\pgfpathlineto{\pgfqpoint{2.827222in}{0.673830in}}%
\pgfpathlineto{\pgfqpoint{2.859656in}{0.670706in}}%
\pgfpathlineto{\pgfqpoint{2.892089in}{0.667686in}}%
\pgfpathlineto{\pgfqpoint{2.924522in}{0.664766in}}%
\pgfpathlineto{\pgfqpoint{2.956956in}{0.661943in}}%
\pgfpathlineto{\pgfqpoint{2.989389in}{0.659213in}}%
\pgfpathlineto{\pgfqpoint{3.021822in}{0.656573in}}%
\pgfpathlineto{\pgfqpoint{3.054256in}{0.654020in}}%
\pgfpathlineto{\pgfqpoint{3.086689in}{0.651551in}}%
\pgfpathlineto{\pgfqpoint{3.119122in}{0.649163in}}%
\pgfpathlineto{\pgfqpoint{3.151556in}{0.646852in}}%
\pgfpathlineto{\pgfqpoint{3.183989in}{0.644617in}}%
\pgfpathlineto{\pgfqpoint{3.216422in}{0.642455in}}%
\pgfpathlineto{\pgfqpoint{3.248856in}{0.640362in}}%
\pgfpathlineto{\pgfqpoint{3.281289in}{0.638338in}}%
\pgfpathlineto{\pgfqpoint{3.313722in}{0.636379in}}%
\pgfpathlineto{\pgfqpoint{3.346156in}{0.634483in}}%
\pgfpathlineto{\pgfqpoint{3.354375in}{0.634018in}}%
\pgfusepath{stroke}%
\end{pgfscope}%
\begin{pgfscope}%
\pgfpathrectangle{\pgfqpoint{0.394722in}{0.512778in}}{\pgfqpoint{2.949653in}{1.625764in}}%
\pgfusepath{clip}%
\pgfsetrectcap%
\pgfsetroundjoin%
\pgfsetlinewidth{1.505625pt}%
\definecolor{currentstroke}{rgb}{0.172549,0.627451,0.172549}%
\pgfsetstrokecolor{currentstroke}%
\pgfsetdash{}{0pt}%
\pgfpathmoveto{\pgfqpoint{0.394722in}{2.064643in}}%
\pgfpathlineto{\pgfqpoint{0.427156in}{2.012591in}}%
\pgfpathlineto{\pgfqpoint{0.459589in}{1.942101in}}%
\pgfpathlineto{\pgfqpoint{0.492022in}{1.876123in}}%
\pgfpathlineto{\pgfqpoint{0.524456in}{1.814191in}}%
\pgfpathlineto{\pgfqpoint{0.556889in}{1.755926in}}%
\pgfpathlineto{\pgfqpoint{0.589322in}{1.701012in}}%
\pgfpathlineto{\pgfqpoint{0.621756in}{1.649182in}}%
\pgfpathlineto{\pgfqpoint{0.654189in}{1.600204in}}%
\pgfpathlineto{\pgfqpoint{0.686622in}{1.553874in}}%
\pgfpathlineto{\pgfqpoint{0.719056in}{1.510011in}}%
\pgfpathlineto{\pgfqpoint{0.751489in}{1.468451in}}%
\pgfpathlineto{\pgfqpoint{0.783922in}{1.429046in}}%
\pgfpathlineto{\pgfqpoint{0.816356in}{1.391661in}}%
\pgfpathlineto{\pgfqpoint{0.848789in}{1.356172in}}%
\pgfpathlineto{\pgfqpoint{0.881222in}{1.322463in}}%
\pgfpathlineto{\pgfqpoint{0.913656in}{1.290429in}}%
\pgfpathlineto{\pgfqpoint{0.946089in}{1.259971in}}%
\pgfpathlineto{\pgfqpoint{0.978522in}{1.230999in}}%
\pgfpathlineto{\pgfqpoint{1.010956in}{1.203427in}}%
\pgfpathlineto{\pgfqpoint{1.043389in}{1.177176in}}%
\pgfpathlineto{\pgfqpoint{1.075822in}{1.152173in}}%
\pgfpathlineto{\pgfqpoint{1.108256in}{1.128349in}}%
\pgfpathlineto{\pgfqpoint{1.140689in}{1.105638in}}%
\pgfpathlineto{\pgfqpoint{1.173122in}{1.083981in}}%
\pgfpathlineto{\pgfqpoint{1.205556in}{1.063322in}}%
\pgfpathlineto{\pgfqpoint{1.237989in}{1.043606in}}%
\pgfpathlineto{\pgfqpoint{1.270422in}{1.024786in}}%
\pgfpathlineto{\pgfqpoint{1.302856in}{1.006812in}}%
\pgfpathlineto{\pgfqpoint{1.335289in}{0.989643in}}%
\pgfpathlineto{\pgfqpoint{1.367722in}{0.973237in}}%
\pgfpathlineto{\pgfqpoint{1.400156in}{0.957555in}}%
\pgfpathlineto{\pgfqpoint{1.432589in}{0.942560in}}%
\pgfpathlineto{\pgfqpoint{1.465022in}{0.928218in}}%
\pgfpathlineto{\pgfqpoint{1.497456in}{0.914498in}}%
\pgfpathlineto{\pgfqpoint{1.529889in}{0.901367in}}%
\pgfpathlineto{\pgfqpoint{1.562322in}{0.888798in}}%
\pgfpathlineto{\pgfqpoint{1.594756in}{0.876763in}}%
\pgfpathlineto{\pgfqpoint{1.627189in}{0.865237in}}%
\pgfpathlineto{\pgfqpoint{1.659622in}{0.854195in}}%
\pgfpathlineto{\pgfqpoint{1.692056in}{0.843615in}}%
\pgfpathlineto{\pgfqpoint{1.724489in}{0.833474in}}%
\pgfpathlineto{\pgfqpoint{1.756922in}{0.823752in}}%
\pgfpathlineto{\pgfqpoint{1.789356in}{0.814430in}}%
\pgfpathlineto{\pgfqpoint{1.821789in}{0.805489in}}%
\pgfpathlineto{\pgfqpoint{1.854222in}{0.796913in}}%
\pgfpathlineto{\pgfqpoint{1.886656in}{0.788683in}}%
\pgfpathlineto{\pgfqpoint{1.919089in}{0.780786in}}%
\pgfpathlineto{\pgfqpoint{1.951522in}{0.773205in}}%
\pgfpathlineto{\pgfqpoint{1.983956in}{0.765926in}}%
\pgfpathlineto{\pgfqpoint{2.016389in}{0.758937in}}%
\pgfpathlineto{\pgfqpoint{2.048822in}{0.752225in}}%
\pgfpathlineto{\pgfqpoint{2.081256in}{0.745777in}}%
\pgfpathlineto{\pgfqpoint{2.113689in}{0.739581in}}%
\pgfpathlineto{\pgfqpoint{2.146122in}{0.733628in}}%
\pgfpathlineto{\pgfqpoint{2.178556in}{0.727906in}}%
\pgfpathlineto{\pgfqpoint{2.210989in}{0.722406in}}%
\pgfpathlineto{\pgfqpoint{2.243422in}{0.717118in}}%
\pgfpathlineto{\pgfqpoint{2.275856in}{0.712034in}}%
\pgfpathlineto{\pgfqpoint{2.308289in}{0.707144in}}%
\pgfpathlineto{\pgfqpoint{2.340722in}{0.702440in}}%
\pgfpathlineto{\pgfqpoint{2.373156in}{0.697916in}}%
\pgfpathlineto{\pgfqpoint{2.405589in}{0.693562in}}%
\pgfpathlineto{\pgfqpoint{2.438022in}{0.689373in}}%
\pgfpathlineto{\pgfqpoint{2.470456in}{0.685342in}}%
\pgfpathlineto{\pgfqpoint{2.502889in}{0.681462in}}%
\pgfpathlineto{\pgfqpoint{2.535322in}{0.677726in}}%
\pgfpathlineto{\pgfqpoint{2.567756in}{0.674130in}}%
\pgfpathlineto{\pgfqpoint{2.600189in}{0.670667in}}%
\pgfpathlineto{\pgfqpoint{2.632622in}{0.667333in}}%
\pgfpathlineto{\pgfqpoint{2.665056in}{0.664121in}}%
\pgfpathlineto{\pgfqpoint{2.697489in}{0.661027in}}%
\pgfpathlineto{\pgfqpoint{2.729922in}{0.658047in}}%
\pgfpathlineto{\pgfqpoint{2.762356in}{0.655176in}}%
\pgfpathlineto{\pgfqpoint{2.794789in}{0.652410in}}%
\pgfpathlineto{\pgfqpoint{2.827222in}{0.649744in}}%
\pgfpathlineto{\pgfqpoint{2.859656in}{0.647175in}}%
\pgfpathlineto{\pgfqpoint{2.892089in}{0.644699in}}%
\pgfpathlineto{\pgfqpoint{2.924522in}{0.642313in}}%
\pgfpathlineto{\pgfqpoint{2.956956in}{0.640012in}}%
\pgfpathlineto{\pgfqpoint{2.989389in}{0.637794in}}%
\pgfpathlineto{\pgfqpoint{3.021822in}{0.635655in}}%
\pgfpathlineto{\pgfqpoint{3.054256in}{0.633593in}}%
\pgfpathlineto{\pgfqpoint{3.086689in}{0.631604in}}%
\pgfpathlineto{\pgfqpoint{3.119122in}{0.629686in}}%
\pgfpathlineto{\pgfqpoint{3.151556in}{0.627837in}}%
\pgfpathlineto{\pgfqpoint{3.183989in}{0.626052in}}%
\pgfpathlineto{\pgfqpoint{3.216422in}{0.624331in}}%
\pgfpathlineto{\pgfqpoint{3.248856in}{0.622671in}}%
\pgfpathlineto{\pgfqpoint{3.281289in}{0.621069in}}%
\pgfpathlineto{\pgfqpoint{3.313722in}{0.619523in}}%
\pgfpathlineto{\pgfqpoint{3.346156in}{0.618032in}}%
\pgfpathlineto{\pgfqpoint{3.354375in}{0.617667in}}%
\pgfusepath{stroke}%
\end{pgfscope}%
\begin{pgfscope}%
\pgfpathrectangle{\pgfqpoint{0.394722in}{0.512778in}}{\pgfqpoint{2.949653in}{1.625764in}}%
\pgfusepath{clip}%
\pgfsetrectcap%
\pgfsetroundjoin%
\pgfsetlinewidth{1.505625pt}%
\definecolor{currentstroke}{rgb}{0.839216,0.152941,0.156863}%
\pgfsetstrokecolor{currentstroke}%
\pgfsetdash{}{0pt}%
\pgfpathmoveto{\pgfqpoint{0.394722in}{2.064643in}}%
\pgfpathlineto{\pgfqpoint{0.427156in}{1.882958in}}%
\pgfpathlineto{\pgfqpoint{0.459589in}{1.743668in}}%
\pgfpathlineto{\pgfqpoint{0.492022in}{1.635071in}}%
\pgfpathlineto{\pgfqpoint{0.524456in}{1.545893in}}%
\pgfpathlineto{\pgfqpoint{0.556889in}{1.470300in}}%
\pgfpathlineto{\pgfqpoint{0.589322in}{1.404804in}}%
\pgfpathlineto{\pgfqpoint{0.621756in}{1.347142in}}%
\pgfpathlineto{\pgfqpoint{0.654189in}{1.295754in}}%
\pgfpathlineto{\pgfqpoint{0.686622in}{1.249523in}}%
\pgfpathlineto{\pgfqpoint{0.719056in}{1.207617in}}%
\pgfpathlineto{\pgfqpoint{0.751489in}{1.169400in}}%
\pgfpathlineto{\pgfqpoint{0.783922in}{1.134371in}}%
\pgfpathlineto{\pgfqpoint{0.816356in}{1.102130in}}%
\pgfpathlineto{\pgfqpoint{0.848789in}{1.072348in}}%
\pgfpathlineto{\pgfqpoint{0.881222in}{1.044753in}}%
\pgfpathlineto{\pgfqpoint{0.913656in}{1.019117in}}%
\pgfpathlineto{\pgfqpoint{0.946089in}{0.995244in}}%
\pgfpathlineto{\pgfqpoint{0.978522in}{0.972967in}}%
\pgfpathlineto{\pgfqpoint{1.010956in}{0.952143in}}%
\pgfpathlineto{\pgfqpoint{1.043389in}{0.932643in}}%
\pgfpathlineto{\pgfqpoint{1.075822in}{0.914357in}}%
\pgfpathlineto{\pgfqpoint{1.108256in}{0.897187in}}%
\pgfpathlineto{\pgfqpoint{1.140689in}{0.881045in}}%
\pgfpathlineto{\pgfqpoint{1.173122in}{0.865853in}}%
\pgfpathlineto{\pgfqpoint{1.205556in}{0.851540in}}%
\pgfpathlineto{\pgfqpoint{1.237989in}{0.838043in}}%
\pgfpathlineto{\pgfqpoint{1.270422in}{0.825304in}}%
\pgfpathlineto{\pgfqpoint{1.302856in}{0.813273in}}%
\pgfpathlineto{\pgfqpoint{1.335289in}{0.801900in}}%
\pgfpathlineto{\pgfqpoint{1.367722in}{0.791142in}}%
\pgfpathlineto{\pgfqpoint{1.400156in}{0.780960in}}%
\pgfpathlineto{\pgfqpoint{1.432589in}{0.771317in}}%
\pgfpathlineto{\pgfqpoint{1.465022in}{0.762180in}}%
\pgfpathlineto{\pgfqpoint{1.497456in}{0.753517in}}%
\pgfpathlineto{\pgfqpoint{1.529889in}{0.745300in}}%
\pgfpathlineto{\pgfqpoint{1.562322in}{0.737503in}}%
\pgfpathlineto{\pgfqpoint{1.594756in}{0.730099in}}%
\pgfpathlineto{\pgfqpoint{1.627189in}{0.723068in}}%
\pgfpathlineto{\pgfqpoint{1.659622in}{0.716386in}}%
\pgfpathlineto{\pgfqpoint{1.692056in}{0.710035in}}%
\pgfpathlineto{\pgfqpoint{1.724489in}{0.703997in}}%
\pgfpathlineto{\pgfqpoint{1.756922in}{0.698253in}}%
\pgfpathlineto{\pgfqpoint{1.789356in}{0.692787in}}%
\pgfpathlineto{\pgfqpoint{1.821789in}{0.687585in}}%
\pgfpathlineto{\pgfqpoint{1.854222in}{0.682633in}}%
\pgfpathlineto{\pgfqpoint{1.886656in}{0.677916in}}%
\pgfpathlineto{\pgfqpoint{1.919089in}{0.673423in}}%
\pgfpathlineto{\pgfqpoint{1.951522in}{0.669142in}}%
\pgfpathlineto{\pgfqpoint{1.983956in}{0.665062in}}%
\pgfpathlineto{\pgfqpoint{2.016389in}{0.661172in}}%
\pgfpathlineto{\pgfqpoint{2.048822in}{0.657463in}}%
\pgfpathlineto{\pgfqpoint{2.081256in}{0.653926in}}%
\pgfpathlineto{\pgfqpoint{2.113689in}{0.650552in}}%
\pgfpathlineto{\pgfqpoint{2.146122in}{0.647332in}}%
\pgfpathlineto{\pgfqpoint{2.178556in}{0.644260in}}%
\pgfpathlineto{\pgfqpoint{2.210989in}{0.641327in}}%
\pgfpathlineto{\pgfqpoint{2.243422in}{0.638528in}}%
\pgfpathlineto{\pgfqpoint{2.275856in}{0.635855in}}%
\pgfpathlineto{\pgfqpoint{2.308289in}{0.633302in}}%
\pgfpathlineto{\pgfqpoint{2.340722in}{0.630864in}}%
\pgfpathlineto{\pgfqpoint{2.373156in}{0.628534in}}%
\pgfpathlineto{\pgfqpoint{2.405589in}{0.626309in}}%
\pgfpathlineto{\pgfqpoint{2.438022in}{0.624182in}}%
\pgfpathlineto{\pgfqpoint{2.470456in}{0.622149in}}%
\pgfpathlineto{\pgfqpoint{2.502889in}{0.620207in}}%
\pgfpathlineto{\pgfqpoint{2.535322in}{0.618349in}}%
\pgfpathlineto{\pgfqpoint{2.567756in}{0.616573in}}%
\pgfpathlineto{\pgfqpoint{2.600189in}{0.614875in}}%
\pgfpathlineto{\pgfqpoint{2.632622in}{0.613251in}}%
\pgfpathlineto{\pgfqpoint{2.665056in}{0.611698in}}%
\pgfpathlineto{\pgfqpoint{2.697489in}{0.610212in}}%
\pgfpathlineto{\pgfqpoint{2.729922in}{0.608790in}}%
\pgfpathlineto{\pgfqpoint{2.762356in}{0.607430in}}%
\pgfpathlineto{\pgfqpoint{2.794789in}{0.606128in}}%
\pgfpathlineto{\pgfqpoint{2.827222in}{0.604883in}}%
\pgfpathlineto{\pgfqpoint{2.859656in}{0.603691in}}%
\pgfpathlineto{\pgfqpoint{2.892089in}{0.602549in}}%
\pgfpathlineto{\pgfqpoint{2.924522in}{0.601457in}}%
\pgfpathlineto{\pgfqpoint{2.956956in}{0.600411in}}%
\pgfpathlineto{\pgfqpoint{2.989389in}{0.599410in}}%
\pgfpathlineto{\pgfqpoint{3.021822in}{0.598451in}}%
\pgfpathlineto{\pgfqpoint{3.054256in}{0.597532in}}%
\pgfpathlineto{\pgfqpoint{3.086689in}{0.596653in}}%
\pgfpathlineto{\pgfqpoint{3.119122in}{0.595810in}}%
\pgfpathlineto{\pgfqpoint{3.151556in}{0.595003in}}%
\pgfpathlineto{\pgfqpoint{3.183989in}{0.594230in}}%
\pgfpathlineto{\pgfqpoint{3.216422in}{0.593490in}}%
\pgfpathlineto{\pgfqpoint{3.248856in}{0.592780in}}%
\pgfpathlineto{\pgfqpoint{3.281289in}{0.592100in}}%
\pgfpathlineto{\pgfqpoint{3.313722in}{0.591449in}}%
\pgfpathlineto{\pgfqpoint{3.346156in}{0.590824in}}%
\pgfpathlineto{\pgfqpoint{3.354375in}{0.590673in}}%
\pgfusepath{stroke}%
\end{pgfscope}%
\begin{pgfscope}%
\pgfsetrectcap%
\pgfsetmiterjoin%
\pgfsetlinewidth{0.803000pt}%
\definecolor{currentstroke}{rgb}{0.000000,0.000000,0.000000}%
\pgfsetstrokecolor{currentstroke}%
\pgfsetdash{}{0pt}%
\pgfpathmoveto{\pgfqpoint{0.394722in}{0.512778in}}%
\pgfpathlineto{\pgfqpoint{0.394722in}{2.138542in}}%
\pgfusepath{stroke}%
\end{pgfscope}%
\begin{pgfscope}%
\pgfsetrectcap%
\pgfsetmiterjoin%
\pgfsetlinewidth{0.803000pt}%
\definecolor{currentstroke}{rgb}{0.000000,0.000000,0.000000}%
\pgfsetstrokecolor{currentstroke}%
\pgfsetdash{}{0pt}%
\pgfpathmoveto{\pgfqpoint{3.344375in}{0.512778in}}%
\pgfpathlineto{\pgfqpoint{3.344375in}{2.138542in}}%
\pgfusepath{stroke}%
\end{pgfscope}%
\begin{pgfscope}%
\pgfsetrectcap%
\pgfsetmiterjoin%
\pgfsetlinewidth{0.803000pt}%
\definecolor{currentstroke}{rgb}{0.000000,0.000000,0.000000}%
\pgfsetstrokecolor{currentstroke}%
\pgfsetdash{}{0pt}%
\pgfpathmoveto{\pgfqpoint{0.394722in}{0.512778in}}%
\pgfpathlineto{\pgfqpoint{3.344375in}{0.512778in}}%
\pgfusepath{stroke}%
\end{pgfscope}%
\begin{pgfscope}%
\pgfsetrectcap%
\pgfsetmiterjoin%
\pgfsetlinewidth{0.803000pt}%
\definecolor{currentstroke}{rgb}{0.000000,0.000000,0.000000}%
\pgfsetstrokecolor{currentstroke}%
\pgfsetdash{}{0pt}%
\pgfpathmoveto{\pgfqpoint{0.394722in}{2.138542in}}%
\pgfpathlineto{\pgfqpoint{3.344375in}{2.138542in}}%
\pgfusepath{stroke}%
\end{pgfscope}%
\begin{pgfscope}%
\definecolor{textcolor}{rgb}{0.000000,0.000000,0.000000}%
\pgfsetstrokecolor{textcolor}%
\pgfsetfillcolor{textcolor}%
\pgftext[x=0.394722in,y=2.262519in,left,base]{\color{textcolor}\sffamily\fontsize{9.600000}{11.520000}\selectfont direct signal (normalized)}%
\end{pgfscope}%
\begin{pgfscope}%
\pgfsetbuttcap%
\pgfsetmiterjoin%
\definecolor{currentfill}{rgb}{1.000000,1.000000,1.000000}%
\pgfsetfillcolor{currentfill}%
\pgfsetfillopacity{0.800000}%
\pgfsetlinewidth{1.003750pt}%
\definecolor{currentstroke}{rgb}{0.800000,0.800000,0.800000}%
\pgfsetstrokecolor{currentstroke}%
\pgfsetstrokeopacity{0.800000}%
\pgfsetdash{}{0pt}%
\pgfpathmoveto{\pgfqpoint{2.427597in}{1.430098in}}%
\pgfpathlineto{\pgfqpoint{3.266597in}{1.430098in}}%
\pgfpathquadraticcurveto{\pgfqpoint{3.288819in}{1.430098in}}{\pgfqpoint{3.288819in}{1.452320in}}%
\pgfpathlineto{\pgfqpoint{3.288819in}{2.060764in}}%
\pgfpathquadraticcurveto{\pgfqpoint{3.288819in}{2.082986in}}{\pgfqpoint{3.266597in}{2.082986in}}%
\pgfpathlineto{\pgfqpoint{2.427597in}{2.082986in}}%
\pgfpathquadraticcurveto{\pgfqpoint{2.405375in}{2.082986in}}{\pgfqpoint{2.405375in}{2.060764in}}%
\pgfpathlineto{\pgfqpoint{2.405375in}{1.452320in}}%
\pgfpathquadraticcurveto{\pgfqpoint{2.405375in}{1.430098in}}{\pgfqpoint{2.427597in}{1.430098in}}%
\pgfpathclose%
\pgfusepath{stroke,fill}%
\end{pgfscope}%
\begin{pgfscope}%
\pgfsetrectcap%
\pgfsetroundjoin%
\pgfsetlinewidth{1.505625pt}%
\definecolor{currentstroke}{rgb}{0.121569,0.466667,0.705882}%
\pgfsetstrokecolor{currentstroke}%
\pgfsetdash{}{0pt}%
\pgfpathmoveto{\pgfqpoint{2.449819in}{1.999653in}}%
\pgfpathlineto{\pgfqpoint{2.672042in}{1.999653in}}%
\pgfusepath{stroke}%
\end{pgfscope}%
\begin{pgfscope}%
\definecolor{textcolor}{rgb}{0.000000,0.000000,0.000000}%
\pgfsetstrokecolor{textcolor}%
\pgfsetfillcolor{textcolor}%
\pgftext[x=2.760931in,y=1.960764in,left,base]{\color{textcolor}\sffamily\fontsize{8.000000}{9.600000}\selectfont aluminum}%
\end{pgfscope}%
\begin{pgfscope}%
\pgfsetrectcap%
\pgfsetroundjoin%
\pgfsetlinewidth{1.505625pt}%
\definecolor{currentstroke}{rgb}{1.000000,0.498039,0.054902}%
\pgfsetstrokecolor{currentstroke}%
\pgfsetdash{}{0pt}%
\pgfpathmoveto{\pgfqpoint{2.449819in}{1.844764in}}%
\pgfpathlineto{\pgfqpoint{2.672042in}{1.844764in}}%
\pgfusepath{stroke}%
\end{pgfscope}%
\begin{pgfscope}%
\definecolor{textcolor}{rgb}{0.000000,0.000000,0.000000}%
\pgfsetstrokecolor{textcolor}%
\pgfsetfillcolor{textcolor}%
\pgftext[x=2.760931in,y=1.805875in,left,base]{\color{textcolor}\sffamily\fontsize{8.000000}{9.600000}\selectfont iron}%
\end{pgfscope}%
\begin{pgfscope}%
\pgfsetrectcap%
\pgfsetroundjoin%
\pgfsetlinewidth{1.505625pt}%
\definecolor{currentstroke}{rgb}{0.172549,0.627451,0.172549}%
\pgfsetstrokecolor{currentstroke}%
\pgfsetdash{}{0pt}%
\pgfpathmoveto{\pgfqpoint{2.449819in}{1.689875in}}%
\pgfpathlineto{\pgfqpoint{2.672042in}{1.689875in}}%
\pgfusepath{stroke}%
\end{pgfscope}%
\begin{pgfscope}%
\definecolor{textcolor}{rgb}{0.000000,0.000000,0.000000}%
\pgfsetstrokecolor{textcolor}%
\pgfsetfillcolor{textcolor}%
\pgftext[x=2.760931in,y=1.650986in,left,base]{\color{textcolor}\sffamily\fontsize{8.000000}{9.600000}\selectfont copper}%
\end{pgfscope}%
\begin{pgfscope}%
\pgfsetrectcap%
\pgfsetroundjoin%
\pgfsetlinewidth{1.505625pt}%
\definecolor{currentstroke}{rgb}{0.839216,0.152941,0.156863}%
\pgfsetstrokecolor{currentstroke}%
\pgfsetdash{}{0pt}%
\pgfpathmoveto{\pgfqpoint{2.449819in}{1.534987in}}%
\pgfpathlineto{\pgfqpoint{2.672042in}{1.534987in}}%
\pgfusepath{stroke}%
\end{pgfscope}%
\begin{pgfscope}%
\definecolor{textcolor}{rgb}{0.000000,0.000000,0.000000}%
\pgfsetstrokecolor{textcolor}%
\pgfsetfillcolor{textcolor}%
\pgftext[x=2.760931in,y=1.496098in,left,base]{\color{textcolor}\sffamily\fontsize{8.000000}{9.600000}\selectfont lead}%
\end{pgfscope}%
\end{pgfpicture}%
\makeatother%
\endgroup%

%% file: figures/MCNP_t.pgf
\begingroup%
\makeatletter%
\begin{pgfpicture}%
\pgfpathrectangle{\pgfpointorigin}{\pgfqpoint{3.500000in}{2.400000in}}%
\pgfusepath{use as bounding box, clip}%
\begin{pgfscope}%
\pgfsetbuttcap%
\pgfsetmiterjoin%
\definecolor{currentfill}{rgb}{1.000000,1.000000,1.000000}%
\pgfsetfillcolor{currentfill}%
\pgfsetlinewidth{0.000000pt}%
\definecolor{currentstroke}{rgb}{1.000000,1.000000,1.000000}%
\pgfsetstrokecolor{currentstroke}%
\pgfsetdash{}{0pt}%
\pgfpathmoveto{\pgfqpoint{0.000000in}{0.000000in}}%
\pgfpathlineto{\pgfqpoint{3.500000in}{0.000000in}}%
\pgfpathlineto{\pgfqpoint{3.500000in}{2.400000in}}%
\pgfpathlineto{\pgfqpoint{0.000000in}{2.400000in}}%
\pgfpathclose%
\pgfusepath{fill}%
\end{pgfscope}%
\begin{pgfscope}%
\pgfsetbuttcap%
\pgfsetmiterjoin%
\definecolor{currentfill}{rgb}{1.000000,1.000000,1.000000}%
\pgfsetfillcolor{currentfill}%
\pgfsetlinewidth{0.000000pt}%
\definecolor{currentstroke}{rgb}{0.000000,0.000000,0.000000}%
\pgfsetstrokecolor{currentstroke}%
\pgfsetstrokeopacity{0.000000}%
\pgfsetdash{}{0pt}%
\pgfpathmoveto{\pgfqpoint{0.451167in}{0.295304in}}%
\pgfpathlineto{\pgfqpoint{2.790194in}{0.295304in}}%
\pgfpathlineto{\pgfqpoint{2.790194in}{2.240000in}}%
\pgfpathlineto{\pgfqpoint{0.451167in}{2.240000in}}%
\pgfpathclose%
\pgfusepath{fill}%
\end{pgfscope}%
\begin{pgfscope}%
\pgfsys@transformshift{0.450000in}{0.300000in}%
\pgftext[left,bottom]{\includegraphics[interpolate=true,width=2.340000in,height=1.940000in]{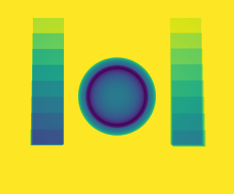}}%
\end{pgfscope}%
\begin{pgfscope}%
\pgfsetbuttcap%
\pgfsetroundjoin%
\definecolor{currentfill}{rgb}{0.000000,0.000000,0.000000}%
\pgfsetfillcolor{currentfill}%
\pgfsetlinewidth{0.803000pt}%
\definecolor{currentstroke}{rgb}{0.000000,0.000000,0.000000}%
\pgfsetstrokecolor{currentstroke}%
\pgfsetdash{}{0pt}%
\pgfsys@defobject{currentmarker}{\pgfqpoint{0.000000in}{-0.048611in}}{\pgfqpoint{0.000000in}{0.000000in}}{%
\pgfpathmoveto{\pgfqpoint{0.000000in}{0.000000in}}%
\pgfpathlineto{\pgfqpoint{0.000000in}{-0.048611in}}%
\pgfusepath{stroke,fill}%
}%
\begin{pgfscope}%
\pgfsys@transformshift{0.452852in}{0.295304in}%
\pgfsys@useobject{currentmarker}{}%
\end{pgfscope}%
\end{pgfscope}%
\begin{pgfscope}%
\definecolor{textcolor}{rgb}{0.000000,0.000000,0.000000}%
\pgfsetstrokecolor{textcolor}%
\pgfsetfillcolor{textcolor}%
\pgftext[x=0.452852in,y=0.198082in,,top]{\color{textcolor}\sffamily\fontsize{8.000000}{9.600000}\selectfont 0}%
\end{pgfscope}%
\begin{pgfscope}%
\pgfsetbuttcap%
\pgfsetroundjoin%
\definecolor{currentfill}{rgb}{0.000000,0.000000,0.000000}%
\pgfsetfillcolor{currentfill}%
\pgfsetlinewidth{0.803000pt}%
\definecolor{currentstroke}{rgb}{0.000000,0.000000,0.000000}%
\pgfsetstrokecolor{currentstroke}%
\pgfsetdash{}{0pt}%
\pgfsys@defobject{currentmarker}{\pgfqpoint{0.000000in}{-0.048611in}}{\pgfqpoint{0.000000in}{0.000000in}}{%
\pgfpathmoveto{\pgfqpoint{0.000000in}{0.000000in}}%
\pgfpathlineto{\pgfqpoint{0.000000in}{-0.048611in}}%
\pgfusepath{stroke,fill}%
}%
\begin{pgfscope}%
\pgfsys@transformshift{0.789888in}{0.295304in}%
\pgfsys@useobject{currentmarker}{}%
\end{pgfscope}%
\end{pgfscope}%
\begin{pgfscope}%
\definecolor{textcolor}{rgb}{0.000000,0.000000,0.000000}%
\pgfsetstrokecolor{textcolor}%
\pgfsetfillcolor{textcolor}%
\pgftext[x=0.789888in,y=0.198082in,,top]{\color{textcolor}\sffamily\fontsize{8.000000}{9.600000}\selectfont 100}%
\end{pgfscope}%
\begin{pgfscope}%
\pgfsetbuttcap%
\pgfsetroundjoin%
\definecolor{currentfill}{rgb}{0.000000,0.000000,0.000000}%
\pgfsetfillcolor{currentfill}%
\pgfsetlinewidth{0.803000pt}%
\definecolor{currentstroke}{rgb}{0.000000,0.000000,0.000000}%
\pgfsetstrokecolor{currentstroke}%
\pgfsetdash{}{0pt}%
\pgfsys@defobject{currentmarker}{\pgfqpoint{0.000000in}{-0.048611in}}{\pgfqpoint{0.000000in}{0.000000in}}{%
\pgfpathmoveto{\pgfqpoint{0.000000in}{0.000000in}}%
\pgfpathlineto{\pgfqpoint{0.000000in}{-0.048611in}}%
\pgfusepath{stroke,fill}%
}%
\begin{pgfscope}%
\pgfsys@transformshift{1.126924in}{0.295304in}%
\pgfsys@useobject{currentmarker}{}%
\end{pgfscope}%
\end{pgfscope}%
\begin{pgfscope}%
\definecolor{textcolor}{rgb}{0.000000,0.000000,0.000000}%
\pgfsetstrokecolor{textcolor}%
\pgfsetfillcolor{textcolor}%
\pgftext[x=1.126924in,y=0.198082in,,top]{\color{textcolor}\sffamily\fontsize{8.000000}{9.600000}\selectfont 200}%
\end{pgfscope}%
\begin{pgfscope}%
\pgfsetbuttcap%
\pgfsetroundjoin%
\definecolor{currentfill}{rgb}{0.000000,0.000000,0.000000}%
\pgfsetfillcolor{currentfill}%
\pgfsetlinewidth{0.803000pt}%
\definecolor{currentstroke}{rgb}{0.000000,0.000000,0.000000}%
\pgfsetstrokecolor{currentstroke}%
\pgfsetdash{}{0pt}%
\pgfsys@defobject{currentmarker}{\pgfqpoint{0.000000in}{-0.048611in}}{\pgfqpoint{0.000000in}{0.000000in}}{%
\pgfpathmoveto{\pgfqpoint{0.000000in}{0.000000in}}%
\pgfpathlineto{\pgfqpoint{0.000000in}{-0.048611in}}%
\pgfusepath{stroke,fill}%
}%
\begin{pgfscope}%
\pgfsys@transformshift{1.463959in}{0.295304in}%
\pgfsys@useobject{currentmarker}{}%
\end{pgfscope}%
\end{pgfscope}%
\begin{pgfscope}%
\definecolor{textcolor}{rgb}{0.000000,0.000000,0.000000}%
\pgfsetstrokecolor{textcolor}%
\pgfsetfillcolor{textcolor}%
\pgftext[x=1.463959in,y=0.198082in,,top]{\color{textcolor}\sffamily\fontsize{8.000000}{9.600000}\selectfont 300}%
\end{pgfscope}%
\begin{pgfscope}%
\pgfsetbuttcap%
\pgfsetroundjoin%
\definecolor{currentfill}{rgb}{0.000000,0.000000,0.000000}%
\pgfsetfillcolor{currentfill}%
\pgfsetlinewidth{0.803000pt}%
\definecolor{currentstroke}{rgb}{0.000000,0.000000,0.000000}%
\pgfsetstrokecolor{currentstroke}%
\pgfsetdash{}{0pt}%
\pgfsys@defobject{currentmarker}{\pgfqpoint{0.000000in}{-0.048611in}}{\pgfqpoint{0.000000in}{0.000000in}}{%
\pgfpathmoveto{\pgfqpoint{0.000000in}{0.000000in}}%
\pgfpathlineto{\pgfqpoint{0.000000in}{-0.048611in}}%
\pgfusepath{stroke,fill}%
}%
\begin{pgfscope}%
\pgfsys@transformshift{1.800995in}{0.295304in}%
\pgfsys@useobject{currentmarker}{}%
\end{pgfscope}%
\end{pgfscope}%
\begin{pgfscope}%
\definecolor{textcolor}{rgb}{0.000000,0.000000,0.000000}%
\pgfsetstrokecolor{textcolor}%
\pgfsetfillcolor{textcolor}%
\pgftext[x=1.800995in,y=0.198082in,,top]{\color{textcolor}\sffamily\fontsize{8.000000}{9.600000}\selectfont 400}%
\end{pgfscope}%
\begin{pgfscope}%
\pgfsetbuttcap%
\pgfsetroundjoin%
\definecolor{currentfill}{rgb}{0.000000,0.000000,0.000000}%
\pgfsetfillcolor{currentfill}%
\pgfsetlinewidth{0.803000pt}%
\definecolor{currentstroke}{rgb}{0.000000,0.000000,0.000000}%
\pgfsetstrokecolor{currentstroke}%
\pgfsetdash{}{0pt}%
\pgfsys@defobject{currentmarker}{\pgfqpoint{0.000000in}{-0.048611in}}{\pgfqpoint{0.000000in}{0.000000in}}{%
\pgfpathmoveto{\pgfqpoint{0.000000in}{0.000000in}}%
\pgfpathlineto{\pgfqpoint{0.000000in}{-0.048611in}}%
\pgfusepath{stroke,fill}%
}%
\begin{pgfscope}%
\pgfsys@transformshift{2.138030in}{0.295304in}%
\pgfsys@useobject{currentmarker}{}%
\end{pgfscope}%
\end{pgfscope}%
\begin{pgfscope}%
\definecolor{textcolor}{rgb}{0.000000,0.000000,0.000000}%
\pgfsetstrokecolor{textcolor}%
\pgfsetfillcolor{textcolor}%
\pgftext[x=2.138030in,y=0.198082in,,top]{\color{textcolor}\sffamily\fontsize{8.000000}{9.600000}\selectfont 500}%
\end{pgfscope}%
\begin{pgfscope}%
\pgfsetbuttcap%
\pgfsetroundjoin%
\definecolor{currentfill}{rgb}{0.000000,0.000000,0.000000}%
\pgfsetfillcolor{currentfill}%
\pgfsetlinewidth{0.803000pt}%
\definecolor{currentstroke}{rgb}{0.000000,0.000000,0.000000}%
\pgfsetstrokecolor{currentstroke}%
\pgfsetdash{}{0pt}%
\pgfsys@defobject{currentmarker}{\pgfqpoint{0.000000in}{-0.048611in}}{\pgfqpoint{0.000000in}{0.000000in}}{%
\pgfpathmoveto{\pgfqpoint{0.000000in}{0.000000in}}%
\pgfpathlineto{\pgfqpoint{0.000000in}{-0.048611in}}%
\pgfusepath{stroke,fill}%
}%
\begin{pgfscope}%
\pgfsys@transformshift{2.475066in}{0.295304in}%
\pgfsys@useobject{currentmarker}{}%
\end{pgfscope}%
\end{pgfscope}%
\begin{pgfscope}%
\definecolor{textcolor}{rgb}{0.000000,0.000000,0.000000}%
\pgfsetstrokecolor{textcolor}%
\pgfsetfillcolor{textcolor}%
\pgftext[x=2.475066in,y=0.198082in,,top]{\color{textcolor}\sffamily\fontsize{8.000000}{9.600000}\selectfont 600}%
\end{pgfscope}%
\begin{pgfscope}%
\pgfsetbuttcap%
\pgfsetroundjoin%
\definecolor{currentfill}{rgb}{0.000000,0.000000,0.000000}%
\pgfsetfillcolor{currentfill}%
\pgfsetlinewidth{0.803000pt}%
\definecolor{currentstroke}{rgb}{0.000000,0.000000,0.000000}%
\pgfsetstrokecolor{currentstroke}%
\pgfsetdash{}{0pt}%
\pgfsys@defobject{currentmarker}{\pgfqpoint{-0.048611in}{0.000000in}}{\pgfqpoint{-0.000000in}{0.000000in}}{%
\pgfpathmoveto{\pgfqpoint{-0.000000in}{0.000000in}}%
\pgfpathlineto{\pgfqpoint{-0.048611in}{0.000000in}}%
\pgfusepath{stroke,fill}%
}%
\begin{pgfscope}%
\pgfsys@transformshift{0.451167in}{2.238315in}%
\pgfsys@useobject{currentmarker}{}%
\end{pgfscope}%
\end{pgfscope}%
\begin{pgfscope}%
\definecolor{textcolor}{rgb}{0.000000,0.000000,0.000000}%
\pgfsetstrokecolor{textcolor}%
\pgfsetfillcolor{textcolor}%
\pgftext[x=0.294945in, y=2.199759in, left, base]{\color{textcolor}\sffamily\fontsize{8.000000}{9.600000}\selectfont 0}%
\end{pgfscope}%
\begin{pgfscope}%
\pgfsetbuttcap%
\pgfsetroundjoin%
\definecolor{currentfill}{rgb}{0.000000,0.000000,0.000000}%
\pgfsetfillcolor{currentfill}%
\pgfsetlinewidth{0.803000pt}%
\definecolor{currentstroke}{rgb}{0.000000,0.000000,0.000000}%
\pgfsetstrokecolor{currentstroke}%
\pgfsetdash{}{0pt}%
\pgfsys@defobject{currentmarker}{\pgfqpoint{-0.048611in}{0.000000in}}{\pgfqpoint{-0.000000in}{0.000000in}}{%
\pgfpathmoveto{\pgfqpoint{-0.000000in}{0.000000in}}%
\pgfpathlineto{\pgfqpoint{-0.048611in}{0.000000in}}%
\pgfusepath{stroke,fill}%
}%
\begin{pgfscope}%
\pgfsys@transformshift{0.451167in}{1.901279in}%
\pgfsys@useobject{currentmarker}{}%
\end{pgfscope}%
\end{pgfscope}%
\begin{pgfscope}%
\definecolor{textcolor}{rgb}{0.000000,0.000000,0.000000}%
\pgfsetstrokecolor{textcolor}%
\pgfsetfillcolor{textcolor}%
\pgftext[x=0.176945in, y=1.862724in, left, base]{\color{textcolor}\sffamily\fontsize{8.000000}{9.600000}\selectfont 100}%
\end{pgfscope}%
\begin{pgfscope}%
\pgfsetbuttcap%
\pgfsetroundjoin%
\definecolor{currentfill}{rgb}{0.000000,0.000000,0.000000}%
\pgfsetfillcolor{currentfill}%
\pgfsetlinewidth{0.803000pt}%
\definecolor{currentstroke}{rgb}{0.000000,0.000000,0.000000}%
\pgfsetstrokecolor{currentstroke}%
\pgfsetdash{}{0pt}%
\pgfsys@defobject{currentmarker}{\pgfqpoint{-0.048611in}{0.000000in}}{\pgfqpoint{-0.000000in}{0.000000in}}{%
\pgfpathmoveto{\pgfqpoint{-0.000000in}{0.000000in}}%
\pgfpathlineto{\pgfqpoint{-0.048611in}{0.000000in}}%
\pgfusepath{stroke,fill}%
}%
\begin{pgfscope}%
\pgfsys@transformshift{0.451167in}{1.564244in}%
\pgfsys@useobject{currentmarker}{}%
\end{pgfscope}%
\end{pgfscope}%
\begin{pgfscope}%
\definecolor{textcolor}{rgb}{0.000000,0.000000,0.000000}%
\pgfsetstrokecolor{textcolor}%
\pgfsetfillcolor{textcolor}%
\pgftext[x=0.176945in, y=1.525688in, left, base]{\color{textcolor}\sffamily\fontsize{8.000000}{9.600000}\selectfont 200}%
\end{pgfscope}%
\begin{pgfscope}%
\pgfsetbuttcap%
\pgfsetroundjoin%
\definecolor{currentfill}{rgb}{0.000000,0.000000,0.000000}%
\pgfsetfillcolor{currentfill}%
\pgfsetlinewidth{0.803000pt}%
\definecolor{currentstroke}{rgb}{0.000000,0.000000,0.000000}%
\pgfsetstrokecolor{currentstroke}%
\pgfsetdash{}{0pt}%
\pgfsys@defobject{currentmarker}{\pgfqpoint{-0.048611in}{0.000000in}}{\pgfqpoint{-0.000000in}{0.000000in}}{%
\pgfpathmoveto{\pgfqpoint{-0.000000in}{0.000000in}}%
\pgfpathlineto{\pgfqpoint{-0.048611in}{0.000000in}}%
\pgfusepath{stroke,fill}%
}%
\begin{pgfscope}%
\pgfsys@transformshift{0.451167in}{1.227208in}%
\pgfsys@useobject{currentmarker}{}%
\end{pgfscope}%
\end{pgfscope}%
\begin{pgfscope}%
\definecolor{textcolor}{rgb}{0.000000,0.000000,0.000000}%
\pgfsetstrokecolor{textcolor}%
\pgfsetfillcolor{textcolor}%
\pgftext[x=0.176945in, y=1.188652in, left, base]{\color{textcolor}\sffamily\fontsize{8.000000}{9.600000}\selectfont 300}%
\end{pgfscope}%
\begin{pgfscope}%
\pgfsetbuttcap%
\pgfsetroundjoin%
\definecolor{currentfill}{rgb}{0.000000,0.000000,0.000000}%
\pgfsetfillcolor{currentfill}%
\pgfsetlinewidth{0.803000pt}%
\definecolor{currentstroke}{rgb}{0.000000,0.000000,0.000000}%
\pgfsetstrokecolor{currentstroke}%
\pgfsetdash{}{0pt}%
\pgfsys@defobject{currentmarker}{\pgfqpoint{-0.048611in}{0.000000in}}{\pgfqpoint{-0.000000in}{0.000000in}}{%
\pgfpathmoveto{\pgfqpoint{-0.000000in}{0.000000in}}%
\pgfpathlineto{\pgfqpoint{-0.048611in}{0.000000in}}%
\pgfusepath{stroke,fill}%
}%
\begin{pgfscope}%
\pgfsys@transformshift{0.451167in}{0.890172in}%
\pgfsys@useobject{currentmarker}{}%
\end{pgfscope}%
\end{pgfscope}%
\begin{pgfscope}%
\definecolor{textcolor}{rgb}{0.000000,0.000000,0.000000}%
\pgfsetstrokecolor{textcolor}%
\pgfsetfillcolor{textcolor}%
\pgftext[x=0.176945in, y=0.851617in, left, base]{\color{textcolor}\sffamily\fontsize{8.000000}{9.600000}\selectfont 400}%
\end{pgfscope}%
\begin{pgfscope}%
\pgfsetbuttcap%
\pgfsetroundjoin%
\definecolor{currentfill}{rgb}{0.000000,0.000000,0.000000}%
\pgfsetfillcolor{currentfill}%
\pgfsetlinewidth{0.803000pt}%
\definecolor{currentstroke}{rgb}{0.000000,0.000000,0.000000}%
\pgfsetstrokecolor{currentstroke}%
\pgfsetdash{}{0pt}%
\pgfsys@defobject{currentmarker}{\pgfqpoint{-0.048611in}{0.000000in}}{\pgfqpoint{-0.000000in}{0.000000in}}{%
\pgfpathmoveto{\pgfqpoint{-0.000000in}{0.000000in}}%
\pgfpathlineto{\pgfqpoint{-0.048611in}{0.000000in}}%
\pgfusepath{stroke,fill}%
}%
\begin{pgfscope}%
\pgfsys@transformshift{0.451167in}{0.553137in}%
\pgfsys@useobject{currentmarker}{}%
\end{pgfscope}%
\end{pgfscope}%
\begin{pgfscope}%
\definecolor{textcolor}{rgb}{0.000000,0.000000,0.000000}%
\pgfsetstrokecolor{textcolor}%
\pgfsetfillcolor{textcolor}%
\pgftext[x=0.176945in, y=0.514581in, left, base]{\color{textcolor}\sffamily\fontsize{8.000000}{9.600000}\selectfont 500}%
\end{pgfscope}%
\begin{pgfscope}%
\pgfsetrectcap%
\pgfsetmiterjoin%
\pgfsetlinewidth{0.803000pt}%
\definecolor{currentstroke}{rgb}{0.000000,0.000000,0.000000}%
\pgfsetstrokecolor{currentstroke}%
\pgfsetdash{}{0pt}%
\pgfpathmoveto{\pgfqpoint{0.451167in}{0.295304in}}%
\pgfpathlineto{\pgfqpoint{0.451167in}{2.240000in}}%
\pgfusepath{stroke}%
\end{pgfscope}%
\begin{pgfscope}%
\pgfsetrectcap%
\pgfsetmiterjoin%
\pgfsetlinewidth{0.803000pt}%
\definecolor{currentstroke}{rgb}{0.000000,0.000000,0.000000}%
\pgfsetstrokecolor{currentstroke}%
\pgfsetdash{}{0pt}%
\pgfpathmoveto{\pgfqpoint{2.790194in}{0.295304in}}%
\pgfpathlineto{\pgfqpoint{2.790194in}{2.240000in}}%
\pgfusepath{stroke}%
\end{pgfscope}%
\begin{pgfscope}%
\pgfsetrectcap%
\pgfsetmiterjoin%
\pgfsetlinewidth{0.803000pt}%
\definecolor{currentstroke}{rgb}{0.000000,0.000000,0.000000}%
\pgfsetstrokecolor{currentstroke}%
\pgfsetdash{}{0pt}%
\pgfpathmoveto{\pgfqpoint{0.451167in}{0.295304in}}%
\pgfpathlineto{\pgfqpoint{2.790194in}{0.295304in}}%
\pgfusepath{stroke}%
\end{pgfscope}%
\begin{pgfscope}%
\pgfsetrectcap%
\pgfsetmiterjoin%
\pgfsetlinewidth{0.803000pt}%
\definecolor{currentstroke}{rgb}{0.000000,0.000000,0.000000}%
\pgfsetstrokecolor{currentstroke}%
\pgfsetdash{}{0pt}%
\pgfpathmoveto{\pgfqpoint{0.451167in}{2.240000in}}%
\pgfpathlineto{\pgfqpoint{2.790194in}{2.240000in}}%
\pgfusepath{stroke}%
\end{pgfscope}%
\begin{pgfscope}%
\pgfsetbuttcap%
\pgfsetmiterjoin%
\definecolor{currentfill}{rgb}{1.000000,1.000000,1.000000}%
\pgfsetfillcolor{currentfill}%
\pgfsetlinewidth{0.000000pt}%
\definecolor{currentstroke}{rgb}{0.000000,0.000000,0.000000}%
\pgfsetstrokecolor{currentstroke}%
\pgfsetstrokeopacity{0.000000}%
\pgfsetdash{}{0pt}%
\pgfpathmoveto{\pgfqpoint{2.937646in}{0.295304in}}%
\pgfpathlineto{\pgfqpoint{3.034881in}{0.295304in}}%
\pgfpathlineto{\pgfqpoint{3.034881in}{2.240000in}}%
\pgfpathlineto{\pgfqpoint{2.937646in}{2.240000in}}%
\pgfpathclose%
\pgfusepath{fill}%
\end{pgfscope}%
\begin{pgfscope}%
\pgfpathrectangle{\pgfqpoint{2.937646in}{0.295304in}}{\pgfqpoint{0.097235in}{1.944696in}}%
\pgfusepath{clip}%
\pgfsetbuttcap%
\pgfsetmiterjoin%
\definecolor{currentfill}{rgb}{1.000000,1.000000,1.000000}%
\pgfsetfillcolor{currentfill}%
\pgfsetlinewidth{0.010037pt}%
\definecolor{currentstroke}{rgb}{1.000000,1.000000,1.000000}%
\pgfsetstrokecolor{currentstroke}%
\pgfsetdash{}{0pt}%
\pgfpathmoveto{\pgfqpoint{2.937646in}{0.295304in}}%
\pgfpathlineto{\pgfqpoint{2.937646in}{0.302901in}}%
\pgfpathlineto{\pgfqpoint{2.937646in}{2.232404in}}%
\pgfpathlineto{\pgfqpoint{2.937646in}{2.240000in}}%
\pgfpathlineto{\pgfqpoint{3.034881in}{2.240000in}}%
\pgfpathlineto{\pgfqpoint{3.034881in}{2.232404in}}%
\pgfpathlineto{\pgfqpoint{3.034881in}{0.302901in}}%
\pgfpathlineto{\pgfqpoint{3.034881in}{0.295304in}}%
\pgfpathlineto{\pgfqpoint{3.034881in}{0.295304in}}%
\pgfpathclose%
\pgfusepath{stroke,fill}%
\end{pgfscope}%
\begin{pgfscope}%
\pgfsys@transformshift{2.940000in}{0.300000in}%
\pgftext[left,bottom]{\includegraphics[interpolate=true,width=0.090000in,height=1.940000in]{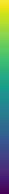}}%
\end{pgfscope}%
\begin{pgfscope}%
\pgfsetbuttcap%
\pgfsetroundjoin%
\definecolor{currentfill}{rgb}{0.000000,0.000000,0.000000}%
\pgfsetfillcolor{currentfill}%
\pgfsetlinewidth{0.803000pt}%
\definecolor{currentstroke}{rgb}{0.000000,0.000000,0.000000}%
\pgfsetstrokecolor{currentstroke}%
\pgfsetdash{}{0pt}%
\pgfsys@defobject{currentmarker}{\pgfqpoint{0.000000in}{0.000000in}}{\pgfqpoint{0.048611in}{0.000000in}}{%
\pgfpathmoveto{\pgfqpoint{0.000000in}{0.000000in}}%
\pgfpathlineto{\pgfqpoint{0.048611in}{0.000000in}}%
\pgfusepath{stroke,fill}%
}%
\begin{pgfscope}%
\pgfsys@transformshift{3.034881in}{0.403159in}%
\pgfsys@useobject{currentmarker}{}%
\end{pgfscope}%
\end{pgfscope}%
\begin{pgfscope}%
\definecolor{textcolor}{rgb}{0.000000,0.000000,0.000000}%
\pgfsetstrokecolor{textcolor}%
\pgfsetfillcolor{textcolor}%
\pgftext[x=3.132103in, y=0.364604in, left, base]{\color{textcolor}\sffamily\fontsize{8.000000}{9.600000}\selectfont 0.2}%
\end{pgfscope}%
\begin{pgfscope}%
\pgfsetbuttcap%
\pgfsetroundjoin%
\definecolor{currentfill}{rgb}{0.000000,0.000000,0.000000}%
\pgfsetfillcolor{currentfill}%
\pgfsetlinewidth{0.803000pt}%
\definecolor{currentstroke}{rgb}{0.000000,0.000000,0.000000}%
\pgfsetstrokecolor{currentstroke}%
\pgfsetdash{}{0pt}%
\pgfsys@defobject{currentmarker}{\pgfqpoint{0.000000in}{0.000000in}}{\pgfqpoint{0.048611in}{0.000000in}}{%
\pgfpathmoveto{\pgfqpoint{0.000000in}{0.000000in}}%
\pgfpathlineto{\pgfqpoint{0.048611in}{0.000000in}}%
\pgfusepath{stroke,fill}%
}%
\begin{pgfscope}%
\pgfsys@transformshift{3.034881in}{0.862369in}%
\pgfsys@useobject{currentmarker}{}%
\end{pgfscope}%
\end{pgfscope}%
\begin{pgfscope}%
\definecolor{textcolor}{rgb}{0.000000,0.000000,0.000000}%
\pgfsetstrokecolor{textcolor}%
\pgfsetfillcolor{textcolor}%
\pgftext[x=3.132103in, y=0.823814in, left, base]{\color{textcolor}\sffamily\fontsize{8.000000}{9.600000}\selectfont 0.4}%
\end{pgfscope}%
\begin{pgfscope}%
\pgfsetbuttcap%
\pgfsetroundjoin%
\definecolor{currentfill}{rgb}{0.000000,0.000000,0.000000}%
\pgfsetfillcolor{currentfill}%
\pgfsetlinewidth{0.803000pt}%
\definecolor{currentstroke}{rgb}{0.000000,0.000000,0.000000}%
\pgfsetstrokecolor{currentstroke}%
\pgfsetdash{}{0pt}%
\pgfsys@defobject{currentmarker}{\pgfqpoint{0.000000in}{0.000000in}}{\pgfqpoint{0.048611in}{0.000000in}}{%
\pgfpathmoveto{\pgfqpoint{0.000000in}{0.000000in}}%
\pgfpathlineto{\pgfqpoint{0.048611in}{0.000000in}}%
\pgfusepath{stroke,fill}%
}%
\begin{pgfscope}%
\pgfsys@transformshift{3.034881in}{1.321580in}%
\pgfsys@useobject{currentmarker}{}%
\end{pgfscope}%
\end{pgfscope}%
\begin{pgfscope}%
\definecolor{textcolor}{rgb}{0.000000,0.000000,0.000000}%
\pgfsetstrokecolor{textcolor}%
\pgfsetfillcolor{textcolor}%
\pgftext[x=3.132103in, y=1.283024in, left, base]{\color{textcolor}\sffamily\fontsize{8.000000}{9.600000}\selectfont 0.6}%
\end{pgfscope}%
\begin{pgfscope}%
\pgfsetbuttcap%
\pgfsetroundjoin%
\definecolor{currentfill}{rgb}{0.000000,0.000000,0.000000}%
\pgfsetfillcolor{currentfill}%
\pgfsetlinewidth{0.803000pt}%
\definecolor{currentstroke}{rgb}{0.000000,0.000000,0.000000}%
\pgfsetstrokecolor{currentstroke}%
\pgfsetdash{}{0pt}%
\pgfsys@defobject{currentmarker}{\pgfqpoint{0.000000in}{0.000000in}}{\pgfqpoint{0.048611in}{0.000000in}}{%
\pgfpathmoveto{\pgfqpoint{0.000000in}{0.000000in}}%
\pgfpathlineto{\pgfqpoint{0.048611in}{0.000000in}}%
\pgfusepath{stroke,fill}%
}%
\begin{pgfscope}%
\pgfsys@transformshift{3.034881in}{1.780790in}%
\pgfsys@useobject{currentmarker}{}%
\end{pgfscope}%
\end{pgfscope}%
\begin{pgfscope}%
\definecolor{textcolor}{rgb}{0.000000,0.000000,0.000000}%
\pgfsetstrokecolor{textcolor}%
\pgfsetfillcolor{textcolor}%
\pgftext[x=3.132103in, y=1.742234in, left, base]{\color{textcolor}\sffamily\fontsize{8.000000}{9.600000}\selectfont 0.8}%
\end{pgfscope}%
\begin{pgfscope}%
\pgfsetbuttcap%
\pgfsetroundjoin%
\definecolor{currentfill}{rgb}{0.000000,0.000000,0.000000}%
\pgfsetfillcolor{currentfill}%
\pgfsetlinewidth{0.803000pt}%
\definecolor{currentstroke}{rgb}{0.000000,0.000000,0.000000}%
\pgfsetstrokecolor{currentstroke}%
\pgfsetdash{}{0pt}%
\pgfsys@defobject{currentmarker}{\pgfqpoint{0.000000in}{0.000000in}}{\pgfqpoint{0.048611in}{0.000000in}}{%
\pgfpathmoveto{\pgfqpoint{0.000000in}{0.000000in}}%
\pgfpathlineto{\pgfqpoint{0.048611in}{0.000000in}}%
\pgfusepath{stroke,fill}%
}%
\begin{pgfscope}%
\pgfsys@transformshift{3.034881in}{2.240000in}%
\pgfsys@useobject{currentmarker}{}%
\end{pgfscope}%
\end{pgfscope}%
\begin{pgfscope}%
\definecolor{textcolor}{rgb}{0.000000,0.000000,0.000000}%
\pgfsetstrokecolor{textcolor}%
\pgfsetfillcolor{textcolor}%
\pgftext[x=3.132103in, y=2.201444in, left, base]{\color{textcolor}\sffamily\fontsize{8.000000}{9.600000}\selectfont 1.0}%
\end{pgfscope}%
\begin{pgfscope}%
\pgfsetrectcap%
\pgfsetmiterjoin%
\pgfsetlinewidth{0.803000pt}%
\definecolor{currentstroke}{rgb}{0.000000,0.000000,0.000000}%
\pgfsetstrokecolor{currentstroke}%
\pgfsetdash{}{0pt}%
\pgfpathmoveto{\pgfqpoint{2.937646in}{0.295304in}}%
\pgfpathlineto{\pgfqpoint{2.937646in}{0.302901in}}%
\pgfpathlineto{\pgfqpoint{2.937646in}{2.232404in}}%
\pgfpathlineto{\pgfqpoint{2.937646in}{2.240000in}}%
\pgfpathlineto{\pgfqpoint{3.034881in}{2.240000in}}%
\pgfpathlineto{\pgfqpoint{3.034881in}{2.232404in}}%
\pgfpathlineto{\pgfqpoint{3.034881in}{0.302901in}}%
\pgfpathlineto{\pgfqpoint{3.034881in}{0.295304in}}%
\pgfpathclose%
\pgfusepath{stroke}%
\end{pgfscope}%
\end{pgfpicture}%
\makeatother%
\endgroup%

%% file: figures/MCNP_s.pgf
\begingroup%
\makeatletter%
\begin{pgfpicture}%
\pgfpathrectangle{\pgfpointorigin}{\pgfqpoint{3.500000in}{2.400000in}}%
\pgfusepath{use as bounding box, clip}%
\begin{pgfscope}%
\pgfsetbuttcap%
\pgfsetmiterjoin%
\definecolor{currentfill}{rgb}{1.000000,1.000000,1.000000}%
\pgfsetfillcolor{currentfill}%
\pgfsetlinewidth{0.000000pt}%
\definecolor{currentstroke}{rgb}{1.000000,1.000000,1.000000}%
\pgfsetstrokecolor{currentstroke}%
\pgfsetdash{}{0pt}%
\pgfpathmoveto{\pgfqpoint{0.000000in}{0.000000in}}%
\pgfpathlineto{\pgfqpoint{3.500000in}{0.000000in}}%
\pgfpathlineto{\pgfqpoint{3.500000in}{2.400000in}}%
\pgfpathlineto{\pgfqpoint{0.000000in}{2.400000in}}%
\pgfpathclose%
\pgfusepath{fill}%
\end{pgfscope}%
\begin{pgfscope}%
\pgfsetbuttcap%
\pgfsetmiterjoin%
\definecolor{currentfill}{rgb}{1.000000,1.000000,1.000000}%
\pgfsetfillcolor{currentfill}%
\pgfsetlinewidth{0.000000pt}%
\definecolor{currentstroke}{rgb}{0.000000,0.000000,0.000000}%
\pgfsetstrokecolor{currentstroke}%
\pgfsetstrokeopacity{0.000000}%
\pgfsetdash{}{0pt}%
\pgfpathmoveto{\pgfqpoint{0.430972in}{0.309125in}}%
\pgfpathlineto{\pgfqpoint{2.764997in}{0.309125in}}%
\pgfpathlineto{\pgfqpoint{2.764997in}{2.249661in}}%
\pgfpathlineto{\pgfqpoint{0.430972in}{2.249661in}}%
\pgfpathclose%
\pgfusepath{fill}%
\end{pgfscope}%
\begin{pgfscope}%
\pgfsys@transformshift{0.430000in}{0.310000in}%
\pgftext[left,bottom]{\includegraphics[interpolate=true,width=2.340000in,height=1.940000in]{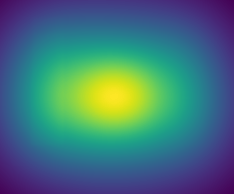}}%
\end{pgfscope}%
\begin{pgfscope}%
\pgfsetbuttcap%
\pgfsetroundjoin%
\definecolor{currentfill}{rgb}{0.000000,0.000000,0.000000}%
\pgfsetfillcolor{currentfill}%
\pgfsetlinewidth{0.803000pt}%
\definecolor{currentstroke}{rgb}{0.000000,0.000000,0.000000}%
\pgfsetstrokecolor{currentstroke}%
\pgfsetdash{}{0pt}%
\pgfsys@defobject{currentmarker}{\pgfqpoint{0.000000in}{-0.048611in}}{\pgfqpoint{0.000000in}{0.000000in}}{%
\pgfpathmoveto{\pgfqpoint{0.000000in}{0.000000in}}%
\pgfpathlineto{\pgfqpoint{0.000000in}{-0.048611in}}%
\pgfusepath{stroke,fill}%
}%
\begin{pgfscope}%
\pgfsys@transformshift{0.432654in}{0.309125in}%
\pgfsys@useobject{currentmarker}{}%
\end{pgfscope}%
\end{pgfscope}%
\begin{pgfscope}%
\definecolor{textcolor}{rgb}{0.000000,0.000000,0.000000}%
\pgfsetstrokecolor{textcolor}%
\pgfsetfillcolor{textcolor}%
\pgftext[x=0.432654in,y=0.211903in,,top]{\color{textcolor}\sffamily\fontsize{8.000000}{9.600000}\selectfont 0}%
\end{pgfscope}%
\begin{pgfscope}%
\pgfsetbuttcap%
\pgfsetroundjoin%
\definecolor{currentfill}{rgb}{0.000000,0.000000,0.000000}%
\pgfsetfillcolor{currentfill}%
\pgfsetlinewidth{0.803000pt}%
\definecolor{currentstroke}{rgb}{0.000000,0.000000,0.000000}%
\pgfsetstrokecolor{currentstroke}%
\pgfsetdash{}{0pt}%
\pgfsys@defobject{currentmarker}{\pgfqpoint{0.000000in}{-0.048611in}}{\pgfqpoint{0.000000in}{0.000000in}}{%
\pgfpathmoveto{\pgfqpoint{0.000000in}{0.000000in}}%
\pgfpathlineto{\pgfqpoint{0.000000in}{-0.048611in}}%
\pgfusepath{stroke,fill}%
}%
\begin{pgfscope}%
\pgfsys@transformshift{0.768969in}{0.309125in}%
\pgfsys@useobject{currentmarker}{}%
\end{pgfscope}%
\end{pgfscope}%
\begin{pgfscope}%
\definecolor{textcolor}{rgb}{0.000000,0.000000,0.000000}%
\pgfsetstrokecolor{textcolor}%
\pgfsetfillcolor{textcolor}%
\pgftext[x=0.768969in,y=0.211903in,,top]{\color{textcolor}\sffamily\fontsize{8.000000}{9.600000}\selectfont 100}%
\end{pgfscope}%
\begin{pgfscope}%
\pgfsetbuttcap%
\pgfsetroundjoin%
\definecolor{currentfill}{rgb}{0.000000,0.000000,0.000000}%
\pgfsetfillcolor{currentfill}%
\pgfsetlinewidth{0.803000pt}%
\definecolor{currentstroke}{rgb}{0.000000,0.000000,0.000000}%
\pgfsetstrokecolor{currentstroke}%
\pgfsetdash{}{0pt}%
\pgfsys@defobject{currentmarker}{\pgfqpoint{0.000000in}{-0.048611in}}{\pgfqpoint{0.000000in}{0.000000in}}{%
\pgfpathmoveto{\pgfqpoint{0.000000in}{0.000000in}}%
\pgfpathlineto{\pgfqpoint{0.000000in}{-0.048611in}}%
\pgfusepath{stroke,fill}%
}%
\begin{pgfscope}%
\pgfsys@transformshift{1.105283in}{0.309125in}%
\pgfsys@useobject{currentmarker}{}%
\end{pgfscope}%
\end{pgfscope}%
\begin{pgfscope}%
\definecolor{textcolor}{rgb}{0.000000,0.000000,0.000000}%
\pgfsetstrokecolor{textcolor}%
\pgfsetfillcolor{textcolor}%
\pgftext[x=1.105283in,y=0.211903in,,top]{\color{textcolor}\sffamily\fontsize{8.000000}{9.600000}\selectfont 200}%
\end{pgfscope}%
\begin{pgfscope}%
\pgfsetbuttcap%
\pgfsetroundjoin%
\definecolor{currentfill}{rgb}{0.000000,0.000000,0.000000}%
\pgfsetfillcolor{currentfill}%
\pgfsetlinewidth{0.803000pt}%
\definecolor{currentstroke}{rgb}{0.000000,0.000000,0.000000}%
\pgfsetstrokecolor{currentstroke}%
\pgfsetdash{}{0pt}%
\pgfsys@defobject{currentmarker}{\pgfqpoint{0.000000in}{-0.048611in}}{\pgfqpoint{0.000000in}{0.000000in}}{%
\pgfpathmoveto{\pgfqpoint{0.000000in}{0.000000in}}%
\pgfpathlineto{\pgfqpoint{0.000000in}{-0.048611in}}%
\pgfusepath{stroke,fill}%
}%
\begin{pgfscope}%
\pgfsys@transformshift{1.441598in}{0.309125in}%
\pgfsys@useobject{currentmarker}{}%
\end{pgfscope}%
\end{pgfscope}%
\begin{pgfscope}%
\definecolor{textcolor}{rgb}{0.000000,0.000000,0.000000}%
\pgfsetstrokecolor{textcolor}%
\pgfsetfillcolor{textcolor}%
\pgftext[x=1.441598in,y=0.211903in,,top]{\color{textcolor}\sffamily\fontsize{8.000000}{9.600000}\selectfont 300}%
\end{pgfscope}%
\begin{pgfscope}%
\pgfsetbuttcap%
\pgfsetroundjoin%
\definecolor{currentfill}{rgb}{0.000000,0.000000,0.000000}%
\pgfsetfillcolor{currentfill}%
\pgfsetlinewidth{0.803000pt}%
\definecolor{currentstroke}{rgb}{0.000000,0.000000,0.000000}%
\pgfsetstrokecolor{currentstroke}%
\pgfsetdash{}{0pt}%
\pgfsys@defobject{currentmarker}{\pgfqpoint{0.000000in}{-0.048611in}}{\pgfqpoint{0.000000in}{0.000000in}}{%
\pgfpathmoveto{\pgfqpoint{0.000000in}{0.000000in}}%
\pgfpathlineto{\pgfqpoint{0.000000in}{-0.048611in}}%
\pgfusepath{stroke,fill}%
}%
\begin{pgfscope}%
\pgfsys@transformshift{1.777913in}{0.309125in}%
\pgfsys@useobject{currentmarker}{}%
\end{pgfscope}%
\end{pgfscope}%
\begin{pgfscope}%
\definecolor{textcolor}{rgb}{0.000000,0.000000,0.000000}%
\pgfsetstrokecolor{textcolor}%
\pgfsetfillcolor{textcolor}%
\pgftext[x=1.777913in,y=0.211903in,,top]{\color{textcolor}\sffamily\fontsize{8.000000}{9.600000}\selectfont 400}%
\end{pgfscope}%
\begin{pgfscope}%
\pgfsetbuttcap%
\pgfsetroundjoin%
\definecolor{currentfill}{rgb}{0.000000,0.000000,0.000000}%
\pgfsetfillcolor{currentfill}%
\pgfsetlinewidth{0.803000pt}%
\definecolor{currentstroke}{rgb}{0.000000,0.000000,0.000000}%
\pgfsetstrokecolor{currentstroke}%
\pgfsetdash{}{0pt}%
\pgfsys@defobject{currentmarker}{\pgfqpoint{0.000000in}{-0.048611in}}{\pgfqpoint{0.000000in}{0.000000in}}{%
\pgfpathmoveto{\pgfqpoint{0.000000in}{0.000000in}}%
\pgfpathlineto{\pgfqpoint{0.000000in}{-0.048611in}}%
\pgfusepath{stroke,fill}%
}%
\begin{pgfscope}%
\pgfsys@transformshift{2.114228in}{0.309125in}%
\pgfsys@useobject{currentmarker}{}%
\end{pgfscope}%
\end{pgfscope}%
\begin{pgfscope}%
\definecolor{textcolor}{rgb}{0.000000,0.000000,0.000000}%
\pgfsetstrokecolor{textcolor}%
\pgfsetfillcolor{textcolor}%
\pgftext[x=2.114228in,y=0.211903in,,top]{\color{textcolor}\sffamily\fontsize{8.000000}{9.600000}\selectfont 500}%
\end{pgfscope}%
\begin{pgfscope}%
\pgfsetbuttcap%
\pgfsetroundjoin%
\definecolor{currentfill}{rgb}{0.000000,0.000000,0.000000}%
\pgfsetfillcolor{currentfill}%
\pgfsetlinewidth{0.803000pt}%
\definecolor{currentstroke}{rgb}{0.000000,0.000000,0.000000}%
\pgfsetstrokecolor{currentstroke}%
\pgfsetdash{}{0pt}%
\pgfsys@defobject{currentmarker}{\pgfqpoint{0.000000in}{-0.048611in}}{\pgfqpoint{0.000000in}{0.000000in}}{%
\pgfpathmoveto{\pgfqpoint{0.000000in}{0.000000in}}%
\pgfpathlineto{\pgfqpoint{0.000000in}{-0.048611in}}%
\pgfusepath{stroke,fill}%
}%
\begin{pgfscope}%
\pgfsys@transformshift{2.450542in}{0.309125in}%
\pgfsys@useobject{currentmarker}{}%
\end{pgfscope}%
\end{pgfscope}%
\begin{pgfscope}%
\definecolor{textcolor}{rgb}{0.000000,0.000000,0.000000}%
\pgfsetstrokecolor{textcolor}%
\pgfsetfillcolor{textcolor}%
\pgftext[x=2.450542in,y=0.211903in,,top]{\color{textcolor}\sffamily\fontsize{8.000000}{9.600000}\selectfont 600}%
\end{pgfscope}%
\begin{pgfscope}%
\pgfsetbuttcap%
\pgfsetroundjoin%
\definecolor{currentfill}{rgb}{0.000000,0.000000,0.000000}%
\pgfsetfillcolor{currentfill}%
\pgfsetlinewidth{0.803000pt}%
\definecolor{currentstroke}{rgb}{0.000000,0.000000,0.000000}%
\pgfsetstrokecolor{currentstroke}%
\pgfsetdash{}{0pt}%
\pgfsys@defobject{currentmarker}{\pgfqpoint{-0.048611in}{0.000000in}}{\pgfqpoint{-0.000000in}{0.000000in}}{%
\pgfpathmoveto{\pgfqpoint{-0.000000in}{0.000000in}}%
\pgfpathlineto{\pgfqpoint{-0.048611in}{0.000000in}}%
\pgfusepath{stroke,fill}%
}%
\begin{pgfscope}%
\pgfsys@transformshift{0.430972in}{2.247979in}%
\pgfsys@useobject{currentmarker}{}%
\end{pgfscope}%
\end{pgfscope}%
\begin{pgfscope}%
\definecolor{textcolor}{rgb}{0.000000,0.000000,0.000000}%
\pgfsetstrokecolor{textcolor}%
\pgfsetfillcolor{textcolor}%
\pgftext[x=0.274750in, y=2.209424in, left, base]{\color{textcolor}\sffamily\fontsize{8.000000}{9.600000}\selectfont 0}%
\end{pgfscope}%
\begin{pgfscope}%
\pgfsetbuttcap%
\pgfsetroundjoin%
\definecolor{currentfill}{rgb}{0.000000,0.000000,0.000000}%
\pgfsetfillcolor{currentfill}%
\pgfsetlinewidth{0.803000pt}%
\definecolor{currentstroke}{rgb}{0.000000,0.000000,0.000000}%
\pgfsetstrokecolor{currentstroke}%
\pgfsetdash{}{0pt}%
\pgfsys@defobject{currentmarker}{\pgfqpoint{-0.048611in}{0.000000in}}{\pgfqpoint{-0.000000in}{0.000000in}}{%
\pgfpathmoveto{\pgfqpoint{-0.000000in}{0.000000in}}%
\pgfpathlineto{\pgfqpoint{-0.048611in}{0.000000in}}%
\pgfusepath{stroke,fill}%
}%
\begin{pgfscope}%
\pgfsys@transformshift{0.430972in}{1.911665in}%
\pgfsys@useobject{currentmarker}{}%
\end{pgfscope}%
\end{pgfscope}%
\begin{pgfscope}%
\definecolor{textcolor}{rgb}{0.000000,0.000000,0.000000}%
\pgfsetstrokecolor{textcolor}%
\pgfsetfillcolor{textcolor}%
\pgftext[x=0.156750in, y=1.873109in, left, base]{\color{textcolor}\sffamily\fontsize{8.000000}{9.600000}\selectfont 100}%
\end{pgfscope}%
\begin{pgfscope}%
\pgfsetbuttcap%
\pgfsetroundjoin%
\definecolor{currentfill}{rgb}{0.000000,0.000000,0.000000}%
\pgfsetfillcolor{currentfill}%
\pgfsetlinewidth{0.803000pt}%
\definecolor{currentstroke}{rgb}{0.000000,0.000000,0.000000}%
\pgfsetstrokecolor{currentstroke}%
\pgfsetdash{}{0pt}%
\pgfsys@defobject{currentmarker}{\pgfqpoint{-0.048611in}{0.000000in}}{\pgfqpoint{-0.000000in}{0.000000in}}{%
\pgfpathmoveto{\pgfqpoint{-0.000000in}{0.000000in}}%
\pgfpathlineto{\pgfqpoint{-0.048611in}{0.000000in}}%
\pgfusepath{stroke,fill}%
}%
\begin{pgfscope}%
\pgfsys@transformshift{0.430972in}{1.575350in}%
\pgfsys@useobject{currentmarker}{}%
\end{pgfscope}%
\end{pgfscope}%
\begin{pgfscope}%
\definecolor{textcolor}{rgb}{0.000000,0.000000,0.000000}%
\pgfsetstrokecolor{textcolor}%
\pgfsetfillcolor{textcolor}%
\pgftext[x=0.156750in, y=1.536794in, left, base]{\color{textcolor}\sffamily\fontsize{8.000000}{9.600000}\selectfont 200}%
\end{pgfscope}%
\begin{pgfscope}%
\pgfsetbuttcap%
\pgfsetroundjoin%
\definecolor{currentfill}{rgb}{0.000000,0.000000,0.000000}%
\pgfsetfillcolor{currentfill}%
\pgfsetlinewidth{0.803000pt}%
\definecolor{currentstroke}{rgb}{0.000000,0.000000,0.000000}%
\pgfsetstrokecolor{currentstroke}%
\pgfsetdash{}{0pt}%
\pgfsys@defobject{currentmarker}{\pgfqpoint{-0.048611in}{0.000000in}}{\pgfqpoint{-0.000000in}{0.000000in}}{%
\pgfpathmoveto{\pgfqpoint{-0.000000in}{0.000000in}}%
\pgfpathlineto{\pgfqpoint{-0.048611in}{0.000000in}}%
\pgfusepath{stroke,fill}%
}%
\begin{pgfscope}%
\pgfsys@transformshift{0.430972in}{1.239035in}%
\pgfsys@useobject{currentmarker}{}%
\end{pgfscope}%
\end{pgfscope}%
\begin{pgfscope}%
\definecolor{textcolor}{rgb}{0.000000,0.000000,0.000000}%
\pgfsetstrokecolor{textcolor}%
\pgfsetfillcolor{textcolor}%
\pgftext[x=0.156750in, y=1.200479in, left, base]{\color{textcolor}\sffamily\fontsize{8.000000}{9.600000}\selectfont 300}%
\end{pgfscope}%
\begin{pgfscope}%
\pgfsetbuttcap%
\pgfsetroundjoin%
\definecolor{currentfill}{rgb}{0.000000,0.000000,0.000000}%
\pgfsetfillcolor{currentfill}%
\pgfsetlinewidth{0.803000pt}%
\definecolor{currentstroke}{rgb}{0.000000,0.000000,0.000000}%
\pgfsetstrokecolor{currentstroke}%
\pgfsetdash{}{0pt}%
\pgfsys@defobject{currentmarker}{\pgfqpoint{-0.048611in}{0.000000in}}{\pgfqpoint{-0.000000in}{0.000000in}}{%
\pgfpathmoveto{\pgfqpoint{-0.000000in}{0.000000in}}%
\pgfpathlineto{\pgfqpoint{-0.048611in}{0.000000in}}%
\pgfusepath{stroke,fill}%
}%
\begin{pgfscope}%
\pgfsys@transformshift{0.430972in}{0.902720in}%
\pgfsys@useobject{currentmarker}{}%
\end{pgfscope}%
\end{pgfscope}%
\begin{pgfscope}%
\definecolor{textcolor}{rgb}{0.000000,0.000000,0.000000}%
\pgfsetstrokecolor{textcolor}%
\pgfsetfillcolor{textcolor}%
\pgftext[x=0.156750in, y=0.864165in, left, base]{\color{textcolor}\sffamily\fontsize{8.000000}{9.600000}\selectfont 400}%
\end{pgfscope}%
\begin{pgfscope}%
\pgfsetbuttcap%
\pgfsetroundjoin%
\definecolor{currentfill}{rgb}{0.000000,0.000000,0.000000}%
\pgfsetfillcolor{currentfill}%
\pgfsetlinewidth{0.803000pt}%
\definecolor{currentstroke}{rgb}{0.000000,0.000000,0.000000}%
\pgfsetstrokecolor{currentstroke}%
\pgfsetdash{}{0pt}%
\pgfsys@defobject{currentmarker}{\pgfqpoint{-0.048611in}{0.000000in}}{\pgfqpoint{-0.000000in}{0.000000in}}{%
\pgfpathmoveto{\pgfqpoint{-0.000000in}{0.000000in}}%
\pgfpathlineto{\pgfqpoint{-0.048611in}{0.000000in}}%
\pgfusepath{stroke,fill}%
}%
\begin{pgfscope}%
\pgfsys@transformshift{0.430972in}{0.566406in}%
\pgfsys@useobject{currentmarker}{}%
\end{pgfscope}%
\end{pgfscope}%
\begin{pgfscope}%
\definecolor{textcolor}{rgb}{0.000000,0.000000,0.000000}%
\pgfsetstrokecolor{textcolor}%
\pgfsetfillcolor{textcolor}%
\pgftext[x=0.156750in, y=0.527850in, left, base]{\color{textcolor}\sffamily\fontsize{8.000000}{9.600000}\selectfont 500}%
\end{pgfscope}%
\begin{pgfscope}%
\pgfsetrectcap%
\pgfsetmiterjoin%
\pgfsetlinewidth{0.803000pt}%
\definecolor{currentstroke}{rgb}{0.000000,0.000000,0.000000}%
\pgfsetstrokecolor{currentstroke}%
\pgfsetdash{}{0pt}%
\pgfpathmoveto{\pgfqpoint{0.430972in}{0.309125in}}%
\pgfpathlineto{\pgfqpoint{0.430972in}{2.249661in}}%
\pgfusepath{stroke}%
\end{pgfscope}%
\begin{pgfscope}%
\pgfsetrectcap%
\pgfsetmiterjoin%
\pgfsetlinewidth{0.803000pt}%
\definecolor{currentstroke}{rgb}{0.000000,0.000000,0.000000}%
\pgfsetstrokecolor{currentstroke}%
\pgfsetdash{}{0pt}%
\pgfpathmoveto{\pgfqpoint{2.764997in}{0.309125in}}%
\pgfpathlineto{\pgfqpoint{2.764997in}{2.249661in}}%
\pgfusepath{stroke}%
\end{pgfscope}%
\begin{pgfscope}%
\pgfsetrectcap%
\pgfsetmiterjoin%
\pgfsetlinewidth{0.803000pt}%
\definecolor{currentstroke}{rgb}{0.000000,0.000000,0.000000}%
\pgfsetstrokecolor{currentstroke}%
\pgfsetdash{}{0pt}%
\pgfpathmoveto{\pgfqpoint{0.430972in}{0.309125in}}%
\pgfpathlineto{\pgfqpoint{2.764997in}{0.309125in}}%
\pgfusepath{stroke}%
\end{pgfscope}%
\begin{pgfscope}%
\pgfsetrectcap%
\pgfsetmiterjoin%
\pgfsetlinewidth{0.803000pt}%
\definecolor{currentstroke}{rgb}{0.000000,0.000000,0.000000}%
\pgfsetstrokecolor{currentstroke}%
\pgfsetdash{}{0pt}%
\pgfpathmoveto{\pgfqpoint{0.430972in}{2.249661in}}%
\pgfpathlineto{\pgfqpoint{2.764997in}{2.249661in}}%
\pgfusepath{stroke}%
\end{pgfscope}%
\begin{pgfscope}%
\pgfsetbuttcap%
\pgfsetmiterjoin%
\definecolor{currentfill}{rgb}{1.000000,1.000000,1.000000}%
\pgfsetfillcolor{currentfill}%
\pgfsetlinewidth{0.000000pt}%
\definecolor{currentstroke}{rgb}{0.000000,0.000000,0.000000}%
\pgfsetstrokecolor{currentstroke}%
\pgfsetstrokeopacity{0.000000}%
\pgfsetdash{}{0pt}%
\pgfpathmoveto{\pgfqpoint{2.910873in}{0.295304in}}%
\pgfpathlineto{\pgfqpoint{3.009282in}{0.295304in}}%
\pgfpathlineto{\pgfqpoint{3.009282in}{2.263481in}}%
\pgfpathlineto{\pgfqpoint{2.910873in}{2.263481in}}%
\pgfpathclose%
\pgfusepath{fill}%
\end{pgfscope}%
\begin{pgfscope}%
\pgfpathrectangle{\pgfqpoint{2.910873in}{0.295304in}}{\pgfqpoint{0.098409in}{1.968177in}}%
\pgfusepath{clip}%
\pgfsetbuttcap%
\pgfsetmiterjoin%
\definecolor{currentfill}{rgb}{1.000000,1.000000,1.000000}%
\pgfsetfillcolor{currentfill}%
\pgfsetlinewidth{0.010037pt}%
\definecolor{currentstroke}{rgb}{1.000000,1.000000,1.000000}%
\pgfsetstrokecolor{currentstroke}%
\pgfsetdash{}{0pt}%
\pgfpathmoveto{\pgfqpoint{2.910873in}{0.295304in}}%
\pgfpathlineto{\pgfqpoint{2.910873in}{0.302993in}}%
\pgfpathlineto{\pgfqpoint{2.910873in}{2.255793in}}%
\pgfpathlineto{\pgfqpoint{2.910873in}{2.263481in}}%
\pgfpathlineto{\pgfqpoint{3.009282in}{2.263481in}}%
\pgfpathlineto{\pgfqpoint{3.009282in}{2.255793in}}%
\pgfpathlineto{\pgfqpoint{3.009282in}{0.302993in}}%
\pgfpathlineto{\pgfqpoint{3.009282in}{0.295304in}}%
\pgfpathlineto{\pgfqpoint{3.009282in}{0.295304in}}%
\pgfpathclose%
\pgfusepath{stroke,fill}%
\end{pgfscope}%
\begin{pgfscope}%
\pgfsys@transformshift{2.910000in}{0.300000in}%
\pgftext[left,bottom]{\includegraphics[interpolate=true,width=0.100000in,height=1.960000in]{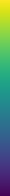}}%
\end{pgfscope}%
\begin{pgfscope}%
\pgfsetbuttcap%
\pgfsetroundjoin%
\definecolor{currentfill}{rgb}{0.000000,0.000000,0.000000}%
\pgfsetfillcolor{currentfill}%
\pgfsetlinewidth{0.803000pt}%
\definecolor{currentstroke}{rgb}{0.000000,0.000000,0.000000}%
\pgfsetstrokecolor{currentstroke}%
\pgfsetdash{}{0pt}%
\pgfsys@defobject{currentmarker}{\pgfqpoint{0.000000in}{0.000000in}}{\pgfqpoint{0.048611in}{0.000000in}}{%
\pgfpathmoveto{\pgfqpoint{0.000000in}{0.000000in}}%
\pgfpathlineto{\pgfqpoint{0.048611in}{0.000000in}}%
\pgfusepath{stroke,fill}%
}%
\begin{pgfscope}%
\pgfsys@transformshift{3.009282in}{0.316873in}%
\pgfsys@useobject{currentmarker}{}%
\end{pgfscope}%
\end{pgfscope}%
\begin{pgfscope}%
\definecolor{textcolor}{rgb}{0.000000,0.000000,0.000000}%
\pgfsetstrokecolor{textcolor}%
\pgfsetfillcolor{textcolor}%
\pgftext[x=3.106504in, y=0.278317in, left, base]{\color{textcolor}\sffamily\fontsize{8.000000}{9.600000}\selectfont 0.02}%
\end{pgfscope}%
\begin{pgfscope}%
\pgfsetbuttcap%
\pgfsetroundjoin%
\definecolor{currentfill}{rgb}{0.000000,0.000000,0.000000}%
\pgfsetfillcolor{currentfill}%
\pgfsetlinewidth{0.803000pt}%
\definecolor{currentstroke}{rgb}{0.000000,0.000000,0.000000}%
\pgfsetstrokecolor{currentstroke}%
\pgfsetdash{}{0pt}%
\pgfsys@defobject{currentmarker}{\pgfqpoint{0.000000in}{0.000000in}}{\pgfqpoint{0.048611in}{0.000000in}}{%
\pgfpathmoveto{\pgfqpoint{0.000000in}{0.000000in}}%
\pgfpathlineto{\pgfqpoint{0.048611in}{0.000000in}}%
\pgfusepath{stroke,fill}%
}%
\begin{pgfscope}%
\pgfsys@transformshift{3.009282in}{0.723251in}%
\pgfsys@useobject{currentmarker}{}%
\end{pgfscope}%
\end{pgfscope}%
\begin{pgfscope}%
\definecolor{textcolor}{rgb}{0.000000,0.000000,0.000000}%
\pgfsetstrokecolor{textcolor}%
\pgfsetfillcolor{textcolor}%
\pgftext[x=3.106504in, y=0.684696in, left, base]{\color{textcolor}\sffamily\fontsize{8.000000}{9.600000}\selectfont 0.04}%
\end{pgfscope}%
\begin{pgfscope}%
\pgfsetbuttcap%
\pgfsetroundjoin%
\definecolor{currentfill}{rgb}{0.000000,0.000000,0.000000}%
\pgfsetfillcolor{currentfill}%
\pgfsetlinewidth{0.803000pt}%
\definecolor{currentstroke}{rgb}{0.000000,0.000000,0.000000}%
\pgfsetstrokecolor{currentstroke}%
\pgfsetdash{}{0pt}%
\pgfsys@defobject{currentmarker}{\pgfqpoint{0.000000in}{0.000000in}}{\pgfqpoint{0.048611in}{0.000000in}}{%
\pgfpathmoveto{\pgfqpoint{0.000000in}{0.000000in}}%
\pgfpathlineto{\pgfqpoint{0.048611in}{0.000000in}}%
\pgfusepath{stroke,fill}%
}%
\begin{pgfscope}%
\pgfsys@transformshift{3.009282in}{1.129629in}%
\pgfsys@useobject{currentmarker}{}%
\end{pgfscope}%
\end{pgfscope}%
\begin{pgfscope}%
\definecolor{textcolor}{rgb}{0.000000,0.000000,0.000000}%
\pgfsetstrokecolor{textcolor}%
\pgfsetfillcolor{textcolor}%
\pgftext[x=3.106504in, y=1.091074in, left, base]{\color{textcolor}\sffamily\fontsize{8.000000}{9.600000}\selectfont 0.06}%
\end{pgfscope}%
\begin{pgfscope}%
\pgfsetbuttcap%
\pgfsetroundjoin%
\definecolor{currentfill}{rgb}{0.000000,0.000000,0.000000}%
\pgfsetfillcolor{currentfill}%
\pgfsetlinewidth{0.803000pt}%
\definecolor{currentstroke}{rgb}{0.000000,0.000000,0.000000}%
\pgfsetstrokecolor{currentstroke}%
\pgfsetdash{}{0pt}%
\pgfsys@defobject{currentmarker}{\pgfqpoint{0.000000in}{0.000000in}}{\pgfqpoint{0.048611in}{0.000000in}}{%
\pgfpathmoveto{\pgfqpoint{0.000000in}{0.000000in}}%
\pgfpathlineto{\pgfqpoint{0.048611in}{0.000000in}}%
\pgfusepath{stroke,fill}%
}%
\begin{pgfscope}%
\pgfsys@transformshift{3.009282in}{1.536008in}%
\pgfsys@useobject{currentmarker}{}%
\end{pgfscope}%
\end{pgfscope}%
\begin{pgfscope}%
\definecolor{textcolor}{rgb}{0.000000,0.000000,0.000000}%
\pgfsetstrokecolor{textcolor}%
\pgfsetfillcolor{textcolor}%
\pgftext[x=3.106504in, y=1.497452in, left, base]{\color{textcolor}\sffamily\fontsize{8.000000}{9.600000}\selectfont 0.08}%
\end{pgfscope}%
\begin{pgfscope}%
\pgfsetbuttcap%
\pgfsetroundjoin%
\definecolor{currentfill}{rgb}{0.000000,0.000000,0.000000}%
\pgfsetfillcolor{currentfill}%
\pgfsetlinewidth{0.803000pt}%
\definecolor{currentstroke}{rgb}{0.000000,0.000000,0.000000}%
\pgfsetstrokecolor{currentstroke}%
\pgfsetdash{}{0pt}%
\pgfsys@defobject{currentmarker}{\pgfqpoint{0.000000in}{0.000000in}}{\pgfqpoint{0.048611in}{0.000000in}}{%
\pgfpathmoveto{\pgfqpoint{0.000000in}{0.000000in}}%
\pgfpathlineto{\pgfqpoint{0.048611in}{0.000000in}}%
\pgfusepath{stroke,fill}%
}%
\begin{pgfscope}%
\pgfsys@transformshift{3.009282in}{1.942386in}%
\pgfsys@useobject{currentmarker}{}%
\end{pgfscope}%
\end{pgfscope}%
\begin{pgfscope}%
\definecolor{textcolor}{rgb}{0.000000,0.000000,0.000000}%
\pgfsetstrokecolor{textcolor}%
\pgfsetfillcolor{textcolor}%
\pgftext[x=3.106504in, y=1.903830in, left, base]{\color{textcolor}\sffamily\fontsize{8.000000}{9.600000}\selectfont 0.10}%
\end{pgfscope}%
\begin{pgfscope}%
\pgfsetrectcap%
\pgfsetmiterjoin%
\pgfsetlinewidth{0.803000pt}%
\definecolor{currentstroke}{rgb}{0.000000,0.000000,0.000000}%
\pgfsetstrokecolor{currentstroke}%
\pgfsetdash{}{0pt}%
\pgfpathmoveto{\pgfqpoint{2.910873in}{0.295304in}}%
\pgfpathlineto{\pgfqpoint{2.910873in}{0.302993in}}%
\pgfpathlineto{\pgfqpoint{2.910873in}{2.255793in}}%
\pgfpathlineto{\pgfqpoint{2.910873in}{2.263481in}}%
\pgfpathlineto{\pgfqpoint{3.009282in}{2.263481in}}%
\pgfpathlineto{\pgfqpoint{3.009282in}{2.255793in}}%
\pgfpathlineto{\pgfqpoint{3.009282in}{0.302993in}}%
\pgfpathlineto{\pgfqpoint{3.009282in}{0.295304in}}%
\pgfpathclose%
\pgfusepath{stroke}%
\end{pgfscope}%
\end{pgfpicture}%
\makeatother%
\endgroup%

%% file: figures/MCNP_lineouts.pgf
\begingroup%
\makeatletter%
\begin{pgfpicture}%
\pgfpathrectangle{\pgfpointorigin}{\pgfqpoint{3.500000in}{2.400000in}}%
\pgfusepath{use as bounding box, clip}%
\begin{pgfscope}%
\pgfsetbuttcap%
\pgfsetmiterjoin%
\definecolor{currentfill}{rgb}{1.000000,1.000000,1.000000}%
\pgfsetfillcolor{currentfill}%
\pgfsetlinewidth{0.000000pt}%
\definecolor{currentstroke}{rgb}{1.000000,1.000000,1.000000}%
\pgfsetstrokecolor{currentstroke}%
\pgfsetdash{}{0pt}%
\pgfpathmoveto{\pgfqpoint{0.000000in}{0.000000in}}%
\pgfpathlineto{\pgfqpoint{3.500000in}{0.000000in}}%
\pgfpathlineto{\pgfqpoint{3.500000in}{2.400000in}}%
\pgfpathlineto{\pgfqpoint{0.000000in}{2.400000in}}%
\pgfpathclose%
\pgfusepath{fill}%
\end{pgfscope}%
\begin{pgfscope}%
\pgfsetbuttcap%
\pgfsetmiterjoin%
\definecolor{currentfill}{rgb}{1.000000,1.000000,1.000000}%
\pgfsetfillcolor{currentfill}%
\pgfsetlinewidth{0.000000pt}%
\definecolor{currentstroke}{rgb}{0.000000,0.000000,0.000000}%
\pgfsetstrokecolor{currentstroke}%
\pgfsetstrokeopacity{0.000000}%
\pgfsetdash{}{0pt}%
\pgfpathmoveto{\pgfqpoint{0.465972in}{0.317222in}}%
\pgfpathlineto{\pgfqpoint{3.371512in}{0.317222in}}%
\pgfpathlineto{\pgfqpoint{3.371512in}{2.280000in}}%
\pgfpathlineto{\pgfqpoint{0.465972in}{2.280000in}}%
\pgfpathclose%
\pgfusepath{fill}%
\end{pgfscope}%
\begin{pgfscope}%
\pgfsetbuttcap%
\pgfsetroundjoin%
\definecolor{currentfill}{rgb}{0.000000,0.000000,0.000000}%
\pgfsetfillcolor{currentfill}%
\pgfsetlinewidth{0.803000pt}%
\definecolor{currentstroke}{rgb}{0.000000,0.000000,0.000000}%
\pgfsetstrokecolor{currentstroke}%
\pgfsetdash{}{0pt}%
\pgfsys@defobject{currentmarker}{\pgfqpoint{0.000000in}{-0.048611in}}{\pgfqpoint{0.000000in}{0.000000in}}{%
\pgfpathmoveto{\pgfqpoint{0.000000in}{0.000000in}}%
\pgfpathlineto{\pgfqpoint{0.000000in}{-0.048611in}}%
\pgfusepath{stroke,fill}%
}%
\begin{pgfscope}%
\pgfsys@transformshift{0.598042in}{0.317222in}%
\pgfsys@useobject{currentmarker}{}%
\end{pgfscope}%
\end{pgfscope}%
\begin{pgfscope}%
\definecolor{textcolor}{rgb}{0.000000,0.000000,0.000000}%
\pgfsetstrokecolor{textcolor}%
\pgfsetfillcolor{textcolor}%
\pgftext[x=0.598042in,y=0.220000in,,top]{\color{textcolor}\sffamily\fontsize{8.000000}{9.600000}\selectfont 0}%
\end{pgfscope}%
\begin{pgfscope}%
\pgfsetbuttcap%
\pgfsetroundjoin%
\definecolor{currentfill}{rgb}{0.000000,0.000000,0.000000}%
\pgfsetfillcolor{currentfill}%
\pgfsetlinewidth{0.803000pt}%
\definecolor{currentstroke}{rgb}{0.000000,0.000000,0.000000}%
\pgfsetstrokecolor{currentstroke}%
\pgfsetdash{}{0pt}%
\pgfsys@defobject{currentmarker}{\pgfqpoint{0.000000in}{-0.048611in}}{\pgfqpoint{0.000000in}{0.000000in}}{%
\pgfpathmoveto{\pgfqpoint{0.000000in}{0.000000in}}%
\pgfpathlineto{\pgfqpoint{0.000000in}{-0.048611in}}%
\pgfusepath{stroke,fill}%
}%
\begin{pgfscope}%
\pgfsys@transformshift{0.979197in}{0.317222in}%
\pgfsys@useobject{currentmarker}{}%
\end{pgfscope}%
\end{pgfscope}%
\begin{pgfscope}%
\definecolor{textcolor}{rgb}{0.000000,0.000000,0.000000}%
\pgfsetstrokecolor{textcolor}%
\pgfsetfillcolor{textcolor}%
\pgftext[x=0.979197in,y=0.220000in,,top]{\color{textcolor}\sffamily\fontsize{8.000000}{9.600000}\selectfont 100}%
\end{pgfscope}%
\begin{pgfscope}%
\pgfsetbuttcap%
\pgfsetroundjoin%
\definecolor{currentfill}{rgb}{0.000000,0.000000,0.000000}%
\pgfsetfillcolor{currentfill}%
\pgfsetlinewidth{0.803000pt}%
\definecolor{currentstroke}{rgb}{0.000000,0.000000,0.000000}%
\pgfsetstrokecolor{currentstroke}%
\pgfsetdash{}{0pt}%
\pgfsys@defobject{currentmarker}{\pgfqpoint{0.000000in}{-0.048611in}}{\pgfqpoint{0.000000in}{0.000000in}}{%
\pgfpathmoveto{\pgfqpoint{0.000000in}{0.000000in}}%
\pgfpathlineto{\pgfqpoint{0.000000in}{-0.048611in}}%
\pgfusepath{stroke,fill}%
}%
\begin{pgfscope}%
\pgfsys@transformshift{1.360351in}{0.317222in}%
\pgfsys@useobject{currentmarker}{}%
\end{pgfscope}%
\end{pgfscope}%
\begin{pgfscope}%
\definecolor{textcolor}{rgb}{0.000000,0.000000,0.000000}%
\pgfsetstrokecolor{textcolor}%
\pgfsetfillcolor{textcolor}%
\pgftext[x=1.360351in,y=0.220000in,,top]{\color{textcolor}\sffamily\fontsize{8.000000}{9.600000}\selectfont 200}%
\end{pgfscope}%
\begin{pgfscope}%
\pgfsetbuttcap%
\pgfsetroundjoin%
\definecolor{currentfill}{rgb}{0.000000,0.000000,0.000000}%
\pgfsetfillcolor{currentfill}%
\pgfsetlinewidth{0.803000pt}%
\definecolor{currentstroke}{rgb}{0.000000,0.000000,0.000000}%
\pgfsetstrokecolor{currentstroke}%
\pgfsetdash{}{0pt}%
\pgfsys@defobject{currentmarker}{\pgfqpoint{0.000000in}{-0.048611in}}{\pgfqpoint{0.000000in}{0.000000in}}{%
\pgfpathmoveto{\pgfqpoint{0.000000in}{0.000000in}}%
\pgfpathlineto{\pgfqpoint{0.000000in}{-0.048611in}}%
\pgfusepath{stroke,fill}%
}%
\begin{pgfscope}%
\pgfsys@transformshift{1.741505in}{0.317222in}%
\pgfsys@useobject{currentmarker}{}%
\end{pgfscope}%
\end{pgfscope}%
\begin{pgfscope}%
\definecolor{textcolor}{rgb}{0.000000,0.000000,0.000000}%
\pgfsetstrokecolor{textcolor}%
\pgfsetfillcolor{textcolor}%
\pgftext[x=1.741505in,y=0.220000in,,top]{\color{textcolor}\sffamily\fontsize{8.000000}{9.600000}\selectfont 300}%
\end{pgfscope}%
\begin{pgfscope}%
\pgfsetbuttcap%
\pgfsetroundjoin%
\definecolor{currentfill}{rgb}{0.000000,0.000000,0.000000}%
\pgfsetfillcolor{currentfill}%
\pgfsetlinewidth{0.803000pt}%
\definecolor{currentstroke}{rgb}{0.000000,0.000000,0.000000}%
\pgfsetstrokecolor{currentstroke}%
\pgfsetdash{}{0pt}%
\pgfsys@defobject{currentmarker}{\pgfqpoint{0.000000in}{-0.048611in}}{\pgfqpoint{0.000000in}{0.000000in}}{%
\pgfpathmoveto{\pgfqpoint{0.000000in}{0.000000in}}%
\pgfpathlineto{\pgfqpoint{0.000000in}{-0.048611in}}%
\pgfusepath{stroke,fill}%
}%
\begin{pgfscope}%
\pgfsys@transformshift{2.122660in}{0.317222in}%
\pgfsys@useobject{currentmarker}{}%
\end{pgfscope}%
\end{pgfscope}%
\begin{pgfscope}%
\definecolor{textcolor}{rgb}{0.000000,0.000000,0.000000}%
\pgfsetstrokecolor{textcolor}%
\pgfsetfillcolor{textcolor}%
\pgftext[x=2.122660in,y=0.220000in,,top]{\color{textcolor}\sffamily\fontsize{8.000000}{9.600000}\selectfont 400}%
\end{pgfscope}%
\begin{pgfscope}%
\pgfsetbuttcap%
\pgfsetroundjoin%
\definecolor{currentfill}{rgb}{0.000000,0.000000,0.000000}%
\pgfsetfillcolor{currentfill}%
\pgfsetlinewidth{0.803000pt}%
\definecolor{currentstroke}{rgb}{0.000000,0.000000,0.000000}%
\pgfsetstrokecolor{currentstroke}%
\pgfsetdash{}{0pt}%
\pgfsys@defobject{currentmarker}{\pgfqpoint{0.000000in}{-0.048611in}}{\pgfqpoint{0.000000in}{0.000000in}}{%
\pgfpathmoveto{\pgfqpoint{0.000000in}{0.000000in}}%
\pgfpathlineto{\pgfqpoint{0.000000in}{-0.048611in}}%
\pgfusepath{stroke,fill}%
}%
\begin{pgfscope}%
\pgfsys@transformshift{2.503814in}{0.317222in}%
\pgfsys@useobject{currentmarker}{}%
\end{pgfscope}%
\end{pgfscope}%
\begin{pgfscope}%
\definecolor{textcolor}{rgb}{0.000000,0.000000,0.000000}%
\pgfsetstrokecolor{textcolor}%
\pgfsetfillcolor{textcolor}%
\pgftext[x=2.503814in,y=0.220000in,,top]{\color{textcolor}\sffamily\fontsize{8.000000}{9.600000}\selectfont 500}%
\end{pgfscope}%
\begin{pgfscope}%
\pgfsetbuttcap%
\pgfsetroundjoin%
\definecolor{currentfill}{rgb}{0.000000,0.000000,0.000000}%
\pgfsetfillcolor{currentfill}%
\pgfsetlinewidth{0.803000pt}%
\definecolor{currentstroke}{rgb}{0.000000,0.000000,0.000000}%
\pgfsetstrokecolor{currentstroke}%
\pgfsetdash{}{0pt}%
\pgfsys@defobject{currentmarker}{\pgfqpoint{0.000000in}{-0.048611in}}{\pgfqpoint{0.000000in}{0.000000in}}{%
\pgfpathmoveto{\pgfqpoint{0.000000in}{0.000000in}}%
\pgfpathlineto{\pgfqpoint{0.000000in}{-0.048611in}}%
\pgfusepath{stroke,fill}%
}%
\begin{pgfscope}%
\pgfsys@transformshift{2.884969in}{0.317222in}%
\pgfsys@useobject{currentmarker}{}%
\end{pgfscope}%
\end{pgfscope}%
\begin{pgfscope}%
\definecolor{textcolor}{rgb}{0.000000,0.000000,0.000000}%
\pgfsetstrokecolor{textcolor}%
\pgfsetfillcolor{textcolor}%
\pgftext[x=2.884969in,y=0.220000in,,top]{\color{textcolor}\sffamily\fontsize{8.000000}{9.600000}\selectfont 600}%
\end{pgfscope}%
\begin{pgfscope}%
\pgfsetbuttcap%
\pgfsetroundjoin%
\definecolor{currentfill}{rgb}{0.000000,0.000000,0.000000}%
\pgfsetfillcolor{currentfill}%
\pgfsetlinewidth{0.803000pt}%
\definecolor{currentstroke}{rgb}{0.000000,0.000000,0.000000}%
\pgfsetstrokecolor{currentstroke}%
\pgfsetdash{}{0pt}%
\pgfsys@defobject{currentmarker}{\pgfqpoint{0.000000in}{-0.048611in}}{\pgfqpoint{0.000000in}{0.000000in}}{%
\pgfpathmoveto{\pgfqpoint{0.000000in}{0.000000in}}%
\pgfpathlineto{\pgfqpoint{0.000000in}{-0.048611in}}%
\pgfusepath{stroke,fill}%
}%
\begin{pgfscope}%
\pgfsys@transformshift{3.266123in}{0.317222in}%
\pgfsys@useobject{currentmarker}{}%
\end{pgfscope}%
\end{pgfscope}%
\begin{pgfscope}%
\definecolor{textcolor}{rgb}{0.000000,0.000000,0.000000}%
\pgfsetstrokecolor{textcolor}%
\pgfsetfillcolor{textcolor}%
\pgftext[x=3.266123in,y=0.220000in,,top]{\color{textcolor}\sffamily\fontsize{8.000000}{9.600000}\selectfont 700}%
\end{pgfscope}%
\begin{pgfscope}%
\pgfsetbuttcap%
\pgfsetroundjoin%
\definecolor{currentfill}{rgb}{0.000000,0.000000,0.000000}%
\pgfsetfillcolor{currentfill}%
\pgfsetlinewidth{0.803000pt}%
\definecolor{currentstroke}{rgb}{0.000000,0.000000,0.000000}%
\pgfsetstrokecolor{currentstroke}%
\pgfsetdash{}{0pt}%
\pgfsys@defobject{currentmarker}{\pgfqpoint{-0.048611in}{0.000000in}}{\pgfqpoint{-0.000000in}{0.000000in}}{%
\pgfpathmoveto{\pgfqpoint{-0.000000in}{0.000000in}}%
\pgfpathlineto{\pgfqpoint{-0.048611in}{0.000000in}}%
\pgfusepath{stroke,fill}%
}%
\begin{pgfscope}%
\pgfsys@transformshift{0.465972in}{0.317222in}%
\pgfsys@useobject{currentmarker}{}%
\end{pgfscope}%
\end{pgfscope}%
\begin{pgfscope}%
\definecolor{textcolor}{rgb}{0.000000,0.000000,0.000000}%
\pgfsetstrokecolor{textcolor}%
\pgfsetfillcolor{textcolor}%
\pgftext[x=0.158972in, y=0.278667in, left, base]{\color{textcolor}\sffamily\fontsize{8.000000}{9.600000}\selectfont 0.00}%
\end{pgfscope}%
\begin{pgfscope}%
\pgfsetbuttcap%
\pgfsetroundjoin%
\definecolor{currentfill}{rgb}{0.000000,0.000000,0.000000}%
\pgfsetfillcolor{currentfill}%
\pgfsetlinewidth{0.803000pt}%
\definecolor{currentstroke}{rgb}{0.000000,0.000000,0.000000}%
\pgfsetstrokecolor{currentstroke}%
\pgfsetdash{}{0pt}%
\pgfsys@defobject{currentmarker}{\pgfqpoint{-0.048611in}{0.000000in}}{\pgfqpoint{-0.000000in}{0.000000in}}{%
\pgfpathmoveto{\pgfqpoint{-0.000000in}{0.000000in}}%
\pgfpathlineto{\pgfqpoint{-0.048611in}{0.000000in}}%
\pgfusepath{stroke,fill}%
}%
\begin{pgfscope}%
\pgfsys@transformshift{0.465972in}{0.645326in}%
\pgfsys@useobject{currentmarker}{}%
\end{pgfscope}%
\end{pgfscope}%
\begin{pgfscope}%
\definecolor{textcolor}{rgb}{0.000000,0.000000,0.000000}%
\pgfsetstrokecolor{textcolor}%
\pgfsetfillcolor{textcolor}%
\pgftext[x=0.158972in, y=0.606771in, left, base]{\color{textcolor}\sffamily\fontsize{8.000000}{9.600000}\selectfont 0.02}%
\end{pgfscope}%
\begin{pgfscope}%
\pgfsetbuttcap%
\pgfsetroundjoin%
\definecolor{currentfill}{rgb}{0.000000,0.000000,0.000000}%
\pgfsetfillcolor{currentfill}%
\pgfsetlinewidth{0.803000pt}%
\definecolor{currentstroke}{rgb}{0.000000,0.000000,0.000000}%
\pgfsetstrokecolor{currentstroke}%
\pgfsetdash{}{0pt}%
\pgfsys@defobject{currentmarker}{\pgfqpoint{-0.048611in}{0.000000in}}{\pgfqpoint{-0.000000in}{0.000000in}}{%
\pgfpathmoveto{\pgfqpoint{-0.000000in}{0.000000in}}%
\pgfpathlineto{\pgfqpoint{-0.048611in}{0.000000in}}%
\pgfusepath{stroke,fill}%
}%
\begin{pgfscope}%
\pgfsys@transformshift{0.465972in}{0.973430in}%
\pgfsys@useobject{currentmarker}{}%
\end{pgfscope}%
\end{pgfscope}%
\begin{pgfscope}%
\definecolor{textcolor}{rgb}{0.000000,0.000000,0.000000}%
\pgfsetstrokecolor{textcolor}%
\pgfsetfillcolor{textcolor}%
\pgftext[x=0.158972in, y=0.934875in, left, base]{\color{textcolor}\sffamily\fontsize{8.000000}{9.600000}\selectfont 0.04}%
\end{pgfscope}%
\begin{pgfscope}%
\pgfsetbuttcap%
\pgfsetroundjoin%
\definecolor{currentfill}{rgb}{0.000000,0.000000,0.000000}%
\pgfsetfillcolor{currentfill}%
\pgfsetlinewidth{0.803000pt}%
\definecolor{currentstroke}{rgb}{0.000000,0.000000,0.000000}%
\pgfsetstrokecolor{currentstroke}%
\pgfsetdash{}{0pt}%
\pgfsys@defobject{currentmarker}{\pgfqpoint{-0.048611in}{0.000000in}}{\pgfqpoint{-0.000000in}{0.000000in}}{%
\pgfpathmoveto{\pgfqpoint{-0.000000in}{0.000000in}}%
\pgfpathlineto{\pgfqpoint{-0.048611in}{0.000000in}}%
\pgfusepath{stroke,fill}%
}%
\begin{pgfscope}%
\pgfsys@transformshift{0.465972in}{1.301535in}%
\pgfsys@useobject{currentmarker}{}%
\end{pgfscope}%
\end{pgfscope}%
\begin{pgfscope}%
\definecolor{textcolor}{rgb}{0.000000,0.000000,0.000000}%
\pgfsetstrokecolor{textcolor}%
\pgfsetfillcolor{textcolor}%
\pgftext[x=0.158972in, y=1.262979in, left, base]{\color{textcolor}\sffamily\fontsize{8.000000}{9.600000}\selectfont 0.06}%
\end{pgfscope}%
\begin{pgfscope}%
\pgfsetbuttcap%
\pgfsetroundjoin%
\definecolor{currentfill}{rgb}{0.000000,0.000000,0.000000}%
\pgfsetfillcolor{currentfill}%
\pgfsetlinewidth{0.803000pt}%
\definecolor{currentstroke}{rgb}{0.000000,0.000000,0.000000}%
\pgfsetstrokecolor{currentstroke}%
\pgfsetdash{}{0pt}%
\pgfsys@defobject{currentmarker}{\pgfqpoint{-0.048611in}{0.000000in}}{\pgfqpoint{-0.000000in}{0.000000in}}{%
\pgfpathmoveto{\pgfqpoint{-0.000000in}{0.000000in}}%
\pgfpathlineto{\pgfqpoint{-0.048611in}{0.000000in}}%
\pgfusepath{stroke,fill}%
}%
\begin{pgfscope}%
\pgfsys@transformshift{0.465972in}{1.629639in}%
\pgfsys@useobject{currentmarker}{}%
\end{pgfscope}%
\end{pgfscope}%
\begin{pgfscope}%
\definecolor{textcolor}{rgb}{0.000000,0.000000,0.000000}%
\pgfsetstrokecolor{textcolor}%
\pgfsetfillcolor{textcolor}%
\pgftext[x=0.158972in, y=1.591083in, left, base]{\color{textcolor}\sffamily\fontsize{8.000000}{9.600000}\selectfont 0.08}%
\end{pgfscope}%
\begin{pgfscope}%
\pgfsetbuttcap%
\pgfsetroundjoin%
\definecolor{currentfill}{rgb}{0.000000,0.000000,0.000000}%
\pgfsetfillcolor{currentfill}%
\pgfsetlinewidth{0.803000pt}%
\definecolor{currentstroke}{rgb}{0.000000,0.000000,0.000000}%
\pgfsetstrokecolor{currentstroke}%
\pgfsetdash{}{0pt}%
\pgfsys@defobject{currentmarker}{\pgfqpoint{-0.048611in}{0.000000in}}{\pgfqpoint{-0.000000in}{0.000000in}}{%
\pgfpathmoveto{\pgfqpoint{-0.000000in}{0.000000in}}%
\pgfpathlineto{\pgfqpoint{-0.048611in}{0.000000in}}%
\pgfusepath{stroke,fill}%
}%
\begin{pgfscope}%
\pgfsys@transformshift{0.465972in}{1.957743in}%
\pgfsys@useobject{currentmarker}{}%
\end{pgfscope}%
\end{pgfscope}%
\begin{pgfscope}%
\definecolor{textcolor}{rgb}{0.000000,0.000000,0.000000}%
\pgfsetstrokecolor{textcolor}%
\pgfsetfillcolor{textcolor}%
\pgftext[x=0.158972in, y=1.919187in, left, base]{\color{textcolor}\sffamily\fontsize{8.000000}{9.600000}\selectfont 0.10}%
\end{pgfscope}%
\begin{pgfscope}%
\pgfpathrectangle{\pgfqpoint{0.465972in}{0.317222in}}{\pgfqpoint{2.905540in}{1.962778in}}%
\pgfusepath{clip}%
\pgfsetrectcap%
\pgfsetroundjoin%
\pgfsetlinewidth{3.011250pt}%
\definecolor{currentstroke}{rgb}{0.000000,0.000000,0.000000}%
\pgfsetstrokecolor{currentstroke}%
\pgfsetdash{}{0pt}%
\pgfpathmoveto{\pgfqpoint{0.598042in}{0.996141in}}%
\pgfpathlineto{\pgfqpoint{0.639969in}{1.042884in}}%
\pgfpathlineto{\pgfqpoint{0.681896in}{1.092939in}}%
\pgfpathlineto{\pgfqpoint{0.735258in}{1.160475in}}%
\pgfpathlineto{\pgfqpoint{0.796243in}{1.242169in}}%
\pgfpathlineto{\pgfqpoint{0.883908in}{1.365338in}}%
\pgfpathlineto{\pgfqpoint{0.929647in}{1.433301in}}%
\pgfpathlineto{\pgfqpoint{0.944893in}{1.459555in}}%
\pgfpathlineto{\pgfqpoint{0.967762in}{1.504452in}}%
\pgfpathlineto{\pgfqpoint{0.983008in}{1.528080in}}%
\pgfpathlineto{\pgfqpoint{1.009689in}{1.567211in}}%
\pgfpathlineto{\pgfqpoint{1.024935in}{1.587396in}}%
\pgfpathlineto{\pgfqpoint{1.047804in}{1.617376in}}%
\pgfpathlineto{\pgfqpoint{1.066862in}{1.640770in}}%
\pgfpathlineto{\pgfqpoint{1.078297in}{1.654777in}}%
\pgfpathlineto{\pgfqpoint{1.089731in}{1.669252in}}%
\pgfpathlineto{\pgfqpoint{1.101166in}{1.682327in}}%
\pgfpathlineto{\pgfqpoint{1.112601in}{1.695488in}}%
\pgfpathlineto{\pgfqpoint{1.124035in}{1.709250in}}%
\pgfpathlineto{\pgfqpoint{1.135470in}{1.720372in}}%
\pgfpathlineto{\pgfqpoint{1.169774in}{1.760126in}}%
\pgfpathlineto{\pgfqpoint{1.196455in}{1.786678in}}%
\pgfpathlineto{\pgfqpoint{1.280309in}{1.864757in}}%
\pgfpathlineto{\pgfqpoint{1.287932in}{1.870163in}}%
\pgfpathlineto{\pgfqpoint{1.310801in}{1.881262in}}%
\pgfpathlineto{\pgfqpoint{1.333670in}{1.886419in}}%
\pgfpathlineto{\pgfqpoint{1.356539in}{1.900162in}}%
\pgfpathlineto{\pgfqpoint{1.436582in}{1.955698in}}%
\pgfpathlineto{\pgfqpoint{1.470886in}{1.982668in}}%
\pgfpathlineto{\pgfqpoint{1.486132in}{1.996415in}}%
\pgfpathlineto{\pgfqpoint{1.520436in}{2.022476in}}%
\pgfpathlineto{\pgfqpoint{1.535682in}{2.033537in}}%
\pgfpathlineto{\pgfqpoint{1.554740in}{2.049195in}}%
\pgfpathlineto{\pgfqpoint{1.573798in}{2.065199in}}%
\pgfpathlineto{\pgfqpoint{1.596667in}{2.080944in}}%
\pgfpathlineto{\pgfqpoint{1.661463in}{2.129412in}}%
\pgfpathlineto{\pgfqpoint{1.672898in}{2.135857in}}%
\pgfpathlineto{\pgfqpoint{1.684332in}{2.144011in}}%
\pgfpathlineto{\pgfqpoint{1.691955in}{2.149346in}}%
\pgfpathlineto{\pgfqpoint{1.737694in}{2.174612in}}%
\pgfpathlineto{\pgfqpoint{1.749129in}{2.181449in}}%
\pgfpathlineto{\pgfqpoint{1.779621in}{2.194282in}}%
\pgfpathlineto{\pgfqpoint{1.791056in}{2.197444in}}%
\pgfpathlineto{\pgfqpoint{1.832983in}{2.210020in}}%
\pgfpathlineto{\pgfqpoint{1.859663in}{2.214602in}}%
\pgfpathlineto{\pgfqpoint{1.882533in}{2.215234in}}%
\pgfpathlineto{\pgfqpoint{1.916836in}{2.212492in}}%
\pgfpathlineto{\pgfqpoint{1.951140in}{2.207136in}}%
\pgfpathlineto{\pgfqpoint{1.958763in}{2.203166in}}%
\pgfpathlineto{\pgfqpoint{1.989256in}{2.195527in}}%
\pgfpathlineto{\pgfqpoint{2.004502in}{2.187793in}}%
\pgfpathlineto{\pgfqpoint{2.038806in}{2.171513in}}%
\pgfpathlineto{\pgfqpoint{2.084544in}{2.141090in}}%
\pgfpathlineto{\pgfqpoint{2.103602in}{2.125607in}}%
\pgfpathlineto{\pgfqpoint{2.126471in}{2.109440in}}%
\pgfpathlineto{\pgfqpoint{2.153152in}{2.085921in}}%
\pgfpathlineto{\pgfqpoint{2.164587in}{2.075247in}}%
\pgfpathlineto{\pgfqpoint{2.183645in}{2.059650in}}%
\pgfpathlineto{\pgfqpoint{2.214137in}{2.029144in}}%
\pgfpathlineto{\pgfqpoint{2.256064in}{1.983933in}}%
\pgfpathlineto{\pgfqpoint{2.271310in}{1.968802in}}%
\pgfpathlineto{\pgfqpoint{2.305614in}{1.931329in}}%
\pgfpathlineto{\pgfqpoint{2.355164in}{1.879408in}}%
\pgfpathlineto{\pgfqpoint{2.385656in}{1.846057in}}%
\pgfpathlineto{\pgfqpoint{2.423772in}{1.808907in}}%
\pgfpathlineto{\pgfqpoint{2.522872in}{1.718831in}}%
\pgfpathlineto{\pgfqpoint{2.541930in}{1.702667in}}%
\pgfpathlineto{\pgfqpoint{2.572422in}{1.673458in}}%
\pgfpathlineto{\pgfqpoint{2.621972in}{1.623588in}}%
\pgfpathlineto{\pgfqpoint{2.740130in}{1.495006in}}%
\pgfpathlineto{\pgfqpoint{2.843042in}{1.375275in}}%
\pgfpathlineto{\pgfqpoint{2.877346in}{1.333187in}}%
\pgfpathlineto{\pgfqpoint{2.907838in}{1.293779in}}%
\pgfpathlineto{\pgfqpoint{2.942142in}{1.251156in}}%
\pgfpathlineto{\pgfqpoint{3.025996in}{1.151339in}}%
\pgfpathlineto{\pgfqpoint{3.079357in}{1.090645in}}%
\pgfpathlineto{\pgfqpoint{3.128908in}{1.036799in}}%
\pgfpathlineto{\pgfqpoint{3.174646in}{0.989864in}}%
\pgfpathlineto{\pgfqpoint{3.220385in}{0.945566in}}%
\pgfpathlineto{\pgfqpoint{3.239442in}{0.927934in}}%
\pgfpathlineto{\pgfqpoint{3.239442in}{0.927934in}}%
\pgfusepath{stroke}%
\end{pgfscope}%
\begin{pgfscope}%
\pgfpathrectangle{\pgfqpoint{0.465972in}{0.317222in}}{\pgfqpoint{2.905540in}{1.962778in}}%
\pgfusepath{clip}%
\pgfsetrectcap%
\pgfsetroundjoin%
\pgfsetlinewidth{1.505625pt}%
\definecolor{currentstroke}{rgb}{0.121569,0.466667,0.705882}%
\pgfsetstrokecolor{currentstroke}%
\pgfsetstrokeopacity{0.500000}%
\pgfsetdash{}{0pt}%
\pgfpathmoveto{\pgfqpoint{0.598042in}{1.336865in}}%
\pgfpathlineto{\pgfqpoint{3.239442in}{1.212360in}}%
\pgfpathlineto{\pgfqpoint{3.239442in}{1.212360in}}%
\pgfusepath{stroke}%
\end{pgfscope}%
\begin{pgfscope}%
\pgfpathrectangle{\pgfqpoint{0.465972in}{0.317222in}}{\pgfqpoint{2.905540in}{1.962778in}}%
\pgfusepath{clip}%
\pgfsetrectcap%
\pgfsetroundjoin%
\pgfsetlinewidth{1.505625pt}%
\definecolor{currentstroke}{rgb}{1.000000,0.498039,0.054902}%
\pgfsetstrokecolor{currentstroke}%
\pgfsetstrokeopacity{0.500000}%
\pgfsetdash{}{0pt}%
\pgfpathmoveto{\pgfqpoint{0.598042in}{1.167125in}}%
\pgfpathlineto{\pgfqpoint{0.659027in}{1.230817in}}%
\pgfpathlineto{\pgfqpoint{0.720012in}{1.291364in}}%
\pgfpathlineto{\pgfqpoint{0.780996in}{1.348764in}}%
\pgfpathlineto{\pgfqpoint{0.841981in}{1.403018in}}%
\pgfpathlineto{\pgfqpoint{0.899154in}{1.451024in}}%
\pgfpathlineto{\pgfqpoint{0.956327in}{1.496264in}}%
\pgfpathlineto{\pgfqpoint{1.013501in}{1.538740in}}%
\pgfpathlineto{\pgfqpoint{1.070674in}{1.578450in}}%
\pgfpathlineto{\pgfqpoint{1.127847in}{1.615395in}}%
\pgfpathlineto{\pgfqpoint{1.185020in}{1.649575in}}%
\pgfpathlineto{\pgfqpoint{1.238382in}{1.678981in}}%
\pgfpathlineto{\pgfqpoint{1.291743in}{1.705979in}}%
\pgfpathlineto{\pgfqpoint{1.345105in}{1.730568in}}%
\pgfpathlineto{\pgfqpoint{1.398466in}{1.752748in}}%
\pgfpathlineto{\pgfqpoint{1.451828in}{1.772519in}}%
\pgfpathlineto{\pgfqpoint{1.505190in}{1.789882in}}%
\pgfpathlineto{\pgfqpoint{1.558551in}{1.804835in}}%
\pgfpathlineto{\pgfqpoint{1.611913in}{1.817380in}}%
\pgfpathlineto{\pgfqpoint{1.665275in}{1.827516in}}%
\pgfpathlineto{\pgfqpoint{1.718636in}{1.835243in}}%
\pgfpathlineto{\pgfqpoint{1.771998in}{1.840562in}}%
\pgfpathlineto{\pgfqpoint{1.821548in}{1.843344in}}%
\pgfpathlineto{\pgfqpoint{1.871098in}{1.844049in}}%
\pgfpathlineto{\pgfqpoint{1.920648in}{1.842676in}}%
\pgfpathlineto{\pgfqpoint{1.970198in}{1.839227in}}%
\pgfpathlineto{\pgfqpoint{2.019748in}{1.833701in}}%
\pgfpathlineto{\pgfqpoint{2.073110in}{1.825427in}}%
\pgfpathlineto{\pgfqpoint{2.126471in}{1.814744in}}%
\pgfpathlineto{\pgfqpoint{2.179833in}{1.801653in}}%
\pgfpathlineto{\pgfqpoint{2.233195in}{1.786153in}}%
\pgfpathlineto{\pgfqpoint{2.286556in}{1.768243in}}%
\pgfpathlineto{\pgfqpoint{2.339918in}{1.747926in}}%
\pgfpathlineto{\pgfqpoint{2.393279in}{1.725199in}}%
\pgfpathlineto{\pgfqpoint{2.446641in}{1.700063in}}%
\pgfpathlineto{\pgfqpoint{2.500003in}{1.672519in}}%
\pgfpathlineto{\pgfqpoint{2.553364in}{1.642566in}}%
\pgfpathlineto{\pgfqpoint{2.606726in}{1.610204in}}%
\pgfpathlineto{\pgfqpoint{2.663899in}{1.572858in}}%
\pgfpathlineto{\pgfqpoint{2.721072in}{1.532746in}}%
\pgfpathlineto{\pgfqpoint{2.778245in}{1.489869in}}%
\pgfpathlineto{\pgfqpoint{2.835419in}{1.444227in}}%
\pgfpathlineto{\pgfqpoint{2.892592in}{1.395820in}}%
\pgfpathlineto{\pgfqpoint{2.949765in}{1.344648in}}%
\pgfpathlineto{\pgfqpoint{3.010750in}{1.287016in}}%
\pgfpathlineto{\pgfqpoint{3.071734in}{1.226238in}}%
\pgfpathlineto{\pgfqpoint{3.132719in}{1.162314in}}%
\pgfpathlineto{\pgfqpoint{3.193704in}{1.095244in}}%
\pgfpathlineto{\pgfqpoint{3.239442in}{1.042876in}}%
\pgfpathlineto{\pgfqpoint{3.239442in}{1.042876in}}%
\pgfusepath{stroke}%
\end{pgfscope}%
\begin{pgfscope}%
\pgfpathrectangle{\pgfqpoint{0.465972in}{0.317222in}}{\pgfqpoint{2.905540in}{1.962778in}}%
\pgfusepath{clip}%
\pgfsetrectcap%
\pgfsetroundjoin%
\pgfsetlinewidth{1.505625pt}%
\definecolor{currentstroke}{rgb}{0.172549,0.627451,0.172549}%
\pgfsetstrokecolor{currentstroke}%
\pgfsetstrokeopacity{0.500000}%
\pgfsetdash{}{0pt}%
\pgfpathmoveto{\pgfqpoint{0.598042in}{1.011532in}}%
\pgfpathlineto{\pgfqpoint{0.666650in}{1.117010in}}%
\pgfpathlineto{\pgfqpoint{0.731446in}{1.212766in}}%
\pgfpathlineto{\pgfqpoint{0.792431in}{1.299191in}}%
\pgfpathlineto{\pgfqpoint{0.853416in}{1.381795in}}%
\pgfpathlineto{\pgfqpoint{0.910589in}{1.455575in}}%
\pgfpathlineto{\pgfqpoint{0.967762in}{1.525644in}}%
\pgfpathlineto{\pgfqpoint{1.021124in}{1.587559in}}%
\pgfpathlineto{\pgfqpoint{1.074485in}{1.645996in}}%
\pgfpathlineto{\pgfqpoint{1.124035in}{1.697052in}}%
\pgfpathlineto{\pgfqpoint{1.173585in}{1.744942in}}%
\pgfpathlineto{\pgfqpoint{1.223135in}{1.789595in}}%
\pgfpathlineto{\pgfqpoint{1.268874in}{1.827886in}}%
\pgfpathlineto{\pgfqpoint{1.314613in}{1.863322in}}%
\pgfpathlineto{\pgfqpoint{1.360351in}{1.895863in}}%
\pgfpathlineto{\pgfqpoint{1.402278in}{1.923119in}}%
\pgfpathlineto{\pgfqpoint{1.444205in}{1.947891in}}%
\pgfpathlineto{\pgfqpoint{1.486132in}{1.970160in}}%
\pgfpathlineto{\pgfqpoint{1.528059in}{1.989909in}}%
\pgfpathlineto{\pgfqpoint{1.569986in}{2.007127in}}%
\pgfpathlineto{\pgfqpoint{1.611913in}{2.021805in}}%
\pgfpathlineto{\pgfqpoint{1.653840in}{2.033937in}}%
\pgfpathlineto{\pgfqpoint{1.691955in}{2.042757in}}%
\pgfpathlineto{\pgfqpoint{1.730071in}{2.049472in}}%
\pgfpathlineto{\pgfqpoint{1.768186in}{2.054088in}}%
\pgfpathlineto{\pgfqpoint{1.806302in}{2.056609in}}%
\pgfpathlineto{\pgfqpoint{1.844417in}{2.057045in}}%
\pgfpathlineto{\pgfqpoint{1.882533in}{2.055407in}}%
\pgfpathlineto{\pgfqpoint{1.920648in}{2.051707in}}%
\pgfpathlineto{\pgfqpoint{1.958763in}{2.045961in}}%
\pgfpathlineto{\pgfqpoint{2.000690in}{2.037301in}}%
\pgfpathlineto{\pgfqpoint{2.042617in}{2.026214in}}%
\pgfpathlineto{\pgfqpoint{2.084544in}{2.012732in}}%
\pgfpathlineto{\pgfqpoint{2.126471in}{1.996890in}}%
\pgfpathlineto{\pgfqpoint{2.168398in}{1.978726in}}%
\pgfpathlineto{\pgfqpoint{2.210325in}{1.958281in}}%
\pgfpathlineto{\pgfqpoint{2.256064in}{1.933429in}}%
\pgfpathlineto{\pgfqpoint{2.301802in}{1.905980in}}%
\pgfpathlineto{\pgfqpoint{2.347541in}{1.876001in}}%
\pgfpathlineto{\pgfqpoint{2.397091in}{1.840755in}}%
\pgfpathlineto{\pgfqpoint{2.446641in}{1.802727in}}%
\pgfpathlineto{\pgfqpoint{2.500003in}{1.758784in}}%
\pgfpathlineto{\pgfqpoint{2.553364in}{1.711879in}}%
\pgfpathlineto{\pgfqpoint{2.610538in}{1.658509in}}%
\pgfpathlineto{\pgfqpoint{2.667711in}{1.602112in}}%
\pgfpathlineto{\pgfqpoint{2.728695in}{1.538853in}}%
\pgfpathlineto{\pgfqpoint{2.793492in}{1.468430in}}%
\pgfpathlineto{\pgfqpoint{2.862099in}{1.390633in}}%
\pgfpathlineto{\pgfqpoint{2.938330in}{1.300812in}}%
\pgfpathlineto{\pgfqpoint{3.025996in}{1.193912in}}%
\pgfpathlineto{\pgfqpoint{3.132719in}{1.059993in}}%
\pgfpathlineto{\pgfqpoint{3.239442in}{0.923785in}}%
\pgfpathlineto{\pgfqpoint{3.239442in}{0.923785in}}%
\pgfusepath{stroke}%
\end{pgfscope}%
\begin{pgfscope}%
\pgfpathrectangle{\pgfqpoint{0.465972in}{0.317222in}}{\pgfqpoint{2.905540in}{1.962778in}}%
\pgfusepath{clip}%
\pgfsetrectcap%
\pgfsetroundjoin%
\pgfsetlinewidth{1.505625pt}%
\definecolor{currentstroke}{rgb}{0.839216,0.152941,0.156863}%
\pgfsetstrokecolor{currentstroke}%
\pgfsetstrokeopacity{0.500000}%
\pgfsetdash{}{0pt}%
\pgfpathmoveto{\pgfqpoint{0.598042in}{0.972364in}}%
\pgfpathlineto{\pgfqpoint{0.792431in}{1.262911in}}%
\pgfpathlineto{\pgfqpoint{0.800054in}{1.274042in}}%
\pgfpathlineto{\pgfqpoint{0.807677in}{1.287249in}}%
\pgfpathlineto{\pgfqpoint{0.883908in}{1.396097in}}%
\pgfpathlineto{\pgfqpoint{0.948704in}{1.484942in}}%
\pgfpathlineto{\pgfqpoint{1.009689in}{1.564811in}}%
\pgfpathlineto{\pgfqpoint{1.066862in}{1.635911in}}%
\pgfpathlineto{\pgfqpoint{1.070674in}{1.640510in}}%
\pgfpathlineto{\pgfqpoint{1.074485in}{1.647982in}}%
\pgfpathlineto{\pgfqpoint{1.127847in}{1.710160in}}%
\pgfpathlineto{\pgfqpoint{1.177397in}{1.764444in}}%
\pgfpathlineto{\pgfqpoint{1.226947in}{1.817298in}}%
\pgfpathlineto{\pgfqpoint{1.230759in}{1.823784in}}%
\pgfpathlineto{\pgfqpoint{1.276497in}{1.867042in}}%
\pgfpathlineto{\pgfqpoint{1.318424in}{1.903768in}}%
\pgfpathlineto{\pgfqpoint{1.322236in}{1.906964in}}%
\pgfpathlineto{\pgfqpoint{1.329859in}{1.915006in}}%
\pgfpathlineto{\pgfqpoint{1.371786in}{1.948016in}}%
\pgfpathlineto{\pgfqpoint{1.417524in}{1.981384in}}%
\pgfpathlineto{\pgfqpoint{1.421336in}{1.986820in}}%
\pgfpathlineto{\pgfqpoint{1.474697in}{2.020774in}}%
\pgfpathlineto{\pgfqpoint{1.505190in}{2.038724in}}%
\pgfpathlineto{\pgfqpoint{1.509001in}{2.043390in}}%
\pgfpathlineto{\pgfqpoint{1.543305in}{2.059521in}}%
\pgfpathlineto{\pgfqpoint{1.550928in}{2.062797in}}%
\pgfpathlineto{\pgfqpoint{1.554740in}{2.066115in}}%
\pgfpathlineto{\pgfqpoint{1.600478in}{2.084017in}}%
\pgfpathlineto{\pgfqpoint{1.615724in}{2.089615in}}%
\pgfpathlineto{\pgfqpoint{1.619536in}{2.092828in}}%
\pgfpathlineto{\pgfqpoint{1.623348in}{2.093900in}}%
\pgfpathlineto{\pgfqpoint{1.627159in}{2.096459in}}%
\pgfpathlineto{\pgfqpoint{1.657651in}{2.103719in}}%
\pgfpathlineto{\pgfqpoint{1.661463in}{2.107380in}}%
\pgfpathlineto{\pgfqpoint{1.672898in}{2.109509in}}%
\pgfpathlineto{\pgfqpoint{1.676709in}{2.112891in}}%
\pgfpathlineto{\pgfqpoint{1.680521in}{2.113508in}}%
\pgfpathlineto{\pgfqpoint{1.688144in}{2.117188in}}%
\pgfpathlineto{\pgfqpoint{1.699578in}{2.121250in}}%
\pgfpathlineto{\pgfqpoint{1.730071in}{2.123793in}}%
\pgfpathlineto{\pgfqpoint{1.737694in}{2.124108in}}%
\pgfpathlineto{\pgfqpoint{1.745317in}{2.126702in}}%
\pgfpathlineto{\pgfqpoint{1.771998in}{2.127306in}}%
\pgfpathlineto{\pgfqpoint{1.779621in}{2.130133in}}%
\pgfpathlineto{\pgfqpoint{1.798679in}{2.129407in}}%
\pgfpathlineto{\pgfqpoint{1.802490in}{2.131892in}}%
\pgfpathlineto{\pgfqpoint{1.806302in}{2.132947in}}%
\pgfpathlineto{\pgfqpoint{1.817736in}{2.131357in}}%
\pgfpathlineto{\pgfqpoint{1.821548in}{2.132468in}}%
\pgfpathlineto{\pgfqpoint{1.829171in}{2.131163in}}%
\pgfpathlineto{\pgfqpoint{1.832983in}{2.131974in}}%
\pgfpathlineto{\pgfqpoint{1.852040in}{2.128982in}}%
\pgfpathlineto{\pgfqpoint{1.855852in}{2.134190in}}%
\pgfpathlineto{\pgfqpoint{1.871098in}{2.130145in}}%
\pgfpathlineto{\pgfqpoint{1.874909in}{2.131375in}}%
\pgfpathlineto{\pgfqpoint{1.882533in}{2.130127in}}%
\pgfpathlineto{\pgfqpoint{1.890156in}{2.127660in}}%
\pgfpathlineto{\pgfqpoint{1.893967in}{2.127886in}}%
\pgfpathlineto{\pgfqpoint{1.905402in}{2.124681in}}%
\pgfpathlineto{\pgfqpoint{1.909213in}{2.127667in}}%
\pgfpathlineto{\pgfqpoint{1.916836in}{2.125483in}}%
\pgfpathlineto{\pgfqpoint{1.924460in}{2.126459in}}%
\pgfpathlineto{\pgfqpoint{1.935894in}{2.121607in}}%
\pgfpathlineto{\pgfqpoint{1.939706in}{2.122260in}}%
\pgfpathlineto{\pgfqpoint{1.947329in}{2.118630in}}%
\pgfpathlineto{\pgfqpoint{1.954952in}{2.117171in}}%
\pgfpathlineto{\pgfqpoint{1.966387in}{2.113762in}}%
\pgfpathlineto{\pgfqpoint{1.970198in}{2.111681in}}%
\pgfpathlineto{\pgfqpoint{1.974010in}{2.113691in}}%
\pgfpathlineto{\pgfqpoint{1.981633in}{2.109321in}}%
\pgfpathlineto{\pgfqpoint{1.989256in}{2.108185in}}%
\pgfpathlineto{\pgfqpoint{1.993067in}{2.105852in}}%
\pgfpathlineto{\pgfqpoint{1.996879in}{2.104993in}}%
\pgfpathlineto{\pgfqpoint{2.004502in}{2.105432in}}%
\pgfpathlineto{\pgfqpoint{2.012125in}{2.100377in}}%
\pgfpathlineto{\pgfqpoint{2.019748in}{2.098372in}}%
\pgfpathlineto{\pgfqpoint{2.023560in}{2.101818in}}%
\pgfpathlineto{\pgfqpoint{2.031183in}{2.101016in}}%
\pgfpathlineto{\pgfqpoint{2.034994in}{2.098199in}}%
\pgfpathlineto{\pgfqpoint{2.046429in}{2.093576in}}%
\pgfpathlineto{\pgfqpoint{2.050241in}{2.091392in}}%
\pgfpathlineto{\pgfqpoint{2.054052in}{2.090754in}}%
\pgfpathlineto{\pgfqpoint{2.057864in}{2.087661in}}%
\pgfpathlineto{\pgfqpoint{2.069298in}{2.083920in}}%
\pgfpathlineto{\pgfqpoint{2.073110in}{2.080636in}}%
\pgfpathlineto{\pgfqpoint{2.080733in}{2.080061in}}%
\pgfpathlineto{\pgfqpoint{2.088356in}{2.077079in}}%
\pgfpathlineto{\pgfqpoint{2.092168in}{2.073728in}}%
\pgfpathlineto{\pgfqpoint{2.095979in}{2.072560in}}%
\pgfpathlineto{\pgfqpoint{2.099791in}{2.069750in}}%
\pgfpathlineto{\pgfqpoint{2.107414in}{2.066446in}}%
\pgfpathlineto{\pgfqpoint{2.118848in}{2.060603in}}%
\pgfpathlineto{\pgfqpoint{2.122660in}{2.060071in}}%
\pgfpathlineto{\pgfqpoint{2.126471in}{2.056085in}}%
\pgfpathlineto{\pgfqpoint{2.134094in}{2.052057in}}%
\pgfpathlineto{\pgfqpoint{2.141718in}{2.046123in}}%
\pgfpathlineto{\pgfqpoint{2.149341in}{2.044205in}}%
\pgfpathlineto{\pgfqpoint{2.153152in}{2.044988in}}%
\pgfpathlineto{\pgfqpoint{2.156964in}{2.040572in}}%
\pgfpathlineto{\pgfqpoint{2.191268in}{2.023170in}}%
\pgfpathlineto{\pgfqpoint{2.198891in}{2.015719in}}%
\pgfpathlineto{\pgfqpoint{2.202702in}{2.013021in}}%
\pgfpathlineto{\pgfqpoint{2.206514in}{2.011880in}}%
\pgfpathlineto{\pgfqpoint{2.214137in}{2.006181in}}%
\pgfpathlineto{\pgfqpoint{2.217948in}{2.006980in}}%
\pgfpathlineto{\pgfqpoint{2.225572in}{2.001903in}}%
\pgfpathlineto{\pgfqpoint{2.233195in}{1.994689in}}%
\pgfpathlineto{\pgfqpoint{2.248441in}{1.983070in}}%
\pgfpathlineto{\pgfqpoint{2.256064in}{1.979310in}}%
\pgfpathlineto{\pgfqpoint{2.275122in}{1.962181in}}%
\pgfpathlineto{\pgfqpoint{2.278933in}{1.964340in}}%
\pgfpathlineto{\pgfqpoint{2.294179in}{1.949291in}}%
\pgfpathlineto{\pgfqpoint{2.305614in}{1.940240in}}%
\pgfpathlineto{\pgfqpoint{2.309426in}{1.941804in}}%
\pgfpathlineto{\pgfqpoint{2.313237in}{1.937341in}}%
\pgfpathlineto{\pgfqpoint{2.317049in}{1.934961in}}%
\pgfpathlineto{\pgfqpoint{2.324672in}{1.926999in}}%
\pgfpathlineto{\pgfqpoint{2.332295in}{1.920171in}}%
\pgfpathlineto{\pgfqpoint{2.336106in}{1.919707in}}%
\pgfpathlineto{\pgfqpoint{2.385656in}{1.874784in}}%
\pgfpathlineto{\pgfqpoint{2.389468in}{1.873671in}}%
\pgfpathlineto{\pgfqpoint{2.431395in}{1.832235in}}%
\pgfpathlineto{\pgfqpoint{2.435206in}{1.830191in}}%
\pgfpathlineto{\pgfqpoint{2.439018in}{1.832380in}}%
\pgfpathlineto{\pgfqpoint{2.446641in}{1.822780in}}%
\pgfpathlineto{\pgfqpoint{2.454264in}{1.817012in}}%
\pgfpathlineto{\pgfqpoint{2.473322in}{1.794839in}}%
\pgfpathlineto{\pgfqpoint{2.477133in}{1.791928in}}%
\pgfpathlineto{\pgfqpoint{2.484757in}{1.781953in}}%
\pgfpathlineto{\pgfqpoint{2.488568in}{1.782317in}}%
\pgfpathlineto{\pgfqpoint{2.500003in}{1.769012in}}%
\pgfpathlineto{\pgfqpoint{2.507626in}{1.760934in}}%
\pgfpathlineto{\pgfqpoint{2.511437in}{1.756180in}}%
\pgfpathlineto{\pgfqpoint{2.519060in}{1.749134in}}%
\pgfpathlineto{\pgfqpoint{2.522872in}{1.743406in}}%
\pgfpathlineto{\pgfqpoint{2.526684in}{1.739894in}}%
\pgfpathlineto{\pgfqpoint{2.530495in}{1.734375in}}%
\pgfpathlineto{\pgfqpoint{2.534307in}{1.733563in}}%
\pgfpathlineto{\pgfqpoint{2.545741in}{1.719934in}}%
\pgfpathlineto{\pgfqpoint{2.549553in}{1.722653in}}%
\pgfpathlineto{\pgfqpoint{2.560987in}{1.712128in}}%
\pgfpathlineto{\pgfqpoint{2.564799in}{1.706552in}}%
\pgfpathlineto{\pgfqpoint{2.580045in}{1.692947in}}%
\pgfpathlineto{\pgfqpoint{2.587668in}{1.682661in}}%
\pgfpathlineto{\pgfqpoint{2.599103in}{1.669978in}}%
\pgfpathlineto{\pgfqpoint{2.602914in}{1.664734in}}%
\pgfpathlineto{\pgfqpoint{2.606726in}{1.661710in}}%
\pgfpathlineto{\pgfqpoint{2.614349in}{1.650344in}}%
\pgfpathlineto{\pgfqpoint{2.618161in}{1.649489in}}%
\pgfpathlineto{\pgfqpoint{2.633407in}{1.629317in}}%
\pgfpathlineto{\pgfqpoint{2.652464in}{1.605571in}}%
\pgfpathlineto{\pgfqpoint{2.656276in}{1.604822in}}%
\pgfpathlineto{\pgfqpoint{2.663899in}{1.595067in}}%
\pgfpathlineto{\pgfqpoint{2.671522in}{1.589089in}}%
\pgfpathlineto{\pgfqpoint{2.679145in}{1.577113in}}%
\pgfpathlineto{\pgfqpoint{2.682957in}{1.571636in}}%
\pgfpathlineto{\pgfqpoint{2.686768in}{1.568408in}}%
\pgfpathlineto{\pgfqpoint{2.694391in}{1.558281in}}%
\pgfpathlineto{\pgfqpoint{2.702015in}{1.551979in}}%
\pgfpathlineto{\pgfqpoint{2.713449in}{1.535741in}}%
\pgfpathlineto{\pgfqpoint{2.724884in}{1.522148in}}%
\pgfpathlineto{\pgfqpoint{2.728695in}{1.515199in}}%
\pgfpathlineto{\pgfqpoint{2.732507in}{1.514403in}}%
\pgfpathlineto{\pgfqpoint{2.740130in}{1.502052in}}%
\pgfpathlineto{\pgfqpoint{2.743942in}{1.505511in}}%
\pgfpathlineto{\pgfqpoint{2.751565in}{1.494383in}}%
\pgfpathlineto{\pgfqpoint{2.770622in}{1.470728in}}%
\pgfpathlineto{\pgfqpoint{2.774434in}{1.470078in}}%
\pgfpathlineto{\pgfqpoint{2.782057in}{1.458854in}}%
\pgfpathlineto{\pgfqpoint{2.789680in}{1.449360in}}%
\pgfpathlineto{\pgfqpoint{2.801115in}{1.432829in}}%
\pgfpathlineto{\pgfqpoint{2.804926in}{1.430876in}}%
\pgfpathlineto{\pgfqpoint{2.808738in}{1.424443in}}%
\pgfpathlineto{\pgfqpoint{2.812549in}{1.420841in}}%
\pgfpathlineto{\pgfqpoint{2.835419in}{1.389477in}}%
\pgfpathlineto{\pgfqpoint{2.843042in}{1.382663in}}%
\pgfpathlineto{\pgfqpoint{2.858288in}{1.360614in}}%
\pgfpathlineto{\pgfqpoint{2.881157in}{1.334022in}}%
\pgfpathlineto{\pgfqpoint{2.884969in}{1.338207in}}%
\pgfpathlineto{\pgfqpoint{2.892592in}{1.325338in}}%
\pgfpathlineto{\pgfqpoint{2.900215in}{1.317396in}}%
\pgfpathlineto{\pgfqpoint{2.904026in}{1.311667in}}%
\pgfpathlineto{\pgfqpoint{2.907838in}{1.308088in}}%
\pgfpathlineto{\pgfqpoint{2.915461in}{1.295721in}}%
\pgfpathlineto{\pgfqpoint{2.919273in}{1.292063in}}%
\pgfpathlineto{\pgfqpoint{2.926896in}{1.280324in}}%
\pgfpathlineto{\pgfqpoint{2.934519in}{1.271812in}}%
\pgfpathlineto{\pgfqpoint{2.942142in}{1.261041in}}%
\pgfpathlineto{\pgfqpoint{2.961200in}{1.236094in}}%
\pgfpathlineto{\pgfqpoint{2.980257in}{1.214190in}}%
\pgfpathlineto{\pgfqpoint{2.987880in}{1.202297in}}%
\pgfpathlineto{\pgfqpoint{2.991692in}{1.200063in}}%
\pgfpathlineto{\pgfqpoint{2.999315in}{1.187587in}}%
\pgfpathlineto{\pgfqpoint{3.010750in}{1.172572in}}%
\pgfpathlineto{\pgfqpoint{3.018373in}{1.161394in}}%
\pgfpathlineto{\pgfqpoint{3.022184in}{1.160188in}}%
\pgfpathlineto{\pgfqpoint{3.045054in}{1.126364in}}%
\pgfpathlineto{\pgfqpoint{3.048865in}{1.134087in}}%
\pgfpathlineto{\pgfqpoint{3.052677in}{1.128354in}}%
\pgfpathlineto{\pgfqpoint{3.056488in}{1.130002in}}%
\pgfpathlineto{\pgfqpoint{3.071734in}{1.110903in}}%
\pgfpathlineto{\pgfqpoint{3.079357in}{1.104790in}}%
\pgfpathlineto{\pgfqpoint{3.083169in}{1.097252in}}%
\pgfpathlineto{\pgfqpoint{3.086981in}{1.093414in}}%
\pgfpathlineto{\pgfqpoint{3.090792in}{1.086703in}}%
\pgfpathlineto{\pgfqpoint{3.094604in}{1.083509in}}%
\pgfpathlineto{\pgfqpoint{3.098415in}{1.077451in}}%
\pgfpathlineto{\pgfqpoint{3.102227in}{1.074557in}}%
\pgfpathlineto{\pgfqpoint{3.106038in}{1.070025in}}%
\pgfpathlineto{\pgfqpoint{3.109850in}{1.063317in}}%
\pgfpathlineto{\pgfqpoint{3.113661in}{1.060384in}}%
\pgfpathlineto{\pgfqpoint{3.125096in}{1.042554in}}%
\pgfpathlineto{\pgfqpoint{3.128908in}{1.041380in}}%
\pgfpathlineto{\pgfqpoint{3.136531in}{1.029014in}}%
\pgfpathlineto{\pgfqpoint{3.140342in}{1.025627in}}%
\pgfpathlineto{\pgfqpoint{3.144154in}{1.018764in}}%
\pgfpathlineto{\pgfqpoint{3.151777in}{1.008952in}}%
\pgfpathlineto{\pgfqpoint{3.155588in}{1.010977in}}%
\pgfpathlineto{\pgfqpoint{3.159400in}{1.003634in}}%
\pgfpathlineto{\pgfqpoint{3.163211in}{0.999837in}}%
\pgfpathlineto{\pgfqpoint{3.170834in}{0.989574in}}%
\pgfpathlineto{\pgfqpoint{3.174646in}{0.990252in}}%
\pgfpathlineto{\pgfqpoint{3.178458in}{0.995668in}}%
\pgfpathlineto{\pgfqpoint{3.189892in}{0.979118in}}%
\pgfpathlineto{\pgfqpoint{3.197515in}{0.969340in}}%
\pgfpathlineto{\pgfqpoint{3.201327in}{0.963928in}}%
\pgfpathlineto{\pgfqpoint{3.205138in}{0.961599in}}%
\pgfpathlineto{\pgfqpoint{3.224196in}{0.936168in}}%
\pgfpathlineto{\pgfqpoint{3.228008in}{0.935136in}}%
\pgfpathlineto{\pgfqpoint{3.231819in}{0.929756in}}%
\pgfpathlineto{\pgfqpoint{3.235631in}{0.926495in}}%
\pgfpathlineto{\pgfqpoint{3.239442in}{0.919919in}}%
\pgfpathlineto{\pgfqpoint{3.239442in}{0.919919in}}%
\pgfusepath{stroke}%
\end{pgfscope}%
\begin{pgfscope}%
\pgfsetrectcap%
\pgfsetmiterjoin%
\pgfsetlinewidth{0.803000pt}%
\definecolor{currentstroke}{rgb}{0.000000,0.000000,0.000000}%
\pgfsetstrokecolor{currentstroke}%
\pgfsetdash{}{0pt}%
\pgfpathmoveto{\pgfqpoint{0.465972in}{0.317222in}}%
\pgfpathlineto{\pgfqpoint{0.465972in}{2.280000in}}%
\pgfusepath{stroke}%
\end{pgfscope}%
\begin{pgfscope}%
\pgfsetrectcap%
\pgfsetmiterjoin%
\pgfsetlinewidth{0.803000pt}%
\definecolor{currentstroke}{rgb}{0.000000,0.000000,0.000000}%
\pgfsetstrokecolor{currentstroke}%
\pgfsetdash{}{0pt}%
\pgfpathmoveto{\pgfqpoint{3.371512in}{0.317222in}}%
\pgfpathlineto{\pgfqpoint{3.371512in}{2.280000in}}%
\pgfusepath{stroke}%
\end{pgfscope}%
\begin{pgfscope}%
\pgfsetrectcap%
\pgfsetmiterjoin%
\pgfsetlinewidth{0.803000pt}%
\definecolor{currentstroke}{rgb}{0.000000,0.000000,0.000000}%
\pgfsetstrokecolor{currentstroke}%
\pgfsetdash{}{0pt}%
\pgfpathmoveto{\pgfqpoint{0.465972in}{0.317222in}}%
\pgfpathlineto{\pgfqpoint{3.371512in}{0.317222in}}%
\pgfusepath{stroke}%
\end{pgfscope}%
\begin{pgfscope}%
\pgfsetrectcap%
\pgfsetmiterjoin%
\pgfsetlinewidth{0.803000pt}%
\definecolor{currentstroke}{rgb}{0.000000,0.000000,0.000000}%
\pgfsetstrokecolor{currentstroke}%
\pgfsetdash{}{0pt}%
\pgfpathmoveto{\pgfqpoint{0.465972in}{2.280000in}}%
\pgfpathlineto{\pgfqpoint{3.371512in}{2.280000in}}%
\pgfusepath{stroke}%
\end{pgfscope}%
\end{pgfpicture}%
\makeatother%
\endgroup%